\documentclass{elsart}
\usepackage{amsmath}
\usepackage{amssymb}
\usepackage{psfig}
\usepackage[dvips]{color}
\usepackage{longtable}
\definecolor{titlecol}{named}{DarkOrchid}
\makeatletter
\@addtoreset{equation}{subsection}
\@addtoreset{equation}{section}
\makeatother

\newcommand{\cal}{\mathcal}
\newcommand{\sss}{\scriptscriptstyle}
\newcommand{\bm}[1]{{\mbox{\boldmath$#1$}}}

\def\Quadrat#1#2{{\vcenter{\hrule height #2
  \hbox{\vrule width #2 height #1 \kern#1
    \vrule width #2}
  \hrule height #2}}}
\def\dAl{\mathop{\kern 1pt\hbox{$\Quadrat{8pt}{0.4pt}$} \kern1pt}}

\newcommand{\nn}[1]{\sum\limits_{#1}^\infty}
\newcommand{\nna}{\frac{(2l-1)!!}{l!}}
\newcommand{\nnb}[2]{{\varphi}(#1,{\bm #2})}
\newcommand{\ffrac}[2]{\frac{{\displaystyle#1}}{{\displaystyle#2}}}
\newcommand{\nnbb}[3]{\hat\varphi^{\sss(#1)}(#2, {\bm #3})}

\newcommand{\nnd}[1]{\stackrel{{\sss(#1)}}{\zeta}}
\newcommand{\nnw}{{\bm w}}
\newcommand{\nne}[3]{\stackrel{{\sss(#1)}}{h}_{#2#3}}
\newcommand{\nnt}[3]{\stackrel{{\sss(#1)}}{T}_{#2#3}}
\newcommand{\nnfa}[5]{\hat h^{{\sss(#1)}}_{#2#3}(#4,
 {\bm #5})}
\newcommand{\nnc}{\hat U^{{\sss(B)}}(u, {\bm w})}
\newcommand{\nng}[4]{\Phi^{{\sss(#1)}}_{#2}(#3,
 {\bm #4})}
\newcommand{\nngk}[1]{\hat{\Phi}^{{\sss(B)}}_{#1}(u,
 {\bm w})}
\newcommand{\nnff}[1]{\Phi_{#1}(t, {\bm x})}
\newcommand{\nnh}[2]{{\rm\bf I}^{{\sss(#1)}}_{-1}\left\{#2\right\}}
\newcommand{\nnhh}[2]{\hat{\rm\bf I}^{{\sss(#1)}}_{-1}\left\{#2\right\}}
\newcommand{\nnj}{U^{{\sss(B)}}(t, {\bm x})}
\newcommand{\nnxe}{({\bm x}_{{\sss B}})}

\newcommand{\nnx}{{\bm x}}

\newcommand{\slch}{\epsilon}
\newcommand{\be}{\begin{equation}}
\newcommand{\en}{\end{equation}}
\newcommand{\br}{\begin{eqnarray}}
\newcommand{\er}{\end{eqnarray}}
\newcommand{\Dd}{\displaystyle}
\newcommand{\nag}[1]{\hat h_{#1}(u,{\bm w})}

\newcommand{\nxi}{\hat\varphi(u,{\bm w})}

\newcommand{\nnv}[1]{v^{#1}_{{\sss B}}}
\newcommand{\nnr}[1]{R^{#1}_{{\sss B}}}
\newcommand{\nnaa}[1]{a^{#1}_{{\sss B}}}
\newcommand{\da}[1]{\dot{a}^{#1}_{{\sss B}}}
\newcommand{\dda}[1]{\ddot{a}^{#1}_{{\sss B}}}
\begin{document}
\begin{frontmatter}
\title{Parametrized Post-Newtonian Theory of Reference Frames, Multipolar Expansions and Equations of Motion
in the N-body Problem}
\author{Sergei Kopeikin\corauthref{cor1}}\ead{kopeikins@missouri.edu} \and \author{Igor Vlasov}
\address{Department of Physics and Astronomy,
University of Missouri-Columbia, Columbia, Missouri 65211, USA}
\corauth[cor1]{Corresponding author}
\begin{abstract}
Post-Newtonian relativistic theory of astronomical reference frames based on Einstein's general theory of
relativity was
adopted by General
Assembly of the International Astronomical Union in 2000. This theory is extended in the present paper by taking
into account all relativistic effects caused by the presumable existence of a scalar field and parametrized by
two parameters, $\beta$
and $\gamma$, of the parametrized post-Newtonian (PPN) formalism.
We use a general class of the scalar-tensor (Brans-Dicke type) theories of gravitation to work out
PPN concepts of global and local reference frames for an astronomical N-body system. The global reference frame
is a standard PPN coordinate system.  A local reference
frame is constructed in the vicinity of a weakly self-gravitating body (a sub-system of the bodies) that is
a member of the astronomical N-body system. Such local inertial
frame is required for unambiguous derivation of the equations of motion of the body in the field of other members
of the N-body system and for construction of adequate algorithms for data analysis of various
gravitational experiments conducted in ground-based
laboratories and/or on board of spacecrafts in the solar system.

We assume that the bodies comprising the N-body system
have weak gravitational field and move slowly. At the same time we do not impose any specific limitations
on the distribution of density, velocity and the equation of state of the body's matter. Scalar-tensor equations
of the gravitational field are solved by making use of the post-Newtonian
approximations so that the metric tensor and the scalar field are obtained as functions of the global and local
coordinates. A correspondence between the local and global coordinate frames is found by making use of asymptotic
expansion matching technique.  This technique allows us to find a class of the post-Newtonian
coordinate transformations between the frames as well as equations of translational motion of the origin of the
local frame along with the law of relativistic precession of its spatial axes. These transformations depend on the
PPN parameters $\beta$ and $\gamma$, generalize general relativistic transformations of the IAU 2000 resolutions,
and should be used in the data processing of
the solar system gravitational experiments aimed to detect the presence of the scalar field. These PPN
transformations are also applicable in the precise time-keeping metrology,
celestial mechanics, astrometry, geodesy and navigation.

We consider a multipolar post-Newtonian expansion of the gravitational and scalar fields and construct a set
of internal and external gravitational multipoles depending on the parameters $\beta$ and $\gamma$. These
PPN multipoles generalize the Thorne-Blanchet-Damour multipoles defined in harmonic coordinates of general
theory of relativity. The PPN multipoles of the scalar-tensor theory of gravity are split in three
classes -- {\it active}, {\it conformal}, and {\it scalar} multipoles. Only two of them are algebraically
independent and we chose to work with the {\it conformal} and {\it active} multipoles. We
derive the laws of conservations of the multipole moments and show that they must be formulated in terms
of the {\it conformal} multipoles.  We focus then on the law of conservation of body's linear momentum
which is defined as a time derivative of the {\it conformal} dipole moment of the body in the local coordinates.
We prove that the local force violating the law of conservation of the body's linear momentum depends
exclusively on the {\it active} multipole moments of the body along with a few other terms which depend
on the internal structure of the body and are responsible for the violation of the strong principle of
equivalence (the Nordtvedt effect).

The PPN translational equations of motion of extended bodies in
the global coordinate frame and with all gravitational multipoles
taken into account are derived from the law of conservation of the
body's linear momentum supplemented by the law of motion of the
origin of the local frame derived from the matching procedure. We
use these equations to analyze translational motion of
spherically-symmetric and rigidly rotating bodies having finite
size. Spherical symmetry is defined in the local frame of each
body through a set of conditions imposed on the shape of the body
and the distribution of its internal density, pressure and velocity
field. We prove that our formalism brings about the parametrized
post-Newtonian EIH equations of motion of the bodies if the
finite-size effects are neglected. Analysis of the finite-size
effects reveal that they are proportional to the parameter $\beta$
coupled with the second and higher-order rotational moments of
inertia of the bodies. The finite-size effects in the
translational equations of motion can be appreciably large at the
latest stage of coalescence of binary neutron stars and can be
important in calculations of gravitational waveform templates for
the gravitational-wave interferometers.

The PPN rotational equations of motion for each extended body possessing an arbitrary multipolar structure of
its gravitational field, have been derived in body's local coordinates. Spin of the body is defined
phenomenologically in accordance with the post-Newtonian law of conservation of angular momentum of an
isolated system. Torque consists of a general relativistic part and the PPN contribution due to the
presence of the scalar field. The PPN scalar-field-dependent part is proportional to the difference
between {\it active} and {\it conformal} dipole moments of the body which disappears in general relativity.
Finite-size effects in rotational equations of motion can be a matter of interest for calculating gravitational
wave radiation from coalescing binaries.
\end{abstract}
\maketitle
\tableofcontents
\listoffigures
\end{frontmatter}
\newpage
\newpage
\section{Notations}
\subsection{General Conventions}
Greek indices $\alpha, \beta, \gamma,...$ run from 0 to 3 and mark space-time components of four-dimensional
objects. Roman indices $i,j,k,...$ run from 1 to 3 and denote components of three-dimensional objects
(zero component belongs to time). Repeated indices mean the Einstein summation rule, for instance,
$A^\alpha B_\alpha=A^0B_0+A^1B_1+A^2B_2+A^3B_3$ and $T^k_k=T^1_{\;1}+T^2_{\;2}+T^3_{\;3}$, etc.

Minkowski metric is denoted $\eta_{\alpha\beta}={\rm diag}(-1,+1,+1,+1)$. Kronecker symbol (the unit matrix)
is denoted $\delta_{ij}= {\rm diag}(1,1,1)$. Levi-Civita fully antisymmetric symbol is $\varepsilon_{ijk}$
such that $\varepsilon_{123}=+1$. Kronecker symbol is used to rise and lower Roman indices.
Complete metric tensor $g_{\alpha\beta}$ is used to rise and lower the Greek indices in exact tensor
equations whereas the Minkowski metric $\eta_{\alpha\beta}$ is employed for rising and lowering indices
in the post-Newtonian approximations

Parentheses surrounding a group of Roman indices mean symmetrization, for example, $A_{(ij)}=(1/2)(A_{ij}+A_{ji})$.
Brackets around two Roman indices denote antisymmetrization, that is $A_{[ij]}=(1/2)(A_{ij}-A_{ji})$.
Angle brackets surrounding a group of Roman indices denote the symmetric trace-free (STF) part of the
corresponding three-dimensional object, for instance,
$$A_{<ij>}=A_{(ij)}-\frac{1}{3}\delta_{ij}A_{kk}\;,\quad\quad
A_{<ijk>}=A_{(ijk)} -\frac{1}{5}\delta_{ij}A_{kpp}-\frac{1}{5}\delta_{jk}A_{ipp}-\frac{1}{5}\delta_{ik}A_{jpp} \;.$$
We also use multi-index notations, for example,
$$A_L=A_{i_1i_2...i_l}\;, \qquad\quad B_{P-1}=B_{i_1i_2...i_{p-1}}\;, \qquad\qquad
D_{<L>}=D_{<i_1i_2...i_l>}\;.$$ Sum over multi-indices is understood as
$$A_L Q^L=A_{i_1i_2...i_l}Q^{i_1i_2...i_l}\;,\qquad\quad
P_{aL-1} T^{bL-1}=P_{ai_1i_2...i_{l-1}}T^{bi_1i_2...i_{l-1}}\;.$$
Comma denotes a partial derivative, for example, $\phi_{,\alpha}=\partial\phi/\partial x^\alpha$,
where $\phi_{,0}=c^{-1}\partial\phi/\partial t,\, \phi_{,i}=\partial\phi/\partial x^i,$
and semicolon $T^\alpha_{\,\;;\beta}$ denotes a covariant derivative.
$L$-order partial derivative with respect to spatial coordinates is denoted
$\partial_L=\partial_{i_1}...\partial_{i_l}$. Other conventions are introduced as they appear in the text.
We summarize these particular conventions and notations in the next section for the convenience of the readers.
\subsection{Particular Conventions and Symbols Used in the Paper}
\renewcommand{\arraystretch}{1.5}
\setlength{\arrayrulewidth}{0.1mm}
\begin{longtable}{|c|@{\,}p{7cm}@{\,}|c|}
\hline
\textbf{\color{blue}Symbol}&\multicolumn{1}{c|}{\textbf{\color{blue}Description}}&
\multicolumn{1}{c|}{\textbf{\color{blue}Equation(s)}}
\endhead\hline
$g_{\mu\nu} $&physical (Jordan-Fierz frame) metric tensor&(\ref{10.3})\\\hline
$\tilde g_{\mu\nu} $&conformal (Einstein frame) metric tensor&(\ref{13.18})\\\hline
$ g $&the determinant of $g_{\mu\nu}$ &(\ref{10.1})\\\hline
$\tilde g $&the determinant of $\tilde g_{\mu\nu}$&(\ref{13.20})\\\hline
$\eta_{\mu\nu} $&the Minkowski (flat) metric tensor&(\ref{exp})\\\hline
$\Gamma^\alpha_{\mu\nu} $& the Christoffel symbol  &(\ref{covd})\\\hline
$R_{\mu\nu} $& the Ricci tensor &(\ref{10.2})\\\hline
$R $& the Ricci scalar & (\ref{10.1})\\\hline
$\tilde R_{\mu\nu}$& the conformal Ricci tensor  &(\ref{13.19})\\\hline
$T_{\mu\nu}$& the energy-momentum tensor of matter &(\ref{10.2})\\\hline
$T $& the trace of the energy-momentum tensor &(\ref{10.2})\\\hline
$\phi$& the scalar field &(\ref{10.1})\\\hline
$\phi_0 $& the background value of the scalar field $\phi$& (\ref{aa})\\\hline
$\zeta $& the dimensionless perturbation of the scalar field &(\ref{aa})\\\hline
$\theta(\phi) $& the coupling function of the scalar field  &(\ref{10.1})\\\hline
${\dAl}_g $& the Laplace-Beltrami operator &(\ref{covd})\\\hline
$\dAl $&the D'Alembert operator in the Minkowski  space-time  &(\ref{11.29})\\\hline
$\rho $& the density of matter in the co-moving frame &(\ref{11.1})\\\hline
$\rho^* $& the invariant (Fock) density of matter  &(\ref{11.19})\\\hline
$\Pi $& the internal energy of matter in the co-moving frame&(\ref{11.1})\\\hline
$\pi^{\mu\nu} $& the tensor of (anisotropic) stresses of matter   &(\ref{11.1})\\\hline
$u^\alpha $& the 4-velocity of matter &(\ref{11.1})\\\hline
$v^i $&the 3-dimensional velocity of matter in the global frame &(\ref{11.23})\\\hline
$\omega$& the asymptotic value of the coupling function  $\theta(\phi) $  &(\ref{10.5})\\\hline
$ \omega' $& the asymptotic value of the derivative of the coupling function $\theta(\phi) $ &(\ref{10.5})\\\hline
$c$& the ultimate speed of general and special theories of relativity&(\ref{10.1})\\\hline
$\epsilon $& a small dimensional parameter, $\epsilon=1/c$ &(\ref{exp})\\\hline
$h_{\mu\nu} $& the metric tensor perturbation, $g_{\mu\nu}-\eta_{\mu\nu}$  &(\ref{mtp})\\\hline
$\nne{n}{\mu}{\nu} $& the metric tensor perturbation of order $\epsilon^n$
in the post-Newtonian expansion of the metric tensor   &(\ref{exp})\\\hline
$N $&a shorthand notation for $\nne{2}{0}{0}$  &(\ref{not})\\\hline
$L $&a shorthand notation for $\nne{4}{0}{0}$  &(\ref{not})\\\hline
$N_i $&a shorthand notation for $\nne{1}{0}{i}$  &(\ref{not})\\\hline
$L_i $&a shorthand notation for $\nne{3}{0}{i}$  &(\ref{not})\\\hline
$H_{ij} $&a shorthand notation for $\nne{2}{i}{j}$  &(\ref{not})\\\hline
$H $&a shorthand notation for $\nne{2}{k}{k}$  &(\ref{not})\\\hline
$\tilde{N},\; \tilde{L}$& shorthand notations for perturbations of conformal metric
$\tilde g_{\mu\nu}$ &(\ref{mp4})\\\hline
$\gamma$&the `space-curvature' PPN parameter  &(\ref{11.27})\\\hline
$\beta $&the `non-linearity' PPN parameter  &(\ref{11.28})\\\hline
$\eta $& the Nordtvedt parameter, $\eta =4\beta-\gamma-3$ &(\ref{mp3})\\\hline
$G $&the observed value of the universal gravitational constant  &(\ref{10.6})\\\hline
${\cal G} $&the bare value of the universal gravitational constant  &(\ref{10.7}), (\ref{10.6})\\\hline
$x^\alpha=(x^0,x^i) $&the global coordinates with $x^0=ct$ and $x^i\equiv{\bm x}$  &
\\\hline
$w^\alpha=(w^0,w^i) $&the local coordinates with $w^0=cu$ and $w^i\equiv{\bm w}$  &
\\\hline
$U $&the Newtonian gravitational potential in the global frame  &(\ref{12.5})\\\hline
$U^{\sss(A)} $&the Newtonian gravitational potential of body A in the global frame  &(\ref{12.9a})\\\hline
$U_i $&a vector potential in the global frame  &(\ref{12.8})\\\hline
$U_i^{\sss(A)} $&a vector potential of body A in the global frame  &(\ref{12.9a})\\\hline
$\chi,\;\Phi_1,\ldots,\Phi_4 $&various special gravitational potentials in the global frame  &(\ref{12.6}),
(\ref{12.9ex})
\\\hline
$V,\; V^i $&potentials of the physical metric in the global frame &(\ref{mp1}), \ref{mp2})\\\hline
$\sigma,\;\sigma^i $&the {\it active} mass and current-mass densities in the global frame  &(\ref{13.0}), (\ref{13.1})\\
\hline
$I_{\sss{<L>}} $&the {\it active} Thorne-Blanchet-Damour mass multipole moments in the global frame  &(\ref{13.9})\\
\hline
$S_{\sss{<L>}} $&the {\it active} spin multipole moments in the global frame  &(\ref{13.10})\\\hline
$\bar V $&potential of the scalar field in the global frame  &(\ref{mp3})\\\hline
$\bar\sigma $&scalar mass density in the global frame  &(\ref{13.13})\\\hline
$\bar I_{\sss{<L>}} $&scalar mass multipole moments in the global frame  &(\ref{13.17})\\\hline
$\tilde V $&gravitational potential of the conformal metric in the global frame  &(\ref{mp4})\\\hline
$\tilde\sigma $&the {\it conformal} mass density in the global frame  &(\ref{13.21})\\\hline
$\tilde I_{\sss{<L>}} $&the {\it conformal} mass multipole moments in the global frame  &(\ref{13.30})\\\hline
$\mathbb{M} $&conserved mass of an isolated system   &(\ref{13.35})\\\hline
$\mathbb{P}^i $&conserved linear momentum of an isolated system  &(\ref{13.36})\\\hline
$\mathbb{S}^i $&conserved angular momentum of an isolated system  &(\ref{13.361})\\\hline
$\mathbb{D}^i $&integral of the center of mass of an isolated system  &(\ref{13.35a})\\\hline
$\hat{A} $&symbols with the hat stand for quantities in the local frame  &
\\\hline
$(B) $&sub-index referring to the body and standing for the internal solution in the local frame  &(\ref{1.1}),
(\ref{1.2})\\\hline
$(E) $&sub-index referring to the external with respect to (B) bodies and standing for the external solution in the
local frame&(\ref{1.1}), (\ref{1.2})\\\hline
$(C) $&sub-index standing for the coupling part of the solution in the local frame  &(\ref{1.2})\\\hline
$P_{\sss L} $&external STF multipole moments of the scalar field  &(\ref{1.7a})\\\hline
$Q_{\sss L} $&external STF gravitoelectric multipole moments of the metric tensor  &(\ref{1.8a})\\\hline
$C_{\sss L} $&external STF gravitomagnetic multipole moments of the metric tensor  &(\ref{1.9a})\\\hline
$Z_{\sss L},\; S_{\sss L} $&other sets of STF multipole moments entering the general solution for the
                            space-time part of the external local metric&(\ref{1.9a})\\\hline
$Y_{\sss L}, B_{\sss L},
D_{\sss L}, E_{\sss L},
F_{\sss L}, G_{\sss L}$& STF multipole moments entering the general solution for the space-space part of the external
local metric  &(\ref{1.10a})   \\\hline
${\mathcal V}_i,\;\Omega_i $&linear and angular velocities of kinematic motion of the local frame;
                            we put them to zero throughout the rest of the paper &(\ref{1.8a}), (\ref{1.8aaa})\\\hline
$\nu^i $&3-dimensional velocity of matter in the local frame  &(\ref{1.12})\\\hline
${\cal I}_{\sss{L}} $&active Thorne-Blanchet-Damour  STF mass multipole moments
                        of the body in the local frame  &(\ref{1.31})\\\hline
$\sigma_{\sss B}$&active mass density of body B in the local frame  &(\ref{pz3})\\\hline
$\bar{\cal I}_{\sss{L}} $&scalar STF mass multipole moments of the body in the local frame  &(\ref{1.33})\\\hline
$\bar\sigma_{\sss B} $&scalar mass density of body B in the local frame  &(\ref{pz4})\\\hline
$\tilde{\cal I}_{\sss{L}} $&conformal STF mass multipole moments of the body in the local frame  &(\ref{1.34})\\\hline
$\tilde\sigma_{\sss B} $&conformal mass density of body B in the local frame  &(\ref{pz5})\\\hline
$\sigma^i_{\sss B} $&current mass density of body B in the local frame  &(\ref{pz6})\\\hline
$S_{\sss L} $&spin STF multipole moments of the body in the local frame  &(\ref{1.32})\\\hline
$\xi^0,\;\xi^i $&Relativistic corrections in the post-Newtonian transformation of time and space coordinates
&(\ref{2.2}),
(\ref{2.3})\\\hline
$x^i_{{\sss B}},\;\nnv{i},\;\nnaa{i} $&position, velocity and acceleration of the body's center of mass
                                      with respect to the global frame &(\ref{2.3}), (\ref{2.8}), (\ref{2.10})\\\hline
$\nnr{i} $&$x^i-x^i_{{\sss B}}(t),$ i.e. the spacial coordinates taken with respect
                                    to the center of mass of body B in the global frame  &(\ref{2.3})\\\hline
${\cal A},\;{\cal B}_{\sss{<L>}} $&functions appearing in the relativistic transformation of time  &(\ref{2.8}),
(\ref{eq1})
\\\hline
${\cal D}_{\sss{<L>}},\;{\cal F}_{\sss{<L>}},\;{\cal E}_{\sss{<L>}} $&
                   functions appearing in the relativistic transformation of spacial coordinates  &(\ref{eq2})\\\hline
$\Lambda^{\beta}_{\;\,\alpha} $&matrix of transformation between local and global coordinate bases  &(\ref{cb})\\\hline
$\mathfrak{B},\;\mathfrak{D},\;\mathfrak{B}^i,\;\mathfrak{P}^i,\;\mathfrak{R}^{i}_{\;j} $&
        PN corrections in the matrix of transformation $\Lambda^{\beta}_{\;\,\alpha}$&(\ref{ma00})--(\ref{ma03})\\\hline
$\bar U,\;\bar U^i,\;$ etc.& external gravitational potentials  &(\ref{3.1})--(\ref{3.4})\\\hline
$\bar U_{,L}\nnxe,\;\bar U^i_{,L}\nnxe $& $l$-th spatial derivative of an external potential
                            taken at the center of mass of body B  &(\ref{3.13})\\\hline
$\mathcal{U}^{{\sss(B)}} $& PN correction in the formula of matching of the local Newtonian potential  &(\ref{3.5})\\\hline
${F}^{ik} $&the matrix of relativistic precession of local coordinates with respect to global coordinates&(\ref{5.8})\\
\hline
${\cal M}_*,\;{\cal J}^i_*,\;{\cal P}^i_* $&
Newtonian-type mass, center of mass, and linear momentum of the body in the local frame  &(\ref{a})--(\ref{c})\\\hline
${\rm M} $& general relativistic PN mass of the body in the local frame &(\ref{ij3})\\\hline
${\cal M} $& active mass of the body in the local frame &(\ref{ij2})\\\hline
$\tilde {\cal M} $& conformal mass of the body in the local frame  &(\ref{ij1})\\\hline
${\cal I}^{\sss{(2)}} $&rotational moment of inertia of the body in the local frame  &(\ref{fop})\\\hline
${\cal N}^{\sss L} $& a set of STF multipole moments  &(\ref{aer})\\\hline
${\cal P}^i $& PN linear momentum of the body in the local frame &(\ref{6.2b})\\\hline
$\Delta\dot{\cal P}^i $& scalar-tensor PN correction to $\dot{\cal P}^i$  &(\ref{bvo})\\\hline
$\tilde{\cal M}_{ij} $&conformal anisotropic mass of the body in the local frame  &(\ref{brt})\\\hline
${\mathbb {F}}^i_N,\;\Delta{\mathbb {F}}^i_{N},\;{\mathbb {F}}^i_{pN},\;\Delta{\mathbb {F}}^i_{pN} $&
                gravitational forces in the expression for $Q_i$  &(\ref{6.9})--(\ref{fopm})\\\hline
${\cal S}^i$&the bare post-Newtonian definition of the angular momentum (spin) of a body&(\ref{spin-3}) \\\hline
${\cal T}^i$&the post-Newtonian torque in equations of rotational motion&(\ref{spin-5}) \\\hline
$\Delta{\cal T}^i$&the post-Newtonian correction to the torque ${\cal T}^i$ &(\ref{spin-6}) \\\hline
$\Delta{\cal S}^i$&the post-Newtonian correction to the bare spin ${\cal S}^i$ &(\ref{spin-7}) \\\hline
${\cal R}^i$&velocity-dependent multipole moments &(\ref{spin-8}) \\\hline
 ${\cal S}^i_+$&the (measured) post-Newtonian spin of the body&(\ref{spin-9})\\\hline
$r $& radial space coordinate in the body's local frame, $r=|{\bm w}|$ &(\ref{as7})\\\hline
${\Omega}^j_{\sss B} $& angular velocity of rigid rotation of the body B referred to its local frame &(\ref{qma})\\\hline
$I^{(2l)}_{\sss B} $& $l$-th rotational moment of inertia of the body B  &(\ref{9.6.3})\\\hline
$\mathbb{I}_{\sss B}^L $& multipole moments of the multipolar expansion of the Newtonian potential
                            in the global coordinates    &(\ref{aq9})\\\hline
$R_{\sss B} $&$|{\bm R}_{\sss B}|$, where ${\bm R}_{\sss B}={\bm x}-{\bm x}_{\sss B}$ &(\ref{re4})\\\hline
$R_{\sss{BC}}^i $&$x^i_{\sss C} - x^i_{\sss B}$ &(\ref{pn7})\\\hline
$F^i_{\sss N},\;F^i_{\sss{EIH}},\;F^i_{\sss{\cal S}},\;F^i_{\sss{\cal I}GR},\; \delta F^i_{\sss{\cal I}GR} $&
                    forces from the equation of motion of spherically-symmetric bodies  &(\ref{gko1})--(\ref{gko5})\\
                    \hline
$\mathfrak{M}_{\sss B} $&Nordtvedt's gravitational mass of the body B  &(\ref{dm})\\\hline
\end{longtable}

\section{Introduction}
\subsection{General Outline of the Paper}
This paper consists of 11 sections and 3 appendices. In this section we give a brief introduction
to the problem of relativistic reference frames and describe our motivations for doing this work.
Section 3 outlines the statement of the problem, field equations and the principles of the post-Newtonian
approximations. Section 4 is devoted to the construction of the global (barycentric) reference frame which
is based on the solution of the field equations in entire space.  We make a multipolar expansion of the
gravitational field in the global coordinates and discuss the post-Newtonian conservation laws in section 5.
Section 6 is devoted to the construction of local coordinates in the vicinity of each body being
a member of N-body system. A general structure of the post-Newtonian coordinate transformations
between the global and local coordinate frames is discussed in section 7. This structure is
specified in section 8 where the matching procedure between the global and local coordinates is
employed on a systematic ground. We use results of the matching procedure in section 9 in order to
derive PPN translational equations of motion of the extended bodies in the N-body system. PPN equations
of rotational motion of each body are derived in section 10. These general equations are applied to
the case of motion of spherically-symmetric bodies which is considered in section 11. Appendix A gives
solution of the Laplace equation in terms of scalar, vector and tensor harmonics. Appendix B provides
with explicit expressions for calculation of the Christoffel symbols and Riemann tensor in terms of
the post-Newtonian perturbation of the metric tensor. Appendix C compares our results with those obtained by
Klioner and Soffel \cite{kls} by making use of a different approach.

\subsection{Motivations and Historical Background}

General theory of relativity is the most powerful theoretical tool for experimental gravitational physics both
in the solar
system and outside of its boundaries. It passed all tests with
unparallel degree of accuracy \cite{wilrv,schaefer,dam2000}.
However, alternative theoretical models are required for deeper
understanding of the nature of space-time gravitational physics and for studying possible
violations of general relativistic relationships which may be observed
in
near-future gravitational experiments designed for testing the principle of
equivalence \cite{step},mapping astrometric positions of stars in our galaxy
with micro-arcsecond precision \cite{gaia} and searching
for extra-solar planets \cite{sim}, testing near-zone relativistic effects associated
with finite speed of propagation of gravitational fields
\cite{kop2001,fk-apj,k-cqg}, detection of freely propagating gravitational waves \cite{ligo}
--\cite{lisa}, etc.

Recently, International Astronomical Union (IAU) has adopted new resolutions \cite{iau1997}
--\cite{skpw} which lay down a self-consistent general relativistic
foundation for further applications in modern geodesy, fundamental astrometry, and celestial
mechanics in the solar system. These resolutions
combined two independent approaches to the theory of relativistic
reference frames in the solar system developed in a series of
publications of various authors \footnote{These approaches are called Brumberg-Kopeikin (BK) and Damour-Soffel-Xu (DSX)
formalisms. The reader is invited to review \cite{skpw} for full list of bibliographic references.}.
The goal of the present paper is to incorporate the parametrized post-Newtonian (PPN) formalism \cite{nord1}
--\cite{will} to the
IAU theory of general relativistic reference frames in the solar
system.  This will extend domain of applicability of the resolutions
to more general class of gravity theories. Furthermore, it will make
the IAU resolutions fully compatible with the JPL equations of motion used for calculation of ephemerides of major
planets, Sun and Moon.
These equations depend on two PPN parameters $\beta$ and $\gamma$ \cite{sman} and they
are presently compatible with the IAU resolutions only in the case of $\beta=\gamma=1$.

PPN parameters $\beta$ and $\gamma$ are characteristics of a scalar field which perturbs the metric tensor and makes
it different from general relativity.
Scalar fields has not yet been detected but they already play
significant role in modern physics. This is because scalar fields help us to explain the origin
of masses of elementary particles \cite{higgs}, to solve various
cosmological problems
\cite{star} --\cite{linde}, to disclose the nature of dark
energy in the universe
\cite{quint}, to develop a gauge-invariant theory of cosmological
perturbations \cite{krms,rk} joining in a very natural way the ideas contained in the original gauge-invariant
formulation proposed
by Bardeen \cite{bard} \footnote{See \cite{mfb} for review of more recent results related to the development
of Bardeen's theory of cosmological perturbations.} with a coordinate-based
approach of Lifshitz \cite{lif,lifkh}. In the present paper
we employ a general class of the scalar-tensor theories of gravity initiated in the pioneering works by
Jordan \cite{jo1,jo2}, Fierz \cite{frz} and, especially, Brans
and Dicke \cite{brd} --\cite{dicke1} \footnote{For a
well-written introduction to this theory and other relevant references can be found in \cite{will} and \cite{wein}. }.
This class of theories is
based on the metric tensor $g_{\alpha\beta}$ representing gravitational field and a scalar
field $\phi$ that couples with the metric tensor through the coupling
function $\theta(\phi)$ which we keep arbitrary. We assume that $\phi$ and $\theta(\phi)$ are analytic functions
which can be expanded about their cosmological background values $\bar{\phi}$ and $\bar{\theta}$. Existence of
the scalar field $\phi$
brings about dependence of the universal gravitational constant $G$ on
the background value of the field $\bar{\phi}$ which can be
considered as constant on the time scale much shorter than the Hubble
cosmological time.

Our purpose is to develop a theory of relativistic reference
frames in an N-body problem (solar system) with two parameters $\beta$ and
$\gamma$ of the PPN formalism. There is a principal difficulty in developing such a theory associated with
the problem of construction of a local reference frame in the vicinity of each
self-gravitating body (Sun, Earth, planet) comprising the N-body system.
Standard textbook on the PPN formalism \cite{will} does not contain solution of this problem in the post-Newtonian
approximation  because the
original PPN formalism was constructed in a single, asymptotically flat,
global coordinate chart (PPN coordinates) covering the entire space-time and having the
origin at the barycenter of the solar system. PPN formalism admits existence of several fields which are responsible
for gravity -- scalar, vector, tensor, etc. After imposing boundary conditions on all these fields at infinity the
standard PPN metric tensor combines their contributions all together in a single expression so that they get absorbed
to the Newtonian and other general relativistic potentials and their contributions are strongly
mixed up. It becomes technically impossible to disentangle the fields in order to find out relativistic space-time
transformation between local
frame of a self-gravitating body (Earth) and the global PPN coordinates which would be consistent with the law of
transformation of the fields imposed by each specific theory of gravitation.
Rapidly growing precision of optical and radio astronomical
observations as well as calculation of relativistic equations of motion in gravitational wave astronomy urgently
demands to work out a PPN theory of such relativistic
transformations between the local and global frames.

It is quite straightforward to construct the post-Newtonian Fermi coordinates along a world line of a massless particle
\cite{niz}. Such approach can be directly applied in the PPN formalism to construct the Fermi reference
frame around a world line of, for example, an artificial satellite. However, account for gravitational self-field
of the particle (extended body) changes physics of the problem and introduces new mathematical aspects to the
existing procedure of construction of the Fermi frames as well as to the PPN formalism.
To the best of our knowledge only two papers \cite{kls,ssa} have been published so far by other researchers where
possible
approaches aimed to derive the
relativistic transformations between the local (geocentric,
planetocentric) and the PPN global coordinate frame were discussed in the framework of the PPN formalism.
The approach proposed in \cite{ssa}
is based on the formalism that was originally worked out by Ashby and
Bertotti \cite{ashb1,ashb2} in order to construct a local
inertial frame in the vicinity of a self-gravitating body that is
a member of an N-body system \footnote{Fukushima (see \cite{fuk1} and references therein)  developed similar ideas
independently by making use of a slightly different mathematical technique.}. In the Ashby-Bertotti formalism
the PPN metric
tensor is taken in its standard form \cite{will} and it treats all massive bodies
as point-like monopole massive particles without rotation. Construction of
a local inertial frame in the vicinity of such massive particle
requires to impose some specific restrictions on the world line of the particle. Namely,
the particle is assumed to be moving along a geodesic defined on the ``effective"
space-time manifold which is obtained by elimination of the body under
consideration from the expression for the standard PPN metric tensor. This
way of introduction of the ``effective" manifold is not
defined uniquely
bringing about an ambiguity in the construction of the ``effective"
manifold \cite{kop1988}. Moreover, the assumption that bodies
are point-like and non-rotating is insufficient for modern geodesy and relativistic celestial mechanics.
For example,
planets in the solar system and stars in binary systems have appreciable rotational speeds and noticeable higher-order
multipole moments. Gravitational interaction of the multipole moments of a celestial body with external tidal field does
not allow the body to
move along the geodesic line \cite{kop1988}. Deviation of the body's center-of-mass world line from the geodesic
motion can be significant and important in numerical calculations of planetary ephemerides (see, e.g., \cite{standish}
and
discussion on page 307 in \cite{kop1991b}) and must be taken into account when one constructs a theory of the
relativistic reference frames in the N-body system.

Different approach to the problem of construction of a local
(geocentric) reference frame in the PPN formalism was proposed in
the paper by Klioner and Soffel \cite{kls}. These authors have
used a phenomenological approach which
does not assume that the PPN metric tensor in local coordinates must be a solution of the field equations of a
specific
theory of gravity. The intention was to make the Klioner-Soffel formalism as general as possible. To this end these
authors assumed that the structure of the metric tensor written down in the local
(geocentric) reference frame must have the following properties:
\begin{itemize}
\item[{\bf A.}]
gravitational field of external bodies (Sun, Moon,
planets) is represented in the vicinity of the Earth in the form
of tidal potentials which should reduce in the Newtonian limit to the Newtonian tidal potential,
\item[{\bf B.}]
switching off the tidal potentials must reduce the metric
tensor of the local coordinate system to its standard PPN
form.
\end{itemize}
Direct calculations revealed that under assumptions made in \cite{kls} the
properties (A) and (B) can not be satisfied simultaneously. This
is a direct consequence of the matching procedure applied in \cite{kls} in order
to transform the local geocentric
coordinates to the global barycentric ones. More specifically, at each step of
the matching procedure four kinds of different terms in the metric tensors have been
singling out and equating separately in the corresponding matching equations for the
metric tensor (for more details see page 024019-10 in \cite{kls}):
\begin{itemize}
\item the terms depending on internal potentials of the body under consideration (Earth);
\item the terms which are functions of time only;
\item the terms which are linear functions of the local spatial
coordinates;
\item the terms which are quadratic and higher-order polynomials of the
local coordinates.
\end{itemize}
These matching conditions are implemented
in order to solve the matching equations. It is implicitly assumed in \cite{kls} that their application
will not give rise to contradiction with other principles of the parametrized gravitational theory in the
curved space-time.

We draw attention of the reader to the problem of choosing the right number of the matching equations. In general
theory of relativity the only gravitational field variable is the metric tensor. Therefore, it is necessary and
sufficient to write down the matching equations for the metric tensor only. However, alternative theories of gravity
have additional fields (scalar, vector, tensor) which contribute to the gravitational field as well. Hence, in these
theories one has to work out matching equations not only for the metric tensor but also for the additional fields.
This problem has not been discussed in \cite{kls} which assumed that it will be sufficient to solve the matching
equations merely for the metric tensor in order to obtain complete information about the structure of the
parametrized post-Newtonian transformation from the local to global frames. This might probably work for some
(yet unknown)
alternative theory of gravity but the result of matching would be rather formal whereas the physical content of such
matching and the degree of applicability of such post-Newtonian transformations will have remained unclear. In the
present paper we rely upon quite general class of the scalar-tensor theories of gravity and consistently use the
matching equation for
the metric tensor along with that for the scalar field which are direct consequences of the field equations.
We have found that our results diverge pretty strongly from the results
of Klioner-Soffel's paper \cite{kls}. This divergence is an indication that the phenomenological (no-gravity-field
equations) Klioner - Soffel approach to the
PPN formalism with local frames taken into account has too many degrees of freedom so that the method of
construction of the parametrized metric tensor in the local coordinates along with the PPN coordinate
transformations proposed in \cite{kls} can not fix them uniquely.  Phenomenological restriction of this freedom
can be done
in many different ways {\it ad liberum}, thus leading to additional (researcher-dependent) ambiguity in the
interpretation of relativistic effects in the local (geocentric) reference frame.

We have already commented that in Klioner-Soffel approach \cite{kls} the metric tensor in the local coordinates is
not determined from the field equations \footnote{Observe the presence of a free function $\Psi$ in Eq. (3.33) of
the paper \cite{kls}.} but is supposed
to be found from the four matching conditions indicated
above. However, the first of the matching
conditions requires that all internal potentials generated by the body's (Earth's) matter can be fully segregated
from the other terms in the metric tensor. This can be done, for example, in general relativity and in the
scalar-tensor theory of gravity as we shall show later in the present paper. However, complete separation of the
internal potentials describing gravitational field of a body under consideration from the other terms in matching
equations may not work out in arbitrary alternative theory of gravity. Thus, the class of gravity theories to
which the first of the matching conditions can be applied remains unclear and yet has to be identified.
Further discussion of the results obtained by Klioner and Soffel is rather technical and deferred to appendix
\ref{ksfor}.

Our point of view is that in order to eliminate any inconsistency and undesirable ambiguities in the construction
of the PPN
metric tensor in the local reference frame of the body under consideration and to apply mathematically rigorous
procedure for derivation of the
relativistic coordinate transformations from the local to global
coordinates, a specific theory of gravity must be used. The field equations in such a case are known and the
number of functions entering the PPN metric tensor in the local coordinates is exactly equal to the number
of matching equations. Hence, all of them can be determined unambiguously. Thus, we propose to build a parametrized
theory of relativistic reference
frames in an N-body system by making use of the following procedure:
\begin{enumerate}
\item Chose a class of gravitational theories with a well-defined system of field equations.
\item Impose a specific gauge condition on the metric tensor and other fields to single out a class of global and
local coordinate systems and to reduce the field equations to a solvable form.
\item Solve the reduced field equations in the global coordinate system $x^\alpha=(x^0,x^i)$ by imposing fall-off
boundary conditions at infinity.
\item Solve the reduced field equations in the local coordinate system $w^\alpha=(w^0,w^i)$ defined in the vicinity
of a world line of the center-of-mass of a body. This will give N local coordinate systems.
\item Make use of the residual gauge freedom to eliminate nonphysical degrees of freedom and to find out the most
general structure of the space-time coordinate transformation between the global and local coordinates.
\item Transform the metric tensor and the other fields from the local coordinates to the global ones by making use
of the general form of the coordinate transformations found at the previous step.
\item Derive from this transformation a set of matching (first-order differential and/or algebraic) equations for
all functions entering the metric tensor and the coordinate transformations.
\item Solve the matching equations and determine all functions entering the matching equations explicitly.
\end{enumerate}\noindent
This procedure works perfectly in the case of general relativity \cite{skpw} and is valid also in the class of the
scalar-tensor theories of gravity as we shall show in the present paper. We do not elaborate on this procedure in
the case of vector-tensor and tensor-tensor theories of gravity. This problem is supposed to be solved somewhere else.

The scalar-tensor
theory of gravity employed in this paper operates with one tensor,
$g_{\alpha\beta}$, and one scalar, $\phi$, fields. The tensor
field $g_{\alpha\beta}$ is the metric tensor of the Riemannian
space-time manifold. The scalar field $\phi$ is not fully independent and is
generated by matter of the gravitating bodies comprising an N-body system. We assume that the N-body system
(solar system, binary star) consists of extended bodies which
gravitational fields are weak everywhere and characteristic velocity of motion is slow. These assumptions allow
us to use the post-Newtonian
approximation (PNA) scheme developed earlier by various researchers \footnote{PNA solves the gravity field
equations by making use of expansions with respect to the weak-field and slow-motion parameters. The reader
is referred to the cornerstone works \cite{fock} --\cite{futs} which reflect different aspects of the
post-Newtonian approximations.}
in order to
find solutions of the scalar-tensor field equations with non-singular distribution of matter in space. The method, we
work out in the present paper, is a significant extension and further
improvement of the general relativistic calculations performed in our previous papers
\cite{kop1988,kop1991b,kop1989a,kop1989b,kop1991,bk-rf,bk-kin,bk-nc}. It takes into account the
post-Newtonian definition of multipole moments of an isolated
self-gravitating body (or a system of bodies) introduced by Kip Thorne \cite{thorne} which has been
mathematically elucidated and
further developed by Blanchet and Damour \cite{bld} (see
also \cite{di} and references therein). We do not specify the internal structure of the bodies so that our
consideration is not restricted with the case of a perfect fluid as it is usually done in the PPN formalism.

\section{Statement of the Problem}
\subsection{Field Equations in the Scalar-Tensor Theory of Gravity}

The purpose of this paper is to develop a relativistic theory of reference frames for N-body problem in the
PPN formalism which contains 10 parameters \cite{will}.
Michelson-Morley and Hughes-Drever type experiments strongly
restricted possible violations of the local isotropy of space whereas E\"otv\"os-Dicke-Braginsky type experiments
verified a weak
equivalence principle with very high precision \cite{will}. These
remarkable experimental achievements and  modern theoretical attempts to unify gravity with other fundamental fields
strongly restrict class of viable alternative
theories of gravity and very likely reduce the number of parameters of the standard PPN formalism \cite{will}
to two - $\beta$ and $\gamma$ \footnote{Experimental testing of the Lorentz-invariance of the gravity field
equations (that is Einstein's principle of relativity for gravitational field) requires introducing more
parameters \cite{k-cqg,will}. We assume in this paper that the Lorentz-invariance is not violated.}.
These parameters appear naturally in the class of alternative theories of gravity with one or several scalar
fields \cite{will,dames} which can be taken as a basis for making generalization of the IAU resolutions on
relativistic reference frames.
For this reason, we shall work in this paper
only with the class of scalar-tensor theories of gravity assuming
that additional vector and/or tensor fields do not exist. For
simplicity we focus on the case with one real-valued scalar field $\phi$ loosely coupled with gravity by means
of a coupling function $\theta(\phi)$.

Field equations in such
scalar-tensor theory are derived from the action \cite{will}
\be
     S=\frac{c^3}{16\pi}\int\biggl(\phi R                  \label{10.1}
     - \theta(\phi)\frac{\phi^{,\alpha}\phi_{,\alpha}}
     {\phi} - \frac{16\pi}{ c^4} {\cal L}(g_{\mu\nu},\,\Psi)\biggr)\sqrt{-g}\; d^4 x\; ,
\en where the first, second and third terms in the right side of Eq. (\ref{10.1}) are
the Lagrangian densities of gravitational field, scalar field and matter
respectively, $g={\rm det}[g_{\alpha\beta}]<0$ is the determinant of the metric tensor
$g_{\alpha\beta}$, $R$ is the Ricci scalar, $\Psi$ indicates dependence of the matter Lagrangian ${\cal L}$ on
matter fields, and $\theta(\phi)$ is
the coupling function which is kept arbitrary. This makes the class of the theories we are working with to be
sufficiently large.
For the sake
of simplicity we postulate that the
self-coupling potential of the scalar field is identically zero so that the scalar field does not interact
with itself.
This is because we do not expect that this potential can lead to
measurable relativistic effects within the boundaries of
the solar system. However, this potential can be important in the case of a strong gravitational field and its
inclusion to the theory
can lead to interesting physical consequences \cite{dames}.

Equations of gravitational field are obtained by variation of the
action (\ref{10.1}) with respect to $g_{\alpha\beta}$ and it's spatial derivatives. It yields
 \br\label{10.2}
   R_{\mu\nu}&=&
   \frac{8\pi}{\phi c^2}\left(T_{\mu\nu}-\frac12 g_{\mu\nu}T\right)+\theta(\phi)\frac{\phi_{,\mu}\phi_{,\nu}}{\phi^2}
   + \frac{1}{\phi}\left(\phi_{;\mu\nu}+\frac12 g_{\mu\nu}{\dAl}_g\phi\right)\;,
\er where
\be\label{covd}
{\dAl}_g\equiv g^{\mu\nu}\frac{\partial^2}{\partial x^\mu\partial x^\nu}-g^{\mu\nu}\Gamma^\alpha_{\mu\nu}
\frac{\partial}{\partial x^\alpha}\;
\en
is
the scalar Laplace-Beltrami operator and $T_{\mu\nu}$
is the stress-energy-momentum tensor of matter comprising the N-body system. It is defined
by equation \cite{ll} \be
  \frac{c^2}2\sqrt{-g}\;T_{\mu\nu}\equiv                         \label{10.3}
  \frac{\partial(\sqrt{-g}{\cal L})}{\partial g^{\mu\nu}}-
  \frac{\partial}{\partial x^{\alpha}}
  \frac{\partial(\sqrt{-g}{\cal L})}{\partial g^{\mu\nu}{}{}_{,\alpha}} \;.
\en The field equation for the scalar field is obtained by
variation of the action (\ref{10.1}) with respect  to $\phi$ and it's spatial derivatives. After making use of
use of the contracted form of Eq. (\ref{10.2}) it yields \be
  {\dAl}_g\phi=\frac1{3+2\theta(\phi)}
  \left(\frac{8\pi}{c^2}\; T-\phi_{,\alpha}\;\phi^{,\alpha}\;    \label{10.4}
  \frac{d\theta}{d\phi}\right)\;.
\en In what follows, we shall also utilize another version of the
Einstein equations (\ref{10.2}) which is obtained after conformal
transformation of the metric tensor \be
   \tilde g_{\mu\nu}=\frac{\phi}{\phi_0} g_{\mu\nu}\qquad,\qquad\tilde g^{\mu\nu}=\frac{\phi_0}{\phi} g^{\mu\nu}\;.
                            \label{13.18}
\en
Here $\phi_0$ denotes the background value of the scalar field which will be introduced in (\ref{aa}).
It is worth noting that the determinant $\tilde g$ of the conformal metric tensor relates to the determinant
$g$ of the metric $g_{\mu\nu}$ as $\tilde g=(\phi/\phi_0)^4 g$. Conformal transformation of the metric tensor leads to
the conformal transformation of the
Christoffel symbols and the Ricci tensor. Denoting the conformal Ricci tensor by $\tilde R_{\mu\nu}$ one can
reduce the field equations (\ref{10.2}) to a simpler form \be
   \tilde R_{\mu\nu}=\frac{8\pi}{\phi c^2}
   \Bigl(T_{\mu\nu}-\frac12 g_{\mu\nu}T\Bigr)+               \label{13.19}
   \frac{2\theta(\phi)+3}{2\phi^2}\,\phi_{,\mu}\,
   \phi_{,\nu}\,.
\en The metric tensor $g_{\mu\nu}$ is called the physical (Jordan-Fierz-frame)
metric \cite{will} because it is used in real measurements of time intervals
and space distances. The conformal metric $\tilde g_{\mu\nu}$ is
called the Einstein-frame metric. Its main advantage is that this metric is in many technical
aspects more convenient for doing calculations than the Jordan-Fierz frame
metric. Indeed, if the last (quadratic with respect to the scalar
field) term in Eq. (\ref{13.19}) was omitted, it would make them
look similar to the Einstein equations of general relativity.
Nevertheless, we prefer to construct the parametrized post-Newtonian
theory of reference frames for N-body problem in terms of the Jordan-Fierz-frame metric in order to avoid unnecessary
conformal transformation to convert results of our calculations to physically meaningful form.

\subsection{The Tensor of Energy-Momentum}

In order to find the gravitational field and determine the motion of the bodies comprising the N-body system
one needs: \begin{itemize}
\item[(1)] to specify a model of matter composing of the N-body system,
\item[(2)] to specify the gauge condition on the metric
tensor $g_{\alpha\beta}$,
\item[(3)] to simplify (reduce) the field equations by making use of the
chosen gauge,
\item[(4)] to solve the reduced field equations,
\item[(5)] to derive equations of motion of the bodies by making use of the solutions of the field equations.
\end{itemize}
This program will be completed in the present paper for the case of
an isolated system of N bodies moving slowly and having
weak gravitational field. In principle,
the formalism which will be developed in the present paper allows us to treat
N-body systems consisting of black holes, neutron stars, or other compact
relativistic bodies if the strong field zones are excluded and the appropriate matching of the strong-field
and weak-field zones is done \cite{dam300}. This
problem will be considered somewhere else. The most important
example of the weak-field and slow-motion N-body system represents our
solar system and one can keep this example in mind for future practical applications of the PPN formalism
developed in the present paper.

We assume that the N-body system is isolated which means that we neglect the tidal
influence of other matter in our galaxy on this system. Thus, the space-time very far away outside of the system is
considered as asymptotically-flat so that the barycenter of the N-body
system is either at rest or moves with respect to the asymptotically flat space along a straight line with a constant
velocity. We assume that the matter comprising the bodies
of the N-body system is described by the energy-momentum tensor with some equation of state which we do not specify.
Following Fock \cite{fock} \footnote{See also \cite{papap} which develops similar ideas.} we define the
energy-momentum tensor as \be
c^2 T^{\mu\nu}=\rho\left(c^2+\Pi\right)u^\mu
u^\nu+\pi^{\mu\nu}\;,    \label{11.1} \en
where $\rho$ and
$\Pi$ are the density and the specific internal energy of matter in the
co-moving frame,
$u^\alpha=dx^\alpha/cd\tau$ is the dimensionless 4-velocity of the matter with $\tau$
being the proper time along the world lines of matter, and
$\pi^{\alpha\beta}$ is the anisotropic tensor of stresses
defined in such a way that it is orthogonal to the 4-velocity
\be\label{pz1}
u^\alpha\pi_{\alpha\beta}=0\;.
\en
Original PPN formalism treats the matter of the N-body system as a perfect fluid for which
\be\label{perf}
\pi^{\alpha\beta}=\left(g^{\alpha\beta}+u^\alpha u^\beta\right)p\;,\en
 where $p$ is an isotropic pressure \cite{will}. Perfect-fluid approximation is not sufficient in the Newtonian
 theory of motion of the solar system bodies where tidal phenomena and dissipative forces play
 essential role \cite{zhtr}. It is also inappropriate for consideration of last stages of coalescing binary
 systems for which a full relativistic theory of tidal deformations must be worked out. For this reason we
 abandon the perfect-fluid approximation and incorporate the anisotropic stresses to the PPN formalism.
 General relativistic consideration of the anisotropic stresses has been done in papers
 \cite{dsx1991,dsx1992,dsx1993,sokx}.

Conservation of the energy-momentum tensor $T^{\mu\nu}{}{_{;\nu}}=0$ leads to the equation of continuity
\be\label{pz2}
\left(\rho u^\alpha\right)_{;\alpha}=\frac{1}{\sqrt{-g}}\left(\rho\sqrt{-g} u^\alpha\right)_{,\alpha} =0\;,
\en
and the second law of thermodynamics that is expressed as
a differential relationship between the specific internal energy
and the stress tensor
\be \label{11.2}
\rho u^{\alpha}\Pi_{,\alpha}+
      \pi^{\alpha\beta}
      u_{\alpha;\beta}=0\;.
\en
These equations define the structure of the tensor of energy-momentum and will be employed later for solving
the field equations and derivation of the equations of motion of the bodies.

\subsection{Basic Principles of the Post-Newtonian Approximations}\label{bpri}

Field equations (\ref{10.2}) and (\ref{10.4}) all together represent a system
of eleventh non-linear differential equations in partial
derivatives and one has to find their solutions for the case of an
N-body system. This problem is complicated and can be
solved only by making use of approximation methods. Two basic methods are known as the post-Minkowskian (see
\cite{dam300,bldam1,bldam2} and references therein)  and the
post-Newtonian (see \cite{dam300} and references therein)
approximation schemes. The post-Newtonian approximation (PNA) scheme deals with slowly moving bodies having weak
gravitational field which makes it
very appropriate for constructing the theory of the
relativistic reference frames in the solar system than the post-Minkowskian approximation (PMA) scheme. This is
because PMA does not use the slow-motion assumption and solves the gravity field equations in terms of retarded
gravitational potentials which are not very convenient for description of relativistic celestial mechanics of
isolated systems.  For this reason, we shall mostly use the PNA scheme in this paper though some elements of
the post-Minkowskian approximation (PMA) scheme will be used for definition of the multipole moments of the
gravitational field.

 Small parameters in the PNA scheme are
$\epsilon_1\sim v/c$ and $\epsilon_2\sim U/c^2$, where $v$
is a characteristic velocity of motion of matter, $c$ is the
ultimate speed (which is numerically equal to the speed of light in vacuum), and $U$ is the Newtonian
gravitational potential. Due to validity of the virial theorem for
self-gravitating isolated systems one has
$\epsilon_2\sim\epsilon_1^2$ and, hence, only
one small parameter $\epsilon_1$ can be used. For the sake of simplicity we introduce parameter
$\epsilon\equiv 1/c$ and consider it formally as a primary parameter of the PNA scheme so,
for example, $\epsilon_1=\epsilon v$, $\epsilon_2=\epsilon^2 U$, etc.

One assumes that the scalar field can be expanded in power
series around its background value $\phi_0$, that is
\be                                                         \label{aa}
\phi= \phi_0(1+\zeta)\;,
\en
where $\zeta$ is dimensionless
perturbation of the scalar field around its background value. The
background value $\phi_0$ of the scalar field can depend on time
due to cosmological evolution of the universe but, according to Damour and Nordtvedt
\cite{damn}, such time-dependence is expected to be rather insignificant due to the presumably
rapid decay of the scalar field in the course of cosmological evolution following
immediately after the Big Bang. According to theoretical expectations
\cite{damn} and experimental data \cite{schaefer}, \cite{dam2000},
\cite{will} the variable part $\zeta$ of the scalar field must have a very
small magnitude so that we can expand all quantities depending on
the scalar field in Taylor series using
$\zeta$ as a small parameter. In particular, decomposition of the
coupling function $\theta(\phi)$ can be written as
\be
\theta(\phi)=\omega+\omega'\;\zeta+O(\zeta^2)\;, \label{10.5}
\en
where $\omega\equiv\theta(\phi_0)$, $\omega'\equiv
\left(d\theta/d\zeta\right)_{\phi=\phi_0}$, and we assume that
$\zeta$ approaches zero as the distance from the N-body system grows to infinity.

Accounting for the decomposition of the scalar field and Eq.
(\ref{10.4}) the gravity field equations (\ref{10.2}) assume the
following form
\br                                                         \label{10.7}
  R_{\mu\nu}&=&\frac{8\pi{\cal G}}{(1+\zeta)c^2}\biggl[ T_{\mu\nu}-
  \frac{\omega+1}{2\omega+3}\;g_{\mu\nu}T\Bigl(
  1+\frac{\omega'\,\zeta}{(\omega+1)(2\omega+3)}\Bigr)\biggl]\\\nonumber&&\qquad\qquad-\frac12g_{\mu\nu}
  \frac{\omega'\, \zeta_{,\alpha}\;
  \zeta^{,\alpha}}{2\omega+3}
  +\frac{\omega\,\zeta_{,\mu}\;\zeta_{,\nu}}{(1+\zeta)^2}\;
  +\frac{\zeta_{;\mu\nu}}{1+\zeta}\;,
\er
where ${\cal G}=1/\phi_0$ is the bare value of the universal gravitational constant and we have taken into account only
linear and quadratic terms of the scalar field which is sufficient for developing the post-Newtonian
parametrized theory of the reference frames in the solar system.

We look for solutions of the field equations
in the form of a Taylor expansion of the metric tensor and the
scalar field with respect to the parameter $\epsilon$
 such that
\br                                                 \label{exp}
g_{\alpha\beta}&=&\eta_{\alpha\beta}+\epsilon\nne{1}{\alpha}{\beta}+\epsilon^2\nne{2}{\alpha}{\beta}+\epsilon^3
\nne{3}{\alpha}{\beta}+O(\epsilon^4)\;,
\er
or, more explicitly,
\br
g_{00}&=&-1+\epsilon^2\nne{2}{0}{0}+\epsilon^4\nne{4}{0}{0}+O(\epsilon^5)
\;,       \label{11.4}
\\
g_{0i}&=&\epsilon\nne{1}{0}{i}+\epsilon^3\nne{3}{0}{i}+O(\epsilon^5) \;,    \label{11.5}
\\
g_{ij}&=&\delta_{ij}+ \epsilon^2\nne{2}{i}{j}+                     \label{11.6}
    \epsilon^4\nne{4}{i}{j}+O(\epsilon^5)  \;,
\\
\zeta&=&\epsilon^2 \nnd{2}+\epsilon^4\nnd{4}+O(\epsilon^6) \;, \label{11.7} \er
where
$\nne{n}{\alpha}{\beta}$ and $\nnd{n}$ denote terms of order
$\epsilon^n\;\;(n=1,2,3...)$.
It has been established that the post-Newtonian expansion of the metric tensor in general theory of relativity
is non-analytic \cite{dam300}. However, the non-analytic terms emerge in the approximations of higher
post-Newtonian order and does not affect our results since we restrict ourselves only with the first
post-Newtonian approximation.
The first post-Newtonian
approximation involves explicitly only terms $\nne{2}{0}{0}$,
$\nne{4}{0}{0}$, $\nne{1}{0}{i}$, $\nne{3}{0}{i}$, $\nne{2}{i}{j}$ and $\nnd{2}$.
In what follows we shall use simplified notations for the metric tensor and scalar field perturbations:
\be                                                             \label{not}
N\equiv\nne{2}{0}{0}\;,\quad
L\equiv\nne{4}{0}{0}\;,\quad
N_i\equiv\nne{1}{0}{i}\;,\quad
L_i\equiv\nne{3}{0}{i}\;,\quad
H_{ij}\equiv\nne{2}{i}{j}\;,\quad H\equiv\nne{2}{k}{k}\;,\en
and
\be\label{not1}
\varphi\equiv(\omega+2)\nnd{2}\;.
\en

The post-Newtonian expansion of the metric tensor and scalar field
introduces a corresponding expansion of the energy-momentum tensor
\br
   T_{00}&=&\nnt{0}{0}{0}+\epsilon^2\nnt{2}{0}{0}+O(\epsilon^4) \;,              \label{11.8}
\\
   T_{0i}&=&\epsilon\nnt{1}{0}{i}+\epsilon^3\nnt{3}{0}{i}+O(\epsilon^5) \;,              \label{11.9}
\\
   T_{ij}&=&\epsilon^2\nnt{2}{i}{j}+\epsilon^4\nnt{4}{i}{j}+O(\epsilon^6) \;,              \label{11.10}
\er where again $\nnt{n}{\alpha}{\beta}\;\;(n=1,2,3...)$ denote terms of order
$\epsilon^n$. In the first post-Newtonian approximation we
need only $\nnt{0}{0}{0}$, $\nnt{2}{0}{0}$, $\nnt{1}{0}{i}$ and
$\nnt{2}{i}{j}$ which are given by the following equations \br
\nnt{0}{0}{0}&=&\rho^*\;, \label{11.21}
\\
 \nnt{1}{0}{i}&=&-\;\rho^*\left(v^i+N^i\right)\;,                       \label{11.23}
\\
 \nnt{2}{i}{j}&=&\rho^*\left(v^i+N^i\right)\left(v^j+N^j\right)+\pi^{ij}\;,          \label{11.24}
\\
 \nnt{2}{0}{0}&=&\rho^*\left(\frac{v^2}{2}-v^kN_k-\frac12 N^kN_k+\Pi-                 \label{11.22}
                  N-
                  \frac{H}{2}\right)\;.
\er
Here we have used the invariant
density \cite{fock}
\be
   \rho^*\equiv\sqrt{-g}u^0\rho=\rho+\epsilon^2\rho\left(\frac{1}{2}H+\frac{1}{2}v^2+\frac{1}{2}N_k N^k+
   v^kN_k \right)\;,                               \label{11.19}
\en that replaces density $\rho$ and is more convenient in calculations because it satisfies the exact
Newtonian-like equation of continuity (\ref{pz2}) which can be recast to \cite{will,fock} \br
    c\rho^*_{\,,0}+                      \label{11.20}
    (\rho^*v^i)_{,i}&=&0\;,
\er where ${\bm v}\equiv (v^i)$ is the 3-dimensional velocity of matter such that
$v^i=c u^i/u^0$.

\subsection{The Gauge Conditions and the Residual Gauge Freedom}

The gauge conditions imposed on the components of the metric tensor had been proposed by Nutku and are
chosen as follows \cite{n1,n2}
\be
   \left(\frac{\phi}{\phi_0}\;\sqrt{-g}\;g^{\mu\nu}\right)_{,\nu}=0\;.                   \label{11.3}
\en
By making use of the conformal metric tensor one can recast
Eq. (\ref{11.3}) to the same form as the de Donder
(or harmonic) gauge conditions in general relativity \cite{fock,papap}
\be
   (\sqrt{-\tilde g}\,\tilde g^{\mu\nu})_{,\nu}=0\,.       \label{13.20}
\en
In what follows, we shall use a more convenient form of Eq. (\ref{11.3}) written as
\be                                                 \label{gau}
g^{\mu\nu}\Gamma^\alpha_{\mu\nu}=\left(\ln\frac{\phi}{\phi_0}\right)^{,\alpha}\;,
\en
so the Laplace-Beltrami operator (\ref{covd}) assumes the form
\be                                             \label{co}
{\dAl}_g\equiv g^{\mu\nu}\left(\frac{\partial^2}{\partial x^\mu\partial x^\nu}-\frac{1}{\phi}\frac{\partial\phi}
{\partial x^\mu}\frac{\partial}{\partial x^\nu}\right)\;.
\en
Dependence of this operator on the scalar field is a property of the adopted gauge condition.

Any function $F(x^\alpha)$ satisfying the homogeneous Laplace-Beltrami equation, $\dAl_gF(x^\alpha)=0$,
is called harmonic. Notice that ${\dAl}_g x^\alpha=-(\ln\phi)^{,\alpha}\not=0$, so the coordinates $x^\alpha$
defined by the gauge conditions (\ref{gau}) are not harmonic functions. Therefore,
we shall call the
coordinate systems singled out by the Nutku conditions (\ref{11.3}) as quasi-harmonic. They have many properties
similar to the harmonic coordinates in general relativity. The choice of the quasi-harmonic coordinates for
constructing theory of the relativistic
reference frames in the scalar-tensor theory of gravity is justified
by the following three factors: (1) the quasi-harmonic coordinates become harmonic when
the scalar field $\phi\rightarrow \phi_0$,
(2) the harmonic coordinates are used in the
resolutions of the IAU 2000 \cite{skpw} on relativistic
reference frames, (3) the condition (\ref{11.3}) significantly
simplifies the field equations and makes it easier to
find their solutions.
One could use, of course, the harmonic coordinates too as it has been done, for example, by Klioner and
Soffel \cite{kls}. They are defined by the condition $g^{\mu\nu}\Gamma^
\alpha_{\mu\nu}=0$  but as we found the field equations and the space-time transformations in these coordinates
look more complicated in contrast to the quasi-harmonic coordinates defined by the Nutku conditions (\ref{11.3}).

Post-Newtonian expansion of the gauge conditions (\ref{gau}) yields
\br
    N_{k,k}&=&0\;,                                                                 \label{11.11a}
    \\
   \frac{c}{2}\left(\frac{2\varphi}{\omega+2}+N+H-N_kN^k\right)_{,0}&=& \label{11.11}
    -\frac{N^j}{2}\left(\frac{2\varphi}{\omega+2}+N+H-N_kN^k\right)_{,j}
    \\\nonumber&&\qquad\quad
     +\left(H_{jk}N^j\right)_{,k}-L_{k,k} \;,
\\
    \frac12\left(\frac{2\varphi}{\omega+2}+N+H-N_kN^k\right)_{,i}&=&  \label{11.12}
    N_{,i}+H_{ik,k}-cN_{i,0}+N^k\left(N_{i,k}-2N_{k,i}\right)\,.
\er
It is worth noting that in the first PNA the gauge-condition Eqs.
(\ref{11.11a}) -- (\ref{11.12}) do not restrict the metric tensor component
$\nne{4}{0}{0}\equiv L$.

Gauge equations (\ref{11.11a}) -- (\ref{11.12}) do not fix the coordinate system uniquely. Indeed, if one changes
coordinates
\be\label{11.12a}
x^\alpha\longrightarrow w^\alpha=w^\alpha\left(x^\alpha\right)\;,
\en
the gauge condition (\ref{gau}) demands only that the new coordinates $w^\alpha$ must satisfy the homogeneous
wave equation
\be                                                         \label{11.b}
g^{\mu\nu}(x^{\beta})\frac{\partial^2 w^\alpha}{\partial x^\mu\partial x^\nu}=0\;,
\en
which have an infinite set of non-trivial solutions.

Eq. (\ref{11.b}) describe the residual gauge freedom existing in the class of the quasi-harmonic coordinate systems
restricted by the Nutku gauge conditions (\ref{gau}). This residual gauge freedom in the scalar-tensor theory is
described by the same equation (\ref{11.b}) as in the case of the harmonic coordinates in general relativity.
We shall discuss this gauge freedom and its applicability to the theory of astronomical reference frames in
more detail in section 7.

\subsection{The Reduced Field Equations}

Reduced field equations for the scalar field and the metric tensor
  are obtained in the first post-Newtonian approximation from Eqs.
(\ref{10.4}) and (\ref{10.7}) after making use of the
post-Newtonian expansions, given by Eqs. (\ref{11.4}) -- (\ref{11.10}). Taking
into account the gauge conditions (\ref{11.11a}) -- (\ref{11.12}) significantly
simplifies the field equations.

The scalar-tensor theory of gravity with variable coupling
function $\theta(\phi)$ has two additional (constant) parameters $\omega$ and $\omega'$ with respect to general
relativity. They are related to the standard PPN parameters $\gamma$ and $\beta$ as follows \cite{will}
\br \gamma=\gamma(\omega)&=&\frac{\omega+1}{\omega+2}\;,                               \label{11.27}
 \\
 \beta=\beta(\omega)&=&1+ \frac{\omega'}{(2\omega+3)(2\omega+4)^2}\;.        \label{11.28}
 \er
We draw attention of the reader that in the book \cite{will} (equation (5.36)), parameter $\Lambda=\beta-1$
is introduced as $\Lambda=\omega'(2\omega+3)^{-2}(2\omega+4)^{-1}$. The difference with our definition (\ref{11.28})
given in the present paper arises due to different definitions of the derivative of the coupling function $\theta$
with respect to the scalar field, that is $(\omega')_{\rm Will}=\phi^{-1}_0(\omega')_{\rm this\;paper}$ where $\phi_0$
is the asymptotic value of the scalar field \footnote{We thank C.M. Will for pointing out this difference to us.}.
All other parameters of the standard PPN formalism describing possible deviations from general relativity
are identically equal to zero \cite{will}. General relativity is obtained as a limiting case of the scalar-tensor
theory when parameters $\gamma=\beta=1$. In order to obtain this limit parameter $\omega$ must go to infinity
with $\omega'$ growing slower than $\omega^3$. If this was not the case one could get
$\lim_{\omega\rightarrow\infty}\gamma=1$ but $\lim_{\omega\rightarrow\infty}\beta\not=1$ which is not a
general relativistic limit.

One can note also that the scalar field perturbation (\ref{not1}) is expressed in terms of $\gamma$ as
\be
\nnd{2}=(1-\gamma)\varphi\;. \label{10.6aa}
\en
As it was established by
previous researchers (see, for instance, \cite{will}) the background scalar field $\phi_0$ and the
parameter of coupling $\omega$ determine the observed numerical
value of the universal gravitational constant
\be G =\frac
{2\omega+4}{2\omega+3 }\;{\cal G}\;, \label{10.6}
\en
where ${\cal
G}\equiv 1/\phi_0$. Had the background value of the scalar
field driven by cosmological evolution, the measured value of the
universal gravitational constant would depend on time and one could hope
to detect it experimentally. The best upper limit on time
variability of $G$ is imposed by lunar laser ranging (LLR) as $|\dot{G}/G|\le 0.5\times 10^{-11}$ yr$^{-1}$
\cite{schaefer}.

After making use of the definition of the tensor of energy-momentum, Eqs. (\ref{11.21}) -- (\ref{11.22}),
and that of the PPN parameters, Eqs. (\ref{11.27}) -- (\ref{10.6}), one obtains
the final form of the reduced field equations:
\br
\label{11.29}
&&\dAl\varphi=-4\pi G\rho^*\;,
\\                                                           \label{11.33}
&&\dAl\left\{ N+\epsilon^2\left[L+\frac{N^2}{2}+2(\beta-1)\varphi^2 \right]\right\}=
\\\nonumber&&\qquad
-8\pi G\rho^*+\frac12\left(N_{i,k}-N_{k,i}\right)\left(N_{i,k}-N_{k,i}\right)+\epsilon^2\biggl\{ H_{<ij>}N_{,ij}
\\\nonumber&&\qquad
-8\pi G\rho^*\left[(\gamma+\frac12)\,v^2+\Pi+
       \gamma\,\frac{\pi^{kk}}{\rho^*}-\frac{H}{6}-(2\beta-\gamma-1)\varphi\right]\biggr\}\,,
\\
&&\dAl N_i=0\;, \label{11.290}
\\\nonumber\\
 &&\dAl L_i= 8\pi G                    \label{11.31}
   \rho^*\left[(1+\gamma)v^i+N^i\right]-2cN^kN_{i,0k}\;,
\\\nonumber\\                                   \label{11.32}
 &&  \dAl H_{ij}=-8\pi G\gamma\rho^*\delta_{ij}+N_{k,i}\left(N_{k,j}-N_{j,k}\right)-
 N_{i,k}\left(N_{j,k}+N_{k,j}\right)  \;,
\er
where
$\dAl\equiv\eta^{\mu\nu}\partial_\mu\partial_\nu $ is the D'Alembert (wave) operator of the Minkowski space-time,
and $H_{<ij>}\equiv H_{ij}-\delta_{ij}H/3$ is the symmetric trace-free (STF) part of the spatial components of
the metric tensor. In these field equations we keep the terms quadratic on $N_i$, but cubic terms and ones proportional
to the products of $N_i$ and perturbations of the metric are omitted.

Equations (\ref{11.29}) -- (\ref{11.32}) are valid in any coordinate system which is admitted by the residual
gauge freedom defined by the gauge conditions (\ref{11.3}). We shall study this residual gauge freedom in full
details when constructing the global coordinates for the entire N-body system and the local coordinates for
each of the bodies. Global coordinates in the solar system are identified with the barycentric reference frame
and the local coordinates are associated with planets. The most interesting case of practical applications
is the geocentric coordinate frame attached to Earth.

\section{Global PPN Coordinate System}

\subsection{Dynamic and Kinematic Properties of the Global Coordinates}

We assume that the gravitational and scalar fields are brought about by the only one system comprising of N extended
bodies
which matter occupies a finite domain in space. Such an astronomical
system is called isolated \cite{fock,papap,dixon} and
the solar system consisting of Sun, Earth, Moon, and other
planets is its particular example. Astronomical systems like a
galaxy, a globular cluster, a binary star, etc. typify other specimens of the isolated systems.
A number of
bodies in the N-body system which must be taken into account depends on the accuracy of astronomical observations
and is determined mathematically by
 the magnitude of residual terms which one must retain in calculations to construct relativistic theory of reference
frames being compatible with the accuracy of the observations. Since we ignore other gravitating
bodies residing outside of the N-body system the space-time can be considered on the global scale as
asymptotically-flat so the metric tensor $g_{\alpha\beta}$ at infinity is
the Minkowski metric $\eta_{\alpha\beta}={\rm diag}(-1,+1,+1,+1)$.

\begin{figure}[p]
\centerline{\psfig{figure=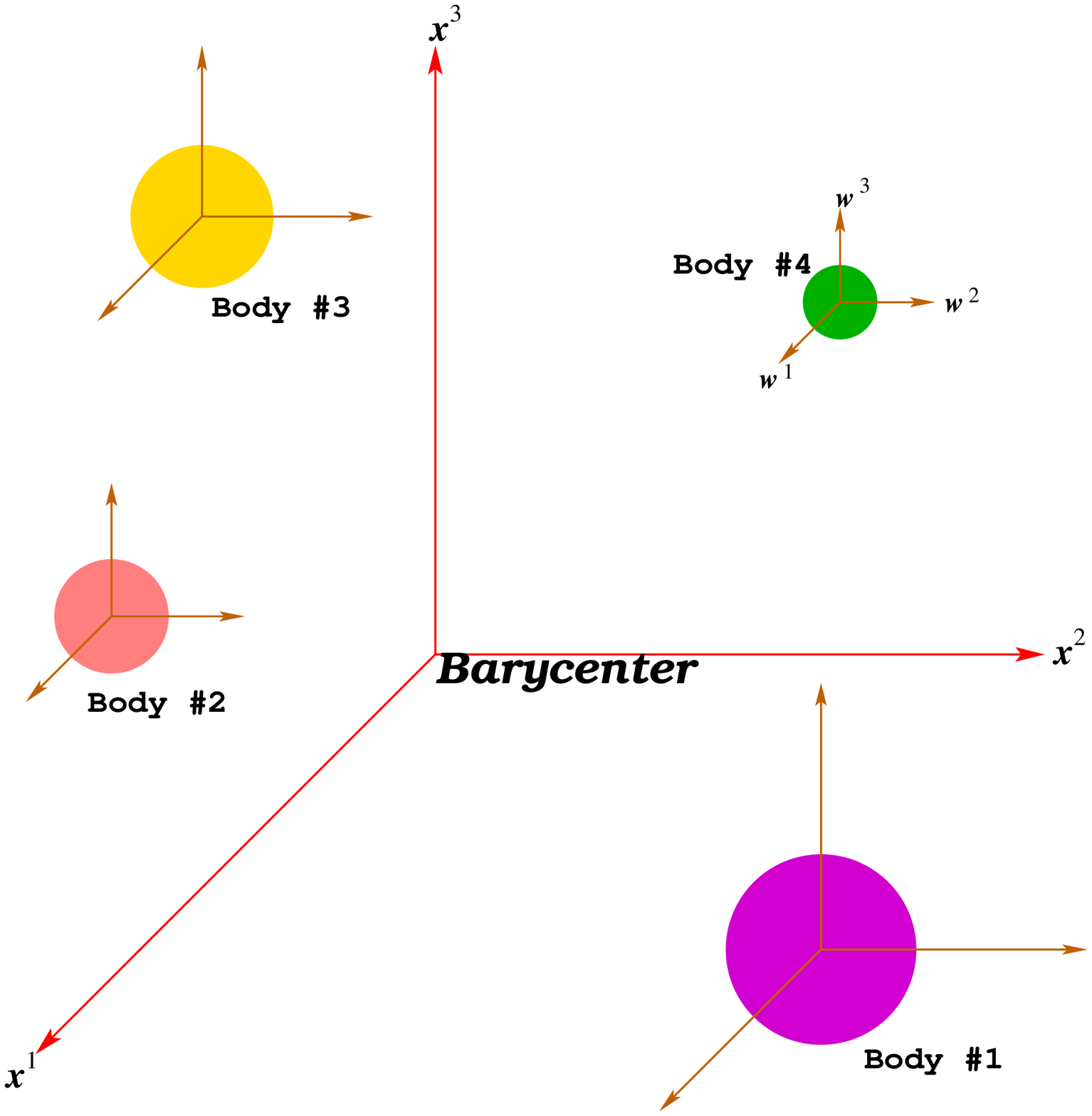,angle=-90,height=14cm,width=18cm}}
\vspace{1cm} \caption[Astronomical N-body System]{The picture illustrates
an astronomical N-body system and coordinate charts associated with
it. Global coordinates, $x^\alpha=(ct, x^i)$, cover the entire
space, have origin at the barycenter of the system, and are Minkowskian at infinity.
Each body has its own local coordinate chart, $w^\alpha=(cu,
w^i)$, having origin at the center of mass of the body under
consideration. Local coordinates are not asymptotically
Minkowskian far away from the body and do not cover the entire space.} \label{fig1}
\end{figure}

\begin{figure}[p]
\centerline{\psfig{figure=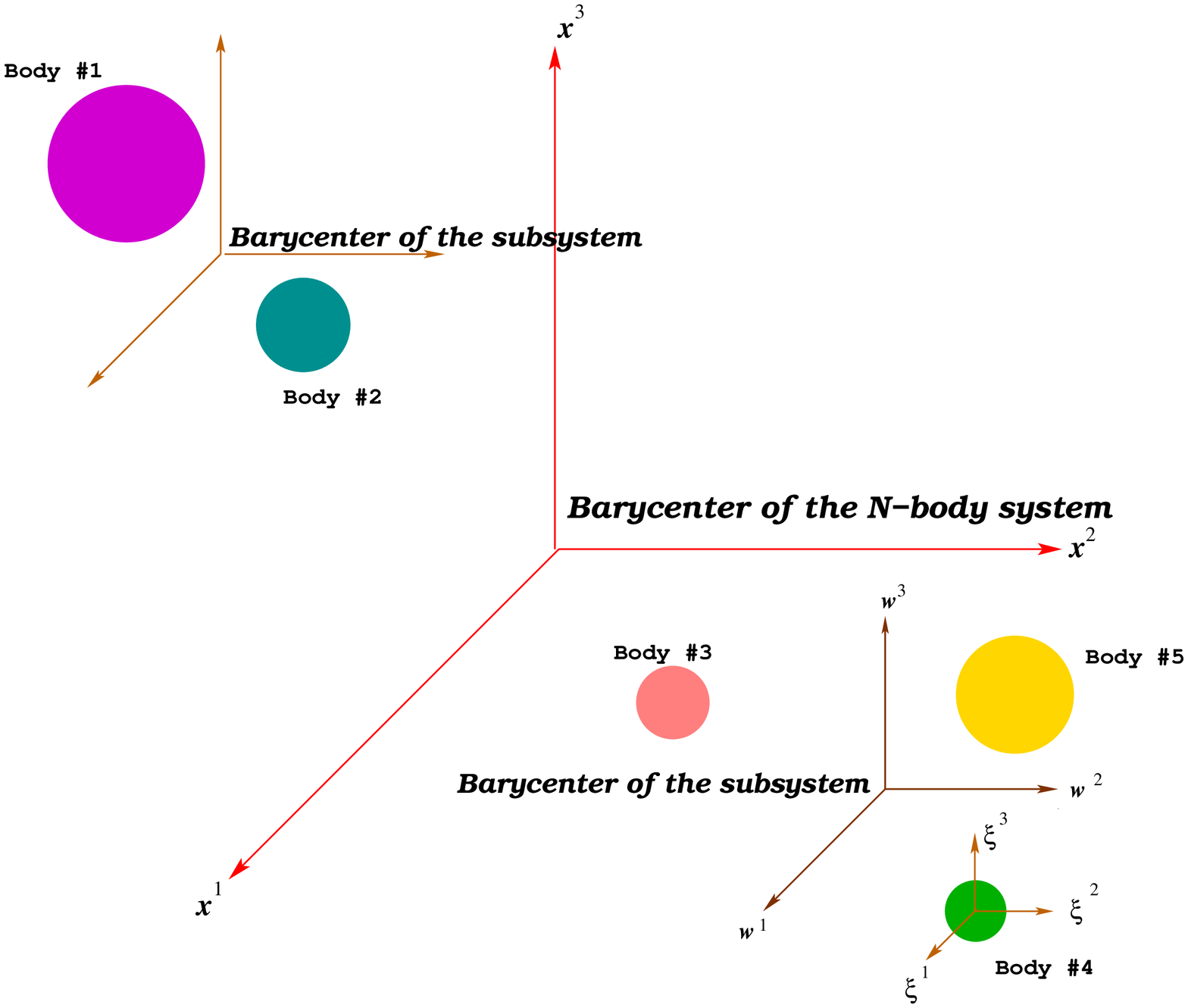,angle=-90,height=14cm,width=18cm}}
\vspace{1cm} \caption[Hierarchy of Coordinate Frames]{The picture
illustrates a hierarchy of coordinate charts existing in N-body
astronomical system (solar system) consisting of several sub-systems: $M_1$, $M_2$,...,$M_p$. One global
coordinate chart, $x^\alpha=(ct, x^i)$, covers the entire space and
has origin at the barycenter of the N-body system. Each
sub-system consists of several gravitationally-bounded bodies (a planet and its moons) and has its
own local coordinate chart, $w^\alpha=(cu, w^i)$, having origin at
the center of mass of the sub-system. At the
same time, each body from the sub-system possesses its own local
coordinate chart, $\xi^\alpha=(cs, \xi^i)$. The hierarchy can have as many levels as necessary for adequate
description of motion of the bodies.} \label{fig1a}
\end{figure}

In the simplest case the N-body system can be comprised of several solitary bodies, like it is shown in Fig \ref{fig1},
but in the most general case it has more complicated hierarchic structure which consists of a sequence
of sub-systems each being comprised of M$_p$ bodies where $p$ is a serial number of the sub-system
(see Fig \ref{fig1a}).
In its own turn each of the sub-systems can contain several sub-sub-systems, and so on. In order to describe
dynamical behavior of the entire N-body system one needs to introduce a global 4-dimensional coordinate system.
We denote such global coordinates $x^\alpha=(x^0,x^i)$, where $x^0=ct$ is time coordinate and $x^i\equiv{\bm x}$
are spatial coordinates. Adequate description of dynamical behavior of the sub-systems of bodies and /or solitary
celestial bodies requires introducing a set of local coordinates attached to each of the sub-systems (or a body)
under consideration. Hence, a hierarchic structure of the coordinate charts in the N-body system repeats that
of the N-body system itself and is fully compatible with mathematical notion of differentiable manifold
\cite{mtw,dfn,schouten}. We shall discuss local coordinates later in section \ref{lcs}.

Let us define the metric tensor perturbation with respect to the Minkowski metric (c.f. Eq. (\ref{exp}))
\be\label{mtp}
h_{\alpha\beta}(t,{\bm x})\equiv g_{\alpha\beta}(t,{\bm x})-\eta_{\alpha\beta}\;.\en
We demand that quantities $rh_{\alpha\beta}$ and $r^2 h_{\alpha\beta,\gamma}$
are bounded, and
\be                                                         \label{12.1}
\lim_{\substack{r\rightarrow\infty\\t+r/c={\rm const.}}}h_{\alpha\beta}(t,{\bm x})=0\;, \en where
$r=|{\bm x}|$. Additional boundary condition must be imposed on the
derivatives of the metric tensor to prevent appearance of non-physical radiative solutions associated with the
advanced wave potentials \cite{fock}. It is written as
\be                                                         \label{12.2}
\lim_{\substack{r\rightarrow\infty\\t+r/c={\rm const.}}}\left[\left(rh_{\alpha\beta}\right),_r+
\left(rh_{\alpha\beta}\right),_0\right]=0\;.
\en
Eq. (\ref{12.2}) is known as a "no-incoming-radiation"
boundary condition \cite{fock,dam1983}. In the case of an
isolated astronomical system this condition singles out a causal
solution of the D'Alembert wave equation depending on the
retarded time $t-r/c$ only.
Similar boundary conditions are imposed on the perturbation of the scalar field defined in Eq. (\ref{10.5})
\be\label{12.3}
\lim_{\substack{r\rightarrow\infty\\t+r/c={\rm const.}}}\zeta(t,{\bm x})=0\;,
\en
\be\label{12.3a}
\lim_{\substack{r\rightarrow\infty\\t+r/c={\rm const.}}}
\left[\left(r\zeta\right),_r+\left(r\zeta\right),_0\right]=0\;.
\en
In principle, the boundary conditions (\ref{12.2}) and (\ref{12.3a}) are not explicitly required in the first
post-Newtonian approximation for solving equations (\ref{11.29}) -- (\ref{11.32}) because the gravitational
potentials in this approximation are time-symmetric. However, they are convenient for doing calculations and
are physically motivated. Therefore, we shall use the (radiative) boundary conditions (\ref{12.2}) and (\ref{12.3a})
later on for giving precise definitions of the multipole moments of the gravitational field of the isolated
astronomical system.

The global coordinates $x^\alpha$ cover the entire space-time and they set up a primary basis for
construction of the theory of relativistic reference frames in the
N-body system \cite{kop1988}. In what follows, we shall assume that the origin of the
global coordinates coincides with the barycenter of the N-body system at any
instant of time. This condition can be satisfied after choosing a suitable
definition of the post-Newtonian dipole moment $\mathbb{D}^i$ of the N-body system and equating its numerical
value to zero along with its first time derivative (see section \ref{pncs}). This can be always done in general
relativity in low orders of the post-Newtonian approximation scheme if one neglects the octuple
and higher-order multipole gravitational radiation \cite{bluc}. In the scalar-tensor theory of gravity one has to take
into account gravitational wave emission in the form of scalar modes \cite{dames} but it does not affect the
first post-Newtonian approximation which is our main concern in the present paper. There are alternative
theories of gravity which violate the third Newton's law so that the dipole moment $\mathbb{D}^i$ of an N-body
system is not conserved even in the first post-Newtonian approximation \cite{will} but we do not consider such
extreme cases.

We shall also assume that spatial axes of the global coordinates do not rotate in
space either kinematically or dynamically \cite{bk-rf}. Spatial
axes of a coordinate system are called kinematically non-rotating \footnote{Angular velocities of dynamic and
kinematic rotations of a reference frame in classic celestial mechanics are equal. However, they have different
values already in the first post-Newtonian approximation due to the presence of the relativistic geodetic
precession caused by the orbital motion of the body.}
if their orientation is kept fixed with respect to a Minkowski
coordinate system defined at the infinite past and at the infinite
distance from the solar system \footnote{At the relativistic language the domain of the asymptotically-flat space-time
is located both at the infinite distance and infinite past. This boundary consisting of null rays is called past null
infinity \cite{mtw}.}. Such kinematically
non-rotating coordinate system can be built on the stellar sky by making use of quasars as reference objects with
accuracy better than $100$ $\mu$arcsec (see
\cite{ma} and references therein). Quasars are
uniformly distributed all over the sky and have negligibly small
parallaxes and proper motions \footnote{Proper motion of an astronomical object in the sky is defined as its
transverse motion in the plane of the sky being orthogonal to the line of sight of observer located at the
barycenter of the solar system.}. Thus, kinematically
non-rotating coordinate system can be determined only through the
experimental analysis of global properties of the space-time manifold
including its global topology. This consideration reveals that the theory of reference frames in N-body system
based on the assumption that the space-time is asymptotically-flat may be corrupted by the influence of some
cosmological effects. Hence, a more appropriate approach to the reference frames taking into account that the
background space-time is the Friedmann-Robertson-Walker universe is to be developed. We have done a
constructive work in this direction in papers \cite{krms,rk} but the results of these papers are still
to be matched with the post-Newtonian approximations.

Dynamically non-rotating coordinate system is defined by the
condition that equations of motion of test particles moving with respect to these coordinates do not have any
terms that might be interpreted as the Coriolis or centripetal
forces \cite{bk-rf}. This definition operates only with local properties of the
space-time and does not require observations of distant celestial
objects like stars or quasars. Dynamical definition of spatially non-rotating coordinates
is used in construction of modern ephemerides of the solar system
bodies which are based primarily on radar and laser ranging measurements to
planets and Moon (see \cite{sman,ma,reframe} and references therein). Because of the assumption that the N-body
system under consideration is isolated, we can
postulate that the global coordinates does not rotate at all in any sense.

\subsection{The Metric Tensor and the Scalar Field in the Global Coordinates}

The metric tensor $g_{\alpha\beta}(t,{\bm x})$ is obtained by
solving the field equations (\ref{11.29}) -- (\ref{11.32}) after
imposing the boundary conditions (\ref{12.1}) -- (\ref{12.3}). We
chose solution of the homogeneous equation (\ref{11.290}) as $N_i=0$. This is because $N_i$ describes rotation of
spatial axes of the coordinate system but we assumed in the previous section that the global coordinates are not
rotating.
It yields solution of the other field equations in the following form \br
   \nnb{t}{x}&=&U(t,\nnx)\,,                                             \label{12.5}
\\
   N(t,{\bm x})&=&2\,U(t,\nnx)\,,                                        \label{12.6}
\\
   L(t,\nnx)&=&2\Phi(t,\nnx)-2\beta U^2(t,\nnx)-c^2\chi_{,00}(t,\nnx)\,, \label{12.9}
\\
   L_i(t,\nnx)&=& -2(1+\gamma)\,U_i(t,\nnx)\,,                           \label{12.8}
\\
   H_{ij}(t,\nnx)&= &2\gamma\, \delta_{ij}\,U(t,\nnx)\,,                 \label{12.7}
\er
where
\be                                                     \label{12.9ex}
\Phi(t,\nnx)\equiv(\gamma+\frac12)\nnff{1}+(1-2\beta)\nnff{2}
         +\nnff{3}+\gamma\nnff{4}\;,
\en and the gravitational potentials $U,\,U^i,\,\chi$, and
$\Phi_k\,\,$ $(k=1,\,...\,,4)$ can be represented as linear
combinations of the gravitational potentials of each body, that is
\be\label{12.9a} U=\sum_A U^{\sss(A)}\,,\qquad U_i=\sum_A
U_i^{\sss(A)}\,,\qquad\Phi_k=\sum_A\Phi^{\sss(A)}_k\:,\qquad
      \chi=\sum_A\chi^{\sss(A)}\:.
¨\en
Herein, the gravitational potentials of body A are defined as integrals taken over the volume of this body
\br
   U^{{\sss(A)}}(t,\nnx)&=&                 \label{12.10}
   G\nnh{A}{\rho^*}\,,
\\\nonumber\\
   U_i^{{\sss(A)}}(t,\nnx)&=&               \label{12.11}
   G\nnh{A}{\rho^*v^i}\,,
\\\nonumber\\
   \chi^{{\sss(A)}}(t,\nnx)&=&              \label{12.12}
   -\,G{\rm\bf I}_1
      ^{{\sss(A)}}\{\rho^*\}\,,
\\\nonumber\\
   \nng{A}{1}{t}{x}&=&G\nnh{A}{\rho^*v^2}\,,                  \label{12.13}
\\\nonumber\\
   \nng{A}{2}{t}{x}&=&G\nnh{A}{\rho^*U}                         \label{12.14a}
\\\nonumber\\
   \nng{A}{3}{t}{x}&=&G\nnh{A}{\rho^*\Pi}\,,                  \label{12.15}
\\\nonumber\\
   \nng{A}{4}{t}{x}&=&G\nnh{A}{\pi^{kk}}\,,                \label{12.16}
\er where notation ${\rm\bf
I}_n^{{\sss(A)}}\{f\}\;\;(n=1,2,3...)$ is used to define the volume
integral \be
   {\rm\bf I}_n^{{\sss(A)}}\{f\}\,
   (t,\nnx)=                                            \label{12.17}
      \int\limits_{V_A}f(t,{\bm x}')
      |{\bm x}-{\bm x}'|^n\,d^3x'\;,
\en with $n$ being an integer, and $V_A$ -- the volume of
integration. Potential $\chi$ is determined as a particular
solution of the inhomogeneous equation \be
   \nabla^2\chi=-2U\,,                                      \label{12.18}
\en with the right side defined in a whole space. Nevertheless, it
proves out that its solution (see Eq. (\ref{12.12})) is spread out over volumes
of the bodies only. It is worthwhile to emphasize that all integrals
defining the metric tensor in the global coordinates are taken over the hypersurface
of (constant) coordinate time $t$. Space-time transformations can
change the time hypersurface, hence transforming
the corresponding integrals. This important issue will be discussed in
section \ref{zza2}.

\section{Multipolar Decomposition of the Metric Tensor and the Scalar Field in the Global Coordinates}\label{miopa}

\subsection{General Description of Multipole Moments}

In what follows a set of certain parameters
describing properties of gravitational and scalar fields and depending on integral characteristics of the N-body system
will be indispensable.
These parameters are called multipole moments. In the Newtonian
approximation they are uniquely defined as coefficients in
Taylor expansion of the Newtonian gravitational potential in
powers of $1/R$ where $R=|{\bm x}|$ is the radial distance from the origin of a coordinate system to a
field point. All Newtonian multipole moments can be functions of time in the most general astronomical situations.
However, very often one assumes that mass is conserved and the center of mass of the system is located at the origin
of the coordinate system under consideration. Provided that these assumptions are satisfied the monopole and
dipole multipole moments
must be constant.

 General relativistic multipolar expansion of
gravitational field is in many aspects
similar to the Newtonian multipolar decomposition. However, due to the non-linearity and tensorial
character of gravitational interaction proper definition of relativistic multipole moments is much more complicated
in contrast to the Newtonian theory. Furthermore, the gauge
freedom existing in the general theory of relativity clearly indicates that
any multipolar decomposition of gravitational field will be
coordinate-dependent. Hence, a great care is required for unambiguous
physical interpretation of various relativistic effects associated with certain multipoles \footnote{See,
for example, section \ref{wep} where we have shown how an appropriate choice of coordinate system allows us
to eliminate a number of coordinate-dependent terms in equations of motion of spherically-symmetric bodies
depending on the "quadrupoles" defined in the global coordinate system. }. It was shown by many researchers
\footnote{For a comprehensive historical review see papers by \cite{thorne}, \cite{bldam2}, \cite{blan} and references
therein} that in general relativity the multipolar expansion of
the gravitational field of an isolated gravitating system is
characterized by only two independent sets -- mass-type and
current-type multipole moments. In particular, Thorne \cite{thorne} had systematized and significantly
perfected works of previous researchers \footnote{Some of the most important of these works
are \cite{sa,pir,bro,ew,wag}.} and defined two sets of the post-Newtonian multipole moments as
follows (see Eqs. (5.32a) and (5.32b) from \cite{thorne})
\begin{eqnarray}\label{tron1}
{\cal I}^L_{\rm Thorne}&=&\int\left(\tau_{00}x^L+A^{l0}r^2x^{<L-2}\tau^{a_{l-1}a_l>}+B^{l0}x^{j<L-1}
\tau^{a_l>j}+C^{l0}x^L\tau^{jj}\right)d^3x \;,\\
\label{tron2}
{\cal S}^L_{\rm Thorne}&=&\varepsilon^{pq<a_l}\int\left(x^{L-1>p}\tau^{0q}+E^{l0}r^2x^{L-2}\partial_t
\tau^{a_{l-1}>j}x^j+F^{l0}x^{L-1>kp}\partial_t\tau^{kq}\right)d^3x\;,
\end{eqnarray}
where numerical coefficients
\begin{eqnarray}
A^{l0}&=&\frac{l(l-1)(l+9)}{2(l+1)(2l+3)}\;,\qquad B^{l0}=-\frac{6l(l-1)}{(l+1)(2l+3)}\;,\\
C^{l0}&=&\frac{2l(l-1)}{(l+1)(2l+3)}\;,\qquad E^{l0}=\frac{(l-1)(l+4)}{2(l+2)(2l+3)}\;,\\
F^{l0}&=&-\frac{l-1}{(l+2)(2l+3)}\;,
\end{eqnarray}
and the multipolar integer-valued index $l$ runs from 0 to infinity.
In these expressions $\tau^{\alpha\beta}$ is the effective stress-energy tensor evaluated at post-Newtonian
order in the post-Newtonian harmonic gauge \cite{thorne}
\begin{equation}\label{amok}
\tau^{\alpha\beta}=\left(1+4\epsilon^2 U\right)T^{\alpha\beta}+\frac{c^4}{16\pi G}N^{\alpha\beta}\;,
\end{equation}
where
\begin{eqnarray}\label{ahv1}
N^{00}&=&-\frac{14}{c^4}U,_{p}U,_{p}\;,\\\label{ahv2}
N^{0i}&=&\frac{4}{c^5}\left[4U,_{p}\left(U^p,_{i}-U^i,_{p}\right)-3U,_{i}U^p,_{p}\right]  \;,\\\label{ahv3}
N^{ij}&=&\frac{2}{c^4}\left(2U,_{i}U,_{j}-\delta_{ij}U,_{p}U,_{p}\right)\;,
\end{eqnarray}
and $U$, $U^i$ are gravitational potentials of the isolated astronomical system defined in Eqs. (\ref{12.9a}).
Thorne \cite{thorne} systematically neglected all surface terms in solution of the boundary-value problem of
gravitational field equations. However,
the effective stress-energy tensor $\tau^{\alpha\beta}$ falls off as distance from the isolated system
grows as $1/R^4$. For this reason,
the multipole moments defined in Eqs. (\ref{tron1}), (\ref{tron2}) are to be formally divergent. This
divergency can be completely eliminated if one makes use of more rigorous mathematical technique developed
by Blanchet and Damour \cite{bld}  for the mass multipole moments and used later on by Damour and
Iyer \cite{dyr2} to define the spin multipoles.  This technique is based on the theory of distributions
\cite{gelfsh} and consists in the replacement in Eqs. (\ref{tron1}), (\ref{tron2}) of the stress-energy
pseudo-tensor $\tau^{\alpha\beta}$ defined in the entire space with the effective source $\tau_c^{\alpha\beta}$
which has a compact support inside the region occupied by matter of the isolated system \cite{bld,dyr2}.
Blanchet and Damour proved \cite{bld} that formal integration by parts of the integrands of Thorne's multipole
moments (\ref{tron1}), (\ref{tron2}) with subsequent discarding of all surface terms recovers the multipole moments
derived by Blanchet and Damour by making use of the compact-support effective source $\tau_c^{\alpha\beta}$.
It effectively demonstrates that Thorne's post-Newtonian multipole moments are physically (and computationally)
meaningful provided that one takes care and operates only with compact-support terms in the integrands
of Eqs. (\ref{tron1}), (\ref{tron2}) after their rearrangement with the proper use of integration by parts
of the non-linear source of gravitational field $N^{\alpha\beta}$ given by Eqs. (\ref{ahv1})--(\ref{ahv3}).
This transformation was done by Blanchet and Damour \cite{bld} who extracted the non-divergent core of
Thorne's multipole moments. We shall use their results in this paper.

In the scalar-tensor theory of gravity
the multipolar series gets more involved because of the presence of the
scalar field. This brings about an additional set of multipole moments which
are intimately related with the multipolar decomposition of
the scalar field outside of the gravitating system. We emphasize that
definition of the multipole moments in the scalar-tensor theory of
gravity depends not only on the choice of the gauge conditions but also on the freedom of conformal transformation
of the metric tensor as was pointed out by Damour and Esposito-Far\'ese \cite{dames} who also derived
(in global coordinates) the set of multipole moments for an isolated astronomical system in two-parametric
class of scalar-tensor theories of gravity.
In this and subsequent sections we shall study the problem of the multipolar decomposition of gravitational
and scalar fields both of the whole N-body system and of each body comprising the system in the framework of the
scalar-tensor theory of gravity under discussion.
In this endeavor we shall follow the line of study outlined and elucidated in \cite{will,thorne,bld,dames}.
The multipole moments under discussion will include the sets of {\it active}, {\it conformal}
and {\it scalar} multipole moments. These three sets are constrained by one identity (see Eq. (\ref{13.31})).
Hence, only two of the sets are algebraically (and physically) independent. The multipole moments we shall
work with will be defined in different reference frames associated both with an isolated astronomical
system and with a single body (or sub-system of the bodies) comprising the isolated system. We call all
these post-Newtonian moments as Thorne-Blanchet-Damour multipoles after the names of the researchers who
strongly stimulated and structured this field by putting it on firm physical and rigorous mathematical bases.
Let us now consider the multipole
moments of the scalar-tensor theory of gravity in more detail.

\subsection{Thorne-Blanchet-Damour Active Multipole Moments}\label{tbz}

Let us introduce the metric tensor potentials
\br\label{mp1}
V&=&\frac12\left\{N+\epsilon^2\left[L+\frac{N^2}{2}+2(\beta-1)\varphi^2\right]\right\}\;,\\\label{mp2}
V^i&=&-\frac{L_i}{2(1+\gamma)}\;,
\er
which enter $g_{00}(t,{\bm x})$ and $g_{0i}(t,{\bm x})$ components of the metric tensor respectively.
Furthermore, throughout this chapter we shall put $N_i=0$ and assume that the spatial metric
component $H_{ij}$ is isotropic,
that is $H_{<ij>}=0$. Then,
the field equations for these potentials follow from Eqs. (\ref{11.33}), (\ref{11.31}) and  read
\br
\label{13.0}
  && \dAl V=-4\pi G\sigma\,,\\\nonumber\\
 &&\dAl V_i=-4\pi G\sigma^i\;,\label{13.1}
\er
where we have introduced the {\it active} mass density
\be                                                         \label{13.2}
   \sigma=\rho^*\left\{1+\epsilon^2\left[
       (\gamma+\frac12)\,v^2+\Pi+
       \gamma\,\frac{\pi^{kk}}
       {\rho^*}-\frac{H}{6}-
       (2\beta-\gamma-1)\varphi\right]\right\}\;,
\en
and the {\it active} current mass density
\be
  \sigma^i=\rho^*\,v^i\;.                                         \label{13.5}
\en
It is worthwhile to observe that in the global coordinates one has $H=6\gamma U(t,{\bm x})$ and $\varphi(t,{\bm x})=
U(t,{\bm x})$. Hence, the expression (\ref{13.2}) for the {\it active} mass density in these coordinates is
simplified and reduced to
\be                                                         \label{rru}
\sigma=\rho^*\left\{1+\epsilon^2\left[
       (\gamma+\frac12)\,v^2+\Pi+
       \gamma\,\frac{\pi^{kk}}
       {\rho^*}-
       (2\beta-1)U\right]\right\}\;.
\en

Solutions of Eqs. (\ref{13.0}) and (\ref{13.1}) are retarded wave potentials \cite{ll} determined up to the solution
of a homogeneous wave equation and satisfying the boundary conditions (\ref{12.1}) -- (\ref{12.2}).
Taking into account that potentials $V$ and $V_i$ are in fact components of the metric
tensor, solutions of Eqs. (\ref{13.0}) and (\ref{13.1}) can be written down as
\br
   V(t,\nnx)&=&G\int_{\cal D}\frac{                           \label{13.3}
   \sigma
   (t-\epsilon|{\bm x}-{\bm x'}|,\,{\bm x'})}{
   |{\bm x}-{\bm x'}|}\, d^3x'+c^2\, \xi^0_{\,\,,0}\;,
\\\nonumber\\
   V^i(t,\nnx)&=&G\int_{\cal D}\frac{                           \label{13.4}
   \sigma^i
   (t-\epsilon|{\bm x}-{\bm x'}|,\,{\bm x'})}{
   |{\bm x}-{\bm x'}|}\, d^3x'+\frac{c^3}{2(1+\gamma)}\Bigl[\xi^i_{\,\,,0}-\xi^0_{\,\,,i}\Bigr]\,,
\er where ${\cal D}$ designates a domain of
integration going over entire space, and the gauge functions $\xi^0$ and $\xi^i$
are solutions of the homogeneous wave equation. We notice that because the densities $\sigma$ and $\sigma^i$
vanish outside the bodies the integration in Eqs. (\ref{13.3}) and (\ref{13.4}) is performed only over the
volume occupied by matter of the bodies.

We take a special choice of the gauge functions as proposed in \cite{bld}
(the only difference is the factor $2(1+\gamma)$ instead of $4$ in \cite{bld}, coming from the field
equation for $g_{0i}$ component), namely
\br                                         \label{gf1}
\xi^0&=&2(1+\gamma)\epsilon^3G\sum\limits_{l=0}^{\infty}\frac{(-1)^l}{(l+1)!}\frac{2l+1}{2l+3}
   \left[\frac1r\int_{\cal D}\sigma^k(t-\epsilon r,\nnx')\,x'^{<kL>}
   \,d^3x'\right]_{,L}\;,\\\nonumber\\                   \label{gf2}
   \xi^i&=&0\;.
   \er
Such gauge transformation preserves the gauge conditions (\ref{11.3}) and also does not change the
post-Newtonian form of the scalar multipole moments which will be discussed in the next section.
Then one can show that potentials $V$ and $V^i$ can be expanded outside of the N-body system in a multipolar series
as follows \cite{bld}
\br
   V(t,\nnx)&=&G\sum\limits_{l=0}^{\infty}                 \label{13.9}
   \frac{(-1)^l}{l!}\left[\frac{I_{<L>}(t-\epsilon r)}{r}\right]_{,L}\,,
\\\nonumber\\
   V^i(t,\nnx)&=&G\sum\limits_{l=0}^{\infty}              \label{13.10}
   \frac{(-1)^l}{(l+1)!}\left\{\left[\frac{\dot{I}_{<iL>}
   (t-\epsilon r)}{r}\right]_{,L}
   \right.\\\nonumber&&\left.
  -\frac{l+1}{l+2}\varepsilon_{ipq}\left[\frac{S_{<pL>}
   (t-\epsilon r)}{r}\right]_{,qL}\right\}\,,
\er
where overdot denotes differentiation with respect to
time $t$. Eqs. (\ref{13.9}) and (\ref{13.10}) define
the {\it active} Thorne-Blanchet-Damour mass multipole moments, $I_{L}$, and the spin
moments, $S_{L}$, which can be expressed in the first PNA in terms of integrals over the N-body system's matter
as follows
\br
   I_{<L>}(t)&= &\int_{\cal D}\sigma
   (t,\nnx')x'^{<L>}\,d^3x'+
   \frac{\epsilon^2}{2(2l+3)}\left[\frac{d^2}{dt^2}                      \label{13.11}
   \int_{\cal D}\sigma
   (t,\nnx')x'^{<L>}x'^2\,d^3x'\right.\\\nonumber\\ &-&\left.
   4(1+\gamma)\,\frac{2l+1}{l+1}
   \frac{d}{dt}\int_{\cal D}\sigma^i(t,\nnx')x'^{<iL>}
   \,d^3x'\right],
  \\\nonumber\\\nonumber\\
   S_{<L>}(t)&=& \int_{\cal D}\varepsilon^{pq<a_l}\hat x'^{L-1>p}             \label{13.12}
   \sigma^q(t,\nnx')\, d^3x'\,.
\er As one can see the mass and current multipole moments of the
scalar-tensor theory define gravitational field of the metric tensor outside
of the N-body system as well as in general relativity \cite{thorne,bld}. When $\beta= \gamma=1$ these multipole
moments coincide with their general relativistic expressions \cite{bld}.
However, in order to complete the multipole decomposition of
gravitational field in the scalar-tensor theory one needs to
obtain a multipolar expansion of the scalar field as well.

\subsection{Thorne-Blanchet-Damour Scalar Multipole Moments}

In order to find out the post-Newtonian definitions of the
multipole moments of the scalar field we again shall use the same
technique as in \cite{thorne,bld}. We take Eq. (\ref{10.4}) and
write it down with the post-Newtonian accuracy by making use of a new (scalar)
potential
\be
\label{mp3}
\bar V=c^2\zeta+\frac{\epsilon^2}{2}\biggl[\eta-(\gamma-1)(\gamma-2)\biggr]\varphi^2\;.
\en
Then, Eq. (\ref{10.4}) assumes the form
\be
\label{13.13}
   \dAl\bar V=-4\pi G \bar{\sigma}\,,
\en
where the conventional notation $\eta\equiv 4\beta-\gamma-3$ for the Nordtvedt parameter \cite{will} has been
used and the {\it scalar} mass density $\bar{\sigma}$ is defined as
\be                                                         \label{13.14}
   \bar{\sigma}=(1-\gamma)\rho^*\left\{1-\epsilon^2\left[
       \frac12\,v^2-\Pi+\frac{\pi^{kk}}{\rho^*}+\frac{H}{6}\right]\right\}
       -\epsilon^2\Bigl[\eta+\gamma(\gamma-1)\Bigr]\rho^*\varphi\;.
\en

We can easily check out that in the global coordinates, where $H=6\gamma U(t,{\bm x})$
and $\varphi(t,{\bm x})=U(t,{\bm x})$,
the {\it scalar} mass density is simplified and is given by
\be                                                         \label{13.14aa}
   \bar{\sigma}=
   (1-\gamma)\rho^*\left[1-\epsilon^2\left(
    \frac12\,v^2-\Pi+\frac{\pi^{kk}}{\rho^*}\right)\right]-\epsilon^2\eta\rho^*U\;.
\en
Solution of Eq. (\ref{13.13}) is the retarded scalar potential
\be
   \bar{V}(t,\nnx)=G\int_{\cal D}\frac{                     \label{13.15}
   \bar{\sigma}
   (t-\epsilon|{\bm x}-{\bm x'}|,\,{\bm x'})}{
   |{\bm x}-{\bm x'}|}\, d^3x'\,.
\en
Multipolar decomposition of the potential (\ref{13.15}) has the
same form as in Eq. (\ref{13.9}) with the {\it scalar} mass
multipole moments defined as integrals over a volume of matter of the N-body system
\be
   \bar{I}_{<L>}(t)= \int_{\cal D}\bar{\sigma}
   (t,\nnx')x'^{<L>}\,d^3x'+
   \frac{\epsilon^2}{2(2l+3)}\frac{d^2}{dt^2}                      \label{13.17}
   \int_{\cal D}\bar{\sigma}
   (t,\nnx')x'^{<L>}x'^2\,d^3x'\,.
\en
We conclude that in the scalar-tensor theory of gravity the
multipolar decomposition of gravitational field requires
introduction of three sets of multipole moments -- the {\it
active} mass moments $I_L$, the {\it scalar} mass moments
$\bar{I}_L$, and the spin moments $S_L$. Neither the {\it
active} nor the {\it scalar} mass multipole moments alone lead to
the laws of conservation of energy, linear momentum, etc. of
an isolated system; only their linear combination does. This
linear combination of the multipole moments can be derived after
making conformal transformation of the metric tensor, solving the
Einstein equations for the conformal metric, and finding its
multipolar decomposition in the similar way as it was done in section \ref{tbz}.

\subsection{Thorne-Blanchet-Damour Conformal Multipole Moments}

Let us now define the {\it conformal} metric potential
\be                         \label{mp4}
\tilde V=\frac{1}{1+\gamma}\left[\tilde N+\epsilon^2\left(\tilde L+\frac{\tilde N^2}{2}\right)\right]\;.
\en
The conformal field equations (\ref{13.19}) in the quasi-harmonic gauge of Nutku (\ref{13.20})
yield
\br\label{13.21}
   \dAl\tilde V&=&-4\pi G \tilde\sigma\,,
\er
where we have introduced a {\it conformal} mass density
\be
   \tilde\sigma=\rho^*\left\{1+\epsilon^2\left[
       \frac32\,v^2+\Pi+
       \frac{\pi^{kk}}
       {\rho^*}-\frac{H}{6}-(1-\gamma)\varphi\right]\right\}\;,           \label{13.24}
\en
which has been calculated directly from Eq. (\ref{13.19}) by making use of the definition of the conformal metric
(\ref{13.18}) and the post-Newtonian
expansions of corresponding quantities described in section \ref{bpri}. Remembering that in the global coordinates
$H=6\gamma U(t,{\bm x})$ and $\varphi(t,{\bm x})=U(t,{\bm x})$ one can simplify expression for the {\it conformal}
mass density which assumes the form
\be
   \tilde\sigma=\rho^*\left[1+\epsilon^2\left(
       \frac32\,v^2+\Pi+
       \frac{\pi^{kk}}
       {\rho^*}-U\right)\right]\;.                       \label{13.24aa}
\en
This equation
coincides precisely  with the post-Newtonian mass density as it is defined in general
relativity (see \cite{will,fock} and \cite{dames} for more detail).
The {\it conformal} current density $\tilde\sigma^i$ is defined in the approximation under consideration by the
same equation as Eq. (\ref{13.5}), that is $\tilde\sigma^i=\sigma^i$.
The field equation for the conformal vector potential $\tilde V^i$ has the form (\ref{13.1}),
therefore $\tilde V^i=V^i$.

Solution of Eq. (\ref{13.21})
gives the retarded {\it conformal} potential
\be                                                     \label{13.29}
  \tilde V(t,\nnx)=G\int_{\cal D}\frac{
  \tilde \sigma
   (t-\epsilon|{\bm x}-{\bm x'}|,\,{\bm x'})}
   {|{\bm x}-{\bm x'}|}\, d^3x'\,.
\en
Multipolar expansion of {\it conformal} potentials $\tilde V$ and $\tilde V^i$ is done in the
same way as it was done previously in section \ref{tbz}. It turns out that the conformal spin moments
coincide with the active spin moments (\ref{13.12}), and the expansion of the potential $\tilde V(t,\nnx)$
acquires the same form as that given in Eq. (\ref{13.9}) but with all
active multipole moments replaced with the conformal multipoles, $\tilde I_{<L>}$, defined as follows
\br\label{13.30}
  \tilde I_{<L>}(t)&=& \int_{\cal D}\tilde\sigma
   (t,\nnx')x'^{<L>}d^3x'+
   \frac{\epsilon^2}{2(2l+3)}\left(\frac{d^2}{dt^2}
   \int_{\cal D}\tilde\sigma
   (t,\nnx')x'^{<L>}x'^2d^3x'\right.
   \\\nonumber \\\nonumber &&\left.
   -8\,\frac{(2l+1)}{l+1}
   \frac{d}{dt}\int_{\cal D}\sigma^i(t,\nnx')x'^{<iL>}d^3x'\right)\;.
\er
These {\it conformal} mass multipole moments coincide exactly
with those introduced by Blanchet and Damour \cite{bld} who also proved (see appendix A in \cite{bld}) that
their definition coincides precisely
(after formal discarding of all surface integrals which have no physical meaning) with the mass multipole
moments introduced originally in the first  post-Newtonian approximation in general relativity by Thorne \cite{thorne}.

There is a simple algebraic relationship between the three mass multipole moments,
$I_L$, $\bar{I}_L$ and $\tilde{I}_L$ in the global frame. Specifically, one has
\be
\label{13.31}
   I_{<L>}=\frac{1+\gamma}{2}\tilde I_{<L>}+\frac12\bar{I}_{<L>}\,.
\en
We shall show later in section \ref{mdloc} that relationship (\ref{13.31}) between the multipole moments obtained
in the global coordinates
for the case of an isolated  astronomical N-body system preserves its form in the local coordinates for each
gravitating body (a sub-system of the
bodies) as well.

\subsection{Post-Newtonian Conservation Laws}\label{pncs}

It is crucial for the following analysis to discuss the laws of
conservation for an isolated astronomical system in the framework
of the scalar-tensor theory of gravity. These laws will allow us to
formulate the post-Newtonian definitions of mass, the center of mass,
the linear and the angular momenta for the isolated system which are
used in derivation of equations of motion of the bodies
comprising the system. In order to derive the laws of conservation
we shall employ a general relativistic approach developed in
\cite{ll} and extended to the Brans-Dicke theory by Nutku \cite{n2}.

To this end it is convenient to recast the field equations
(\ref{10.2}) to the form
\br
 \Theta^{\mu\nu} &\equiv& (-g)\frac{\phi}{\phi_0}\Bigl[c^2 T^{\mu\nu}+  t^{\mu\nu}\Bigr]
   =
    \frac{c^4}{16\pi\phi_0}\Bigl[(-g)\phi^2                      \label{13.32}
   (g^{\mu\nu}g^{\alpha\beta}-
    g^{\mu\alpha}g^{\nu\beta})\Bigr]_{,\alpha\beta}\,,
\er where $t^{\mu\nu}$ is an analog of the Landau-Lifshitz
pseudo-tensor of the gravitational field in the scalar-tensor
theory of gravity. This pseudotensor is defined by the
equation
\br
\label{13.33}
    t^{\mu\nu}&= & \frac{c^4}{16\pi}\,\frac{\phi^3}{\phi_0^2}\,
    \tilde\tau^{\mu\nu}_{LL} +\frac{c^4}{16\pi}
    \frac{2\theta(\phi)+3}{\phi}
    \Bigl(\phi^{,\mu}\phi^{,\nu}-
    \frac{1}{2}g^{\mu\nu}\phi_{,\lambda}\phi^{,\lambda}
    \Bigr)\;,
\er where $\tilde{\tau}^{\mu\nu}_{LL}$ is the (standard) Landau-Lifshitz
pseudotensor \cite{ll} expressed in terms of the conformal metric
$\tilde{g}_{\alpha\beta}$ and its derivatives.

The conservation laws are now obtained from Eq. (\ref{13.32})
\be
  \Theta^{\mu\nu}{}{}_{,\nu}\equiv                           \label{13.34}
  \Bigl[(-g)\frac{\phi}{\phi_0}(c^2 T^{\mu\nu}+t^{\mu\nu})
  \Bigr]_{,\nu}=0\,.
\en
They are a direct consequence of anti-symmetry of the right
side of Eq. (\ref{13.32}) with respect to the upper indices $\nu$ and
$\alpha$. In what follows, we concentrate on the laws of conservation in the first
post-Newtonian approximation only. Hence, we neglect the energy,
linear and angular momenta taken away from the system by
gravitational waves (see \cite{dames} where this problem has been tackled). For this reason, the conserved
mass $\mathbb{M}$,
the linear momentum $\mathbb{P}^i$, and spin  $\mathbb{S}^i$ of the isolated gravitating N-body system are
defined as
\br
  \mathbb{M}&=&\epsilon^2\int_{\cal D}\Theta^{00}\,d^3x\;,                       \label{13.35}
\\\nonumber\\
\mathbb{P}^i&=&\epsilon\int_{\cal D}\Theta^{0i}\,d^3x\,,                      \label{13.36}
\\\nonumber\\
\mathbb{S}^i&=&\epsilon\int_{\cal D}\varepsilon^i_{\;jk}w^j\Theta^{0k}\,d^3x\,.                      \label{13.361}
\er
In these definitions integration is performed over the whole space. Let us remark that the integrals
are finite since in the first PNA $\Theta^{00}$ and $\Theta^{0i}$ are of $O(r^{-4})$ for large $r$.
Moreover, in this approximation the domain of integration can be reduced to the volume of the bodies
comprising the system -- observe that in (\ref{13.37}) -- (\ref{13.38}) the functions under the integrals
are compactly supported. Taking into account the asymptotic behavior of $\Theta^{00}$
one can prove that the linear
momentum $\mathbb{P}^i$ can be represented as the time derivative of the function
\be
\mathbb{D}^i=\epsilon^2\int_{\cal D}\Theta^{00}x^i\,d^3x\;,                                \label{13.35a}
\en
which is
interpreted as the integral of the center of mass. Hence,
\be\label{13.35b} \mathbb{D}^i(t)=\mathbb{P}^i\,t+\mathbb{K}^i\;,\en where $\mathbb{K}^i$ is a constant
vector defining displacement of the barycenter of the N-body system
from the origin of the global coordinate frame. One can chose
$\mathbb{K}^i=0$ and $\mathbb{P}^i=0$. In such case $\mathbb{D}^i=0$, and the center of mass
of the N-body system will always coincide with the origin of the
global reference frame. Such global reference frame is called barycentric. It is used in description of
ephemerides of the solar system bodies, navigation of spacecrafts in deep space and reduction of astronomical
observations of various types.

Direct calculations of the pseudotensor (\ref{13.33}) with
subsequent comparison with the conformal multipole moments
(\ref{13.30}) reveal that for the isolated system the post-Newtonian conserved quantities are
\br
   \mathbb{M}&\equiv&\tilde I =\int_{\cal D}\rho^*\left[
   1+\epsilon^2\left(\Pi+\frac{v^2}{2}-\frac{U}{2}\right)\right]\,d^3x    \label{13.37}
   +O(\epsilon^4)\,,
\\\nonumber\\                                                               \label{13.37a}
\mathbb{D}^i&\equiv&\tilde I^i=\int_{\cal D}\rho^*x^i\left[
   1+\epsilon^2\left(\Pi+\frac{v^2}{2}-\frac{U}{2}\right)\right]\,d^3x
   +O(\epsilon^4)\,,
\\\nonumber\\
   \mathbb{P}^i&=&\int_{\cal D} \Biggl\{
   \rho^*v^i \left[1+\epsilon^2\left(\Pi+                    \label{13.38}
   \frac{v^2}{2}-\frac{U}{2}\right)\right]+
   \epsilon^2\pi^{ik}v^k-\frac{\epsilon^2}{2}
   \rho^*W^i\Biggr\}\,d^3x+O(\epsilon^4)\,,
\er
where by definition
\be
   W^i(t,\nnx)=G\int_{\cal D}\frac{                                 \label{13.39}
   \rho^*(t,\nnx'){\bm v'}\cdot(\nnx-\nnx')(x^i-x'^i)}{|{\bm x}-{\bm x}'|^3}\, d^3x'\,, \en
and the integration is
performed over the hypersurface of constant global coordinate time $t$.  It is evident from Eqs.
(\ref{13.38}) and (\ref{13.39}) that it is the {\it conformal} moments, $\tilde I$ and $\tilde I^i$, which
define the conserved mass $\mathbb{M}$ and linear momentum  $\mathbb{D}^i$ of the N-body system. The {\it active}
monopole and dipole moments defined by Eq. (\ref{13.11}) for $l=0,1$ are not consistent with the laws of
conservation and, hence, can not serve to define the conserved quantities. We fix position of the center
of mass (barycenter) of the N-body
system in the global coordinates by equating {\it conformal} dipole moment of the system to zero,
that is $\tilde I^i=0$.

Now we are prepared to begin construction of a local coordinate
system in the vicinity of a gravitating body or a sub-system of
bodies which are members of the entire N-body system. For concreteness and for the sake of simplicity
we shall focus on the construction of the local coordinate
system around one body (Earth, planet, etc.).

\section{Local  PPN Coordinate System}\label{lcs}

\subsection{Dynamic and Kinematic Properties of the Local Coordinates}

Local coordinate system (local coordinates) is constructed in
the vicinity of each body comprising the N-body system
\footnote{Precise definition of body's center of mass will be given in subsequent sections along with derivation
of its equations of motion.}. Thus, in principle, N local coordinate systems $w^\alpha$ must be introduced
in addition to one global coordinate system $x^\alpha$ (see Fig. \ref{fig1}). In the case of the N-body
system which is divided on sub-systems of bodies the number of the local coordinates increases in accordance
with the underlying hierarchic structure of the N-body system (see Fig. \ref{fig1a}).
The principles of construction of the local coordinates are the same for any weakly gravitating body
(a sub-system of bodies). For this reason, it is sufficient to work out description of only one local
coordinate system, $w^\alpha=(cu, {\bm w})$, as the other local coordinate charts must have a similar
structure \cite{kop1991}. For practical applications in the solar system the most important local coordinates
are associated with the Earth and they are called geocentric coordinates. Local coordinates are not
asymptotically Minkowskian far away from the body because the gravitational field of the body under
consideration must smoothly match with the gravitational field of external bodies.

We assume that each body consists of matter which admits continuous
distribution of mass density, anisotropic stresses and internal velocity field. If one had ``turned off"
gravitational field of all external bodies (Moon, Sun, planets) but the gravitational field of the body
under consideration (Earth), it would be described by a set of the (internal) multipole moments defined
by equations given in previous
section. However, we can not neglect gravitational field of the external bodies if one wants to take into
account classic \cite{mtw,cltide} and relativistic effects associated with tides \cite{sokx,masha}. The
tidal deformation of the
body will be comprehensively large, for example, at the latest stage of coalescence of neutron stars in
binary systems emitting gravitational radiation and is to be taken into account in calculations of the
templates of gravitational waves being emitted by such systems. We also know that already in the Newtonian
limit this external
gravitational field reveals itself in the vicinity of the Earth
as a classical tidal force \cite{mtw}. Gravitational potential of the tidal force is represented
as a Taylor series with respect to the local geocentric
coordinates with time-dependent coefficients which are called external (tidal)
multipole moments \cite{th}. This series usually starts in the Newtonian approximation from the second order
(quadratic) term because the monopole and dipole external multipole moments are not physically associated with
the tidal force. In general relativity this monopole-dipole effacing property
of the external gravitational field is retained in the post-Newtonian approximation as a consequence of the Einstein
principle of equivalence (EEP)
\cite{will,ll,mtw}. In particular, EEP suggests that it is possible to chose such (local) coordinates that all
first derivatives of the metric tensor (i.e., the
Christoffel symbols) vanish along a geodesic world line of a freely falling particle \cite{niz}. This is equivalent
to making a suitable coordinate transformation on the space-time manifold from the global to local frame
\cite{dfn,schouten}.
In general relativity this property of EEP is also valid for a self-gravitating body moving in external
gravitational field of other bodies. The original proof was given in \cite{th,das1,das2} and elaborated
on later in a series of papers by other researchers \cite{ashb1,ashb2,kop1988,bk-nc,dsx1991,dsx1992}.

As contrasted with general relativity the scalar-tensor theory of
gravity has a scalar  (helicity-0) component of the gravitational field which
can not be eliminated by a coordinate transformation to the local frame of the body
being in a free fall. This is
because the scalar field does not change its numerical value under pointwise coordinate transformations
and can not be eliminated if it has a non-zero value on space-time manifold. It
means that scalar fields do not obey the principle
of equivalence and the gravitational field in the scalar-tensor theory can
not be reduced in the local coordinate system to the tidal field only. In particular,
this was the reason why Einstein
had rejected a theory of gravity based exceptionally on a scalar field (for more detail see
\cite{will,mtw}).

This argument makes it clear that in
order to incorporate a local coordinate system to the standard
PPN formalism \cite{nord1} --\cite{will} one needs
to know the nature of the fundamental fields (scalar, vector, tensor,
spinor, etc.) present in the theory because these fields have different behavior
under coordinate transformations. To construct a local coordinate system, solution of the field equations
for all fundamental fields must be found directly in the local coordinate system. Then, this solution must
be matched to the solution of the same equations in the global coordinates and the transformation
laws of the additional fields must be used along with the transformation law
of the metric tensor in order to find relativistic space-time transformation between the global and local
coordinate systems. Nonetheless, one has to keep in mind that the scalar field is not observed directly but
is organically incorporated to the metric tensor which obeys to EEP. It means that the scalar field and its
first derivative at each point of the manifold can be absorbed in the metric tensor and its first derivatives.
Thus, the metric tensor in the origin of the local coordinates can be reduced to the Minkowski metric as long
as the body's gravitational field is not to be a matter of concern.

We demand that the origin of the local coordinates
coincides with the body's center of mass at any instant of time.
This requires a precise definition of the center of mass of each body with respect to its local coordinates.
But when one takes into account the post-Newtonian corrections, the notion of the body's
center of mass becomes ambiguous because it can be chosen in several different ways depending on what kind of
definition of the internal dipole moment of the body in the multipolar
expansion of the local metric tensor is chosen. We have proven by straightforward calculations that it is the
{\it conformal} dipole moment
(\ref{13.37a}) which gives a physically correct definition of the body's center of mass because only this moment
allows to derive equations of translational motion of the body which does not contain self-accelerated terms
violating the Newton's third law of action-counteraction. This property of the {\it conformal} moment is closely
related to its conservation for an isolated
system of N bodies as demonstrated in section \ref{pncs}.

In general, the body (Earth) as a part of the N-body system is not isolated
and interacts gravitationally with other bodies (Moon, Sun, etc.). For this reason, the second
and higher order time derivatives of the conformal dipole moment of the body
are not equal to zero by themselves. It means that there is a local force exerted on the body by external
gravitational field which prevents its linear momentum (the first time derivative of the body's dipole moment)
to conserve. Nevertheless, it is possible
to prove that all time derivatives of the body's dipole moment can be kept equal to zero if one chose the
origin of the local coordinates to move not along a geodesic world line. This goal is achieved by
making use of a specific choice of external dipole moment in the multipolar
expansion of the homogeneous solution of the gravitational field equations (see section \ref{xru}).
The correct choice of the body's center of mass allows us to
eliminate the ill-behaved coordinate-dependent terms in the
equations of motion of the body and facilitates discussion of the strong equivalence principle's violation
(the Nordtvedt effect) for extended bodies.

We admit that the local coordinates can be, in general, dynamically rotating. It
means that translational equations of motion of a test particle written down in the
local coordinates can include the Coriolis and centrifugal forces. If one excludes the
dynamical rotation of the local coordinates, their spatial axes will slowly rotate in the
kinematic sense with respect to the spatial axes of the global coordinates
\cite{iau1997,iau2000}. This effect is called a geodetic precession and it obeys the law of parallel transport of
vectors on curved space-time manifold \cite{mtw}. Nowadays,
the IAU recommends to use a kinematically-nonrotating geocentric system which
spatial axes are anchored to distant quasars used as reference points of the
international celestial reference system (ICRS) (see \cite{mignar}
for more detail). The metric tensor of the kinematically-nonrotating geocentric coordinates has an external
dipole moment in
$\hat{g}_{0i}(u,{\bm w})$ component of the geocentric metric tensor describing the dynamical rotation of the
spatial axes of the geocentric coordinates. This term would be zero if the geocentric coordinates were chosen
to be dynamically nonrotating. The angular velocity of the dynamical rotation is equal to that of the geodetic
precession and is fixed by the corresponding IAU resolution.
At this step of development of our formalism we shall not specify the angular velocity
of the dynamical rotation in order to keep the formalism as general as possible.

\subsection{The Metric Tensor and the Scalar Field in the Local Coordinates}\label{qop}

We denote the local (for example, geocentric) coordinates by
$w^\alpha=(w^0,w^i)=(cu, w^i)$ where $u$ stands for the local
coordinate time. All quantities related to the (central) body around which the local coordinate frame is
constructed will be labelled by
subindex B standing for ``body". We are looking for the solution of the field
equations (\ref{11.29}) -- (\ref{11.33}) inside a world tube containing the world line of the
body's center of mass and spreading up to the nearest external body, so
that the only source of matter inside the region covered by the local frame is the matter of the central body.
Thus, the right side of equations (\ref{11.29})
-- (\ref{11.32}) contains the energy-momentum tensor of the body's
matter only. Spatial domain of
applicability of the local coordinates can be extended after finding the space-time transformation
from local to global coordinates \cite{bk-nc}.

Solution of the differential equations
(\ref{11.29}) -- (\ref{11.33}) is a linear combination of general
solution of the homogeneous equation and a particular
solution of the inhomogeneous equation. For example, solution for a
scalar field in the local coordinates is written as
\be
\nxi=\,\nnbb{B}{u}{w}+\nnbb{E}{u}{w}\;,                                 \label{1.1}
\en
whereas the metric tensor, $\hat{g}_{\mu\nu}(u,{\bm w})=\eta_{\mu\nu}+\hat{h}_{\mu\nu}(u,{\bm w})$,
is given in the form
\be                                                                 \label{1.2}
\nag{\mu\nu}=\nnfa{B}{\mu}{\nu}{u}{w}+
\nnfa{E}{\mu}{\nu}{u}{w}+\nnfa{C}{\mu}{\nu}{u}{w}\;,
\en
where terms with sub-index B refer to the central body (Earth) and describe
the (internal) solution of the inhomogeneous equations, terms with
sub-index E refer to the external bodies (Moon, Sun, etc.) and describe the
(external) solution of the homogeneous equations, and terms with
sub-index C (which stands for coupling) arise because of the non-linearity of the gravity field equations
for the metric tensor. One notices that in the first post-Newtonian approximation the coupling
terms appear only in $\hat g_{00}(u,\nnw)$ component of the metric tensor.
We do not impose any other specific limitations on the structure of the
metric tensor in local coordinates. All
information about its structure can be obtained from
the solution of the field equations (\ref{11.29}) -- (\ref{11.32}). We draw attention of the reader that we
put a hat over all quantities referred to the local coordinates $w^\alpha$. This is because functional
dependence of one and the same quantity looks different in different coordinates. For example, for any
scalar function $F(x)$ and coordinate transformation $x=x(w)$ one has $F(x)=F[x(w)]\equiv\hat{F}(w)$ while
$F(w)$ differs from $F(x)$ \cite{dfn,schouten}.

\subsubsection{The Scalar Field: Internal and External Solutions}

Eq. (\ref{11.29}) gives internal, $\nnbb{B}{u}{w}$, and external, $\nnbb{E}{u}{w}$, solutions for the scalar
field in the following form
\br                                                             \label{1.7}
 \nnbb{B}{u}{w}&=&\nnc\;,\\\nonumber\\
  \nnbb{E}{u}{w}&=& \nn{l=0}\frac1{l!} P_{L}w^L\;.        \label{1.7a}
\er
Here $P_{L}\equiv P_{L}(u)$ are external STF multipole moments in the multipolar decomposition of the scalar
field generated by the bodies which are external with respect to the central body. These external moments
are functions of the local time $u$ only. The internal solution $\nnbb{B}{u}{w}$ describes the scalar field which
is generated by the central body only.

\subsubsection{The Metric Tensor: Internal Solution }

The boundary conditions imposed on the internal solution for the metric tensor
are identical with those given in Eqs. (\ref{12.1}) -- (\ref{12.2}). For
this reason the internal solution for the metric tensor has a form which is similar with that obtained
in the global coordinates where all quantities must be referred now only to the central
body. We obtain
\br
\label{1.8}
\hat N^{{\sss(B)}}(u,\nnw)&=&
2\nnc\;,
\\\nonumber\\                \label{1.8aa}
\hat L^{{\sss(B)}}(u,\nnw)&=&2\hat \Phi^{{\sss(B)}}(u,\nnw)-
    2\beta\left[\nnc\right]^2-c^2\hat{\chi}^{{\sss(B)}}_{,00}(u,\nnw)\;,
    \\\nonumber\\                                                           \label{1.9}
\hat L_i^{{\sss(B)}}(u,\nnw)&=& -2(1+\gamma)
          \hat U_i^{{\sss(B)}}(u,\nnw)\;,
 \\\nonumber\\                                                              \label{1.10}
\hat H_{ij}^{{\sss(B)}}(u,\nnw)&=&2\gamma\delta_{ij}\nnc\;, \er
where all gravitational potentials of the central body are taken over the
volume of the body's matter defined as a cross-section of the body's world tube with the hypersurface of
constant local coordinate time $u$. Specifically, one has \br
  \nnc&=&\;\;G\nnhh{B}{\rho^*}\;,                                    \label{1.11}
  \\\nonumber\\
  \hat{U}_i^{{\sss(B)}}(u,\nnw)&=&\;\;             \label{1.12}
           G\nnhh{B}{\rho^*\nu^i}\;,
  \\\nonumber\\
\label{1.12a}
 \hat\Phi^{{\sss(B)}}(u,\nnw)&=&
(\gamma+\frac12)\nngk{1}+
    (1-2\beta)\nngk{2}
    \\\nonumber&&
    +\nngk{3} +\gamma\nngk{4},
\er
where
\br
\nngk{1}&=&\;\;G\nnhh{B}{\rho^*\nu^2}\;,                           \label{1.14}
  \\\nonumber\\
  \nngk{2}&=&\;\;G\nnhh{B}{\rho^*\hat U^{{\sss(B)}}}\;,     \label{1.15}
  \\\nonumber\\
  \nngk{3}&=&\;\;G\nnhh{B}{\rho^*\Pi}\;,                             \label{1.16}
  \\\nonumber\\
  \nngk{4}&=&\;\;G\nnhh{B}{\pi^{kk}}\;,                           \label{1.17}
  \\\nonumber\\
\hat{\chi}^{{\sss(B)}}(u,\nnw)&=&            \label{1.13}
          -\,G\hat{\rm\bf I}_1^{{\sss(B)}}\{\rho^*\}\;,
\er
the symbol $\nu^i=dw^i/du$ is the velocity of the body's matter with respect to the origin of the local coordinates,
and we have introduced a special notation
\br
  \hat{\rm\bf I}_n^{{\sss(B)}}\{f\}(u,\nnw)&=&\,  \label{1.18}
           \int\limits_{V_{\sss B}}f(u,{\bm w}')
           |{\bm w}-{\bm w}'|^n\,d^3w'\;,
\er
for integrals over the body's volume. We emphasize once again that the integrand of $\hat{\rm\bf I}_n^{{\sss(B)}}
\{f\}(u,\nnw)$ is a function which is taken over
the hypersurface of constant time $u$.

The local metric given by Eqs. (\ref{1.7}), (\ref{1.8}) -- (\ref{1.10})
must obey the gauge condition (\ref{11.3}) which yields
\be
         \frac{1}{c}\frac{\partial\hat U^{\sss(B)}}{\partial u}+               \label{1.19}
         \frac{\partial\hat U_{k}^{\sss(B)}}{\partial w^k}=O(\epsilon^2)\,.  \en
This is the only gauge condition which can be imposed on the local metric in the first post-Newtonian approximation.
We note that Eq. (\ref{1.19}) is satisfied due to the validity of the equation of continuity (\ref{11.20}).

\subsubsection{The Metric Tensor: External Solution }\label{mtex}

Solution of the homogeneous field equations for the metric tensor given in this section is based on and extends
the multipolar
formalism for description of vacuum gravitational fields developed
in \cite{th,suen}. Brief introduction to this formalism is given
in Appendix \ref{ap1}. Boundary conditions imposed on the external
solution must ensure its convergency on the world line of
the origin of the local coordinates where ${\bm w}=0$. However, the external solution for the metric tensor
diverges as the radial distance $r=|{\bm w}|$ from the
origin of the local coordinates grows. This is because of the gravitational field of external bodies which
does not asymptotically vanish in the local coordinates for large $r$ \cite{th,suen}.

Explicit form of the external solution for the linearized metric tensor perturbation in local coordinates
is given by
\br
\label{1.8a}
\hat N^{{\sss(E)}}(u,\nnw)&=& 2\nn{l=0}
                     \frac{1}{l!}Q_Lw^L+\Omega^2 w^2-\Omega^p\Omega^q w^p w^q  \;,
    \\\nonumber\\\nonumber\\                                            \label{1.8aaa}
\hat N_i^{{\sss(E)}}(u,\nnw)&=&{\mathcal V}_i+\varepsilon_{ipq}\Omega_pw^q\;,
\\\nonumber\\\nonumber\\                                                \label{1.9a}
\hat L_i^{{\sss(E)}}(u,\nnw)&=&\sum\limits_{l=1}^{\infty}
          \frac{1}{l!}\varepsilon_{ipq}C_{pL-1}w^{<qL-1>}+\sum\limits_{l=0}^{\infty}\frac{1}{l!}Z_{iL}w^{L}+
          \sum\limits_{l=0}^{\infty}\frac{1}{l!}
           S_{L}w^{<iL>}\;,
 \\\nonumber\\\nonumber\\                                               \label{1.10a}
\hat H_{ij}^{{\sss(E)}}(u,\nnw)&=&
          2\delta_{ij}\sum\limits_{l=0}^{\infty}
          \frac{1}{l!}Y_{L}w^{L}+\sum\limits_{l=0}^{\infty}
          \frac{1}{l!}B_{L}w^{<ijL>}+\frac13\left(\delta_{ij}\Omega^2-\Omega^i\Omega^j\right)w^2
                     \\\nonumber
          &&+\sum\limits_{l=1}^{\infty}\frac{1}{l!}
          \Biggl(
          D_{iL-1}w^{<jL-1>}+\varepsilon_{ipq}E_{pL-1}w^{<jqL-1>}
          \Biggr)
          ^{{\rm Sym}(ij)}                            \\\nonumber&&
          +\sum\limits_{l=2}^{\infty}\frac{1}{l!}
          \Biggl(
          F_{ijL-2}w^{L-2}+\varepsilon_{pq(i}
          G_{j)pL-2}w^{<qL-2>}
          \Biggr)\;,                       \er
where $\Omega_i$ is the angular velocity of kinematic rotation of the local frame with respect to the global
coordinates, ${\mathcal V}^i$ is the velocity of the local frame with respect to the local frame moving along
geodesic world line (see below), and either symbol ``${\rm Sym}(ij)$" or the round brackets around
indices denote symmetry with respect to the indices, for instance,
$[T_{ijL}]^{{\rm Sym}(ij)}\equiv T_{(ij)L}=(1/2)[T_{ijL}+T_{jiL}]$. In (\ref{1.8a}) -- (\ref{1.10a})
we keep only the terms of $O(\Omega)$ and $(\Omega^2)$ which are relevant for the discussion of Newtonian
geodesics.
It is worth to notice that external
solutions for the metric tensor in local coordinates contain monopole terms $Q$ and $Y$. Term $Q$ defines
the unit of measurement of the coordinate time $u$ at the origin of the local frame and $Y$ defines the
unit of measurement of spatial distances with respect to the international system of units. Both these
terms could be equated to zero from the very beginning but we prefer to keep them in our equations for generality.
This is because the IAU resolutions \cite{iau2000} explicitly introduce the non-zero values of $Q$ and $Y$
and this is why we are interested in the impact of these functions on the PPN theory of reference frames
\footnote{Functions $Q=Y=L_C=1.48082686741\times 10^{-8}\pm (2\times 10^{-17}$ \cite{irf,fuk2}
(see also \cite{bk-cm,brs} for detailed theoretical review on the relativistic time scales in the solar system.}.

In order to understand physical meaning of various components of the external solution for the metric tensor
in the local coordinates it is instructive to write down the Newtonian equation of motion of a test particle
falling freely in the field defined by the external metric. This equation is a geodesic world line so that
after calculation of the Christoffel symbols one has particle's acceleration
\br                                                               \label{bcn}
\frac{d^2w^i}{du^2}&=&Q_i-\dot{\cal
V}_i-2\varepsilon_{ijk}\Omega^j
{\nu}^k-\varepsilon_{ijk}\dot\Omega^jw^k+\left(\Omega^2\delta_{ij}-\Omega_i\Omega_j\right)w^j\\\nonumber&&+Q_{ij}w^j+
\sum_{l=2}^\infty\frac{1}{!} Q_{iL} w^L+O\left(\epsilon^2\right)\;,
\er
where ${\nu}^i\equiv dw^i/du$ and we have neglected the post-Newtonian corrections. First two terms in the right
side of this equation, $Q_i-\dot{\cal V}_i$, describe kinematic acceleration of the particle with respect to the
coordinate system moving along geodesic. The third term, $2\varepsilon_{ijk}\Omega^j\nu^k$, in the right side of
Eq. (\ref{bcn}) is the famous Coriolis acceleration \cite{l-l-mech} caused by motion of the particle and rotation
of spatial axes of the local frame with angular velocity $\Omega^i$. The forth term, $\varepsilon_{ijk}\dot\Omega^jw^k$,
in the right side of Eq. (\ref{bcn}) is acceleration due to the non-uniform rotation of the local frame.
The fifth term, $\left(\Omega^2\delta_{ij}-\Omega_i\Omega_j\right)w^j$, describes a centrifugal acceleration
of the particle. The sixth term, $Q_{<ij>}w^j$, is a quadrupole tidal acceleration due to the presence
of external gravitational field from other bodies besides the central one. Last term in the right side
of Eq. (\ref{bcn}) is the tidal acceleration due to the higher order multipoles of the external gravitational
field of other bodies. It is interesting to note that the centrifugal and the quadrupole tidal accelerations
have similar structure. The difference, however, is that the matrix of the centrifugal acceleration,
$\Omega^2\delta_{ij}-\Omega_i\Omega_j$,  is not trace-free in contrast to the tidal matrix, $Q_{ij}$.
However, the trace-free part of $\Omega^2\delta_{ij}-\Omega_i\Omega_j$ can be singled out and absorbed to the
definition of $Q_{ij}$.

It is convenient to construct the external part of the metric tensor in such a way that makes it Minkowskian
(orthogonal) at the origin of the local coordinates. This can be achieved if one chooses function ${\cal V}^i=0$.
This condition also allows us to give a unique interpretation of the dipole term $Q_i$ as equal to the inertial
force per unit mass exerted on the free falling particle due to the accelerated motion of the local frame under
consideration with respect to the geodesic world line. In other words, the metric tensor with ${\cal V}^i=0$
and $Q_i\not= 0$ specifies a local coordinate system such that its origin moves with acceleration $Q_i$ with
respect to a geodesic world-line defined on the background space-time which is determined exclusively by the
external part of the
 metric tensor. We also notice that the dipole term $Z_i$ in Eq. (\ref{1.9a}) is just a post-Newtonian
 correction to ${\cal V}^i$ and would also destroy orthogonality of spatial axes of the local frame at
 its origin. Thus, in addition to the condition, ${\cal V}^i=0$, we also demand, $Z^i=0$. Therefore,
 any post-Newtonian corrections to the equations of motion of the origin of the local coordinates are
 hidden in the inertial acceleration $Q_i$ in Eq. (\ref{1.8a}).

A set of eleven external STF multipole moments
$P_{L}$, $Q_{L}$, $C_{L}$, $Z_{L}$, $S_{L}$, $Y_{L}$, $B_{L}$, $D_{L}$, $E_{L}$, $G_{L}$
(we omit $<\, >$ for simplicity, i.e. $P_L\equiv P_{<L>}$, etc.)
is defined on the world line
of the origin of the local coordinates so that these multipoles are functions of the local coordinate time $u$ only.
Furthermore, the external multipole moments are symmetric and
trace-free (STF) objects with respect to any of two indices, and they are transformed
as tensors with respect to linear coordinate transformations. In what follows, we
shall assume that the angular velocity of rotation of the local frame, $\Omega^i$, is so small that the
metric tensor
component $\epsilon\hat N_i^{{\sss(E)}}$ is comparable with that
$\epsilon^3\hat L_i^{{\sss(E)}}$. For this reason, we shall
neglect all terms which are either quadratic with respect to $\hat
N_i^{{\sss(E)}}$ or are products of $\hat N_i^{{\sss(E)}}$ with one of each of the other components,
$\hat N^{{\sss(E)}}$, $\hat L_i^{{\sss(E)}}$ or $\hat
H_{ij}^{{\sss(E)}}$. Only linear with respect to $\hat
N_i^{{\sss(E)}}$ terms and their first derivatives will be
retained in our calculations.

Imposing the gauge conditions (\ref{11.3}) on the metric tensor given by Eqs. (\ref{1.8a}) -- (\ref{1.10a})
reveals that only 7 from 11 external multipole moments are algebraically independent. More specifically,
the gauge condition (\ref{11.12}) leads to the following relationship between the moments
\br
    D_L&=&\frac{2l(2l-1)}{2l+1}                                     \label{1.21}
            \biggl[Y_L+(1-\gamma)P_L-Q_L\biggr],\qquad\qquad(l\ge1)\,.
\er
Three other relationships are obtained after accounting for Eq. (\ref{1.21}) in the gauge condition (\ref{11.11})
which yields
\be
   S_L=\dot Y_L+(1-\gamma)\frac{2l^2+l+1}
     {(l+1)(2l+3)}\,\dot P_L-                        \label{1.22}
        \frac{2l^2-3l-1}{(l+1)(2l+3)}\,
         \dot Q_L,\qquad\qquad(l\ge0)\,,
\en
and
\br \label{1.20a}
        E_i&=&\frac{2}{5}\,\dot\Omega_i\;,\\\nonumber\\\label{1.20b}
         E_L&=&0\;,\qquad\qquad(l\ge2)\\\nonumber\\\label{1.20}
        B_L&=&0\;,\qquad\qquad(l\ge0)\;.
        \er

Eqs. (\ref{1.21}) -- (\ref{1.20}) allow us to eliminate the
external multipole moments $B_{L},\,E_{L},\,D_{L},\,S_{L}$ from
the local metric so that the space-time and space-space components of the external metric tensor assume the
form
\br
                               \label{1.25}
\hat L_i^{{\sss(E)}}(u,\nnw)& =&\sum\limits_{l=1}^{\infty}
          \frac{1}{l!}\varepsilon_{ipq}C_{pL-1}w^{<qL-1>}
         +\sum\limits_{l=1}^{\infty}\frac{1}{l!}Z_{iL}w^{L}
             \\\nonumber&+&
          \sum\limits_{l=0}^{\infty}\frac{1}{l!}
          \biggl[\dot Y_{L}+(1-\gamma)
          \frac{{\Dd2l^2+l+1}}{{\Dd(l+1)(2l+3)}}\dot P_{L}
          -\frac{{\Dd2l^2-3l-1}}{{\Dd(l+1)(2l+3)}}\dot Q_L\biggr]w^{<iL>}\;,
 \\\nonumber\\\nonumber\\                                                   \label{1.26}
\hat
H_{ij}^{{\sss(E)}}(u,\nnw)&=&\frac15\left(\varepsilon_{ipq}\dot\Omega^p w^{<jq>}+\varepsilon_{jpq}
\dot\Omega^p w^{<iq>}\right)+2\delta_{ij}\sum\limits_{l=0}^{\infty}
     \frac{1}{l!}Y_{L}w^{L}\\\nonumber&&
+2\sum\limits_{l=0}^{\infty}\frac{2l+1}{(2l+3)\,l!}
          \biggl[\Bigl(Y_{iL}
          +(1-\gamma)P_{iL}-Q_{iL}\Bigr)w^{<jL>}
          \biggr]
          ^{{\rm Sym}(ij)}\\\nonumber&&
          +\sum\limits_{l=0}^{\infty}
          \frac{1}{(l+2)!}F_{ijL}w^{L}
+\sum\limits_{l=0}^{\infty}\frac{1}{(l+2)!}
          \varepsilon_{pq(i}
          G_{j)pL}w^{<qL>}\;.
\er
Remaining multipole moments $P_L$, $Q_L$, $Z_L$, $C_L$, $Y_L$, $F_{L}$, $G_{L}$ and the angular
velocity of rotation, $\Omega_i$, can not be constrained by imposing the gauge conditions. However,
a residual gauge freedom described by differential Eq. (\ref{11.b}) allows us
to find further limitations on the remaining 7 sets of the
multipole moments which are explicitly shown in the right side of Eqs. (\ref{1.25}) and
(\ref{1.26}). Exclusion of the residual degrees of freedom makes it clear which multipole moments are in fact
physically relevant, that is can not be excluded by infinitesimal
coordinate transformations $w^\alpha=x^\alpha+\xi^\alpha(x^\beta)$.

In order to exclude the external multipoles, which do not carry out information about gravitational
degrees of freedom, we shall
make use of a well-known property of the gauge-invariance of the (linearized) Riemann
tensor under infinitesimally small coordinate transformations (see Box 18.2 in \cite{mtw})
\be \label{rim} \hat
R_{\alpha\beta\gamma\delta}(u,{\bm w})=R_{\alpha\beta\gamma\delta}(u, {\bm w})\;.
\en
Eq. (\ref{rim}) must be understood as invariance of the functional form of the Riemann tensor after
making an infinitesimally small gauge transformation.
Computing all components of the Riemann tensor, which are functions of the
external metric tensor (\ref{1.8a}), (\ref{1.25}) and (\ref{1.26}) only, one
finds (see Appendix \ref{rten}) that the external part of the Riemann tensor depends only on four
sets of the external multipole moments $P_{L}$, $Q_{L}$, $C_{L}$ and
$G_{L}$. However, one can notice that the multipoles $C_{L}$ and $G_{L}$
enter Eq. (\ref{b9}) only in the form of a linear
combination which can not be split algebraically. This means that only three sets of the external
moments have a real physical meaning. In what follows, we shall choose  $P_{L}$, $Q_{L}$, and $C_{L}$
as the primary external
multipoles. Other multipole moments $Y_L$, $Z_L$, $F_G$ and
$G_{L}$ can be chosen arbitrary which reflects the presence of 4 residual gauge degrees of freedom
generated by the coordinate transformation confined by Eq. (\ref{11.b}).

Hereafter we assume
that the angular velocity of rotation of the local frame
\be\label{ttt}
\Omega_i=0\;, \en
which gives $\hat
N_i^{{\sss(E)}}(u,\nnw)=0$. This assumption greatly simplifies
subsequent calculations without missing any significant physics. We have to notice, however,
that rotating local coordinate systems have a great practical value for satellite geodesy and
global positioning system (GPS) \cite{ashn}.

Various authors used the residual gauge freedom
differently. We shall follow the convention accepted in papers \cite{skpw,kop1988,dsx1991,th} and
postulate \footnote{We draw attention of the reader that in the standard PPN formalism \cite{will} and in the
global coordinates the space-space components of the metric tensor are always diagonal in standard gauge.
However, here we discuss the local coordinates which admit more general degrees of freedom than the global
ones. Hence, our postulate is not redundant.}  that the
space-space components $\hat{g}_{ij}(u,{\bm w})$ of the local metric tensor must form a diagonal
matrix proportional to the Kronecker symbol $\delta_{ij}$.
It allows to simplify the non-linear term $H_{<ij>}N_{,ij}$
in Eq. (\ref{11.33}) for the time-time component of the
metric tensor by making use of the Laplace equation for function $N(t,{\bm w})$ which converts
the non-linear term to that having a compact support, that is $H_{<ij>}N_{,ij}\sim \rho^* H$.
In order to diagonalize $\hat{g}_{ij}(u,{\bm w})$ one chooses the external multipoles
$F_{L}$, $G_{L}$ as follows
\be
      F_{L}=0\;,                                              \label{3.35}
      \en
      \be
      G_{L}=0\;,                                                     \label{3.35a}\en
 for all $l\ge0$. Furthermore,  we chose
 \be
     Y_{L}=Q_L+(\gamma-1)P_{L} \;,                     \label{3.36}
\en
for all $Y_L$ with $l\ge1$ but the monopole moment $Y$ which
is left arbitrary \footnote{Restrictions (\ref{3.35}) -- (\ref{3.36}) on the external multipoles
result of the requirement that the metric tensor must be diagonal in the entire domain of validity
of the local coordinates but not in a single point.}.
  We preserve some gauge freedom and do not fix the external multipoles $Z_L$ ($l\ge 2$).

Finally, the external metric tensor assumes the following simple
form
\br\label{1.24b} \hat N^{{\sss(E)}}(u,\nnw)&=&2\nn{l=0}
                     \frac{1}{l!}Q_Lw^{L}\;,
\\\nonumber\\\nonumber\\                                                                    \label{1.25b}
\hat L_i^{{\sss(E)}}(u,\nnw)&=&\left(\dot Y+\frac13\dot
Q+\frac{1-\gamma}{3}\dot P\right)w^i
          +\sum\limits_{l=1}^{\infty}
          \frac{1}{l!}\varepsilon_{ipq}C_{pL-1}w^{<qL-1>}
             \\\nonumber
          &+&
          2\sum\limits_{l=1}^{\infty}\frac{2l+1}{(2l+3)(l+1)!}
          \biggl[2\dot Q_L+(\gamma-1)\dot P_{L}\biggr]w^{<iL>}+\sum\limits_{l=1}^{\infty}\frac{1}{l!}Z_{iL}w^{L}\;,
 \\\nonumber\\\nonumber\\                                               \label{1.26b}
\hat
H_{ij}^{{\sss(E)}}(u,\nnw)&=&2\delta_{ij}\left\{Y+\nn{l=1}\frac{1}{l!}\biggl[Q_L+(\gamma-1)P_{L}\biggr]w^{L}\right\}\;.
\er
where we still keep the (time-dependent) monopole term $Q\not=0$.

Now we can compute $\nnfa{E}{0}{0}{u}{w}$ component of the
external metric tensor up to the post-Newtonian order by making
use of Eq. (\ref{11.33}). Its the most general solution is determined up to
that of a homogeneous wave equation which we shall assume to be
incorporated to the post-Newtonian correction to the multipole
moments $Q_L$ defined by Eq. (\ref{1.24b}). Hence, we obtain
\br \label{1.24c}
\hat L^{{\sss(E)}}(u,\nnw)&=&
-2\left(\nn{l=1}\frac{1}{l!}Q_Lw^{L}\right)^2-
                 2(\beta-1)\left(\nn{l=1}\frac{1}{l!}P_{L}w^{L}\right)^2
\\\nonumber&&
                 +\nn{l=0}\frac{1}{(2l+3)l!}\Ddot{Q}_{L}w^{L}w^2\,.
\er
It is interesting to note that summation in the first two terms in the right side of Eq. (\ref{1.24c})
was originally started from $l=0$ (see Eqs. (\ref{1.7a}) and (\ref{1.8a})). However, product
of a harmonic polynomial with the monopoles $Q$ and $P$ represent a homogeneous solution of
the Laplace equation and, for this reason, can be absorbed by the Newtonian-like polynomial
$Q_L w^L$ in Eq.  (\ref{1.24b}) by means of re-definition of the multipoles $Q_L$. This is
possible because the mathematical structure of the multipoles $Q_L$ as functions of the
parameters of the external bodies has not yet been specified. Our remark helps to realize
that the influence of the scalar field on the external solution of the metric tensor in
the local coordinates starts from quadratic, with respect to coordinates ${\bm w}$, terms only.
External scalar field can not be eliminated by pointwise coordinate transformation but it enters to the
external metric tensor in such a way that it can be absorbed to the multipole moments $Q_L$ of the metric
tensor gravitational field. Hence, the external multipoles $P_L$ do not contribute linearly to the equations
of translational motion of test particles and extended bodies -- only their non-linear combination is
observable (see section \ref{tem}).

\subsubsection{The Metric Tensor: The Coupling Terms }

The coupling terms in the metric tensor in local coordinates are given as a
particular solution of the inhomogeneous equation (\ref{11.33})
with the right side taken as a product of the internal and
external solutions found on previous step of the approximation procedure. Solving Eq. (\ref{11.33})
yields the coupling terms of the metric tensor in the local coordinates
\br \label{1.24o} \hat
L^{{\sss(C)}}(u,\nnw)&=&-2\nnc\times\\\nonumber&&\times\left\{Y+(2\beta-\gamma-1)P+2\nn{l=0}\frac{1}{l!}
\biggl[Q_L+(\beta-1)P_{L}\biggr]w^{L}\right\}\\\nonumber&&
-2G\nn{l=1}\frac{1}{l!}\biggl[Q_L+2(\beta-1)P_{L}\biggr]\nnhh{B}{\rho^*w^{L}}\;.\er
This completes derivation of the metric tensor in the local coordinates.

\subsection{Multipolar Decomposition of the Body's Gravitational Field in the Local Coordinates}\label{mdloc}

The local coordinates are introduced in the vicinity of each of the gravitating body comprising the N-body system.
We consider one of them and call it the "central" body which is indexed by the letter 'B'. This body, for example,
can be the Earth and the local coordinates in such case are called the geocentric coordinates \cite{iau2000,skpw}.
Gravitational field of the central body taken alone, that is when all other (external) bodies are ignored, is
described in the local coordinates in terms of the metric tensor and scalar field which depend on the internal
field potentials $\hat U^{\scriptstyle (B)}$, $\hat U^i_{\scriptstyle
(B)}$, $\hat \Phi_1^{\scriptstyle (B)}$, etc., defined in Eqs.
(\ref{1.11}) -- (\ref{1.17}). Multipolar decomposition of the internal metric tensor of the central body is totally
equivalent to the procedure of multipolar decomposition of the gravitational field of the N-body in global
coordinates described in section \ref{mdgco}.
However, from the point of view of precise theory the central body is not gravitationally isolated from the
other bodies
of the N-body system because it interacts with them gravitationally. This interaction brings about the
coupling terms to the metric tensor in the local coordinates which can contribute to the numerical values
of the body's multipole moments in the multipolar decomposition of the local metric tensor. The presence
of the coupling terms introduces a post-Newtonian ambiguity of the multipolar decomposition of gravitational
field in the local coordinates and puts a question about what specific definition of the multipole moments of
the central body must be used in deriving translational and rotational equations of motion of the body in the
gravitational field of N-body system. This problem was seemingly pointed out for the first time by Thorne and
Hartle \cite{th}.

Solution of this problem can be found only by doing complete calculations of the equations of motion of the
central body with taking into account all its multipoles. One has two possibilities: either to include or to
exclude the contribution of the coupling terms to the multipole moments of the body and we have explored both
of them. It turns out the final form of the equations of motion can be significantly simplified if the coupling
terms are included to the definition of the multipole moments. In fact, if one excludes the contribution of the
coupling terms to the body's multipoles it produces a lot of additional terms in the equations of motion but all
these terms can be eliminated after a suitable re-definition of the multipole moments. By inspection one can check
that the final form of such renormalized equations of motion coincides with that which would be obtained if one
included contribution of the coupling terms in the local metric tensor to the definition of the multipole moments
of the central body from the very beginning. The significantly simple form of the renormalized equations of motion
is a direct indication that the coupling terms must be included to the definition of the multipole moments of the
body in the local coordinates. This resolves the Thorne-Hartle ambiguity \cite{th} in the definition of the multipole
moments.

Thus, the formal procedure of the multipolar decomposition of the gravitational field in the local coordinates is
based on the same field equations (\ref{13.0}), (\ref{13.13}) and (\ref{13.21}) for {\it active}, {\it scalar}
and {\it conformal} potentials whose right sides depends on the {\it active}, {\it scalar} and {\it conformal}
mass densities defined by Eqs. (\ref{13.2}), (\ref{13.14}) and (\ref{13.24}) respectively. All these densities
depend on the trace of the space-space component of the metric tensor, $H$, and the scalar field, $\varphi$.
In accordance with our procedure of definition of the multipole moments in the local coordinates,
these functions must include the contribution of the external gravitational and scalar fields. In other words,
computation of the mass densities in Eqs. (\ref{13.2}), (\ref{13.14}) and (\ref{13.24}) in the local coordinates
must be relied upon the trace of the metric tensor, $H$, defined by the sum of Eqs. (\ref{1.10}) and (\ref{1.26b}),
and the scalar field, $\varphi$, defined by the sum of Eqs. (\ref{1.7}) and (\ref{1.8}). Solving Eqs. (\ref{13.0}),
(\ref{13.13}), (\ref{13.21}) with the mass densities defined herewith and expanding the metric potentials in the
multipolar series yields the multipole moments of the central body in the local coordinates.

In section 4 we have constructed three
sets of the mass multipole moments -- active, scalar, and conformal. The same type of multipoles presents in the
local coordinates as well.
The {\it active} STF mass multipole
moments of the central body are
\br\label{1.31}
   {\cal I}_{L}&= &\int_{V_{\sss B}}\sigma_{\sss B}
   (u,\nnw)w^{<L>}\,d^3w+
   \frac{\epsilon^2}{2(2l+3)}\times\\\nonumber\\\nonumber &\times&\left[\frac{d^2}{du^2}
   \int_{V_{\sss B}}\sigma_{\sss B}
   (u,\nnw)w^{<L>}w^2\,d^3w
   -4(1+\gamma)\,\frac{2l+1}{l+1}
   \frac{d}{du}\int_{V_{\sss B}}\sigma^i_{\sss B}(u,\nnw)w^{<iL>}
   \,d^3w\right]\\\nonumber\\\nonumber &-&\epsilon^2
   \int_{V_{\sss B}}d^3w\,\sigma_{\sss B}(u,\nnw)\times\\\nonumber\\\nonumber&\times&
   \left\{Y+(2\beta-\gamma-1) P+\sum_{k=1}^{\infty}\frac{1}{k!}\biggl[Q_{K}+2(\beta-1) P_{K}\biggr]
   w^{K}\right\}w^{<L>}\;,\er
   where $V_{\sss B}$ denotes the volume of the central body under consideration and the {\it active} mass density
   in the
   body's interior is defined as
\be\label{pz3}
 \sigma_{\sss B}=\rho^*\left\{1+\epsilon^2\left[(\gamma+\frac12)\nu^2+\Pi+\gamma\frac{\pi^{kk}}{\rho^*}-(2\beta-1)
 \hat U^{(\sss B)}  \right]\right\}\;,
 \en
where $\hat U^{(\sss B)}$ is the gravitational potential of the body given by Eq. (\ref{1.11}).

The {\it scalar} STF mass multipole moments of the body are defined as
\br \label{1.33}
   \bar{\cal I}_{L}&=& \int_{V_{\sss B}}\bar{\sigma}_{\sss B}
   (u,\nnw)\left\{1-\epsilon^2\left[Y-\gamma P+\nn{k=1}\frac{1}{k!}Q_Kw^K\right]\right\}w^{<L>}\,d^3w
\\\nonumber\\\nonumber &
+& \frac{\epsilon^2}{2(2l+3)}\frac{d^2}{du^2}\int_{V_{\sss B}}\bar{\sigma}_{\sss B}(u,\nnw)w^{<L>}w^2\,d^3w
\\\nonumber\\\nonumber &
+&\epsilon^2
   \int_{V_{\sss B}}\rho^*(u,\nnw)\left[4(1-\beta)\nn{k=1}\frac{1}{k!}P_Kw^K-\eta P\right]w^{<L>}d^3w\;,
\er
where the {\it scalar} mass density of the body's matter is defined by
\be\label{pz4}
 \bar\sigma_{\sss B}=(1-\gamma)\rho^*\left[1-\epsilon^2\left(\frac12\nu^2-\Pi+\frac{\pi^{kk}}{\rho^*}\right)\right]
 -\epsilon^2\eta\rho^*\hat U^{(\sss B)}\;.
 \en

The {\it conformal} STF mass multipole moments of the body are
\br                                                 \label{1.34}
   \tilde{\cal I}_{L}&= &\int_{V_{\sss B}}\tilde\sigma_{\sss B}
   (u,\nnw)\left\{1-\epsilon^2\left[Y+(1-\gamma)P+\sum_{k=1}^{\infty}\frac{1}{k!}Q_{K}w^{K}\right]\right\}
   w^{<L>}\,d^3w\\\nonumber\\\nonumber&+&
   \frac{\epsilon^2}{2(2l+3)}\left[\frac{d^2}{du^2}
   \int_{V_{\sss B}}\tilde\sigma_{\sss B}
   (u,\nnw)w^{<L>}w^2\,d^3w
   -\frac{8(2l+1)}{l+1}
   \frac{d}{du}\int_{V_{\sss B}}\sigma^i_{\sss B}(u,\nnw)w^{<iL>}
   \,d^3w\right]\;,
\er
with the {\it conformal} mass density of the body's matter defined as
\be\label{pz5}
 \tilde\sigma_{\sss B}=\rho^*\left[1+\epsilon^2\left(\frac32\nu^2+\Pi+\frac{\pi^{kk}}{\rho^*}-\hat U^{(\sss B)}
 \right)\right]\;.
 \en
The {\it conformal} density does not depend on the PPN parameters $\beta$ and $\gamma$.

The current density is defined in the local coordinates by
\be\label{pz6}
 \sigma^i_{\sss B}=\rho^*\nu^i\;,
 \en
and the spin multipole moments of the body are determined by the formula \footnote{We discuss the post-Newtonian
definition of body's spin in section \ref{req}.}
\be
   S_{L}= \int_{V_{\sss B}}\varepsilon^{pq<a_l}\hat w^{L-1>p}             \label{1.32}
   \sigma^q_{\sss B}(u,\nnw)\, d^3w\,.
\en

\begin{figure}[p]
\centerline{\psfig{figure=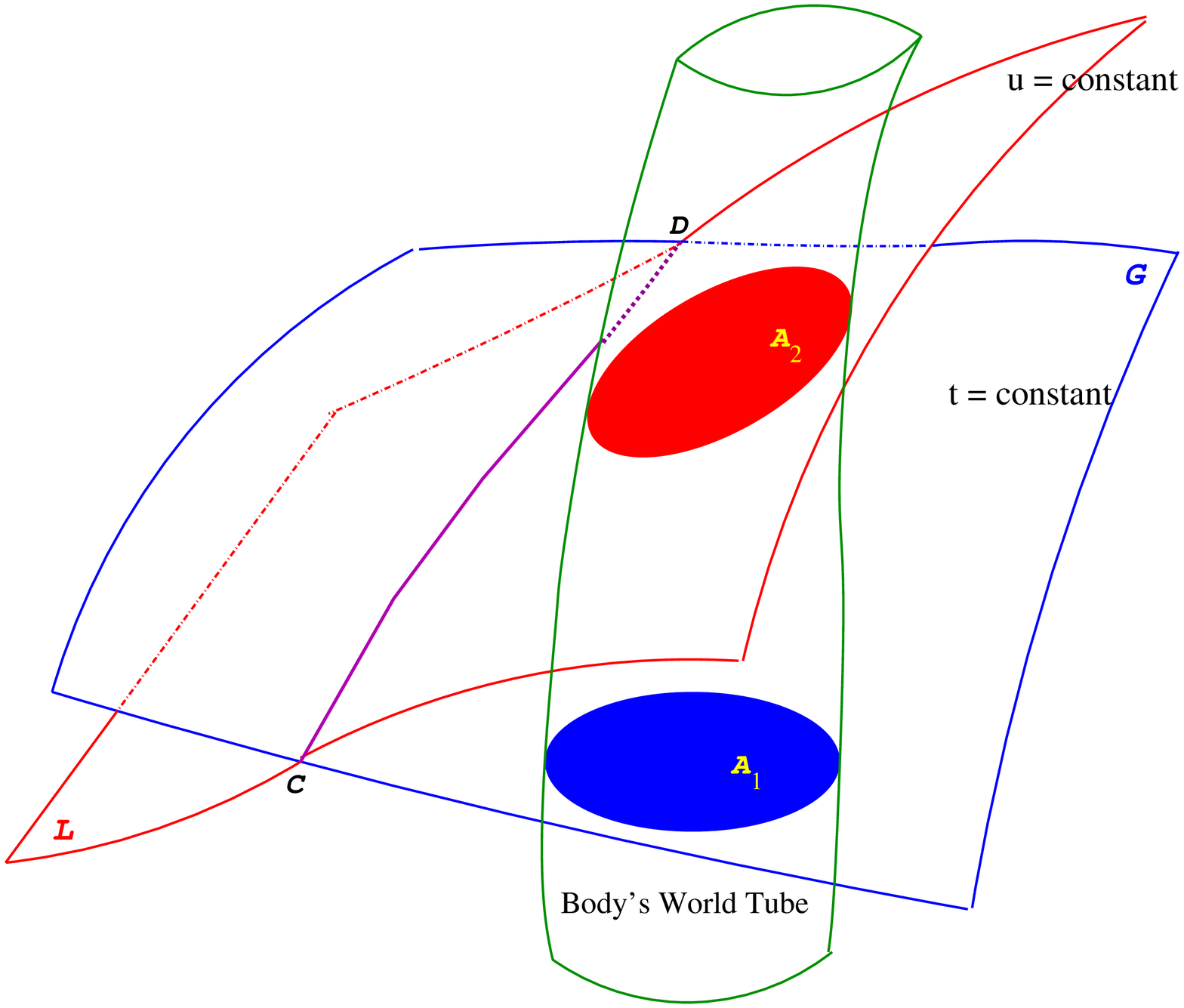,angle=-90,height=14cm,width=18cm}}
\vspace{1cm} \caption[Two Hypersurfaces of Constant Time]{Two
3-dimensional hypersurfaces of constant time related to the global (G) and local
(L) coordinates are shown. The two hypersurfaces do not
coincide because of relativistic transformation of the local
coordinate grid with respect to the global one. A body's world tube
contains a world line of the origin of the local coordinates.
The world tube cross-section, ${\color{blue}A_1}$, is a part of the hypersurface of
constant time $t$ of the global coordinates, and
that, ${\color{red}A_2}$, is a part of the hypersurface of constant time $u$ of the local
coordinates. The line $CD$ marks
intersection of the two hypersurfaces. All integrals
depending on density, velocity and other internal characteristics of the body are performed in
the global coordinates over the cross-section ${\color{blue}A_1}$ whereas in the
local coordinates the integration is over the cross-section ${\color{red}A_2}$.}
\label{fig3}
\end{figure}

It is important to emphasize that the algebraic relationship (\ref{13.31}) preserves its form for the set of
the mass multipole moments taken for each body separately, that is
\be\label{wh1}
{\cal I}_{L}=\frac{1+\gamma}{2}\tilde {\cal I}_{L}+\frac{1}{2}\bar{\cal I}_{L}\,.
\en
Validity of this relationship can be checked out by a straightforward calculation. We also draw attention of the
reader that the hypersurface of the integration in Eqs. (\ref{1.31}), (\ref{1.33}), (\ref{1.34}) is that of
the constant local coordinate time $u$ which does not coincide with the hypersurface of the constant time $t$
in the global coordinates (see Fig. \ref{fig3}).
This remark is of a great importance in the matching procedure of the local and global coordinates discussed
in next section. It will be also important in section \ref{wep} for correct derivation of equations of motion
of the body which requires comparison of the definition of the multipole moments of the body in the local
coordinates used in this paper and that in the global coordinates used by some other authors
\footnote{See section \ref{wep} for more detail.}.

\section{Parametrized Post-Newtonian Transformation from Local to Global Coordinates}\label{zza2}
\subsection{Preliminary Remarks}

General relativistic post-Newtonian coordinate transformation from local, $w^\alpha$, to global, $x^\alpha$,
coordinates are used in standard algorithms of data processing of various astronomical observations to reduce
the observable quantities to the barycentric coordinates of the solar system \cite{sman,reframe} where
they are ``stored" (catalogued) at a certain astrometric epoch recommended for the international usage by
the IAU. The most commonly used epochs are J2000 and B1950 (i.e. instantaneous orientation of the barycentric
coordinates fixed by the position of the vernal equinox as it was located on the sky in 2000AD or 1950AD).
Post-Newtonian coordinate transformation from the local to global frame is an inalienable part in the procedure
of derivation of the general relativistic equations of motion of test and/or extended bodies
in the solar system
\cite{kls,ssa,ashb1,ashb2,kop1988,kop1989a,kop1989b,kop1991,bk-rf,bk-kin,bk-nc,dsx1991,dsx1992,klv,mashn}
as well as compact relativistic stars in binary systems emitting gravitational waves
\cite{dam300,dam1983,das1,das2,kop1985,grkop}.
It is widely accepted that scalar fields must exist since they provide elegant explanation for various
effects and phenomena encountered by modern theoretical physics of fundamental particle interactions,
gravitation and cosmology \cite{scalar}. Therefore, it is natural to generalize the existing form of
the general relativistic post-Newtonian transformation from the local to global coordinates to make
them fully compatible with the scalar-tensor theory of gravity. Solution of this problem in terms of
the PPN parameters $\beta$ and $\gamma$ is displayed in the present section.

The PPN coordinate transformations from the local (geocentric) coordinates to global (barycentric) coordinates
can be found by making use of the asymptotic expansion matching technique \cite{math1} as proposed in \cite{das1,das2}.
The metric tensor and the scalar field are solutions of the field equations which are expressed in the form of two
different post-Newtonian expansions in the global and local coordinates. These two expansions of the metric tensor
and the scalar field must match smoothly in the spatial domain where both coordinate charts overlap. This matching
domain is originally defined as a region in which the post-Newtonian expansion of the metric tensor and scalar field
is not divergent. In the case of a weak gravitational field this domain extends from the origin of the local coordinates
associated with a central body (Earth) up to the next closest celestial body (Moon) \footnote{In the N-body system
comprised of neutron stars and/or black holes the local coordinates would overlap with the global ones in the
so-called {\it buffer} region in which the gravitational field of both the neutron star/black hole and the external
bodies is weak. More details on this subject can be found in papers \cite{th}--\cite{das2}.}. However, after the
structure of the PPN coordinate transformation is found the
domain of applicability of the local coordinates can be stretched out up to much larger distances (see text below after
Eq. (\ref{ma5}) and papers \cite{kop1991b,bk-nc,klv} for more detail).

Special relativistic transformation from local to global coordinates is linear and takes into account only kinematic
aspects of the transformation, i.e. that the local frame moves with respect to the global one with constant velocity.
The special relativistic (Poincar\'e-Lorentz) transformation is described at each instant of time by 10 parameters
characterizing the intrinsic properties of flat space-time which is tangent to curved space-time manifold at any
point on the world line of the origin of the local coordinates \cite{fock,ll,mtw}. These parameters include 4
space and time translations, 3 spatial rotations, and 3 Lorentz boosts depending on the velocity of the origin of
the local frame with respect to the global coordinates. General relativity generalizes the Poincar\'e-Lorentz
transformation by accounting for the presence of dynamic effects (acceleration) in the motion of the local frame
as well as for the effects of the external gravitational field of the other bodies and the background curvature
of the space-time. It was shown
(see, e.g., \cite{ashb2,kop1988,das1,das2} and references therein)
that in general relativity the post-Newtonian coordinate transformation from local to global coordinates is non-linear,
has more than 10 parameters, and all parameters
depend on time. Scalar-tensor theory of gravity brings about additional complications to the relativistic theory of
reference frames caused by the dependence of the post-Newtonian transformation on the scalar field which is described
by two more parameters, $\gamma$ and $\beta$. However, scalar-tensor theory does not change general structure of
the parametrized post-Newtonian (PPN) transformation which remains the same as that used in
general theory of relativity.

The PPN coordinate transformation between the global and local coordinates belongs to the class of transformations
which must comply with the gauge condition (\ref{11.3}). Therefore, one begins with
finding of the most general structure of such coordinate transformations. As soon as this structure is known it can
be further specialized by reducing the number of the gauge degrees of freedom by making use of the residual gauge
transformations and matching technique applied for establishing a one-to-one correspondence between two
asymptotic expansions of the metric tensor and scalar field written down in the global and local coordinates.
This procedure allows us to arrive to a single, ingenious form of the PPN
coordinate transformation. We shall show that the PPN transformation from local to global coordinates
can be represented as a power series
expansion with respect to two small parameters, $\epsilon=1/c$ and $r/R$,
where $r$ and $R$ are distances from the origin of the local coordinates (the central body) to
matching point in space and to the other (external) gravitating body respectively. Coefficients of the power series
expansion are symmetric and trace-free (STF) multi-index functions of time which are
determined simultaneously with
the external multipole moments, $P_L$, $Q_L$, $C_L$, etc., of the local metric in the course of the
matching procedure.
The STF
coefficients of the PPN coordinate transformation are functions of
the local coordinate time $u$ and are ``pinned down" to the origin of
the local coordinates. The matching procedure shows that the STF coefficients of the most general form of
the PPN coordinate transformation couples linearly with the external STF multipole moments of the local metric
tensor entering Eqs. (\ref{1.8a}) -- (\ref{1.10a}) so that the matching allows to derive a set of equations
defining only their algebraic sum. This reflects existence of the residual gauge freedom which can be used in
order to simplify the structure of the PPN
coordinate transformation and/or that of the metric tensor in local coordinates. We have used this gauge freedom
already in section \ref{mtex} to suppress the number of the external multipole moments which have no
physical significance. Elimination of the non-physical multipole moments from the metric tensor leads
to more simple structure of the PPN transformations as well. This issue is discussed in the next sections
in more detail.

\subsection{General Structure of the Coordinate Transformation}\label{gzm}

The most general structure of the coordinate transformation from the local, $w^\alpha=(cu,w^i)$, to global,
$x^\alpha=(ct,x^i)$, coordinates in the weak-field and slow-motion post-Newtonian approximation is given by
two equations:
\br u&=&t +\epsilon^2\xi^0(t,{\bm x})\;,
\label{2.2} \\\nonumber\\
 w^i&=&\nnr{i}+\epsilon^2\xi^i(t,{\bm x})\;,      \label{2.3} \er
 where $\xi^0$ and $\xi^i$ are the post-Newtonian corrections to the Galilean transformation,
$\nnr{i}=x^i-x^i_{{\sss B}}(t)$, and $x^i_{{\sss B}}(t)$ is the position of
the origin of the local frame at time $t$ with respect to the origin of the global coordinates. We shall
prove later on that the origin of the local coordinates can be always chosen at any instant of time at
the center of mass of the body (Earth, Sun, planet) around which the local coordinate chart has been constructed.
In what
follows, we shall denote velocity and acceleration
of the origin of the local coordinates as $\nnv{i}\equiv\dot x^i_{{\sss B}}$ and
$\nnaa{i}\equiv\Ddot x^i_{{\sss B}}$ respectively, where here and everywhere else the overdot
must be understood as a time derivative with respect to time $t$.

Pointwise matching equations for the scalar field, the metric tensor, and
the Christoffel symbols are given by their general laws of a coordinate transformation \cite{mtw,dfn,schouten}
\br
\varphi(t,{\bm x})&=&\nxi\;, \label{2.5}
      \\\nonumber\\
      g_{\mu\nu}(t,{\bm x})&=&                         \label{2.6}
          \hat g_{\alpha\beta}(u, {\bm w})\frac{\partial w^{\alpha}}{\partial x^{\mu}}
          \frac{\partial w^{\beta}}{\partial x^{\nu}}\;,
          \\\nonumber\\
   \Gamma^{\mu}_{\alpha\beta}(t,{\bm x})&=&
    \hat{\Gamma}^{\nu}_{\rho\sigma}(u,\bm w)
     \frac{\partial x^{\mu}}{\partial w^{\nu}}
      \frac{\partial w^{\rho}}{\partial x^{\alpha}}
       \frac{\partial w^{\sigma}}{\partial x^{\beta}}+           \label{2.7}
        \frac{\partial x^{\mu}}{\partial w^{\nu}}
         \frac{{\partial}^2 w^{\nu}}{\partial x^{\alpha}
          \partial x^{\beta}}\,.
\er
One recalls that $g_{0i}$ component of the
metric tensor does not contain terms of order $O(\epsilon)$ because we have assumed that both the
global and the local frames are not rotating. This fact, being used in Eq. (\ref{2.6}), implies that
function $\xi^0(t,{\bm x})$ from Eq. (\ref{2.2}) must be subject to
the following restriction: $\xi^0,_{k}=-\nnv{i}+O(\epsilon^2)$. This is a partial differential equation which
can be integrated so that function $\xi^0$ can be represented
as \be \xi^0(t,\nnx)=-{\cal A}(t)-
       \nnv{k}\nnr{k}                                           \label{2.8}
          +\epsilon^2 \kappa(t,\nnx)+O(\epsilon^4)\;,     \en
where ${\cal A}(t)$ and
$\kappa(t,\nnx)$ are analytic, but in all other aspects yet unspecified, functions \footnote{Notice that
function ${\cal A}(t)$ depends only on time $t$.}.

Let us now use the gauge conditions (\ref{gau}) in order to
impose further restrictions of the PPN functions $\xi^0$ and $\xi^i$
from Eqs. (\ref{2.2}) and (\ref{2.3}). The gauge conditions
can be written in arbitrary coordinates as an exact equality
\be
   g^{\alpha\beta}\Gamma^{\mu}_{\alpha\beta}=
     \frac{\zeta^{,\mu}}{1+\zeta}\;.                                         \label{2.4}
\en
The law of transformation of the Christoffel symbols, Eq.
(\ref{2.7}), being substituted to Eq. (\ref{2.4}) yields a partial
differential equation of the second order
\be
   g^{\alpha\beta}(t,\nnx)\,
    \frac{{\partial}^2 w^{\mu}}{\partial x^{\alpha}             \label{2.9}
     \partial x^{\beta}}=0\;,
\en
which describes any possible freedom of the PPN transformation in the quasi-harmonic gauge of the scalar-tensor
theory of gravity.
Let us now substitute functions $w^0=cu$ and $w^i$ from Eqs.
(\ref{2.2}) and (\ref{2.3}), and $\xi^0$ from Eq. (\ref{2.8}) to
Eq. (\ref{2.9}). One obtains
 \br
   \nabla^2 \kappa(t,\nnx)&=&3\nnv{k}\nnaa{k} -\Ddot {\cal A}- \label{2.10}
        \dot a^k_{{\sss B}}\nnr{k}+ O(\epsilon^2)\;,
\\\nonumber\\
   \nabla^2\xi^i(t,\nnx)&=&-\nnaa{i}+ O(\epsilon^2)\;.              \label{2.11}
\er
General solution of these elliptic-type equations can be written in the form of
the power series expansion in terms of the scalar and vector
spherical harmonics (see appendix \ref{ap1} for more detail). Furthermore, solution for functions
$\kappa(t,\nnx)$ and $\xi^i(t,\nnx)$ in Eqs. (\ref{2.10})
and (\ref{2.11}) consist of two parts -- general solution of the
homogeneous Laplace equation and a particular solution of the inhomogeneous elliptic
equation. We shall omit that part of the solution of the homogeneous
equation which is singular at the origin of the local coordinates, that is at
the point $w^i=0$. In general, such ill-behaved terms can be present in other
alternative theories of gravity which violate the law of conservation of linear momentum. But in the scalar-tensor
theory of gravity such ill-behaved functions do not appear in
the geocentric metric and we do not need singular terms in the
coordinate transformations (\ref{2.2}) and (\ref{2.3}) to match the
scalar field and the metric tensor in two coordinate systems.
Integration of equations (\ref{2.10}) and (\ref{2.11})
results in
\br  \kappa&=& \Bigl(\frac{1}{2}
        \nnv{k}\nnaa{k}-\frac{1}{6}
        \Ddot {\cal A}\Bigr)\nnr{2}-\frac{1}{10}             \label{2.12}
        \dot a^k_{{\sss B}}\nnr{k}\nnr{2}+
        \Xi(t,\nnx)\;,\\\nonumber\\                             \label{2.13}
\xi^i&=&-\frac{1}{6} \nnaa{i} \nnr{2}+\Xi^i(t,\nnx)\,,
\er
where functions $\Xi$ and $\Xi^i$ are solutions of the homogeneous Laplace equation. These solutions can be
written in the form of harmonic polynomials
\br\label{eq1}
\Xi(t,\nnx)&=&\sum\limits_{l=0}^\infty\frac{1}{l!}{\cal B}_{<L>}
        \nnr{L}\;,\\\nonumber\\\label{eq2}
\Xi^i(t,\nnx)&=&\sum\limits_{l=1}^{\infty}
        \frac{1}{l!}{\cal D}_{<iL>}\nnr{L}
        +\sum\limits_{l=0}^{\infty}\frac{\varepsilon_{ipq}}{(l+1)!}
        {\cal F}_{<pL>}\nnr{<qL>}+
        \sum\limits_{l=0}^{\infty}
        \frac{1}{l!}{\cal E}_{<L>}
        \nnr{<iL>}\;,
        \er
where the coefficients ${\cal B}_{<L>}$, ${\cal D}_{<L>}$, ${\cal F}_{<L>}$, and ${\cal E}_{<L>}$ of the
polynomials are different multi-index STF functions. These functions are defined on the world line of the
origin of the local coordinates and depend only on the global time $t$. Explicit form of these functions
will be obtained in the process of matching of the global and local metric tensors as well as the scalar field.
Notice that we have denoted all, yet unknown STF functions
 in Eqs. (\ref{eq1}) and (\ref{eq2}) by capital calligraphic letters, while the STF multipole moments in the
 expressions for the local metric, Eqs.  (\ref{1.24b}) -- (\ref{1.26b}), and the scalar field (\ref{1.7a})
 have been denoted by capital roman letters. This is supposed to help to distinguish the STF functions
 having different origin. We also emphasize that the specific form of functions in the coordinate
 transformations (\ref{2.12}) and (\ref{2.13}) is the most general one which preserves the gauge
 conditions (\ref{2.4}) and allows us to analyze the residual gauge freedom in construction of the local
 coordinates by operating with the same type of functions which have been used in the external solution for
 the metric tensor in the local coordinates.

\subsection{Transformation of the Coordinate Bases}\label{trbs}

Derivation of the PPN coordinate transformation originates from the matching equation (\ref{2.6}) for the
metric tensor applied in the joint domain of validity of the local and global coordinates. This equation
contains the matrix of transformation $\Lambda^{\beta}_{\;\,\alpha}=\partial w^\beta/\partial x^\alpha$
between the two coordinate bases,
$\hat{\bf
e}_{\alpha}\equiv\partial/\partial w^\alpha$ and ${\bf
e}_{\alpha}\equiv\partial/\partial x^\alpha$, in the local, $w^\alpha$, and global, $x^\alpha$, coordinates
respectively.
Transformation between them reads
\be\label{cb} {\bf
e}_{\alpha}=\Lambda^{\beta}_{\;\,\alpha}\hat{\bf e}_{\beta}\;. \en
In that region of space-time where the matrix of transformation $\Lambda^{\beta}_{\;\,\alpha}$ is non-singular
it can be inverted so that one gets the
inverse matrix $\left(\Lambda^{-1}\right)^{\beta}_{\;\alpha}$ defined by a standard rule \cite{mtw}
 \be\label{qaz}
\Lambda^{\alpha}_{\;\,\beta}\left(\Lambda^{-1}\right)^{\beta}_{\;\,\gamma}=\delta^{\alpha}_{\gamma}\;.
\en
The inverse matrix is required to get the inverse transformation between the two bases.
The matrix
$\Lambda^{\alpha}_{\;\,\beta}$ can be expanded in the post-Newtonian series, which is a consequence of the
post-Newtonian expansion of the coordinate transformation described in a previous section. The post-Newtonian
expansion of the matrix of transformation is as follows
\br\label{ma00}
\Lambda^{0}_{\;\,0}&=&1+\epsilon^2\mathfrak{B}(t,\nnx)+\epsilon^4\mathfrak{D}(t,\nnx)+O(\epsilon^5)\;,\\
\nonumber\\\label{ma01}
\Lambda^{
0}_{\;\,i}&=&-\epsilon\nnv{i}+\epsilon^3\mathfrak{B}^i(t,\nnx)+O(\epsilon^5)\;,\\\nonumber\\\label{ma02}
\Lambda^{ i}_{\;\,0}&=&-\epsilon\nnv{ i}+\epsilon^3\mathfrak{P}^{
i}(t,\nnx)+O(\epsilon^5)\;,\\\nonumber\\\label{ma03} \Lambda^{
i}_{\;j}&=&\delta^{ i}_{j}+\epsilon^2\mathfrak{R}^{
i}_{\;j}(t,\nnx)+O(\epsilon^4)\;, \er
where the coefficients of the expansion are the following functions of the global coordinates:
\br
\label{zz1}
\mathfrak{B}(t,\nnx)&=&\nnv{2}-\nnaa{k}\nnr{k}-\dot{\cal A}\;,\\
\nonumber\\                                             \label{zz2}
\mathfrak{D}(t,\nnx)&=&\left(\frac13\ddot{\cal A}-\nnv{k}\nnaa{k}+\frac15\da{k}\nnr{k}\right)
\left(\nnv{j}\nnr{j}\right)\\\nonumber&&+\left(\frac12\nnaa{2}+\frac35\nnv{k}\da{k}-\frac{1}{10}\dda{k}
\nnr{k}-\frac16\dddot{\cal A}\right)\nnr{2}\\\nonumber&&+\sum_{l=0}^\infty\frac{1}{l!}
\left(\dot{\cal B}_{<L>}-\nnv{k}{\cal B}_{<kL>}\right)\nnr{L}\;,\\
\nonumber\\                                                     \label{zz3}
\mathfrak{B}^i(t,\nnx)&=&\left(\nnv{k}\nnaa{k}-\frac13\Ddot{\cal A}\right)\nnr{i}-\frac{1}{10}\da{i}\nnr{2}-
\frac15\da{k}\nnr{k}\nnr{i}+\sum_{l=0}^\infty\frac{1}{l!}{\cal B}_{<iL>}\nnr{L}\;,\\
\nonumber\\\label{zz4}
\mathfrak{P}^i(t,\nnx)&=&\frac13\nnaa{i}\nnv{k}\nnr{k}-\frac16\da{i}\nnr{2}+\sum\limits_{l=1}^{\infty}
        \frac{1}{l!}\left(\dot{\cal D}_{<iL>}-\nnv{k}{\cal D}_{<ikL>}\right)\nnr{L}-\nnv{k}{\cal D}_{<ik>}
        \\\nonumber&&+\varepsilon_{ipq}\sum\limits_{l=1}^{\infty}\frac{1}{l!}\left(\dot{\cal F}_{<pL-1>}-
        \frac{l}{l+1}\nnv{k}{\cal F}_{<pkL-1>}\right)\nnr{qL-1}\\\nonumber&&-\nnv{k}\varepsilon_{ipk}\sum
        \limits_{l=0}^{\infty}\frac{1}{(l+1)!}
        {\cal F}_{<pL>}\nnr{L}+\sum\limits_{l=0}^{\infty}\frac{1}{l!}\dot{\cal E}_{<L>}\nnr{<iL>}\\
        \nonumber&&-\sum\limits_{l=0}^{\infty}\frac{l+1}{l!}{\cal E}_{<L>}\nnv{<i}\nnr{L>}\;,
\er\br                                                         \label{zz5}
\mathfrak{R}^i_{\;j}(t,\nnx)&=&-\frac13\nnaa{i}\nnr{j}+\sum\limits_{l=0}^{\infty}
        \frac{1}{l!}\left({\cal D}_{<ijL>}+\delta_{ij}{\cal E}_{<L>}+\frac{1}{l+1}
        \varepsilon_{ipj}{\cal F}_{<pL>}\right)\nnr{L}\\\nonumber&&+\varepsilon_{ipq}\sum\limits_{l=0}^{\infty}
        \frac{l+1}{(l+2)!}{\cal F}_{<jpL>}\nnr{qL}+\sum\limits_{l=0}^{\infty}
        \frac{1}{l!}{\cal E}_{<jL>}\nnr{<iL>}\\\nonumber&&-2\sum\limits_{l=0}^{\infty}
        \frac{1}{(2l+3)l!}{\cal E}_{<iL>}\nnr{<jL>}\;.
\er
Elements of the inverse matrix $\left(\Lambda^{-1}\right)^{\alpha}_{\;\beta}$
can be deduced from Eqs. (\ref{zz1}) -- (\ref{zz5}) by applying relationship (\ref{qaz}).

Formulas (\ref{zz1}) -- (\ref{zz5}) allow us to evaluate the range
of applicability of the local coordinates. Radius of this
range is determined by the condition that determinant of the coordinate transformation matrix
$\Lambda^{\alpha}_{\;\beta}$ is zero. Calculating the
determinant of the matrix one obtains
\br\label{ma5}
{\rm
det}\left(\Lambda^{\alpha}_{\;\beta}\right)&=&1+\epsilon^2\left[-\dot{\cal
A}+3\,{\cal E}-\frac43\left(\nnaa{k}-\frac52{\cal
E}_k\right)\nnr{k}\right.\\\nonumber&&\left.+\sum_{l=2}^\infty\frac{(l+1)(2l+3)}{(2l+1)l!}{\cal
E}_{<L>}\nnr{L}\right]+O\left(\epsilon^4\right)\;.
\er
Radius of
convergence of the polynomial in the right side of Eq. (\ref{ma5})
crucially depends on the choice of functions ${\cal E}_L$. In what
follows \footnote{See Eqs. (\ref{3.21a}), (\ref{3.22})}, we shall prove that it is possible to make function
${\cal E}_i=\nnaa{i}$ and all other functions ${\cal E}_L=0$ for any $l\ge2$. Then if one also takes into account
(\ref{3.15}) and (\ref{3.21}) (putting $Q=Y=0$ there) the
determinant (\ref{ma5}) turns to zero when the distance $\nnr{}\approx
c^2/(2\nnaa{})$. In case of the local geocentric frame attached to the Earth and moving around the Sun
with acceleration $\nnaa{}\simeq 0.6 cm/s^2$ this
distance $\nnr{}$ is about $10^{21}$ cm.  Hence, the local geocentric frame covers a region which includes the
entire solar system.
In case of a binary pulsar with a
characteristic size of the orbit $\sim 10^{10}$ cm the local
coordinate system attached to the pulsar is spread out of the binary
system at the distance about $10^{14}$ cm which also significantly
exceeds the distance between the pulsar and its companion. This remark can be important for researchers
doing analysis of physical processes going on in pulsar's magnetosphere \cite{mestel}.

This
consideration suggests that the metric tensor defined originally in the
local coordinates only in the domain restricted by the distance to the nearest external gravitating body can
be extrapolated far beyond this boundary. Such extrapolation can
be accomplished by choosing another form of the solution of the
homogeneous field equations describing background gravitational field of external bodies. Research in this direction
has been pursued in \cite{klv}.

\section{Matching the Post-Newtonian Expansions of the Metric Tensor and a Scalar Field }

\subsection{Historical Background}

The method of matched asymptotic expansions has been developed for finding solutions
of ordinary and/or partial differential equations and is well known in
mathematics for a long time (see, for instance, \cite{math1},
\cite{math2} and references therein). The idea of implying
this method in general relativity goes back to earlier work of
Einstein and Rosen \cite{einros} where the authors discussed
the problem of motion of gravitationally interacting particles by
treating them as topological
structures (`bridges') in the space-time manifold being a
regular solution of the Einstein field equations. Fock \cite{fock}
applied the matching technique
to join the metric tensor expansions of the `near' and
`far-radiative' zones of an isolated astronomical system emitting
quadrupole gravitational radiation (see also \cite{blan} where this procedure is discussed in more detail).
Manasse \cite{manas} studied
radial fall of a small black hole onto a massive
gravitating body and calculated distortion of the shape of the black hole's horizon
by making use of the matching technique. Thorne \cite{t69} and Burke \cite{b}
suggested to use the matching technique for imposing an
outgoing-wave radiation condition on the post-Newtonian metric tensor
for an isolated system emitting gravitational waves. This method
helps to chose a causal solution of the homogeneous Einstein
equations in the post-Newtonian approximation scheme and to
postpone appearance of ill-defined (divergent) integrals, at least, up to the
fourth PNA \cite{a1982,kmad,a1983}. Demiansky \& Grishchuk
\cite{dg} used the matching technique to show that a
black hole orbits its companion of a comparable mass in accordance
with the Newtonian equations of motion. At about the same time,
D'Eath \cite{das1,das2}  explored the idea proposed by
Hawking and worked out a detailed analysis of the problem of motion
of two Kerr black holes comprising a binary system by making use of
matching of the internal (local coordinates) and external (global coordinates) solutions of the
Einstein equations. D'Eath derived general relativistic equations of
motion of the black holes in the first post-Newtonian (1 PN) approximation. Kates \cite{katesa,katesb}
extended his analysis and obtained
the gravitational radiation-reaction force (2.5 PNA) for the black
holes in a binary system. He has also elaborated on a
rigorous mathematical treatment of the matched asymptotic expansions
technique for various applications in general relativity \cite{katesc}.
Damour \cite{dam1983} used the asymptotic matching to solve the
problem of motion of two spherically-symmetric and non-rotating
compact bodies with the gravitational radiation reaction force
taken into account. He proved that mass
of each of the bodies, which appears in the external solution of the two-body problem as a constant parameter,
is the same as that characterizing the
Schwarzschild metric of a non-rotating black hole. Thorne and Hartle \cite{th} applied the matching technique
to study the problem of translational
motion and precession of compact bodies having quadrupole gravitational fields. Their method combined with
the mathematical technique of
D'Eath \cite{das1} was employed in \cite{kop1988}
to derive the post-Newtonian equations of motion of
extended bodies making up an N-body system in the weak-field and slow-motion approximation. The paper
\cite{kop1988} also demonstrates for the case of N-body problem how to construct
a local coordinate system with the origin moving exactly along the world line
of the center of mass of an extended body which has arbitrary shape and is rotating.
The matching technique used in \cite{kop1988} led to the development of the Brumberg-Kopeikin (BK) formalism in the
theory of astronomical reference frames and have been improved
in a series of subsequent publications \cite{kop1991,bk-rf,bk-kin,bk-nc,bk-cm,klv}. Similar matching technique and
Damour-Soffel-Xu (DSX) formalism were formulated in \cite{dsx1991,dsx1992} to describe the post-Newtonian celestial
mechanics of isolated astronomical systems. Both BK and DSX formalisms were used
later as a basis of the resolutions on the relativistic reference frames and time scales in the
solar system adopted by the 20-th General Assembly of the IAU \cite{iau2000}. In the present paper
we extend the general relativistic theory of reference frames and apply the matched asymptotic expansions technique
to the case of the scalar-tensor theory of gravity.
This will allow us to incorporate unambiguously the adopted IAU resolutions on reference frames in the solar system
to the
parametrized post-Newtonian (PPN) formalism \cite{will}, thus, making more close the link between the experimental
gravity and modern observational astronomy.

For the benefit of readers we emphasize here the differences between the results obtained in
BK formalism \cite{kop1988,kop1991,bk-rf,bk-nc} and in this paper and those contained in DSX
formalism \cite{dsx1991,dsx1992} and in \cite{kls,dames}.
These differences are as follows:
\begin{itemize}
\item Original BK formalism deals with the Newtonian definitions of the multipole moments while DSX formalism
operates with the post-Newtonian multipoles. This paper extends the BK formalism and incorporates the
post-Newtonian multipoles into the matching technique and equations of motion.
\item BK and DSX formalisms have been developed in the framework of the general theory of relativity only.
The present paper extends the BK formalism to the class of scalar-tensor theories of gravity.
\item Damour and Esposito-Far\'ese \cite{dames} generalized definitions of the post-Newtonian multipole moments
for an isolated N-body system for the case of the scalar-tensor theory of gravity and concentrated on discussion
of experimental tests of this theory for binary pulsars and gravitational wave astronomy. They did not work out
any matching procedure for construction of local frames of reference and derivation of equations of motion of
each body in the local coordinates. The present paper develops the matching procedure in the scalar-tensor theory
of gravity and constructs a set of global and local coordinate frames for description of both global and local
dynamics of the N-body system  as a whole and each body from the system separately. In the present paper we also
construct the post-Newtonian definition of the multipole moments for each body being a member of the N-body system.
\item Klioner and Soffel \cite{kls} used DSX formalism to supplement two-parametric PPN formalism by the DSX matching
technique and construct a set of global and local coordinate frames for description of the dynamics
of the N-body system. They also used PPN-parametrized definitions of the post-Newtonian multipoles given
in  \cite{dames}. Klioner and Soffel did not rely upon a particular class of alternative gravitational
theory and abandoned the use of gravitational field equations. The present paper makes use of the field
equations of the scalar-tensor theory of gravity and applies the matching technique of the BK formalism.
We show that Klioner-Soffel results \cite{kls} do not match with the scalar-tensor theory of gravity which
makes domain of applicability of the phenomenological theory of reference frames constructed in \cite{kls}
uncertain. Further discussion of the Klioner-Soffel approach and comparison of results of their
paper \cite{kls} with those obtained in the present review article are deferred to appendix \ref{ksfor}.
\end{itemize}

\subsection{Method of the Matched Asymptotic Expansions in the PPN Formalism}

Method of the matched asymptotic expansions \footnote{It is also called the boundary layer method \cite{zwill}.}
is one of the powerful mathematical tools for solving differential equations with a small parameter present
for which a regular perturbation series method is inadequate. It occurs as often as a solitary solution can
not match all the boundary conditions in a differential equation \cite{math1}, \cite{math2}. If a regular
perturbation series can not be applied, there may be one or more regions where the solution
can be represented in the
form of asymptotic expansion with respect to one or more small parameters which satisfies at least one of
the boundary conditions. Matching the asymptotic expansions in the buffer region (boundary layer) where
at least two of them are valid and convergent, allows us to find the law of transformation from one expansion
to another and to retrieve coefficients of these expansions.

In the present paper the main asymptotic post-Newtonian expansions, which are used in the matching procedure,
are solutions of the
gravity field equations for the metric tensor
and scalar field found in the global and local coordinates that are subject to different boundary conditions
imposed respectively at infinite distance and/or at the origin of the local coordinates.
These solutions are shown in Eqs. (\ref{12.5})
-- (\ref{12.9}) and (\ref{1.7}), (\ref{1.7a}), (\ref{1.24b}) -- (\ref{1.26b}). The solution for the
metric tensor and scalar field in the global coordinates is valid
everywhere inside and outside of the N-body system up to infinity. This is because we assumed that the
gravity field is weak everywhere and there are no singularities in the space-time manifold
\footnote{Had one worked with the tensor of energy-momentum of the point-like massive particles one would
have singularities on the particle's world lines \cite{infeld}. We argue that in such a case the method
of the matched asymptotic expansions is the only way to derive equations of motion of these particles
without ambiguities present in the higher-order post-Newtonian approximations \cite{blanchuk}.}.

Because we do not deal with space-time singularities the reader may think that one could use the global
coordinates alone to describe relativistic celestial dynamics of the bodies from the N-body system.
However, this idea does not work out for two reasons. First, the local coordinates are still required
to give physically meaningful definition of the multipole moments of each body and one must know how
these definitions correlate with the definitions of these multipoles in the global coordinates. This relationship
between the two definitions of the multipole moments is a key element in the procedure of derivation of equations
of motion of extended bodies having finite size. We discuss importance of this issue in section \ref{wep}
in more detail. Second,
the global coordinate frame is an inappropriate reference for
analysis of gravitational experiments conducted in the vicinity of the Earth. This is because the Earth is
both moving and embedded to the gravitational field of other bodies of the solar system.  Simple translation
of the origin of the global coordinates to the geocenter (the Galilean-Newtonian transformation) frequently
used in early publications \footnote{See, for example, \cite{will,fock,brr,spyr1,capor} and references therein.
One may notice, however, that Will understood this problem fairly clear long time ago as follows from the
discussion in section 6.2 of his book \cite{will}.}  does not take into account relativistic aspects of the
coordinate transformations on the space-time manifold and, hence, can not eliminate a large number of
coordinate-dependent (and for this reason nonphysical) effects which will be present in such treatment of
astronomical observations \cite{kop1988,bk-rf,mashn,srs}. The number of the coordinate-dependent effects
is much smaller if one uses correct relativistic procedure to transform the global to local coordinates.
Such post-Newtonian transformation simplifies drastically analysis of astronomical observations and
description of the dynamics of lunar motion and/or that of artificial satellites \cite{bk-nc,brberg,dsx1994}.

The internal solution for the metric
tensor and scalar field in the local coordinates contain the external multipole moments which can not be
found as explicit functions of time without
matching of the local solution of the gravity field equations to the global one. The matching allows us to
express the external multipole moments in terms of the gravitational potentials (\ref{12.9a})
characterizing the global metric tensor and scalar
field. At the same time the matching procedure determines the
structure of the PPN coordinate transformation between the global and
local coordinates.

Matching of the local and global solutions of the metric tensor and scalar field is based on Eqs. (\ref{2.5}) and
(\ref{2.6}), and consists of the following steps \footnote{Solution of some problems (for instance,
in cosmology \cite{dmyx}) requires to match not only the metric tensor but also its first derivatives.
In this paper this requirement is redundant as the metric tensor and scalar field are smooth differentiable
functions in the matching domain and their derivatives of any order have no jumps. Therefore, Eq. (\ref{2.7})
is a consequence of Eq. (\ref{2.6}).}:
\begin{enumerate}
\item[\bf Step 1.] One re-writes the local metric tensor and scalar field in the right side of Eqs. (\ref{2.5})
and (\ref{2.6}) in terms of the global coordinates $(t,{\bm x})$. This is achieved by making use of a Taylor
expansion of $\nxi$ and $\hat{g}_{\alpha\beta}(u,{\bm w})$ around the point $x^\alpha=(ct,{\bm x})$.
\item[\bf Step 2.] One calculates the partial derivatives of the local coordinates with respect to the global
ones, that is the matrix of transformation of the coordinate bases given in section (\ref{trbs}).
\item[\bf Step 3.] One separates the `global' gravitational potentials in the left side of Eqs. (\ref{2.5})
and (\ref{2.6}) relating to the central body (Earth) and to the external bodies (Moon, Sun, etc.) respectively:
\br
   U(t,\nnx)&=&\nnj+\bar{U}(t,\nnx)\,,                     \label{3.1}
\\\nonumber\\
   U^i(t,\nnx)&=&U_i^{{\sss(B)}}(t,\nnx)+    \label{3.2}
        {\bar{U}}^i(t,\nnx)\,,
\\\nonumber\\
   \chi(t,\nnx)&=&\chi^{{\sss(B)}}(t,\nnx)+  \label{3.3}
         \bar{\chi}(t,\nnx)\,,
\\\nonumber\\
   \nnff{k}&=&\nng{B}{k}{t}{x}+ \bar{\Phi}_k(t,\nnx)\;,       \label{3.4}
    \qquad\qquad(k=1,\,...\,,4)\,,
\er where functions with index $({\rm B})$ are given by integrals (\ref{12.10})--(\ref{12.17}) taken over the
volume of the central body only, and the bar over other functions indicates here and hereafter
that the corresponding sum in the definitions (\ref{12.9a}) of
these functions excludes the central body with the index (B), that is the sum takes into account only external bodies
\be\label{omap}
\bar{U}=\sum_{A\not=B} U^{\sss(A)}\,,\qquad \bar{U}_i=\sum_{A\not=B}
U_i^{\sss(A)}\,,\qquad\bar{\Phi}_k=\sum_{A\not=B} \Phi^{\sss(A)}_k\:,\qquad
      \bar{\chi}=\sum_{A\not=B} \chi^{\sss(A)}\:.
¨\en
\item[\bf Step 4.] One expands the gravitational potentials of the external masses (that is functions with
bars in Eqs. (\ref{3.1}) -- (\ref{omap})) in Taylor's series in powers of $R^i_{\sss B}=x^i-x^i_{\sss B}$
in the vicinity of the origin of the local coordinates, that is the point $x^i=x^i_{\sss B}$.
\item[\bf Step 5.] One equates similar terms of these Taylor expansions from the left side of the matching
equations (\ref{2.5}) and (\ref{2.6}) with the corresponding Taylor expansions entering their right side.
\item[\bf Step 6.] One separates symmetric and anti-symmetric tensor parts in the matching equations and
determine all coefficients in the local metric tensor and scalar field, which remained undetermined so far,
as well as coefficients in the coordinate transformations. This fixes the residual gauge freedom and gives
equations of translational and rotational motion of the local reference frame.
\end{enumerate}
Let us now explain each step of the matching procedure in more detail.

\subsection{Transformation of Gravitational Potentials from the Local to Global Coordinates}
\subsubsection{Transformation of the Internal Potentials}

\begin{figure}[p]
\centerline{\psfig{figure=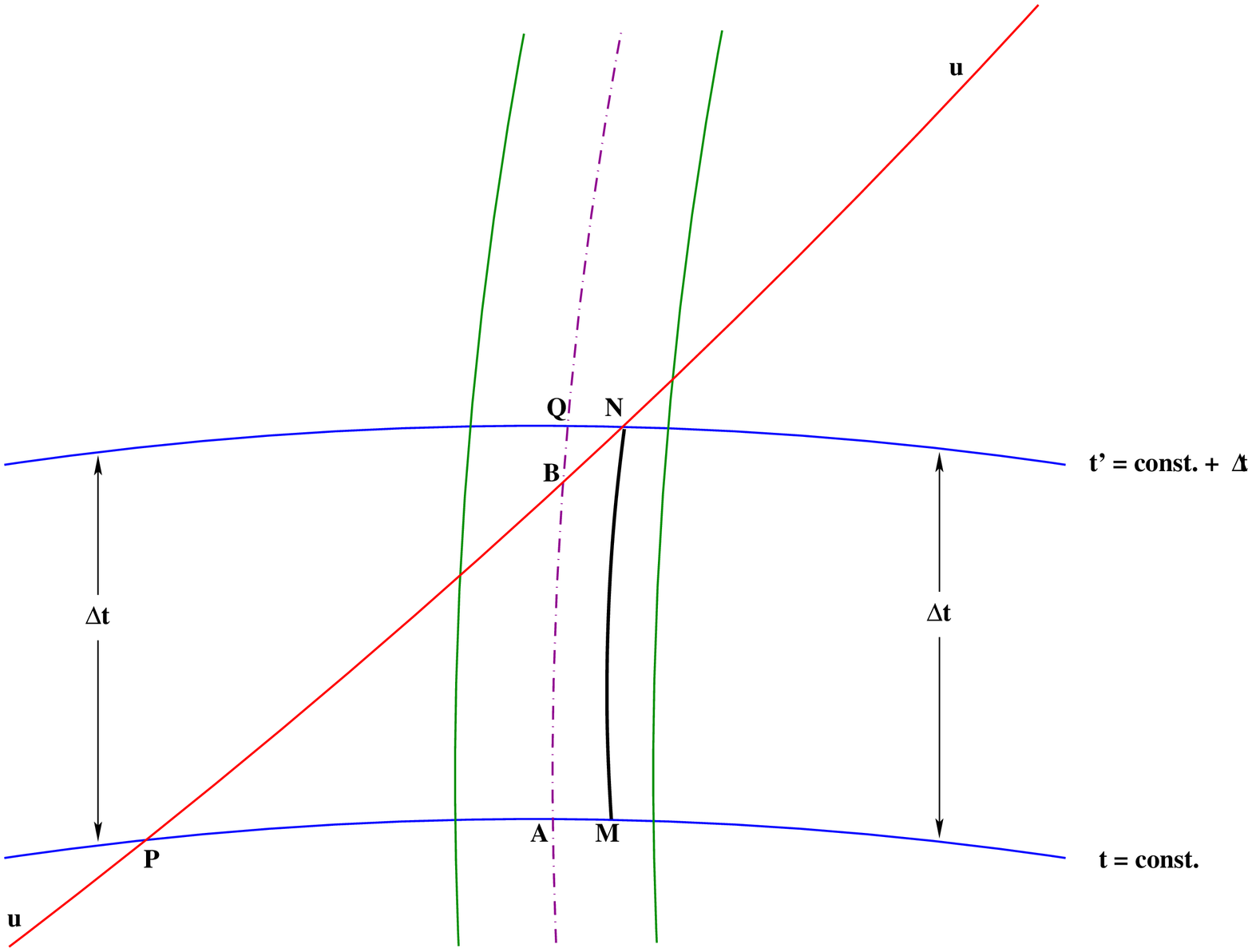,angle=-90,height=14cm,width=18cm}}
\vspace{1cm}
\caption[Lie transfer]{Lie Transfer} Matching the local and global coordinates requires to calculate integrals
of various functions depending on the internal structure of body over two different hypersurfaces of constant
time, $u$ and $t$, as shown in Fig. \ref{fig3}. Relationship between the integrals taken on the two hypersurfaces
is set up by making use of the Lie transfer from one hypersurfaces to another. The integral curves used in the
Lie transfer are the world lines of the four-velocity of body's matter. Only one of such lines, MN, is shown
in the figure. Dashed line is the world line of the origin of the local coordinates which coincides with the
center of mass of the body under consideration. The center of mass is not associated with four-velocity of
body's matter.
\label{fig4}
\end{figure}

At the first step of the matching procedure
one has to transform the metric tensor and the scalar field in the
right side of matching equations (\ref{2.5}) and (\ref{2.6}) from the local, $w^\alpha=(cu,{\bm w})$,
to global, $x^\alpha=(ct,{\bm x})$, coordinates
\footnote{It is also conceivable to make a reciprocal transform of all functions in the left side of
equations (\ref{2.5}) and (\ref{2.6}) to the local coordinates $w^\alpha=(cu,{\bm w})$. However, it
is more convenient and simpler to transform the metric tensor and scalar field from the right side
of Eqs. (\ref{2.5}) and (\ref{2.6}) to the global coordinates $x^\alpha=(ct,{\bm x})$ in accordance
to the transformations (\ref{2.2}), (\ref{2.3}), (\ref{2.12}) -- (\ref{eq2}) which are already displayed
in terms of the global
 coordinates.}. We remind that
the internal gravitational potentials associated with the scalar field, Eq.
(\ref{1.7}), and the metric tensor, Eqs. (\ref{1.11}) -- (\ref{1.13}), are
defined in the local coordinates $w^\alpha=(cu,{\bm w})$ as integrals over hypersurface of a constant
coordinate time $u$. On the other hand, the corresponding gravitational potentials, Eqs.
(\ref{12.10}) -- (\ref{12.16}) are defined in the global coordinates $x^\alpha=(t,{\bm x})$ as
integrals over hypersurface of a
constant coordinate time $t$. These two hypersurfaces do not
coincide and can intersect only at the points that form a
2-dimensional sub-hypersurface  (see Fig. \ref{fig3}). For this reason, in order to
transform the internal potentials defined in the local coordinates, $w^\alpha=(cu, {\bm
w})$, to those defined in the global coordinates, $x^\alpha=(ct, {\bm x})$, one needs to make
a pointwise transformation given by Eqs. (\ref{2.2}) -- (\ref{2.3}) along with a
Lie transform of integrands of the integrals which moves the integrands
from the hypersurface of constant time $u$ to that
of constant time $t$ (see Fig \ref{fig4}).
This procedure was
worked out in \cite{kop1991} \footnote{See also \cite{bk-nc,wwo}.} and is described below.

Let us assume that the field point ${\sf P}$, at which matching of the internal and external solutions of
the metric tensor and the scalar field is done, has global coordinates
$x^\alpha({\sf P})=(ct, {\bm x})$ and local coordinates
$w^\alpha({\sf P})=(cu, {\bm w})$ (see Fig. \ref{fig4}). These
coordinates are related by the pointwise transformation given by Eqs.
(\ref{2.2}) -- (\ref{2.3}). By definition, the matching
point ${\sf P}$ belongs simultaneously both to the hypersurface of
constant time $u$ and to that of the constant time $t$. Let us
consider a matter element of the central body B located at the point ${\sf N}$ and lying on
the same hypersurface of the constant time $u$ inside of the body's world
tube. We assume that the point
${\sf N}$ has local coordinates, $w^\alpha({\sf N})=(cu, {\bm
w}'(u))$, and global coordinates, $x^\alpha({\sf N})=(ct', {\bm
x}'(t'))$. We emphasize that the time coordinate $u$ of
the points ${\sf P}$ and ${\sf N}$ is the same because they
are located on the same hypersurface of the constant time $u$. However, the time coordinate $t'$ of the point
$N$ is different from the time coordinate $t$ of the point $P$, that is $t'\not=t$, because the hypersurfaces
of the
constant global time passing through the points ${\sf P}$ and
${\sf N}$ are different. Let us consider a world line of the element of
the body's matter passing through the point ${\sf N}$ and intersecting a
hypersurface of the constant time $t$ at the point ${\sf M}$. This
world line allows us to map coordinates of the element of the body's matter from the
hypersurface $t'$ to the hypersurface $t$. By the construction, the point ${\sf
M}$ must have the global coordinates $x^\alpha({\sf M})=(ct, {\bm x}'(t))$.

One can expand the spatial coordinates of the world line of the chosen element of the body's matter in a Taylor
series with respect to time
\be                         \label{plk}
x'^i(t')=x'^i(t)+v'^i(t)(t'-t)+O\left(\Delta t^2\right)\;, \en
where $v'^i$ is the spatial velocity of the body's element at the
point ${\sf M}$ and $\Delta t\equiv t'-t$. One denotes by letters ${\sf O}$ and ${\sf Q}$  position of
the origin of the local coordinates on two hypersurfaces, $t$
and $t'$, respectively (see  Fig. \ref{fig4}). Global coordinates of the origin of the local coordinates,
taken on two different
hypersurfaces, are related by equation
\be \label{lko} x_{\sss
B}^i(t')=x_{\sss B}^i(t)+v_{\sss B}^i(t)(t'-t)+O\left(\Delta
t^2\right)\;. \en

Now we shall find the time interval $\Delta t=t'-t$ separating the
two hypersurfaces of the constant global coordinate time, $t$ and $t'$, under condition that the matching
point ${\sf P}$ is fixed. At the point ${\sf N}$
the relationship between the local time $u$ and the global time $t',$ according to Eq. (7.2.1), is
\be
\label{poi} u=t'+\epsilon^2\xi^0(t',{\bm
x}')+O\left(\epsilon^4\right)\;. \en
Subtracting Eq. (\ref{2.2})
from Eq. (\ref{poi}) and accounting for the fact that the space-time
interval between points ${\sf N}$ and ${\sf P}$ is small we obtain
\be                                         \label{tyu}
\Delta t=t'-t=\epsilon^2\left[\xi^0(t,{\bm
x})-\xi^0(t,{\bm x}')\right]+O\left(\epsilon^4\right)\;, \en
where the point
${\bm x}'\equiv{\bm x}'(t)$, that is, it belongs to the hypersurface
$t$.

Local coordinates of the point ${\sf N}$ are transformed to the
global coordinates as follows
\be\label{mnh} w'^i=x'^i(t')-x_{\sss
B}(t')+\epsilon^2\xi^i\left(t',{\bm x}'(t')\right)\;. \en
Expanding functions in the right side of Eq. (\ref{mnh}) in the
vicinity of the time instant, $t$, and taking into account Eqs.
(\ref{plk}) -- (\ref{tyu}) yields
\be \label{sot} w'^i=R'^i_{\sss
B}+\epsilon^2\left[\xi'^i-(v'^i-v^i_{\sss
B})\left(\xi'^0-\xi^0\right)\right]+O\left(\epsilon^4\right)\;,
\en
where $\xi'^i\equiv\xi^i(t,{\bm x}')$,
$\xi'^0\equiv\xi^0(t,{\bm x}')$,
and $R'^i_{\sss B}= x'^i(t)-x^i_{\sss B}(t)$.

Transformation (\ref{sot}) is used for deriving relationship
between the absolute values of functions $|{\bm w}'-{\bm w}|$ and
$|{\bm x}'-{\bm x}|$ which enter denominators of all integrands in the integrals defining the
internal gravitational potentials. Subtracting Eq. (\ref{2.3}) from
Eq. (\ref{sot}) and taking the absolute value of the difference gives
\be \label{gqm} |{\bm w}'-{\bm w}|=|{\bm x}'-{\bm
x}|+\epsilon^2\left[n^k\left(\xi'^k-\xi^k\right)-
n^k\left(v'^k-v_{\sss
B}^k\right)\left(\xi'^0-\xi^0\right)\right]+O\left(\epsilon^4\right)\;,
\en where $\xi'^i\equiv\xi^i(t,{\bm x}')$, $\xi^i\equiv\xi^i(t,{\bm x})$,
and $n^i\equiv(x'^i-x^i)/|{\bm x}'-{\bm x}|$.

We must also perform a Lie transform to find a relationship between the volume elements
$d^3w'$ and $d^3x'$ taken at the points ${\sf N}$ and ${\sf M}$
respectively. We note that the invariant density $\rho^*$
introduced in Eq. (\ref{11.19}) has one more remarkable
property in addition to Eq. (\ref{11.20}) \cite{kop1991}.
Specifically, the Lie derivative of the product of the invariant
density and the volume element is zero, which means that
\be\label{li} \rho^*\left(t, {\bm x}'(t)\right) d^3x'({\sf
M})=\rho^*\left(t', {\bm x}'(t')\right) d^3x'({\sf
N})=\rho^*\left(u,{\bm w}'\right) d^3w'({\sf N})\;. \en

Locally measurable velocity ${\nu'}^i$ of the body's element at the
point ${\sf N}$ is defined with respect to the origin of the local
coordinate system. It relates to velocity ${v'}^i$ of the
same element of the body taken at point ${\sf M}$ in global
coordinates by equation:
\be\label{v}
{\nu'}^i(u)={v'}^i(t)-v^i_{\sss B}(t)+O\left(\epsilon^2\right)\;,
\en which can be derived from Eq. (\ref{sot}) by direct
differentiation with respect to time.

By making use of Eqs. (\ref{gqm}) -- (\ref{v}) we finally obtain
transformations of the internal gravitational potentials from the local to
global coordinates. These transformations are given by the following equations
\br\label{3.5} \nnc&=&
\nnj+\epsilon^2\mathcal{U}^{{\sss(B)}}(t,\nnx)+O(\epsilon^4)\;,
     \\\nonumber\\
     \hat{U}_i^{{\sss(B)}}(u,\nnw)&=&
     U_i^{{\sss(B)}}(t,\nnx)-                \label{3.6}
     \nnv{i}\nnj+O(\epsilon^2)\;,
\\\nonumber\\
     \hat{\chi}^{{\sss(B)}}(u,\nnw)&=&         \label{3.7}
     \chi^{{\sss(B)}}(t,\nnx)+O(\epsilon^2)\;,
\\\nonumber\\
     \nngk{1}&=&\nng{B}{1}{t}{x}+\nnv{2}\nnj-
     2\nnv{i}U^i_{{\sss(B)}}(t,\nnx)+        \label{3.8}
     O(\epsilon^2)\;,
\\\nonumber\\
     \nngk{2}&=&\nng{B}{2}{t}{x}-                                \label{3.9}
     G\nnh{B}{\rho^*\bar U(t\,,x)}+O(\epsilon^2)\;,
\\\nonumber\\
     \nngk{3}&=&\nng{B}{3}{t}{x}+O(\epsilon^2)\;,                \label{3.10}
\\\nonumber\\
     \nngk{4}&=&\nng{B}{4}{t}{x}+O(\epsilon^2)\;,                  \label{3.11}
\er where the post-Newtonian correction,  $\mathcal{U}^{{\sss(B)}}(t,\nnx)$, to the Newtonian potential,
$\nnj$, reads  \be \label{ozs}
\mathcal{U}^{{\sss(B)}}(t,\nnx)=G{\rm\bf
I}^{{\sss(B)}}_{-2}\left\{\rho^*n^k\left(v'^k-v_{\sss
B}^k\right)\left(\xi'^0-\xi^0\right)-\rho^* n^k
\left(\xi'^k-\xi^k\right)\right\}\;. \en
This correction is the result of the post-Newtonian coordinate transformation (\ref{2.2}), (\ref{2.3})
and the Lie-transfer of the integrand of the Newtonian gravitational potential from local to global
coordinates. Transformation of all other internal gravitational potentials from local to global
coordinates does not require to take into account their relativistic corrections as it would exceed
the accuracy of the first post-Newtonian approximation which is redundant.

The matching procedure also requires to derive transformation of
the second time derivative of the potential $\chi$ in explicit form. This
transformation can be directly obtained from the definition
of the potential given by Eq. (\ref{12.12})  and the mapping equation (\ref{3.7}). After straightforward
calculation one gets
\be \frac{\partial\hat {\chi}^{{\sss(B)}}(u,\nnw)}{\partial u^2}
             =
             c^2\chi^{{\sss(B)}}_{,00}
             (t,\nnx)+
             \nnaa{k}\chi^{{\sss(B)}}_{,k}
             (t,\nnx)                                      \label{3.12}
             +2c\nnv{k} \chi^{{\sss(B)}}_{,0k}
             (t,\nnx)+
             \nnv{i}\nnv{j} \chi^{{\sss(B)}}_{,ij}
             (t,\nnx)+O(\epsilon^2)\;,
        \en
where $\chi^{{\sss(B)}}_{,0}$ and $\chi^{{\sss(B)}}_{,i}$ denote partial derivatives of $\chi^{{\sss(B)}}$ with
respect to the global time coordinate $x^0=ct$ and the spatial coordinate $x^i$
respectively.

\subsubsection{Transformation of the External Potentials}

External potentials in the internal solution for the metric tensor in local coordinates $w^\alpha=(cu,{\bm w})$
depend on the external multipole moments $Q_L =Q_L(u)$, $C_L=C_L(u)$,
$P_L=P_L(u)$ depend only on time $u$ and  are `pinned down' to the origin of the local coordinates
located at the point
${\sf B}$ where the hypersurface of the constant time $u$ intersects with the world line of the
origin (see Fig \ref{fig4}). On the other hand, all functions entering the left side of matching
equations (\ref{2.5}), (\ref{2.6}) are defined on the hypersurface of constant global coordinate
time $t$.  Hence, before doing the pointwise coordinate transformation of the external potentials
we must perform a Lie transfer of the potentials
along the world line of the origin of the local coordinates
from point ${\sf B}$ to
point ${\sf A}$ located
on the hypersurface of the constant global coordinate time $t$. Time shift $\Delta t$ along
this world line is determined by Eq. (\ref{tyu}) where one has to
associate the point $\nnx'$ with the
origin of the local coordinates, that is $\nnx'=\nnx_{\sss B}$, under condition that the matching
point ${\sf P}$ is taken the same as defined in previous section.
Keeping in mind that the external potentials are scalars with respect to the Lie transport, one obtains
\be \label{n1} Q_L({\sf
B})=Q_L({\sf A})+\dot Q_L({\sf A})\Delta t+O\left(\Delta
t^2\right)\;, \en
where overdot means differentiation with
respect to time $t$. After making use of Eqs. (\ref{2.2}) and
(\ref{tyu}), and accounting that local coordinates of the point ${\sf
B}$ are, $w^\alpha({\sf B})=(cu,0)$, while the global coordinates of the
point ${\sf A}$ are, $x^\alpha({\sf B})=(ct,\nnx_{\sss B}(t))$, one
gets \be \label{n2} Q_L(u)=Q_L(t)-\epsilon^2\dot
Q_L(t)\left[\xi^0(t,\nnx_{\sss
B})-\xi^0(t,\nnx)\right]+O\left(\epsilon^4\right)\;. \en Similar
formulas are valid for the multipole moments $C_L$ and $P_L$ as well.

Now we can do the pointwise coordinate transformation of the space coordinates given by Eq. (\ref{2.3}).
For the STF product of the local coordinates one has: \be
\label{n3} w^{<i_1i_2...i_l>}=R^{<i_1i_2...i_l>}_{\sss
B}+l\epsilon^2 R^{<i_1i_2...i_{l-1}}_{\sss
B}\xi^{i_l>}+O\left(\epsilon^4\right)\;, \en
where we have used Eq. (\ref{2.3}).
After combining Eqs.
(\ref{n2}) -- (\ref{n3}) all together the post-Newtonian transformation of the
Newtonian part of the external potential of the internal solution of the metric tensor assumes the following form
\br
\label{n4}
\sum_{l=0}^\infty\frac{1}{l!}
Q_L(u)w^L&=&\sum_{l=0}^\infty Q_L(t)R^{L}_{\sss
B}+\epsilon^2\biggl[\xi^0(t,\nnx)-\xi^0(t,\nnx_{\sss
B})\biggr]\sum_{l=0}^\infty\frac{1}{l!}\,
\dot Q_L(t)R^{L}_{\sss
B} \\\nonumber\\\nonumber&&+\epsilon^2\sum_{l=1}^\infty\frac{1}{(l-1)!}\,
Q_{kL-1}(t)R^{<L-1}_{\sss
B}\xi^{k>}+O\left(\epsilon^4\right)\;. \er
This is the
most complicated transformation of the external potential that we need. It takes into account the
post-Newtonian nature of the PPN coordinate transformation and supplements Eq. (\ref{3.5}) for the
internal Newtonian gravitational potential. All other external potentials present in
the local metric (\ref{1.24b}) -- (\ref{1.26b}) are transformed without taking into account the
post-Newtonian corrections by
making use of only the very first term in the right side of Eq.
(\ref{n3}).

\subsection{Matching of a Scalar Field}

Scalar field is used in the post-Newtonian terms only. For this reason
matching of its asymptotic expansions derived in local and global coordinates is quite straightforward.
We operate with
external and internal solutions of the scalar field given by Eqs.
(\ref{12.5}) and (\ref{1.7}), (\ref{1.7a}) respectively. Matching equation
(\ref{2.5}) reveals that the internal potentials referred to the central body B cancel out in its left and
right sides due to Eq. (\ref{3.5}) while the potentials $P_L$ match to the Newtonian potential and its
derivatives referred to the external bodies.  More specifically, for any number $l\geq 0$ the matching yields
\be
       P_L=\bar U_{,L}\nnxe+O(\epsilon^2)\,,                     \label{3.13}
\en
where the external Newtonian potential $\bar{U}$ is defined in Eq. (\ref{omap}) and taken at the origin of the
local coordinates, that is the point $x^i=x^i_{\sss B}(t)$, at the instant of time $t$. Thus, the STF multipole
moment $P_{L}$ of the external scalar field is fully determined
as $l$-th spatial derivative of the Newtonian gravitational potential $\bar U$. We remind that the scalar field
was normalized to the factor $\gamma-1$ \footnote{See Eq. (\ref{not1}) where $\omega+2=1/(1-\gamma)$ in
accordance with definition of the PPN parameter $\gamma$.}, so that physically observed scalar
field $\zeta=(1-\gamma)\varphi$. It vanishes in the limit of general relativity where the multipole
moments $P_L$ play no role.

\subsection{Matching of the Metric Tensor}
\subsubsection{Matching of $g_{00}(t,\nnx)$ and $\hat{g}_{\alpha\beta}(u,{\bm w})$ in the Newtonian approximation}

Newtonian approximation of the matching equation (\ref{2.6}) for  $g_{00}(t,\nnx)$ component of the
metric tensor in its left side yields
\be                                                             \label{j1}
     \hat N(u,\nnw) = N(t,\nnx)+2\mathfrak{B}(t,\nnx)-\nnv{2}+O(\epsilon^2)\,.
\en
Function $\mathfrak{B}(t,\nnx)$ is taken from Eq. (\ref{zz1}) while the global and
local metric tensors are taken from Eqs. (\ref{12.6}) and (\ref{1.24b})
respectively. One finds that after making use of Eq.
(\ref{3.5}) the internal gravitational potentials
$\hat{U}^{\scriptstyle(B)}(u,{\bm w})$ and
${U}^{\scriptstyle(B)}(t,{\bm x})$ drop out of the left and right sides of Eq. (\ref{j1}). Expanding
the external gravitational potential $\bar U(t,{\bm x})$ in a Taylor series around the origin of the
local coordinates, $x^i_{\sss B}$, and
equating similar terms with the same power of $R^i_E$ specifies the matching conditions
\br
    Q+\dot {\cal A}&=&\frac12\nnv{2}+\bar U\nnxe+O(\epsilon^2)\,,   \label{3.15}
\\\nonumber\\
    Q_i&=&\bar U_{,i}\nnxe-\nnaa{i}+O(\epsilon^2)\,,            \label{3.16}
\\\nonumber\\
    Q_L&=&\bar U_{,L}\nnxe +O(\epsilon^2),\qquad\qquad                        \label{3.17}
           (l\ge2)\,.
\er

Equation (\ref{3.15}) makes it evident that two functions ${\cal
A}$ and $Q$ can not be determined from the matching procedure
separately, only their linear combination $Q+\dot{\cal A}$ can be determined. Hence, one of these functions
can be chosen arbitrary. The most preferable choice is to take $Q=0$ as it was done, for example,
in \cite{kop1988,dsx1991,th}. This choice is also consistent with recommendations of the IAU \cite{iau2000,skpw}
and it makes $\hat{g}_{00}(u, {\bm w})$ component of the local
metric tensor equal to $-1$ at the origin of the local coordinates if
gravitational field of the central body (Earth) is neglected. However, if one chooses $Q=0$ the rate of the
coordinate time $u$ can be different from that of the coordinate time $t$ because the average value of
functions $v_B^2$ and $\bar U$ is not zero for elliptic orbits. Hence, the choice of $Q=0$ can be inconvenient
for astronomical data reduction programs in the solar system. Therefore, two time scales, $TDB=k_B t$
and $TDT=k_E u$, have been introduced in such a way that their rate at the origin of the local coordinate
system is the same \footnote{See \cite{iau2000,skpw,irf,bk-cm} for more detail.}. This makes function
\br
\label{pak}
Q(t)=-<\dot{\cal A}>=a+bt+ct^2+...\;,
\er
to be a polynomial of time which numerical coefficients are calculated by means of numerical integration
of Eq. (\ref{3.15}) over sufficiently long interval of time \cite{irf,bk-cm}. Time-rate adjustment
coefficients $k_B$ and $k_E$ relate to each other as  \cite{irf,bk-cm}
\be
\frac{k_B}{k_E}=1+c^{-2}Q(t^*)\;,
\en
where $t^*$ is a certain astronomical epoch chosen by convention.

\begin{figure}[p]
\centerline{\psfig{figure=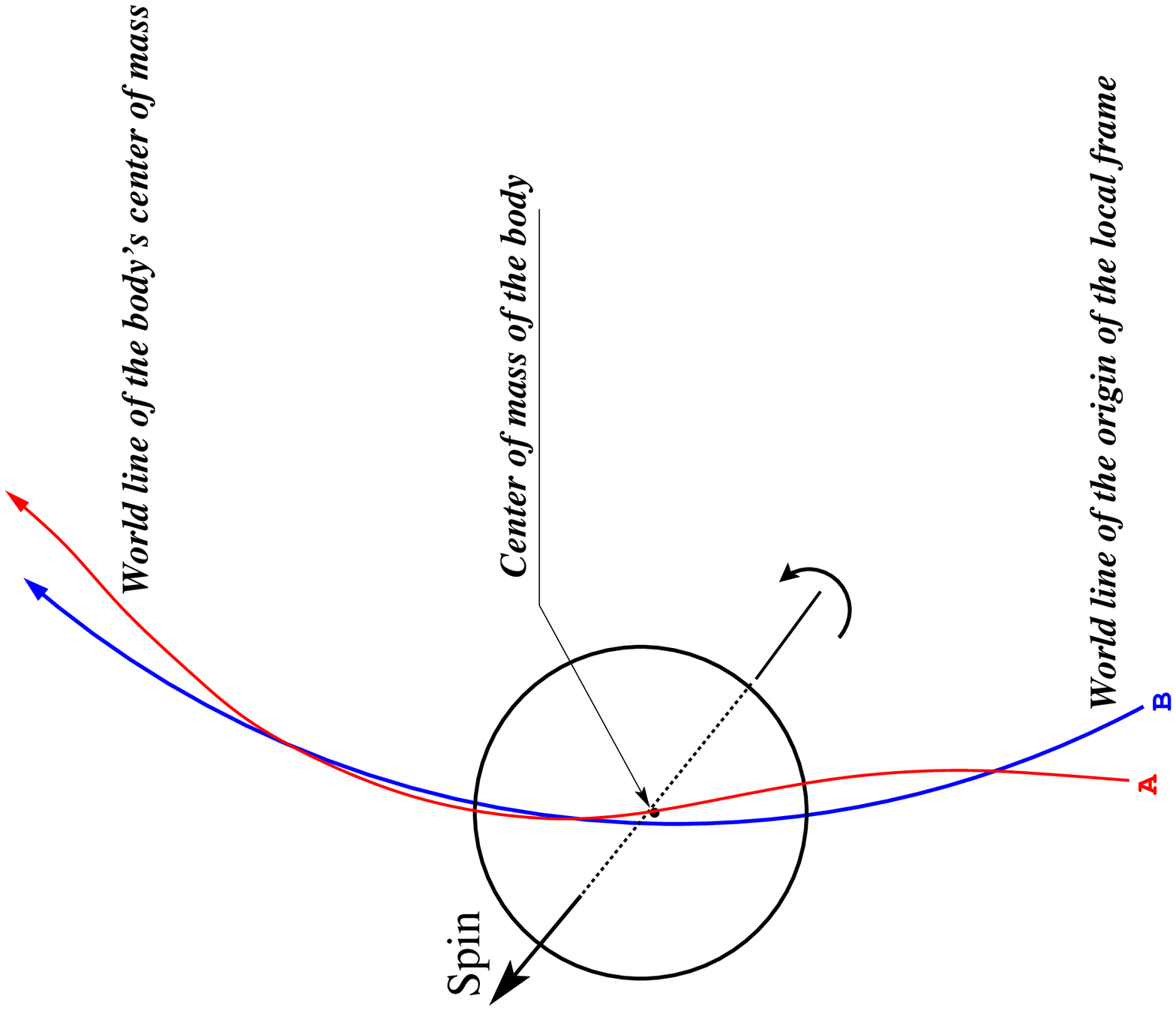,angle=-90,height=14cm,width=18cm}}
\vspace{1cm} \caption[Two World Lines ]{World lines of the origins of
two local coordinates are shown. The
line (A) indicates the world line of the origin of the local
coordinates ${\cal A}$ falling freely in the background
space-time described by the external metric tensor
$g^{(E)}_{\alpha\beta}$. The line (B) depicts motion of the
origin of the local coordinates ${\cal B}$ always located at the
center of mass of the body under consideration. In general, the two world lines do not coincide
due to the existence of the inertial force, ${\cal M}Q_i$, in the local
coordinate system ${\cal B}$. This force arises due to
the gravitational coupling between mass and spin multipole moments
of the body with the gravitoelectric and gravitomagnetic tidal
gravitational fields \cite{mashn,mash2} of the background metric produced
by other bodies of the N-body system.} \label{fig2}
\end{figure}

In accordance with the interpretation given in \cite{mtw,th}, function
$Q_i$ from (\ref{3.16}) must be understood as acceleration measured by
accelerometer being rigidly fixed at rest in the origin of the local coordinates under condition that the
internal gravitational field of the central body B is neglected.
The choice, $Q_i=0$, (see, e.g., \cite{ashb1,th}) leads to construction of a freely falling local coordinate frame
which origin moves along a geodesic world line in the
background space-time manifold defined by the gravitational potentials of
all the celestial bodies of the N-body system but the central body B that is located near the origin of
the local coordinates. Such choice of $Q_i$, however,
disengage the world line of the center of mass of the central body B from that of the origin of the local
coordinates (see Fig. \ref{fig2}).
Indeed, the center of mass of the central body (Earth) does not
move along the geodesic world line due to the interaction of its internal
quadrupole and higher-order multipole moments with the tidal gravitational field of external bodies
\cite{kop1988,kop1991b,kop1991}.  For this reason, a special procedure must be applied for
finding function $Q_i$ which will make the origin of the local coordinates be following the same world
line as the body's center of mass. This procedure is described later in section \ref{xru} in full details.

STF external multipoles $Q_L$ are defined in the Newtonian approximation by Eq. (\ref{3.17}).
They describe gravitoelectric \cite{mashn,mash2} tidal gravitational field of external bodies at
the origin of the local coordinates. Post-Newtonian corrections to the Newtonian value of the
multipoles $Q_L$ can be important for lunar laser ranging and other experimental tests of relativistic
gravity theory in the near-Earth space. The post-Newtonian corrections to the external multipoles can be
also important in construction of the relativistic theory of accretion disc around a star in a close
binary system \cite{cherep}.  These corrections are derived later and shown explicitly in Eq. (\ref{5.9}).

\subsubsection{Matching of $g_{ij}(t,\nnx)$ and $\hat{g}_{\alpha\beta}(u,{\bm w})$ to the order of $O(\epsilon^2)$}

We substitute $g_{ij}(t,\nnx)$ component of the metric tensor in the global coordinates to the left side
of the matching Eq. (\ref{2.6}) and write this equation by
taking into account all post-Newtonian terms of order $O(\epsilon^2)$. We find that in this approximation
the matching equations reads
\be                                                                 \label{3.18}
\hat H_{ij}(u,{\nnw})=H_{ij}(t,{\nnx})-\mathfrak{R}^i_{\;j}(t,\nnx)-\mathfrak{R}^j_{\;i}(t,\nnx)+\nnv{i}\nnv{j} +
          O(\epsilon^2)\;,
\en
where the matrix of transformation $\mathfrak{R}^i_{\;j}(t,\nnx)$ from the local,
$w^\alpha$, to global, $x^\alpha$, coordinates is given in Eq. (\ref{zz5}) whereas components of the metric
tensor are defined by Eq. (\ref{12.7}) in the global coordinates
and by Eq. (\ref{1.26b}) in the local coordinates.  Comparing
similar terms we find that the internal gravitational potentials of the central body B,
$\hat{U}^{\sss(B)}(u,{\bm w})$ and ${U}^{\sss(B)}(t,{\bm x})$, entering the left and right sides of
the matching equation,
cancel each other and drop out of the matching equation completely. The remaining terms belong to the
background gravitational field of external bodies and their matching gives the following set of equations
\br\label{3.21}
 Y+{\cal E}&=&\frac16\nnv{2}+\gamma\bar U\nnxe
                +O(\epsilon^2)\;,\\\nonumber\\\label{3.21a}
{\cal E}_i&=&\nnaa{i}+O(\epsilon^2)\;,
\\\nonumber\\ \label{3.22}
{\cal E}_{L}&=&O(\epsilon^2)\qquad\qquad\qquad\qquad(l\ge2)\;,     \\\nonumber\\\label{3.19}
{\cal D}_{ij}&=&\frac12\nnv{<i}\nnv{j>}
                                      +O(\epsilon^2)\;,
\\\nonumber\\\label{3.20}
{\cal D}_{L}&=&O(\epsilon^2)\qquad\qquad\qquad\qquad(l\ge3)\;,
                                     \\\nonumber\\\label{3.23}
{\cal F}_{L}&=&O(\epsilon^2)\qquad\qquad\qquad\qquad(l\ge2)\;,\er
which define the structure of the PPN transformation between spatial coordinates of the global and local coordinates.

Function $Y=Y(t)$ defines the unit of measurement of spatial distances in the local coordinates.
It would be convenient to chose $Y=0$ as it was done, for example, in \cite{kop1988,th}. However,
introduction of TDB and TDT time scales in ephemeris astronomy must be accompanied by the choice of $Y\not=0$
to compensate effects of function $Q\not=0$ from Eq. (\ref{3.15}) on spatial barycentric (global)
and geocentric (local) coordinates \cite{bk-cm,brs}. In view of this complexity, related to a pure
practical application of the PPN theory of reference frames in the solar system, we do not impose
any restriction on function $Y$ at this step of the matching procedure. It will allow us to trace
how functions $Q$ and $Y$ affect equations of motion of extended bodies.

\subsubsection{Matching of $g_{0i}(t,\nnx)$ and $\hat{g}_{\alpha\beta}(u,{\bm w})$ to the order of $O(\epsilon^3)$}

Matching equation (\ref{2.6}) for $g_{0i}(t,\nnx)$ taken in its left side, reads
\br \label{3.24}
\hat L_i(u,\nnw)&
=&L_i(t,\nnx)+\mathfrak{B}_i(t,\nnx)-\mathfrak{P}_i(t,\nnx)
+\nnv{i}\biggl[\mathfrak{B}(t,\nnx)+N(t,\nnx)\biggr]\\\nonumber\\\nonumber&&+\nnv{j}\biggl[H_{ij}(t,\nnx)-
\mathfrak{R}^j_{\;i}(t,\nnx)\biggr]
           +O(\epsilon^2)\,,
\er where we have employed Eqs.
(\ref{j1}) and (\ref{3.18}) for doing simplification. Subsequent substitution of the
metric tensor given in the global, Eqs. (4.2.2), (4.2.4) and (4.2.5), and the local, Eqs. (6.2.7) and (6.2.37),
coordinates, to Eq. (\ref{3.24}) along with Eqs. (\ref{3.2}) and (\ref{3.6})
for vector-potentials $U_i(t,{\bm x})$ and $\hat{U}^{(\sss B)}_i(u,{\bm w})$ respectively, reveals that all
gravitational potentials depending on the internal structure of
the central body B (Earth) cancel each other. Remaining terms in the
matching equation (\ref{3.24}) depend only on the background values of the gravitational potentials of
external bodies. They yield a number of relationships which allow us to express the
external multipole moments from the metric tensor in local coordinates and functions present
in the PPN coordinate transformations (7.2.11)--(7.2.14) in terms of
the gravitational potentials of external bodies.
These relationships are as follows \br
       {\cal B}_i&=&
        2(1+\gamma){\bar U}^i\nnxe
         -(1+2\gamma)\nnv{i}\bar U\nnxe      \label{3.25}
          -\frac{1}{2}\nnv{i}\nnv{2}-\nnv{i}Q+O(\epsilon^2)\,,
          \\\nonumber\\
          {\cal B}_{<ik>}&=&
        2(1+\gamma)\left[\bar U^{<i,k>}\nnxe
         -\nnv{<i}\bar U^{,k>}\nnxe\right]+               \label{3.26}
          2\nnv{<i}\nnaa{k>}+Z_{<ik>}+O(\epsilon^2)\,,
           \\\nonumber\\
       {\cal B}_{<iL>}&=&
        2(1+\gamma)\left[\bar U^{<i,L>}\nnxe
         -\nnv{<i}\bar U^{,L>}\nnxe\right]+Z_{iL}+               \label{3.27}
          O(\epsilon^2)\,, \qquad(l\ge2),\er
          and\br\hspace{-2cm}
       \varepsilon_{ipk}\left(C_p+\dot{\cal F}_p
        \right)&=&
         -2(1+\gamma)\bar U^{[i,k]}\nnxe
          +(1+2\gamma)                                         \label{3.28}
           \nnv{[i}\bar U^{,k]}\nnxe
            +\nnv{[i}Q^{k]}+O(\epsilon^2)\,,\\\nonumber\\
                                                          \label{3.29}
        \varepsilon_{ipj}C_{pL-1}&=&
         \frac{4l(1+\gamma)}{l+1}\times\\\nonumber\\\nonumber&\times&\left[
          \nnv{{}^{[i}}\bar U^{{}^{,j]L-1}}\nnxe-
           \bar U^{[i,j]L-1}\nnxe
           -\frac{1}{l}\sum\limits_{k=1}^{l-1}
             \delta^{{}^{\scriptstyle a_k[i}}{\dot{\bar U}}^{,j]L-2}\nnxe
              \right]+O(\epsilon^2)\,.
\er

As one can see the matching equation (\ref{3.24})
gives rise to two sets of equations, Eqs. (\ref{3.25}) -- (\ref{3.27}), and
Eqs. (\ref{3.28} -- (\ref{3.29}) which have different properties of
symmetry. More specifically, Eqs. (\ref{3.25}) -- (\ref{3.27})
depend only on objects which are symmetric and trace-free with respect to the
entire set of indices $\{i,a_1,a_2,...,a_l\}$. On the other hand,
Eqs. (\ref{3.28}), (\ref{3.29}) are symmetric with respect to
the set of indices $\{a_1,a_2,...,a_l\}$ but they are anti-symmetric with
respect to any pair of indices consisting of the index $i$ and either one off the set
$\{a_1, a_2,..., a_l\}$. This separation of the matching equation (\ref{3.24}) in the symmetric and
antisymmetric components is due to the fact that this
equation has one free (vector-type) index $i$ and, hence,
can be split in fully symmetric and antisymmetric parts in
accordance with the Clebsch-Gordan decomposition of the vector
spherical harmonics into irreducible representations (see
\cite{thorne,bldam2,gelf} for more details).

Let us now discuss a freedom in choosing angular speed of rotation of spatial axes of the
local coordinates. From the discussion in previous sections one knows that there are two possible
types of the local coordinates -
either dynamically or kinematically non-rotating.
Dynamically non-rotating local coordinates are defined by the condition that the
external dipole moment $C_i=0$ in Eq.
(\ref{1.25}). On the other hand, the kinematically non-rotating local coordinates are realized if
one chooses function ${\cal F}_i=0$ in the coordinate transformation, Eq.
(\ref{2.13}), for spatial axes. If we do not fix the choice of $C_i$, then, Eq. (\ref{3.28}) can be
rewritten as
\be
      \dot{\cal F}_i=\varepsilon_{ijk}\left[
         (1+\gamma)\bar U^{i,k}\nnxe
          -(\gamma+\frac{1}{2})                                       \label{3.43}
           \nnv{i}\bar U^{,k}\nnxe
            -\frac{1}{2}\nnv{i}Q^{k}\right]-C_i  +O(\epsilon^2).
\en
The first term in Eq. (\ref{3.43})
describes the Lense-Thirring gravitomagnetic precession, the
second term describes the de Sitter precession, and the third term
describes the Thomas precession \cite{will,kop1989b,mtw}. We emphasize that in the
scalar-tensor theory both the Lense-Thirring and the de Sitter
precessions depend on the PPN parameter $\gamma$ while the Thomas
precession does not. The reason is that the Thomas precession is a special relativistic effect \cite{mtw}
and, as such can not depend at all on
a particular choice of a specific gravitational theory. If function $C_i=0$, spatial axes of the local
coordinates are kinematically rotating. On the other hand, the choice $\dot{\cal F}_i=0$ gives
kinematically non-rotating local coordinate frame which spatial axes are fixed with respect to distant
quasars with neglected proper motions \cite{ma}.

Functions ${\cal B}_L$ enter the PPN time coordinate transformation \footnote{See Eq. (\ref{eq1}).}
and depend on the gauge-dependent
multipole moments $Z_L$ which can be chosen arbitrary. There are two of the most preferable options:
\begin{enumerate}
\item[1)] One simplifies the time transformation, Eq. (\ref{2.12}), from the local coordinates to the global
ones as much as possible. In this case the moments $Z_L$ have to be chosen such that
functions ${\cal B}_L$ in the time transformation assume the most simple form
\br
     {\cal B}_{<ik>}&=&2\nnv{<i}\nnaa{k>}+O(\epsilon^2)               \label{3.44}
\\\nonumber\\
     {\cal B}_{<iL>}&=&O(\epsilon^2),\qquad\qquad\quad(l\ge2)\,.                  \label{3.45}
\er Here the choice of ${\cal B}_{ij}$ in Eq. (\ref{3.44})
eliminates all terms
explicitly depending on the velocity of the origin of the local coordinates
$\nnv{i}$. Substituting Eqs.
(\ref{3.44}) and (\ref{3.45}) into Eqs. (\ref{3.26}) and (\ref{3.27})
yields
\be
    Z_{iL}=-2(1+\gamma)\left[\bar U^{<i,L>}\nnxe
     -\nnv{<i}\bar U^{,L>}\nnxe\right]+O(\epsilon^2)\;,\qquad\qquad(l\ge1).   \label{3.46}
\en
This makes the metric tensor and the scalar field in the local coordinates
to be determined by four sets of the independent external
multipoles $P_L$, $Q_L$, $C_L$, and $Z_L$. However, the multipole moments $Z_L$ are not physically significant
and describe gauge-dependent coordinate effects.

\item[2)] One removes from the
metric tensor in the local coordinates all physically insignificant multipole moments
$Z_L$. The cost of this choice is a more complicated form of
the time transformation, Eq. (\ref{2.12}), from the local to global coordinates that
involves now the infinite number of coefficients ${\cal B}_L$. Assuming that all $Z_{iL}=0$ for $l\geq 1$
one obtains from Eqs. (\ref{3.26}) and
(\ref{3.27})
\br
       {\cal B}_{<ik>}&
        =&2(1+\gamma)\left[\bar U^{<i,k>}\nnxe
         -\nnv{<i}\bar U^{,k>}\nnxe\right]+               \label{3.47}
          2\nnv{<i}\nnaa{k>}+O(\epsilon^2)\;,
          \\\nonumber\\
       {\cal B}_{<iL>}&
        =&2(1+\gamma)\left[\bar U^{<i,L>}\nnxe
         -\nnv{<i}\bar U^{,L>}\nnxe\right]+               \label{3.48}
          O(\epsilon^2)\;,\qquad(l\ge2).
                                \er
\end{enumerate}
At the present step of the matching procedure we prefer to keep the multipole moments $Z_L$ unspecified to
preserve some freedom in making the gauge transformations.

\subsubsection{Matching of $g_{00}(t,\nnx)$ and $\hat{g}_{\alpha\beta}(u,{\bm w})$ to the order of $O(\epsilon^4)$}

Matching of the metric tensor at the post-Newtonian order of $O(\epsilon^4)$ allows us to infer the
post-Newtonian equations of motion of the origin of the local coordinate system as well as the post-Newtonian
corrections to the external multipole moments $Q_L$ and the remaining part of the time transformation formula (\ref{eq1})
between the local and global coordinates. Expansion of the matching equation (\ref{2.6}) for the metric
tensor component, $g_{00}(t,\nnx)$,  to the post-Newtonian order, generalizes Eq. (\ref{j1})
\br\hspace{-0.9cm}                                                    \label{5.1}
\hat N(u,\nnw)+\epsilon^2\hat L(u,\nnw)&=&
    N(t,\nnx)+2\mathfrak{B}(t,\nnx)-\nnv{2}\\\nonumber\\\nonumber&
    +\epsilon^2&\Bigl[L(t,\nnx)-3{\mathfrak{B}}^2(t,\nnx)+2\mathfrak{D}(t,\nnx)-2\mathfrak{B}(t,\nnx)N(t,\nnx)
    \\\nonumber\\\nonumber&
    +&4\nnv{2}\mathfrak{B}(t,\nnx)+2\nnv{2}N(t,\nnx)
   +2\nnv{i} L_i(t,\nnx)+2\nnv{i}\mathfrak{B}_i(t,\nnx)\\\nonumber\\\nonumber&-&
   \frac23\nnv{2}\mathfrak{R}^k_k(t,\nnx)+
   2\nnv{i}\nnv{j}\mathfrak{R}^i_j(t,\nnx)-\frac53\nnv{4}+\frac13\nnv{2}H(t,\nnx)\Bigr]
    +O(\epsilon^3)\,,
    \er
where the gravitational potentials in the right side of this equations are determined by Eqs.
(\ref{12.6}) -- (\ref{12.7}) and those in the left side are
given by Eqs. (6.2.5), (6.2.6), (6.2.36), (6.2.39) and (6.2.40).
Solution of Eq. (\ref{5.1}) is done in several steps.

First of all, one substitutes components of the matrix of the PPN coordinate transformation, given by Eqs.
(\ref{zz1}) -- (\ref{zz4}), to Eq. (\ref{5.1}). Then, one matches separately the internal gravitational
potentials referred to the central body B and those referred to the external bodies.   The internal gravitational
potentials have to be transformed from the local to global coordinates by making use
of  Eqs. (\ref{3.5}) -- (\ref{3.12}). One notices that transformation equation (\ref{ozs}) for the Newtonian
gravitational potential of the central body B can be written explicitly in terms of the functions
coming about the matching procedure at lowest orders. Taking definitions of
functions $\xi^0$ and $\xi^i$ from section \ref{gzm} and elaborating them by making use of the results
of previous steps of the matching procedure, one obtains explicit form of the relativistic correction,
$\mathcal{U}{{\sss(B)}}(u,{\bm w})$, from Eq. (\ref{ozs}) which describes transformation of the
Newtonian potential from the local to global coordinates. It reads
\br\label{zal}
\mathcal{U}{{\sss(B)}}(u,{\bm w})&=& \nnj\left(\frac{1}{2}\nnv{2}-
          \gamma\bar U\nnxe- \nnaa{k}\nnr{k}+Y\right)
          \\\nonumber\\\nonumber&&
          +\frac{1}{2}\nnv{i}\nnv{j}
          \chi^{{\sss(B)}}_{,ij}(t,\nnx)+
          c\nnv{k}\chi^{{\sss(B)}}_{,0k}
          (t,\nnx)\\\nonumber\\\nonumber&&
          -\frac{1}{2}\nnaa{k}\chi^{{\sss(B)}}_{,k}
          (t,\nnx)
          -\nnv{k}U^k_{{\sss(B)}}(t,\nnx)+
          O(\epsilon^2)\;.
\er
Employing this formula in the matching procedure of the internal Newtonian potential of the central body B
along with transformation equations for other internal potentials, one gets a
remarkable result -- both the Newtonian and the post-Newtonian terms depending on the internal structure of
the body B cancel out and, hence, completely vanish from the matching equation (\ref{5.1}). This
effacing-internal structure effect can be explained in terms of the laws of conservation of intrinsic
linear and angular momenta of the body B which are valid
in the scalar-tensor
theory of gravity as well as in general relativity \cite{will,dames}. In other words, the presence
of a scalar field in
the scalar-tensor theory of gravity does not result in the net self-force or self-torque exerted on
the body which existence would violate classic Newton's laws. The internal-structure effacing principle
for spherically-symmetric extended bodies was extrapolated to the 2.5 post-Newtonian approximation by one
of the authors of the present paper \cite{kop1985,kop1987} by applying the Fock-Papapetrou \cite{fock,papap}
method for derivation of relativistic equations of motion for a binary system with the conservative and
gravitational radiation-reaction forces. Validity of the effacing principle was also confirmed by Damour
\cite{dam1983} who derived equations of motion of two Schwarzschild black holes in the 2.5 post-Newtonian
approximations by the Einstein-Infeld-Hoffmann technique \cite{eih} supplemented by the method of analytic
continuation of the energy-momentum tensor of point-like particles understood in terms of the
generalized functions (distributions) \cite{gelf}.

At the third step, one equates in Eq. (\ref{5.1}) gravitational potentials generated by all external bodies
(Moon, Sun, etc.) but the central body B (Earth). This step requires to know how function,
$\sum\frac{1}{l!}Q_L(u)w^L$,  is transformed from the local to global coordinates within the post-Newtonian
accuracy. General formula of transformation of this function is given by Eq. (\ref{n4}). Substituting to
this formula the explicit form of functions $\xi^0$ and $\xi^i$ displayed in Eqs. (\ref{2.8}),
(\ref{2.12}) -- (\ref{eq2}), one gets
\br
      \nn{l=0} \frac{1}{l!} Q_L(u)w^L&=&   \label{5.3}
      \nn{l=0}\frac{1}{l!} Q_L(t)\nnr{L}
      \Bigl[1+l\epsilon^2\left(\gamma\bar U\nnxe-Y\right)\Bigr]\\\nonumber \\\nonumber  & -&
      \epsilon^2\left\{
      \nn{l=1}\frac{1}{(l-1)!}Q_{kL-1}{F}^{jk}\nnr{<jL-1>}-\dot Q\nnv{k}\nnr{k}\right. \\\nonumber \\\nonumber  &+&\left.
      \nn{l=1}\frac{1}{(l-1)!}
      \Bigl[\frac{1}{2}\nnv{j}\nnv{k}Q_{jL-1}
      \nnr{<kL-1>}+\left(\nnaa{k}Q_L-\frac{1}{l}
      \nnv{k}\dot Q_L\right)\nnr{<kL>}\Bigr]\right.\\\nonumber \\\nonumber &+&
      \left.
          \nn{l=0}\frac{1}{(2l+3)l!}\Bigl[
          \frac{1}{2}
          \nnv{j}\nnv{k}Q_{jkL}
          -\frac{1}{2}\nnaa{k}Q_{kL}-
          \nnv{k}\dot Q_{kL}\Bigr]\nnr{L}\nnr{2}\right\}
          +O(\epsilon^4)\,.
      \er

Matching equation (\ref{5.1}) requires to calculate function $\bar{\chi}_{,00}(t,\nnx)$ generated by the
external bodies other than the body B (Earth) and entering $g_{00}(t,\nnx)$
component of the metric tensor in the global coordinates as shown in Eqs. (\ref{12.9}) and (\ref{3.3}).
Contrary to other potentials, like $\bar U(t,\nnx)\,,
\bar U^i(t,\nnx)\,, \bar{\Phi}_k(t,\nnx)$, which are solutions of the
homogeneous Laplace equation in the vicinity of the body B, function
$\bar{\chi}(t,\nnx)$ is subject to the Poisson
equation \cite{fock} \be
   \nabla^2\bar{\chi}(t,\nnx)=-2\bar U(t,\nnx)\;.   \label{5.4}
\en
After solving this equation and expanding its solution into STF harmonics (see appendix \ref{ap1}) one obtains
\be                                                          \label{5.5}
  \bar{\chi}(t,\nnx)=
  \nn{l=0}\frac{1}{l!}\bar{\chi}_{,<L>}\nnxe\nnr{L}-
  \nn{l=0}\frac{1}{(2l+3)l!}
  \bar U_{,L}\nnxe\nnr{L}\nnr{2}\:.
\en       \\
Differentiating the left and right sides of Eq. (\ref{5.5}) two times with respect to the global
coordinate time, $t$, yields
\br                                       \label{5.6}
  \bar{\chi}_{,tt}(t,\nnx)&=&
  \nn{l=0}\frac{1}{l!}\bar{\chi}_{,tt<L>}\nnxe\nnr{L}+
  \nn{l=0}\frac{1}{(2l+3)l!}\times\\\nonumber\\\nonumber&&\times\Bigl[
  \nnaa{k}\bar U_{,kL}\nnxe+2\nnv{k}\dot{\bar U}_{,kL}\nnxe-
  \nnv{j}\nnv{k}\bar U_{,jkL}\nnxe-\Ddot{\bar U}_{,L}\nnxe
  \Bigr]\nnr{L}\nnr{2}\:.
     \er

Finally, one expands all functions in both sides of Eq.
(\ref{5.1}) in the Taylor series with respect to the distance, $\nnr{i}$, from the central body B, and reduce all
similar terms. One finds that those terms which do not depend on
$\nnr{i}$ (that is, functions of time, $t$, only) form an ordinary differential equation of a first order
for function
${\cal B}(t)$ in the coordinate time transformation given by Eq. (\ref{2.12}). This differential equation reads
\br
   \dot {\cal B}(t)&=&-                                         \label{5.7}
        \frac{1}{8}\nnv{4}-
        (\gamma+\frac{1}{2})\nnv{2}\bar U\nnxe+
        (\beta-\frac{1}{2})\bar U^2\nnxe+Q\left[\frac12\nnv{2}+\frac32Q-\bar U\nnxe\right]\\\nonumber\\\nonumber&&
        +2(1+\gamma)\nnv{k}\bar U^k\nnxe
        -\bar\Phi\nnxe+\frac12\bar\chi_{,tt}\nnxe
        +O(\epsilon^2)\;.
   \er

   Terms which are linear with respect to $\nnr{i}$ give us the post-Newtonian equation of translational
   motion of the origin of the local coordinates, $x^i_B(t)$, in global (barycentric) coordinates.
   It generalizes equation (\ref{3.16}) derived solely in the Newtonian approximation. Barycentric
   acceleration of the origin of the local coordinates with respect to the barycenter of the N-body system is
\br
 \nnaa{i}&=&
           \bar U_{,i}\nnxe-Q_i+\epsilon^2                   \label{5.8}
           \Biggl[F^{ik}Q_k+\bar\Phi_{,i}\nnxe-\frac{1}{2}\bar{\chi}_{,itt}\nnxe+Q_i\Bigl(Y-2Q\Bigr)
           \\\nonumber\\\nonumber  &
           &
           +2(1+\gamma)\dot{\bar{U}}^i\nnxe-2(1+\gamma)
           \nnv{k}\bar U^{k,i}\nnxe-(1+2\gamma)
           \nnv{i}\dot{\bar U}\nnxe\\\nonumber\\\nonumber&&+
           (2-2\beta-\gamma)\bar U\nnxe \bar U_{,i}\nnxe+(1+\gamma)\nnv{2}
           \bar U_{,i}\nnxe-
           \frac{1}{2}\nnv{i}\nnv{k}\bar U_{,k}\nnxe\\\nonumber\\\nonumber  &
           &
           -\frac{1}{2}\nnv{i}\nnv{k}\nnaa{k}-\nnv{2}\nnaa{i}-(2+\gamma)
           \nnaa{i}\bar U\nnxe\Biggr]+O(\epsilon^4)\;.
   \er
This equation contains the external dipole moment, $Q_i$, which must be calculated with the post-Newtonian
accuracy in order to complete derivation of the post-Newtonian equation of translational motion of the origin
of the local coordinates. A simple choice of $Q_i=0$ does not allow us to keep the origin of the local coordinates
at the center of mass of the central body B (Earth) for arbitrary long interval of time. This is because the body
is, in general, interacting with tidal gravitational field of external bodies (Moon, Sun, etc.) and does not move
along a geodesic world line while the choice of $Q_i=0$ makes the origin of the local coordinates moving along a
geodesic world line \cite{kop1988,dsx1991,th}. Thus, function $Q_i$ must be defined in such a way that the body's
center of mass and the origin of the local coordinates coincide at any instant of time. This problem is equivalent
to solution of the problem of motion of the body's center of mass with respect to the origin of the local coordinates
and will be discussed in the next section.

Terms which are quadratic, cubic, etc., with respect to the distance,$\nnr{i}$, determine the post-Newtonian
corrections to the external STF multipole moments $Q_L=Q_L^{\sss{N}}+\epsilon^2 Q_L^{\sss{pN}}$, where the
Newtonian-like term $Q_L^{\sss{N}}$ is shown in equation (\ref{3.17}). The post-Newtonian corrections are
\br\label{5.9}
   Q_L^{\sss{pN}}(t)&=&X_{<L>}+
           \dot Z_{L}+\bar\Phi_{,<L>}\nnxe-\frac12\bar\chi_{,tt<L>}\nnxe+\Bigl(lY-2Q\Bigr)\bar U_{,L}\nnxe
           \\\nonumber\\\nonumber  &&+2(1+\gamma)\dot{\bar U}^{<i_l,L-1>}\nnxe-2(1+\gamma)
           \nnv{k}\bar U^{k,L}\nnxe\\\nonumber\\\nonumber  &&   +(l-2\gamma-2)
           \nnv{<i_l}\dot{\bar U}^{,L-1>}\nnxe+(1+\gamma)\nnv{2}
           \bar U_{,L}\nnxe \\\nonumber\\\nonumber  &&-
           \frac{l}{2}\nnv{k}\nnv{<i_l}\bar U^{,L-1>k}\nnxe+(2-2\beta-l\gamma)\bar U\nnxe\bar U_{,L}\nnxe
           \\\nonumber\\\nonumber & &-
           (l^2-l+2+2\gamma)\nnaa{<i_l}\bar U^{,L-1>}\nnxe
           -lF^{k<i_l}\bar U^{,L-1>k}\nnxe
           \;,\qquad(l\ge2)
   \er
where we used notations
\br
    X_{<ij>}&=&3\nnaa{<i}\nnaa{j>}\,,                      \label{5.10}
\\\nonumber\\
    X_{<L>}&=&0\,,\qquad\qquad\qquad (l\ge3).                          \label{5.11}
\er
These equations finalize description of the STF multipole moments entering external solution of the metric
tensor in local coordinates $w^\alpha=(cu,{\bm w})$ and
the parametrized post-Newtonian transformation between the local and global coordinates.

\subsection{Final Form of the PPN Coordinate Transformation from Local to Global Coordinates}

For the sake of convenience we summarize the final form of the parametrized post-Newtonian coordinate
transformation from local to global coordinates which is given by two equations:
\br                                                 \label{5.12}
u&=&t-\epsilon^2\left({\cal
A}+\nnv{k}\nnr{k}\right)\\\nonumber&+& \epsilon^4\left[{\cal B}+
\biggl(\frac{1}{3}\nnv{k}\nnaa{k}-
   \frac{1}{6}\dot{\bar U}\nnxe\biggr)\nnr{2}-
   \frac{1}{10}\Dot{a}^{k}_{\scriptstyle B}\nnr{k}\nnr{2}+
   \sum\limits_{l=1}^{\infty}\frac{1}{l!}
   {\cal B}_{<L>}\nnr{L}\right]+O(\epsilon^5)\;,
   \\\nonumber\\                                         \label{5.13}
   w^i&=&\nnr{i}+\epsilon^2\left[
   \biggl(\frac{1}{2}\nnv{i}\nnv{k}+\gamma
   \delta^{ik}\bar U\nnxe+{F}^{ik}\biggr)\nnr{k}+
   \nnaa{k}\nnr{i}\nnr{k}-\frac{1}{2}\nnaa{i}\nnr{2}
   \right]+O(\epsilon^4)\;.
   \er
Here functions ${\cal A}$ and ${\cal B}$ are solutions of the ordinary differential equations
\br
   \dot{\cal A}&=&\frac12\nnv{2}+\bar U\nnxe-Q\,,             \label{5.14}
\\\nonumber\\
   \dot {\cal B}&=&-                                         \label{5.7a}
        \frac{1}{8}\nnv{4}-
        (\gamma+\frac{1}{2})\nnv{2}\bar U\nnxe+Q\Bigl[\frac12\nnv{2}+\frac32Q-\bar U\nnxe\Bigr]
        \\\nonumber&+&
        (\beta-\frac{1}{2})\bar U^2\nnxe+2(1+\gamma)\nnv{k}\bar U^k\nnxe
        -\bar\Phi\nnxe+\frac12\bar\chi_{,tt}\nnxe\;,
\er
while the other functions are defined as follows
\br
{\cal B}_i&=&2(1+\gamma)\bar U^i\nnxe-                   \label{5.15}
   (1+2\gamma)\nnv{i}\bar U\nnxe-\frac12
   \nnv{i}\nnv{2}-\nnv{i}Q\,,
\\\nonumber\\
       {\cal B}_{<ik>}&=&\!\!
       Z_{ik}+
        2(1+\gamma)\bar U^{<i,k>}\nnxe
         -2(1+\gamma)\nnv{<i}\bar U^{,k>}\nnxe+               \label{5.16}
          2\nnv{<i}\nnaa{k>}\,,
           \\\nonumber\\
        {\cal B}_{<iL>}&=&\!\!
       Z_{iL}+
        2(1+\gamma)\bar U^{<i,L>}\nnxe
         -2(1+\gamma)\nnv{<i}\bar U^{,L>}\nnxe\;               \label{5.17}
          , \qquad (l\ge2),
           \\\nonumber\\                            \label{5.18}
    \dot{F}^{ik}&=&
    (1+2\gamma)\nnv{[i}\bar U^{,k]}\nnxe-2(1+\gamma)\bar U^{[i,k]}\nnxe
    +\nnv{[i}Q^{k]}\;.
\er
These equations will be used in subsequent sections for derivation of equations of motion of extended bodies.
They are also convenient for
comparison with the relativistic transformations derived by other researchers
\cite{kls,skpw,ashb2,fuk1,kop1988,bk-nc,dsx1991,dsx1992}.

\section{Translational Equations of Motion of Extended Bodies}\label{tem}

\subsection{Introduction}

In the Newtonian theory of gravity definitions of mass and the center of
mass of an extended body, which is a member of a many-body system, are quite simple and
straightforward concepts. Because of the simplicity, they have been directly extrapolated without any change
to the relativistic
theory of gravity by Fock \cite{fock} by making use of
the invariant density $\rho^*$ which obeys the Newtonian-like equation
of continuity, Eq. (\ref{11.20}). The invariant density is defined as mass of baryons per unit of the proper
volume and after integration over volume of a body gives its total baryon mass which is constant \cite{will}.
The invariant density
is used to introduce newtonian definitions of the body's
center of mass, and its linear momentum in the body's local coordinates. The baryon mass, center of mass
and the linear momentum of body B are given \footnote{We skip in this section label B for all quantities
referred to body B as it does not cause misinterpretation. We shall label the bodies with indices A,B,C, etc.
every time when it may cause confusion.}  by integrals \cite{fock,brr}
\br\label{a} {\cal M}_*&=&\int_{V_{\sss B}}\rho^*(u,{\bm w})d^3 w\;,
\\\nonumber\\\label{b}
{\cal J}^i_*&=&\int_{V_{\sss B}}\rho^*(u,{\bm w})w^i d^3 w\;,\\\nonumber\\\label{c}
{\cal P}^i_*&=&\int_{V_{\sss B}}\rho^*(u,{\bm w})\nu^i d^3 w\;.
\er
These
definitions were used
by Fock \cite{fock}, Papapetrou \cite{papap}, Petrova \cite{petr}, Brumberg \cite{brberg},
and some other researchers (see, e.g., \cite{spyr1,capor,spyr2,spyr3}
and references therein) for derivation of the post-Newtonian
equations of translational motion of extended, spherically-symmetric bodies. Physical intuition tells us that
equations of motion of such bodies, in principle, have to depend only on masses of the bodies which are supposed
to be the only parameters characterizing the magnitude of their internal gravitational field. This is indeed true
in the Newtonian theory. It was found, however, that the post-Newtonian equations of motion of spherically-symmetric
bodies, unlike the
Newtonian theory, depend not only on the baryon masses of the bodies, Eq. (\ref{a}), but also
on some other characteristics which are the internal kinetic
and gravitational energy, elastic energy and moments of inertia \cite{fock,brr,spyr1,capor,brberg,spyr2,spyr3,dall}.
Appearance of such terms significantly complicates interpretation of the post-Newtonian
equations of motion and is unsatisfactory from the physical point of view. For instance, due to
the dependence of the post-Newtonian equations of motion on other parameters rather than masses of the bodies,
it is possible to argue
that motion of a binary system consisting of ordinary stars is different from that of a binary system consisting
of black holes having the same value of the mass as the stars.

This point of view is incompatible with the Newtonian equations of motion of two black holes as it was shown
by Demiansky and Grishchuk \cite{dg} who proved that motion of black holes in the Newtonian approximation obeys
to the same laws of gravitational physics as for ordinary spherically-symmetric stars. In addition,
the Einstein-Infeld-Hoffmann method of derivation of the post-Newtonian equations of motion operates
with vacuum Einstein's equations and does not admit appearance of any terms in the equations of motion
of spherical bodies which would depend on the internal structure of the bodies. Thus, one has to expect
that the Newtonian definitions of mass, the dipole moment, and the center of mass given in
Eqs. (\ref{a}) -- (\ref{c}) are not quite appropriate for calculation of the post-Newtonian equations
of motion of extended bodies. Indeed, our study of relativistic translational motion of two
spherically-symmetric compact stars in a binary system \cite{kop1985,grkop} revealed that if the Newtonian
mass and the center of mass are replaced with their
corresponding relativistic definitions all
terms in the equations of motion
depending on the internal structure of the bodies are effectively eliminated via renormalization of masses.
Damour \cite{dam1983} called this property "the effacing principle" and confirmed its validity for
spherically-symmetric and compact relativistic bodies using the matched asymptotic expansion technique.

Newtonian theory predicts that if celestial bodies are not spherically-symmetric and rotate, their equations
of motion must depend on additional parameters which are the mass multipole moments of the bodies.
It is natural to expect that the post-Newtonian equations of motion of such bodies will contain both the mass
and current multipole moments given, for example, by Eqs. (\ref{1.31}) and (\ref{1.32}) respectively.
However, it is not evident whether some other parameters have to appear in the relativistic equations of motion
in addition to these two sets of the internal multipoles. Scrutiny analysis of this problem in general
relativity elucidated that the post-Newtonian equations of motion of extended bodies do contain only the mass
and current multipoles \cite{dsx1991,dsx1992,dsx1993} and does not depend on any other internal characteristic
of the bodies. PPN formalism operates with a class of alternative theories of gravity and does not obey to
the "effacing principle" even for spherically-symmetric bodies because of the violation of the strong principle
of equivalence \cite{will,dicke1,n-1,n-2}. This violation makes two masses -- inertial and gravitational --
be different and, hence, two parameters appear in the post-Newtonian equations of motion of spherical bodies
as contrasted to general relativistic case. It is interesting to answer the question how many parameters have
to be introduced to characterize motion of extended, non-spherically symmetric and rotating bodies in the PPN
formalism, thus, extending general relativistic results of Damour, Soffel, and Xu \cite{dsx1991,dsx1992,dsx1993}.

Like in general relativity, solution of this problem in the PPN formalism could not be achieved in the framework
of a standard {\it a la} Fock-Papapetrou post-Newtonian approach which basically operates with a single (global)
coordinate system. The global coordinates can not be used to define multipole moments of each body in N-body
system in a physically consistent way. A local coordinate system should be constructed around each of the bodies
in order to strongly suppress coordinate-dependent contributions to definitions of the multipole moments caused
by the Lorentz contraction and presence of the background gravitational field caused by external bodies.
Consistent relativistic concept of the local coordinate system was developed in \cite{kop1988,bk-nc,dsx1991,mashn}.
The concept of a local frame of reference associated with a moving body has been also discussed in papers
\cite{ssa,ashb1,ashb2,fuk1}. However, the authors of these papers always assumed that the center of mass of
the body moves along a geodesic world line in the background space-time. This assumption is not valid for
a non-spherical and rotating body as it follows from the Mathisson-Papapetrou-Dixon equations \cite{papap,dixon}
and other arguments present in \cite{kop1988,dsx1991}. For this reason, the method of construction of the
local coordinates in the vicinity of a self-gravitating body suggested in \cite{ashb1,ashb2} is not general enough.
Thorne \cite{thorne}, Blanchet and Damour \cite{bld} invented a post-Newtonian definition of the STF mass
multipole moments which were fruitfully employed by Damour, Soffel and Xu
\cite{dsx1991,dsx1992} for derivation of the
post-Newtonian translational equations of motion of
self-gravitating extended bodies in general relativity. These authors had also proved that the only parameters
present in the equations are Tolman masses of the bodies and their STF multipole moments referred to the local
frame of reference of each body.

In this section we derive the parametrized post-Newtonian
equations of translational motion of extended bodies in the scalar-tensor theory of gravity and
prove that these equations depend only on the inertial and gravitational masses of the bodies and a set of
STF {\it active} multipole moments defined in section \ref{mdloc}. In case of spherically-symmetric bodies the
only parameters present in the equations are the inertial and gravitational masses which are different due
to the Nordtvedt effect \cite{n-1,n-2}.

Post-Newtonian Thorne-Blanchet-Damour definition of the conserved
mass and center of mass of a single isolated body
 are given by Eqs.
(\ref{13.37}) and (\ref{13.37a}).
One might think that if the body is, in fact, a member of N-body system its gravitational
interaction with other bodies of
the system would violate these conservation laws \cite{th}. The law of conservation of mass is indeed violated
if the body is not spherically-symmetric (see below). However, the law of conservation of the center of mass
and linear momentum of the body can be retained.

There are three types of multipole moments in the scalar-tensor theory -- active, conformal and scalar --
which can be
used for definition of body's mass and its center of mass. These moments
were introduced in section \ref{mdloc} and are defined by equations
(\ref{13.11}), (\ref{13.17}) and (\ref{13.30}). By direct calculation we shall demonstrate that the active
and scalar dipole moments of the body are not efficient in derivation of the translational equations of motion.
This is because if one uses either active or scalar dipole and derive equations of motion of the body in its
local coordinates the equations, we shall obtain, will contain a significant number of terms which can be
treated as self-accelerations and they can not be removed by simple translation of the origin of the local
coordinate system to another point. Self-accelerated terms in the equations of motion violate Newton's third
law and are unacceptable. On the other hand, when we use the conformal dipole moment for defining the body's
center of mass and, then, derive equations of motion of the body, we find that self-acceleration terms do not
appear, and the equations have remarkably simple structure of the post-Newtonian force which is a function
of the {\it active} multipole moments of the body coupled with external multipoles.

Let us discuss derivation and specific of the translational equations of motion in more detail starting from
the explicit form of the local (macroscopic) equations of motion of the body's matter in local coordinates.

\subsection{Macroscopic post-Newtonian Equations of Motion}\label{qmez-1}

The macroscopic post-Newtonian equations of motion of matter consist of three groups: (1) the equation of
continuity, (2) the thermodynamic equation relating the elastic energy $\Pi$ and the stress tensor
$\pi_{\alpha\beta}$, and (3) the Euler equation.

The equation of continuity in the local coordinates $(u, {\bm w})$ has the most simple form for the invariant
density $\rho^*$ and reads
\be\label{kp1}
\frac{\partial\rho^*}{\partial u}+\frac{\partial\left(\rho^*\nu^i\right)}{\partial w^i}=0\;.
\en
This equation is exact, that is takes into account all post-Newtonian corrections, as follows from the
definition of the invariant density $\rho^*$ and Eq. (\ref{11.20}).

The thermodynamic equation relating the internal elastic energy $\Pi$ and the stress tensor $\pi_{\alpha\beta}$
is required only in a linearized approximation where the stress tensor is completely characterized by its
spatial components $\pi_{ij}$. Hence, one has from Eq. (\ref{11.2}) the following differential equation
\be\label{kp2}
\frac{d\Pi}{du}+\frac{\pi_{ij}}{\rho^*}\frac{\partial\nu^i}{\partial w^j}=O(\epsilon^2)\;,
\en
where the operator of convective time derivative is $d/du\equiv\partial/\partial u+\nu^i\partial/\partial w^i$.

The Euler equation \footnote{In fact, this is the Navier-Stokes equation because the stress tensor is taken in
its the most general form.}  follows from the spatial part of the law of conservation
of the energy-momentum tensor $T^{i\nu}_{\;\;\;;\nu}=0$. It yields
\br\label{kp3}&&
\rho^*\frac{d}{du}\left\{\nu^i+\epsilon^2\left[\left(\frac12\nu^2+\Pi+\frac12\hat N+\frac13\hat H\right)\nu^i+
\hat L_i\right]\right\}+\epsilon^2\,\frac{\partial\left(\pi_{ij}\nu^j\right)}{\partial u}=\\\nonumber\\
\nonumber&&\frac12\rho^*\frac{\partial\hat N}{\partial w^i}-\frac{\partial\pi_{ij}}{\partial w^j}+\epsilon^2
\left\{\rho^*\left[\frac12\frac{\partial\hat L}{\partial w^i}+\frac14\left(\nu^2+2\Pi+\hat N\right)
\frac{\partial\hat N}{\partial w^i}+\frac16\,\nu^2\,\frac{\partial\hat H}{\partial w^i}+\nu^k\,
\frac{\partial\hat L_k}{\partial w^i}\right]\right.\\\nonumber\\\nonumber&&\left.
+\frac16\,\pi_{kk}\,\frac{\partial\hat H}{\partial w^i}+
\pi_{ik}\left(\frac{\partial \hat N}{\partial w^k}-\frac53\frac{\partial \hat H}{\partial w^k}\right)
+\frac12\left(\hat N-\frac53\hat H\right)\frac{\partial\pi_{ik}}{\partial w^k}
\right\}+O(\epsilon^4)\;,
\er
where gravitational potentials $\hat{N}$, $\hat{H}$, $\hat{L}$, $\hat{L}_i$ are the metric tensor components
in the local coordinates and they have been defined in section \ref{qop}.

\subsection{Definitions of Mass, the Center of Mass and the Linear Momentum of an Extended Body in the N-body System}

There are two algebraically independent definitions of the post-Newtonian mass in the scalar-tensor theory --
the {\it active} mass and the {\it conformal} mass which are derived from Eqs. (\ref{1.31}) and (\ref{1.34})
respectively for index $l=0$. As discussed in section \ref{mdloc} one must retain contribution of the gravitational
potential of external bodies in the definition of the STF multipole moments of the body's gravitational field.
It will allow us to cancel out in equations of motion all terms depending on the internal structure of the
central body B which are not incorporated to the definition of the STF multipoles. Absence of such terms in
equations of motion extends validity of the effacing principle from general theory of relativity
\cite{dsx1991,dam1983,kop1985} to the scalar-tensor theory of gravity at least in the first post-Newtonian
approximation. The question about whether to keep the contribution of the gravitational potentials of
external bodies in the definition of the STF multipolar decomposition of the body's gravitational field
was discussed by Thorne and Hartle \cite{th} but they did not come up with a definite answer. Our approach
is based on direct calculation of equations of motion and we have tried various possible definitions of the
center of mass and the STF multipoles. After tedious and cumbersome calculations we came to the conclusion
that the equations of motion have the most simple form and the minimal set of parameters if we take the
{\it conformal} definition of mass and the center of mass for each body and include  the gravitational
potential of external bodies to the definition of the body's multipole moments

{\it Conformal} multipoles are given by Eq. (\ref{1.34}) and for index $l=0$ can be reduced to simpler
form such that the {\it conformal} mass of the body located near the origin of the local coordinates is
\be\label{ij1}
\tilde {\cal M}={\rm M}-\epsilon^2\left\{\left[Y+(1-\gamma)P\right]{\rm M}+\sum_{l=1}^\infty\frac{l+1}{l!}
Q_L{\cal I}^L\right\}+O(\epsilon^4)\;.
\en
In what follows, we will need definition of the {\it active} mass of the body as well. It is extracted
from Eq. (\ref{1.31}) when index $l=0$
\br\label{ij2}
{\cal M}&=&{\rm M}+\epsilon^2\biggl\{\frac16(\gamma-1)\Ddot{\cal I}^{(2)}-\frac12\,\eta\int_{V_{\sss B}}
\rho^*\hat U^{(B)}d^3w\\\nonumber\\\nonumber&&-\Bigl[Y+(2\beta-\gamma-1)P\Bigr]{\rm M}-\sum_{l=1}^\infty
\frac{1}{l!}\Bigl[(\gamma l+1)Q_L+2(\beta-1)P_L\Bigr]{\cal I}^L\biggr\}+O(\epsilon^4)\;,
\er
where
\be\label{ij3}
{\rm M}=\int_{V_{\sss B}}\rho^*\left[1+\epsilon^2\left(\frac12\nu^2+\Pi-\frac12\hat U^{(B)}\right)\right]d^3w
+O(\epsilon^4)\;,
\en
is {\it general relativistic} definition of the post-Newtonian mass of the body \cite{will} and
\be\label{fop}
{\cal I}^{(2)}=\int_{V_{\sss B}}\rho^*w^2d^3w\;,
\en
is the second-order rotational moment of inertia of the body.

 It is not difficult to derive a relationship between the {\it active} and {\it conformal} post-Newtonian
 masses by making use of Eqs. (\ref{ij1}) -- (\ref{ij3}). It reads
\br\label{acmass}
{\tilde{\cal M}} &=& {\cal M} +\epsilon^2\biggl\{\frac{1}{2}\eta\int_{V_{\sss B}}\rho^* \hat{U}^{\scriptstyle(B)}d^3w-
\frac{\gamma-1}{6}\Ddot{\cal I}^{(2)}\\\nonumber &&
+2(\beta-1)\Bigl({\cal M} P
+\nn{l=1}\frac{1}{l!}P_L{\cal
I}^{L}\Bigr)+(\gamma-1)\nn{l=1}\frac{1}{(l-1)!}Q_L{\cal
I}^{L}\biggr\} + O(\epsilon^4)\;,  \er
where $\eta=4\beta-\gamma-3$ in Eq. (\ref{6.2d}) is called the {\it Nordtvedt} parameter \cite{will}.
Numerical value of this parameter is known with the precision better than 0.02\% from
the lunar laser ranging experiment \cite{llr} which lasts already for more than 30 years.

One can see that in the scalar-tensor theory of gravity the {\it conformal} mass of the body differs from
its {\it active} mass. If the body is completely isolated the difference can be only due to the {\it Nordtvedt}
effect, that is for $\eta\not=0$, and the time-dependence of the body's rotational moment of inertia
(for example, radial oscillations, etc.). In the case when the presence of external bodies can not be
ignored, one has to account for coupling of the external gravitational field multipoles, $Q_L$ and $P_L$,
with the internal multipole moments ${\cal I}_L$ of the body.

It is important to realize that in general case the {\it general relativistic} post-Newtonian mass of
an individual body is not conserved. Indeed, taking a time derivative of Eq. (\ref{ij3}) and making
use of the macroscopic equations of motion of the body's matter given in Section \ref{qmez-1}, one gets \cite{capor}
\be\label{ij4}
\dot{\rm M}=\epsilon^2\sum_{l=1}^\infty\frac{1}{l!}Q_L\dot{\cal I}^L+O(\epsilon^4)\;,
\en
where overdot means the time derivative with respect to the local coordinate time $u$.
This equation reveals that the {\it general relativistic} mass of the body is constant, if and only if,
the mass multipole moments ${\cal I}_L$ of the body do not depend on time and/or there is no external
tidal fields, that is $Q_L=0$. However, one can notice that the {\it conformal} and {\it active}
post-Newtonian masses are not constant in the presence of the tidal field even if the body's multipole
moments ${\cal I}_L$ are constant. This is because the external multipole moments $Q_L$ enter definitions
of these masses, Eqs. (\ref{ij1}) and (\ref{ij2}), explicitly, and in general case of N-body problem they
depend on time.

Direct calculation of the equations of motion elucidates that definition of the {\it  conformal} mass
dipole moment given by Eq. (6.3.5) for $l=1$ gives the most optimal choice of the
post-Newtonian center of mass for each body. This is because after differentiation with respect to time
only the {\it conformal} dipole moment leads to the law of
conservation of the body's linear momentum when one neglects the influence of other external bodies,
while the post-Newtonian {\it scalar} or
{\it active} dipole moments do not bear such property. Thus,
the post-Newtonian center of mass of the body, ${\cal J}^i\equiv\tilde{\cal I}^i$, is derived from
Eq. (6.3.5) for $l=1$ and reads
\br\label{6.2a}
{\cal J}^i&=&\int_{V_{\sss B}}\rho^*w^i \left[1+\epsilon^2\left(\frac12\nu^2+\Pi-
\frac12\hat U^{(B)}\right)\right]d^3w\\\nonumber\\\nonumber&&-
\epsilon^2\left\{\left[Y+(1-\gamma)P\right]{\cal J}^i_*+\sum_{l=1}^\infty
\frac{l+1}{l!}Q_L{\cal I}^{iL}\right.\\\nonumber\\\nonumber&&\left.\qquad+\frac12\sum_{l=1}^\infty
\frac{1}{(2l+1)(l-1)!}Q_{iL-1}{\cal N}^{L-1} \right\}+O(\epsilon^4)\;,
\er
where here and everywhere else symbol
\be\label{aer}
{\cal N}^{L}=\int_{V_{\sss B}}\rho^*w^2w^{<L>}d^3w\;,\hspace{2cm}(l\ge 0)
\en
denotes a new STF object \cite{kls}.
We call attention of the reader to the fact that for $l=0$ the scalar function ${\cal N}\equiv{\cal I}^{(2)}$,
where ${\cal I}^{(2)}$ has been defined in Eq. (\ref{fop}).

It is worth noting that the post-Newtonian definitions of mass and of the center of mass of the body
depend not only on the internal distribution of matter's density, velocity, and stresses inside the body
but also on terms describing
the coupling of body's gravitational field with that of
external masses. As we shall show later, inclusion of these coupling terms
in definitions (\ref{ij1}), (\ref{ij2}) and (\ref{6.2a}) is absolutely
necessary in order to simplify translational equations of motion
as far as possible and bring them to the form which can be reduced to the Einstein-Infeld-Hoffmann
equations of motion in the limiting case of spherically symmetric bodies. In this sense, the question
about whether the coupling of internal and external gravitational fields should be included in the definitions
of the post-Newtonian mass and the center of mass, which was a matter of concern for Thorne and Hartle
\cite{th}, can be considered as having been resolved.

The post-Newtonian linear momentum of the body, ${\cal P}^i$, is defined as the first time derivative of
the dipole moment given by Eq. (\ref{6.2a}), that is ${\cal P}^i=\dot{\cal J}^i$, where dot indicates
a derivative with respect to the local coordinate time $u$. After taking the derivative one obtains
\br\label{6.2b}
{\cal P}^i&=&\int_{V_{\sss B}}\rho^*\nu^i
\left[1+\epsilon^2\left(\frac12\nu^2+\Pi-
\frac12\hat U^{(B)}\right)\right] d^3w\\\nonumber\\\nonumber&&+
\epsilon^2\int_{V_{\sss B}}\left[\pi_{ik}\nu^k-
\frac12\rho^*\hat W^{(B)}_i\right]d^3w\\\nonumber\\\nonumber&&-
\epsilon^2\frac{d}{du}\left\{\left[Y+(1-\gamma)P\right]
{\cal J}^i_*+\sum_{l=1}^\infty\frac{l+1}{l!}Q_L{\cal I}^{iL}\right.\\\nonumber\\\nonumber&&\left.+
\frac12\sum_{l=1}^\infty\frac{1}{(2l+1)(l-1)!}
Q_{iL-1}{\cal N}^{L-1} \right\}\\\nonumber\\\nonumber&&+\epsilon^2
\sum_{l=1}^\infty\frac{1}{l!}\left[Q_L\dot{\cal I}^{iL}+
\frac{l}{2l+1}Q_{iL-1}\dot{\cal N}^{L-1}\right]-\epsilon^2
\sum_{l=1}^\infty\frac{1}{l!}Q_L\int_{V_{\sss B}}\rho^*\nu^i w^L d^3w\;,
\er
where function
\be\label{wh2}
\hat W^{(\sss B)}_i(u,\nnw)=G\int_{V_{\sss B}}\frac{
   \rho^*(u,\nnw')\nu'^k(w^k-w'^k)(w^i-w'^i)}{|{\bm w}-{\bm w}'|^3}\, d^3w'\;.
\en

Until now the point ${\bm x}_{\sss B}(t)$ represented location of the origin of
the local coordinate system in the global coordinates taken at the time $t$. In general,
the origin of the local coordinates does not coincide with the center of mass of the body which can
move with respect to the local coordinates with some velocity and acceleration. In order to be able
to keep the center of mass of the body at
the origin of the local coordinates one must prove that for
any instant of time the dipole moment defined by Eq. (\ref{6.2a}) and its time
derivative (that is, the linear momentum of the body) given by Eq. (\ref{6.2b}) can be made equal to zero.
This requirement can be achieved, if and only if, the second time
derivative of the dipole moment with respect to the local coordinate time $u$
is identically equal to zero, that is
\be\label{6.2c}
\dot{\cal P}^i=0
\en
It is
remarkable that this equation can be satisfied after making an appropriate
choice of the external dipole moment $Q_i$ that characterizes
a locally measurable acceleration of the origin of the local coordinates with respect to another
local coordinate frame whose origin moves along a geodesic world line in the background space-time.
This statement has been proven
in \cite{kop1988} in the Newtonian approximation and, then, extended up
to the first general relativistic post-Newtonian approximation in
\cite{dsx1991}. In the present paper we shall derive the consequences of Eq. (\ref{6.2c}) in the
first post-Newtonian approximation of the scalar-tensor theory of
gravity characterized by two PPN parameters, $\beta$ and $\gamma$.

\subsection{Translational Equation of Motion in Local Coordinates}\label{xru}

Translational equation of motion of the body in the local coordinates, $w^\alpha=(cu, {\bm w})$, is
derived by making use of the definition of its {\it conformal} linear
momentum, ${\cal P}^i$, displayed in Eq. (\ref{6.2b}). Differentiating Eq. (\ref{6.2b}) one time with
respect to the local coordinate
time $u$, operating with the macroscopic
equations of motion, Eqs. (\ref{kp1}) -- (\ref{kp3}), and integrating by parts to re-arrange some terms,
one obtains
\br\label{6.2d}
    \dot{\cal P}^i&=&{\cal M}
    Q_i(u)+\nn{l=1}\frac{1}{l!}\label{6.2.e} Q_{iL}(u){\cal
    I}^{L}(u) + \epsilon^2\Delta\dot{\cal P}^i \\\nonumber&
    -\epsilon^2&\Biggl\{
    \nn{l=2}\frac{1}{(l+1)!}
    \biggl[ (l^2+l+4)Q_L - 2(1-\gamma)P_{L}\biggr] \Ddot{\cal I}^{iL}
    \\\nonumber&&
    +\nn{l=2}\frac{2l+1}{(l+1)(l+1)!}
    \biggl[ (l^2+2l+5)\dot Q_L - 2(1-\gamma)\dot P_{L}\biggr] \dot{\cal I}^{iL}
    \\\nonumber&&
    +\nn{l=2}\frac{2l+1}{(2l+3)(l+1)!}
    \biggl[ (l^2+3l+6)\Ddot Q_L - 2(1-\gamma)\Ddot P_{L}\biggr] {\cal I}^{iL}
    \\\nonumber&&
+\biggl[3Q_k-(1-\gamma)P_k\biggr]\Ddot{\cal I}^{ik}
+\frac{3}{2}\biggl[4\dot Q_k-(1-\gamma)\dot P_k\biggr]\dot{\cal I}^{ik}
    \\\nonumber&&+\frac{3}{5}\biggl[5\Ddot Q_k-(1-\gamma)\Ddot P_k\biggr]{\cal I}^{ik}
    +\nn{l=2}\frac{1}{l!}
    \dot Z_{iL}{\cal I}^{L}\\\nonumber&& +\nn{l=1}\frac{1}{(l+1)!}\varepsilon_{ipq}
    \biggl[\dot C_{pL}{\cal I}^{<qL>} + \frac{l+2}{l+1}C_{pL}\dot{\cal I}^{qL}\biggr] \\\nonumber&&
    -2\nn{l=1}\frac{l+1}{(l+2)!}\varepsilon_{ipq}\biggl[\Bigl( 2Q_{pL}-(1-\gamma)P_{pL}\Bigr)
    \dot{\cal S}^{qL}
    \\\nonumber&&+ \frac{l+1}{l+2}\Bigl( 2\dot Q_{pL}-
    (1-\gamma)\dot P_{pL}\Bigr) {\cal S}^{qL}\biggr]-\nn{l=1}\frac{l(l+2)}{(l+1)(l+1)!}
    C_{iL}{\cal S}^{L}  \\\nonumber&&
    -\frac{1}{2}\varepsilon_{ikq}\biggl[\Bigl(4Q_k-2(1-\gamma)P_k\Bigr)\dot{\cal S}^q
    +\Bigl(2\dot Q_k-(1-\gamma)\dot P_k\Bigr){\cal S}^q\biggr]
    \\\nonumber&&
    +\Bigl(P_i-Q_i\Bigr)\biggl[\frac{1-\gamma}{6}\Ddot{\cal I}^{(2)}+\frac{1}{2}\eta\int_{V_{\sss B}}\rho^*
    \hat{U}^{\scriptstyle(B)}d^3w
    \\\nonumber&&+\nn{l=2}\frac{1}{l!}\biggl(2(\beta-1)P_{L}-(1-\gamma)lQ_L\biggr){\cal
    I}^{L}
    \biggr]\Biggr\}  +O(\epsilon^4)\;,
\er
where we have shown explicitly all terms proportional to $Q_i$, $\eta=4\beta-\gamma-3$, and the
post-Newtonian correction $\Delta\dot{\cal P}^i$ is given by
\br\label{bvo} \Delta\dot{\cal P}^i&=&
\nn{l=1}\frac{1}{l!}\biggl[\frac{2(1-\gamma)(2l+1)}{(2l+3)(l+1)}
\frac{d}{du} \int_{V_{\sss B}}\rho^*\nu^k w^{<kL>}d^3w \\\nonumber&&+(\gamma-1)
\int_{V_{\sss B}}(\rho^*\nu^2+{\hat\sigma}^{kk})w^{L}d^3w\\\nonumber&&
+
2(1-\beta)\int_{V_{\sss B}}\rho^*\Bigl(\hat{U}^{\scriptstyle(B)}+\nn{n=0}\frac{1}{n!}P_{N}w^{N}\Bigr)w^{L}d^3w\biggr]
\biggl(P_{iL}-Q_{iL}\biggr)  \\\nonumber &&
 +2(\beta-1)P\biggl({\cal
M}Q_i+\nn{l=1}\frac{1}{l!}P_{iL}{\cal I}^{L}\biggr)\\\nonumber &&
+2\biggl[2Q-Y+(\gamma-1)P\biggr]\biggl({\cal
M}Q_i+\nn{l=1}\frac{1}{l!}Q_{iL}{\cal I}^{L}\biggr)\\\nonumber &&
+(Q_i-P_i)\biggl[2(\beta-1)P_k+(\gamma-1)Q_k\biggr]{\cal I}^k
-\frac{1}{3}\biggl[6\Ddot Y+4(1-\gamma)\Ddot P\biggr]{\cal I}^i\\\nonumber && -
\dot Z_{ik}{\cal I}^k
-\varepsilon_{ipq}\biggl(\dot C_p{\cal I}^q+2C_p\dot{\cal
I}^q\biggr)\\\nonumber && - \biggl[\dot Q + 4\dot Y +2(1-\gamma)\dot
P\biggr]\dot{\cal I}^i - 2\biggl[2Q+(\gamma-1)P\biggr]\Ddot{\cal
I}^i\;.
\er
It is worth noticing that mass ${\cal M}$ is the {\it active} mass, and the STF multipole moments
${\cal I}^L$ ($l\ge 1$) of the body, which appear in the right side of Eqs. (\ref{6.2d}) and (\ref{bvo}),
are the {\it active} mass multipole moments depending on time $u$. Function $Q_i$ in Eq. (\ref{6.2d})
has not been yet restricted and can be chosen arbitrary. Its choice determines a world line of the origin
of the local coordinates. If one chooses $Q_i=0$, then the origin of the local frame, ${\bm x}_{\sss B}$,
is moving along a geodesic world line defined in the global frame by Eq. (\ref{5.8}) and the center of
mass of the body under consideration is moving in accordance with the law of motion, Eq. (\ref{6.2d}),
with respect to this geodesic. For practical applications, however, it is more convenient to chose the
origin of the local frame to be always located at the center of mass of the body. This can be accomplished
by imposing condition (\ref{6.2c}), that is $\dot{\cal P}^i\equiv\Ddot{\cal
J}^i=0$. If this condition is satisfied it allows us to chose $\dot{\cal J}^i={\cal J}^i=0$ exactly,
which leads to similar conditions for the {\it active} dipole moment, $\Ddot{\cal
I}^i=\dot{\cal I}^i={\cal I}^i=0$, in the Newtonian approximation. However, in the post-Newtonian
approximation the {\it active} dipole moment of the body, ${\cal I}^i\not=0$, even if the {\it conformal}
multipole moment of the body, ${\cal J}^i=0$, because these two moments are defined via different equations
(see Section \ref{mdloc}).

Fixing the origin of the local frame at the body's center of mass (${\cal J}^i=0$) and noticing
that $Q_L=P_L$ for any $l\ge2$ makes the post-Newtonian function $\Delta\dot{\cal P}^i=0$. Then, one can find
solution of Eq. (\ref{6.2d}) for function $Q_i$ which was considered until now as arbitrary variable.
We remind that the physical meaning of $Q_i$ is
acceleration of the origin of the local coordinates with respect to a geodesic world line in the
background space-time defined by the gravitational field of external bodies. Equating $\dot{\cal P}^i$,
separating acceleration $Q_i$ of the body's center of mass from all other terms in Eq.(\ref{6.2d})
and taking all terms with $Q_i$ to the right side yields the following post-Newtonian equation
\br \label{9.4.3}
\tilde{\cal
M}_{ij}Q_j={\mathbb {F}}^i_N+\epsilon^2\left({\mathbb {F}}^i_{pN} +
\Delta{\mathbb {F}}^i_{pN}\right) + O(\epsilon^4)\;, \er
where the {\it conformal} anisotropic tensor of mass
\br \label{brt}
\tilde{\cal M}_{ij}&=&  {\tilde
{\cal
M}}\delta_{ij}-\epsilon^2\biggl[3\Ddot{\cal I}^{ij}-
2\nn{l=1}\frac{1}{l!}\Bigl(Q_{jL}{\cal I}^{iL} - Q_{iL}{\cal
I}^{jL}\Bigr)\biggr]\;, \er
and the (tidal) gravitational forces
\br
\label{6.9}
{\mathbb {F}}^i_N& = & -\nn{l=1}\frac{1}{l!}\,
Q_{iL}(u){\cal I}^{L}(u)\;,
\\\nonumber\\\label{qopx}
{\mathbb {F}}^i_{pN}& = &
\nn{l=2}\frac{l^2+l+2(1+\gamma)}{(l+1)!}\,
 Q_L\Ddot{\cal I}^{iL}+6\dot Q_k\dot{\cal I}^{ik}+3\Ddot Q_k{\cal I}^{ik} \\\nonumber&&
+\nn{l=2}\frac{(2l+1)}{(l+1)!}\biggl(\frac{l^2+2l+2\gamma+3}{l+1}\,  \dot
Q_L\Dot{\cal I}^{iL}
+\frac{l^2+3l+2\gamma+4}{2l+3}\,\Ddot
Q_L{\cal I}^{iL}\biggr)\\\nonumber&&
+\nn{l=2}\frac{1}{l!}\varepsilon_{ipq}\left( \dot
C_{pL-1}{\cal I}^{qL-1}+ \frac{l+1}{l}\,C_{pL-1}\Dot{\cal
I}^{qL-1}\right) \\\nonumber &&
-2(1+\gamma)\nn{l=2}\frac{l}{(l+1)!}\varepsilon_{ipq}
\left(Q_{pL-1}\Dot{\cal S}^{qL-1}+ \frac{l}{l+1}\,\Dot
Q_{pL-1} {\cal S}^{qL-1}\right)\;,
\\\nonumber &&
-\nn{l=1}\frac{l(l+2)}{(l+1)(l+1)!}\,C_{iL}{\cal S}^{L}+\nn{l=2}\frac{1}{l!}\,\dot
Z_{iL}{\cal I}^{L}-\varepsilon_{ipq}\dot Q_p{\cal S}^q
\\\nonumber \\\label{fopm}
\Delta{\mathbb{F}}^i_{pN} &=&
(1-\gamma)\biggl(\frac{1}{2}\varepsilon_{ikq}\dot P_k{\cal S}^q
-\frac{3}{2}\dot P_k\dot{\cal I}^{ik}-\frac{3}{5}\Ddot P_k{\cal
I}^{ik}\biggr) \\\nonumber&
+&(1-\gamma)\Biggl[\nn{l=1}\frac{1}{l!} \biggl(Q_{kL}{\cal
I}^{iL} - Q_{iL}{\cal I}^{kL}\biggr)-\Ddot{\cal
I}^{ik}\Biggr]P_k\\\nonumber &
+&\biggl[\frac{1}{2}\eta\int_{V_{\sss B}}\rho^*
\hat{U}^{\scriptstyle(B)}d^3w+\frac{1-\gamma}{6}\,\Ddot{\cal
I}^{(2)}
+\nn{l=2}\frac{2(\beta-1)+(\gamma-1)l}{l!}\,Q_L{\cal
I}^{L}\biggr]P_i \;. \er

Eqs. (\ref{9.4.3}) -- (\ref{fopm}) describe the law of translational motion of the body in the local
coordinates in the presence of external bodies which create a force dragging motion of the body's center
of mass from geodesic world line. Newtonian, ${\mathbb {F}}^i_{N}$, and the post-Newtonian,
 ${\mathbb {F}}^i_{pN}$, tidal forces are caused by gravitational coupling of the body's internal ({\it active})
  multipole moments, ${\cal I}_L$ and ${\cal S}_L$, with the external multipole moments, $Q_L$ and $C_L$.
  The post-Newtonian tidal force, ${\mathbb {F}}^i_{pN}$, is reduced in the limit of $\gamma =1$ to general
  relativistic expression derived previously by Damour, Soffel and Xu \cite{dsx1991}.

It is worthwhile to emphasize that summation with respect to index $l$ in Eq. (\ref{6.9}) begins from $l=1$.
The point is that we have defined the center of mass of the body $B$ in terms of the {\it conformal}
dipole moment ${\cal J}^i$ by the condition ${\cal J}^i=0$. However, the force  ${\mathbb {F}}^i_{N}$
depends on the {\it active} multipole moments of the body but the {\it active} dipole moment
${\cal I}^i\not={\cal J}^i$ and, hence  ${\cal I}^i\not=0$. For this reason, one has to take into account
the contribution to the force ${\mathbb {F}}^i_{N}$ coming out of the non-zero {\it active} dipole of the body.
This contribution has a post-Newtonian order of magnitude and can be written down in more explicit form as
\be\label{zaq}
\left({\mathbb {F}}^i_{N}\right)_{\rm dipole}=-Q_{ij}{\cal I}^j\;,
\en where the active dipole moment ${\cal I}^i$ is \footnote{We remind that the conformal dipole moment of the
body, ${\cal J}^i=0$. Hence, Eq. (\ref{zaq}) is, in fact, a difference between the {\it active} and
{\it conformal} dipole moments of the body, ${\cal I}^i-{\cal J}^i$ .}
\br\label{zop}
 {\cal I}^j&=&\epsilon^2\biggl\{-\frac{1}{2}\eta\int_{V_{\sss B}}\rho^*\hat U_{\sss{(B)}}w^jd^3w
 +\frac{1}{5}(\gamma-1)\left[3\dot{\cal
R}^j-\frac12\ddot{\cal N}^j\right]
\\\nonumber &+&
\sum_{l=0}^{\infty}\frac{1}{l!}\Bigl[(1-\gamma)l\,Q_L+2(1-\beta)P_L\Bigr]{\cal I}^{jL}
\\\nonumber &+&
\frac{1}{2}\sum_{l=0}^{\infty}\frac{1}{(2l+3)l!}\Bigl[(\gamma-1)Q_{jL}+
4(1-\beta)P_{jL}\Bigr]{\cal N}^L\biggr\}+
 O\left(\epsilon^4\right)\,,
\er
where
\be\label{hamo}
{\cal R}^i=\int_{V_{\sss B}}\rho^*\nu^k w^{<k}w^{i>}d^3w\;.
\en
If one takes into account explicit relationships between the multipole moments $P_L$ of the scalar
field and gravitational potential of the external bodies, then Eq. (\ref{zop}) can be slightly simplified
\br \label{zaop}
{\cal I}^j &=& \epsilon^2\biggl\{-\frac{1}{2}\eta\int_{V_{\sss B}}\rho^*\hat U_{\sss{(B)}}w^jd^3w
+\frac{1}{5}(\gamma-1)\left[3\dot{\cal
R}^j-\frac12\ddot{\cal N}^j\right]
\\\nonumber &+&
2(1-\beta)\left[\bar U\nnxe{\cal I}^j+ \nnaa{k}{\cal
I}^{jk} +\frac13\nnaa{j}{\cal N}\right]
\\\nonumber&-&
\frac{\eta}{2}\sum_{l=0}^{\infty}\frac{1}{(2l+3)l!}Q_{jL}{\cal
N}^L+
\sum_{l=1}^{\infty}\frac{(1-\gamma)l+2(1-\beta)}{l!}Q_L{\cal
I}^{jL}\biggr\}+O\left(\epsilon^4\right)\,.
\er
It is clear that dipole moment of the central body can contribute to the equations of motion of the body
only in the scalar-tensor theory of gravity because in general relativity $\beta=\gamma=1$. It is important
to notice the presence of the Nordtvedt parameter $\eta=4\beta-\gamma-3$ in the {\it active} dipole moment
of the body. It is likely that we observe a general feature of the relativistic equations of motion in the
scalar-tensor theory. It looks like each {\it active} multipole moment of the body under consideration has
a contribution being proportional to the Nordtvedt parameter $\eta$. This, for example, leads to inequality
between inertial and gravitational masses of the body in case of the monopole mass moment. Presence of the
Nordtvedt parameter-dependent term(s) in dipole moment is insignificant for the bodies whose shape is close
to spherically-symmetric. However, it may play a role in motion of sub-systems, like that of Earth and Moon,
which possess large deviations from spherical symmetry. Experimental study of this problem would be desirable.

Gravitational force $\Delta{\mathbb {F}}^i_{pN}$ is also essential only in the scalar-tensor theory of gravity.
It is proportional to the dipole moment of the scalar field $P_i$ and its time derivatives. Had the scalar field
absent the dipole moment $P_i$ of the scalar field could not exist and the force $\Delta{\mathbb {F}}^i_{pN}$
would be zero. The dipole moment of the scalar field $P_i$ couples with the self-gravitational energy
$\sim \int\rho^*\hat{U}^{\scriptstyle(B)}$ of the body as well as with the energy of external gravitational
field and kinetic energy of the body's internal motion \footnote{Observe the term $\Ddot{\cal I}^{(2)}$}.
In the next section we shall show that it is this coupling that is responsible for inequality of inertial
and gravitational masses of the body and should be treated as a violation of the strong principle of equivalence.

\subsection{Equation of Translational Motion in Global Coordinates}

Equation of translational motion of the body's center of mass in the global coordinates $x^\alpha=(ct, {\bm x})$
can be obtained from the equation of motion, Eq. (\ref{5.8}), of the origin of the local coordinates,
${\bm x}_{\sss B}$, if the acceleration $Q_i$ is subject to obey to the local equation of motion. (\ref{9.4.3}).
The acceleration $Q_i$ depends on scalar multipoles $P_L$ so that after substitution Eq. (\ref{9.4.3})
to Eq. (\ref{5.8}) one has to make use of Eq. (\ref{3.13}), that is $P_L=\bar{U}_{,L}({\bm x}_{\sss B})$.
Moreover, one has to use Eq. (\ref{3.16}) to replace in forces $\Delta{\mathbb {F}}^i_{N}$,
${\mathbb {F}}^i_{pN}$, and $\Delta{\mathbb {F}}^i_{pN}$ all terms depending explicitly on $Q_i$
with a linear combination of the barycentric acceleration $a^i_{\sss B}$ and the gradient
$\bar{U}_{,i}({\bm x}_{\sss B})$ of the gravitational potential of external bodies,
$Q_i=\bar{U}_{,i}({\bm x}_{\sss B})-a^i_{\sss B}$. Taking all terms depending on the gravitational gradient
$\bar{U}_{,i}({\bm x}_{\sss B})$ to the right side of Eq. (\ref{5.8}) and those depending on the barycentric
acceleration $a^i_{\sss B}$ to the left side, bring Eq. (\ref{5.8}) to the following form
\br \label{9.5.1}
{\tilde{\cal M}}_{\sss B}\nnaa{i} &=& {\cal M}_{\sss B}\Biggl\{\bar
U_{,i}\nnxe+\epsilon^2\biggl[\bar\Phi_{,i}\nnxe-\frac{1}{2}\bar{\chi}_{,itt}\nnxe\biggr]\Biggr\} -
{\mathbb{F}}^i_N
 -\epsilon^2{\mathbb{F}}^i_{pN} \\\nonumber & +&\epsilon^2{\cal M}_{\sss B}\Biggl\{\biggl[\gamma
\delta_{ik}\nnv{2}-\nnv{i}\nnv{k} -2(\gamma+\beta)\delta_{ik}\bar
U\nnxe\biggr]\bar U_{,k}\nnxe -{\cal A}\dot{Q}_i\\\nonumber &&
+2(1+\gamma)\dot{\bar
U}^i\nnxe-2(1+\gamma)\nnv{k}\bar U^{k,i}\nnxe
-(1+2\gamma)\nnv{i}\dot{\bar U}\nnxe\Biggr\} \\\nonumber &
+&\epsilon^2\Biggl\{\biggl[2Q-Y-\nnv{2}-(2+\gamma)\bar
U\nnxe\biggr]\delta_{ik} -\frac12\nnv{i}\nnv{k}-{F}^{ik}+3\ddot{\cal
I}^{ik} \\\nonumber &&+
 2\nn{n=1}\frac{1}{n!}\biggl[Q_{iN}{\cal
I}^{kN} - Q_{kN}{\cal I}^{iN}\biggr]\Biggr\}
\nn{l=2}\frac1{l!}\,Q_{kL}{\cal I}^{L}\\\nonumber
& +&\epsilon^2(1-\gamma)\Biggl\{\ddot{\cal I}^{ik}\bar
U_{,k}\nnxe +\frac32\dot{\cal I}^{ik}\dot{\bar U}_{,k}\nnxe
+\frac35{\cal I}^{ik}\ddot{\bar U}_{,k}\nnxe
\\\nonumber &&+\frac12\varepsilon_{ipk}{\cal S}^p\dot{\bar U}_{,k}\nnxe
 +\nn{l=1}\frac1{l!}\biggl(
Q_{iL}{\cal I}^{kL} - Q_{kL}{\cal I}^{iL}\biggr)\bar
U_{,k}\nnxe \Biggr\} + O(\epsilon^4)\;,\nonumber
\er
where the external potentials $\bar U({\bm x}_{\sss B})$, $\bar U^i({\bm x}_{\sss B})$, $\bar\Phi({\bm x}_{\sss B})$,
and $\bar\chi({\bm x}_{\sss B})$ are defined in Eq. (\ref{omap}) and are taken on the world line of the center of
mass of the body B. The external potentials can be expanded in multipolar series so that the translational equation
of motion (\ref{9.5.1}) will depend only on the {\it active} multipole moments of the bodies. We do not present this
general result in the present paper but consider a more simple case of spherically-symmetric and rotating bodies in
section \ref{wep}.

One has to notice that {\it inertial} mass $\tilde{\cal M}_{\sss B}$ of the body B is its {\it conformal} mass.
It is not equal to {\it gravitational} mass ${\cal M}_{\sss B}$ of the body in the right side of this equation
which is its {\it active} mass. Difference between the two masses is given by Eq. (\ref{acmass}) and it causes
violation of the strong principle of equivalence for massive extended bodies. Existence of the possible difference
between the inertial and gravitational masses in alternative theories of gravity was pointed out by Dicke
\cite{brd,dicke,dicke1} and Nordtvedt \cite{nord1,nord2,llr}. In our calculations the inertial-gravitational
mass difference originates from Eq. (\ref{fopm}) which has terms proportional to scalar dipole,
$P_i=\bar{U}_{,i}({\bm x}_{\sss B})$, and they contribute to the gravitational (active) mass only.

Forces ${\mathbb{F}}^i_N$ and ${\mathbb{F}}^i_{pN}$ are given by Eqs. (9.4.5) and (9.4.6).
Terms in the first and second curled brackets of Eq. (\ref{9.5.1})
being proportional to mass ${\cal M}_{\sss B}$ are the post-Newtonian corrections to the Newtonian
force acting on the body B considered as a monopole massive particle . The group of terms in third
and forth curled brackets in Eq. (\ref{9.5.1}) represents the post-Newtonian correction to the Newtonian
tidal force ${\mathbb{F}}^i_N$ and takes into account higher order multipoles of the body B. In particular,
these terms contain time-dependent functions $Q$ and $Y$ which define the unit of time and length in the
local coordinates. This correction also contains the matrix of relativistic precession ${F}^{ik}$
given in Eq. (\ref{5.18}).  Eq. (\ref{9.5.1}) describes a generic case of translational equation of
motion of extended bodies having arbitrary shape and rotation (all multipoles). We derive rotational
equation of motion of the body B in the next section.

\section{Rotational Equations of Motion of Extended Bodies}\label{req}
Rotational equations of motion for each body define orientation of the body's angular momentum (spin) at each
instant of time with respect to the local frame of reference $w^\alpha=(cu,{\bm w})$ which axes are not
dynamically-rotating, that is Fermi-Walker transported \cite{mtw} in accordance with Eq. (\ref{5.18})
describing orientation of the axes of the local coordinates with respect to the global coordinates at
each instant of time. We shall work out relativistic equations of the rotational motion by making use
of the method proposed in \cite{dsx1993}.
\subsection{Post-Newtonian Definition of the Angular Momentum of the Body}
First of all, one needs to introduce definition of the angular momentum of an extended body from the N-body
system. In principle, one had to use the same principle of the multipolar expansion of the metric tensor
applied to its $\hat{g}_{0i}(u,{\bm w})$ component calculated with taking into account all terms of the
next post-Newtonian approximation. This procedure was applied by Damour and Iyer \cite{di,dyr2} for the
post-Newtonian definition of angular momentum of an isolated system. It is not known yet how to apply
the Damour-Iyer procedure to a single body from the N-body system because of its complexity. Therefore,
we shall use the approach proposed in \cite{dsx1993} to bypass this difficulty.

Let us introduce a {\it bare} spin of a single body B by making use of the following post-Newtonian
definition
\be\label{spin-1}
{\cal S}^i =
\frac{1}{c}\int\varepsilon_{ijk}w^j\hat{\Theta}^{0k}d^3w\,,
\en
where
\be\label{spin-2}
\hat{\Theta}^{0k}=(-\hat g)\phi\left(T^{0k}+\hat{t}^{0k}\right)\;,
\en
 is a linear combination of tensor of energy-momentum of matter $T^{\mu\nu}$ and the pseudo-tensor of
 gravitational field $\hat{t}^{\mu\nu}$. We have defined the bare spin of the body B by Eq. (\ref{spin-1})
 because it corresponds to the conserved spin of an isolated system (see Eq. \ref{13.361} for more detail).
 We also assume that the center of mass of the body B is chosen such that its conformal dipole moment ${\cal J}^i$ is zero.

 Integration in Eq. (\ref{spin-1}) is formally performed in local coordinates over entire space.
 However, tensor of energy-momentum $T^{0k}$ includes the matter of the body B only and depends on
 the complete metric tensor $\hat{g}_{\mu\nu}$ in the local coordinates. We assume that the
 pseudo-tensor $\hat{t}^{0k}$ depends only on the internal part of the local metric tensor for the body B.
 Integration by parts allows us to reduce the bare spin of the body B in Eq. (\ref{spin-1}) to the following expression
 \br\label{spin-3}
 {\cal S}^i  &=&\frac{1}{c^2}
\int_{V_{\sss B}}\varepsilon_{ijk}w^j\biggl\{\rho^*\nu^k\Bigl[c^2+\frac12\nu^2+\Pi+(2\gamma+1)\hat
U_{\sss{(B)}} \\\nonumber&+&
\sum_{l=1}^{\infty}\frac1{l!}\Bigl(3Q_L+2(\gamma-1)P_L\Bigr)w^L+3Y+(1-\gamma)P\Bigr]
\\\nonumber&+&
\pi^{kn}\nu^n-\frac{1}{2}\,\rho^*\Bigl[\hat
W_k^{\sss{(B)}} +(3+4\gamma)\hat
U^k_{\sss{(B)}}\Bigr]\biggr\}+O\left(\epsilon^3\right)\,,
\er
where integration is over the volume of the body B, and potential $\hat W_k^{\sss{(B)}}$
is defined by Eq. (\ref{wh2}).
We shall use Eq. (\ref{spin-3}) to derive rotational equations of motion of body's angular momentum.

\subsection{Equations of Rotational Motion in Local Coordinates}

Rotational equations of motion for body's spin are derived by differentiation of Eq. (\ref{spin-3})
with respect to the local coordinate time $u$. After taking the time derivative and making use of the
macroscopic equations of motion in local coordinates given in section \ref{qmez-1}, one makes several
transformations of the integrand to reduce similar terms and to simplify final result. After tedious but
straightforward calculations done in the spirit of paper \cite{dsx1993}, one obtains equations of the
rotational motion of body B in its own local coordinate frame
\be\label{spin-4}
\frac{d {\cal S}^i}{du} = {\cal T}^i + \epsilon^2\left(\Delta{\cal
T}^i - \frac{d}{du}\Delta{\cal S}^i\right)+O\left(\slch^4\right)\,,
\en
where ${\cal T}^i$ is a general-relativistic torque for $\gamma=1$, and $\Delta{\cal
T}^i$ is its post-Newtonian correction due to the presence of the scalar field, while $\Delta{\cal S}^i$
can be considered as a supplementary post-Newtonian contribution to the bare spin ${\cal S}^i$.
The torque and other terms in the right side of Eq. (\ref{spin-4}) read as follows:
\br\label{spin-5} {\cal T}^i &=& \sum_{l=0}^{\infty}\frac1{l!}\varepsilon_{ijk}
\left[{\cal I}^{jL}\biggl(Q_{kL}-\slch^2\dot
Z_{kL}\biggr)+\slch^2{\cal S}^{jL}C_{kL}\right]\,,
\\\nonumber\\\label{spin-6}
\Delta{\cal T}^i &=&
\varepsilon_{ijk}\nnaa{j}\left[\frac{3(1-\gamma)}5\dot{\cal R}^k
+\frac{\gamma-1}{10}\ddot{\cal N}^k +
\frac{\eta}2\int_{V_{\sss B}}\rho^*\hat U_{\sss{(B)}}w^kd^3w
\right.\\\nonumber&+&\left.
\frac{\eta}2\sum_{l=0}^{\infty}\frac1{(2l+3)l!}Q_{kL}{\cal N}^L+
\sum_{l=1}^{\infty}\frac{(\gamma-1)l+2(\beta-1)}{l!}Q_L{\cal
I}^{kL}\right]
\\\nonumber&+&2(\beta-1)\nnaa{n}{\cal I}^{kn} +
\Bigl[Y+Q+(2\beta-\gamma-1)\bar
U\nnxe\Bigr]\sum_{l=0}^{\infty}\frac1{l!}\varepsilon_{ijk}{\cal
I}^{jL}Q_{kL}\,,
\\\nonumber\\\label{spin-7}
\Delta{\cal S}^i &=&
-\sum_{l=1}^{\infty}\frac{1}{l!}{\cal I}^{iL}C_L
+\sum_{l=0}^{\infty}\frac{l+2}{(2l+3)(l+1)!}{\cal
N}^{L}C_{iL}
\\\nonumber&+&
\sum_{l=0}^{\infty}\frac{1}{(2l+5)l!}\varepsilon_{ijk}\left[\frac12\dot{\cal
N}^{jL}Q_{kL}- \frac{l+2(2\gamma+3)}{2(l+2)}{\cal N}^{jL}\dot
Q_{kL}-\frac{2(1+\gamma)(2l+3)}{l+2}{\cal R}^{jL}Q_{kL}\right]
\\\nonumber &+&
\frac{1-\gamma}5\varepsilon_{ijk}\Bigl[3{\cal R}^j\nnaa{k}+{\cal
N}^j\da{k}\Bigr]+ \Bigl[Q-Y+(\gamma-1)\bar U\nnxe\Bigr]{\cal S}^i\,,
\er
where
\be\label{spin-8}
{\cal R}^L=\int_{V_{\sss B}}\rho^*\nu^k w^{<kL>}d^3w\;,
\en
is additional set of multipole moments which has been used already in definition of the multipole moments
in section \ref{miopa}.

General relativistic torque ${\cal T}^i$ depends on the multipole moments $Z_L$ which define the residual
gauge freedom. They can be used to simplify the post-Newtonian correction to the torque, $\Delta{\cal T}^i$.
This correction is, in fact, exactly equivalent to $\Delta{\cal T}^i=\varepsilon_{ijk}\nnaa{j}
\left({\cal I}^i-{\cal J}^i\right)$, where ${\cal I}^i$ and ${\cal J}^i$ are {\it active} and {\it conformal}
dipole moments of the body B respectively. The difference between the two dipole moments taken under
condition that ${\cal J}^i=0$ is given by Eq. (\ref{zaop}) and has been reproduced in Eq. (\ref{spin-6}).
We have taken into account the external monopole moments $Q$ and $Y$ defining the units of measurement
of the local time and spatial coordinates respectively. Their contribution is to the rotational equations
of motion is extremely small and can be omitted.

We re-define the spin of the body as
\be\label{spin-9}
{\cal S}^i_{+} = {\cal S}^i + \epsilon^2\Delta{\cal S}^i\,,
\en
so that equations of rotational motion acquire their final form
\be\label{spin-10}
\frac{d}{du}{\cal S}^i_{+} = {\cal T}^i + \epsilon^2\Delta{\cal T}^i + O\left(\epsilon^4\right)\,.
\en
These equations should be compared with analogous equations derived by Klioner and Soffel \cite{kls}
\footnote{These equations are numbered (9.42) -- (9.47) in \cite{kls}.} First of all, we notice that our
definition of the multipole moments ${C}_L^{\rm Kopeikin-Vlasov}$ differs by a numerical factor $(l+1)/l$
from that, $C_L^{\rm Klioner-Soffel}$, used by Klioner and Soffel, that is
\be\label{ao4}
{C}_L^{\sss\rm Kopeikin-Vlasov}=-2(1+\gamma)\,\frac{l+1}{l}\,C_L^{\sss\rm Klioner-Soffel}\;.
\en
Comparison of spins, ${\cal S}^i_{+}$ -- our notations, and, $S'^i$ -- Klioner-Soffel's notation, shows
that they are equal if sign minus in front of three last terms in Eq. (9.45) of Klioner-Soffel's paper \cite{kls},
is replaced with sign plus. The general relativistic torque in the present paper coincides naturally with that
derived in \cite{kls}. The biggest difference occurs between the post-Newtonian correction $\Delta{\cal T}^i$
to the torque in this paper and a corresponding quantity given in two equations, (9.43) and (9.46),
in the paper \cite{kls}. First three terms in our Eq. (\ref{spin-6}) completely coincides with equation (9.43)
derived by Klioner and Soffel, thus, confirming the presence of the Nordtvedt effect for rotational motion of
the bodies. However, we obtained different terms in the second and third lines of our Eq. (\ref{spin-6})
as contrasted with equation (9.46) of Klioner and Soffel's paper \cite{kls}. We suppose that the difference
may come out of slightly different gauge conditions used in this paper and in \cite{kls}. Additional origin
of the difference is that we used definition of the center of mass of the body which is not reduced to that
used by Klioner and Soffel.

\section{Motion of spherically-symmetric and rigidly-rotating bodies}\label{wep}
\subsection{Definition of spherically-symmetric and rigidly rotating body}

It is well understood that the notion of a spherically-symmetric and rigidly rotating body is not
invariant but coordinate-dependent \cite{ll,mtw}. According to special theory of relativity, coordinate
grid of a moving coordinate frame is linearly deformed and magnitude of this deformation depends on velocity
of the frame with respect to a reference frame being at rest. This consideration assumes that
if one considers a spherically-symmetric body in a static frame it will be not spherically-symmetric in
a moving frame.
Deformation of the body's shape can be calculated by applying the Lorentz transformation to the equation
describing the shape of the body in the static frame \cite{bt}. The Lorentz deformation is  solely coordinate
effect which does not lead to appearance of physical stresses (tensions) inside the moving body. Nonetheless,
the Lorentz deformation of the body's shape has to be taken into account for correct calculation of observed
physical effects associated with motion of the body. Poincar\'e and Lorentz were first who took into account
special relativistic deformation of a moving electron for calculation of the electromagnetic radiation-reaction
force exerted on the electron due to the emission of electromagnetic radiation \cite{jack}.

In general relativity gravitational field causes deformation of coordinate grid of a static frame with respect
to the grid of the same frame taken in the absence of gravitational field \footnote{This represents pure
mathematical comparison. It is not physically possible to turn off or to screen gravitational field.}.
Hence, this distortion of the coordinate grid represent pure mathematical effect and does not cause physical
deformation of the body being at rest with respect to this frame. However, gravity-caused deformations of the
local coordinate's grid must be taken into account in calculation of translational equations of motion of the
body with respect to the global coordinate frame. It is worth mentioning that one has to distinguish the
mathematical deformations of the frame from the physical (tidal) deformations of the body itself \cite{sokx}.
This can be achieved if a precise relativistic theory of reference frames is employed. The PPN formalism
brings new complications due to the presence of additional fields which can cause both coordinate and physical
deformations of the body's volume.

Post-Newtonian definition of the multipole moments of the gravitational field is coordinate-dependent.
Therefore, the explicit structure of the multipolar expansion and the number of terms present in this
expansion crucially depend on the choice of coordinates. Transformation from local coordinates to global
ones will change mathematical description of the multipole moments and we have to be careful in finding
the most precise formulation of the notion of spherically-symmetric body to avoid introduction of non-physical
multipole moments. Any misunderstanding of this concept will lead to inconsistencies in calculation of
equations of motion for N-body problem in the first \cite{kop1985} and higher-order post-Newtonian
approximations \cite{itoh} and/or appearance of spurious coordinate-dependent terms having no physical meaning.

We assume that for each body of N-body system the geometrical center of the body's spherical symmetry is
located at the center of mass of the body that coincides with the origin of the local coordinates associated
with this body. We assume that all functions characterizing internal structure of the body have spherically-symmetric
distribution in the local coordinates. These functions are:
the invariant density $\rho^*$, the internal energy $\Pi$, and the stress tensor $\pi_{ij}$. Spherical symmetry
in the local coordinates means that these functions depend only on the local radial coordinate $r=|{\bm w}|$:
\be\label{as7}
\rho^*(u, {\bm w})=\rho^*(r)\;,\qquad\qquad\Pi(u, {\bm w})=\Pi(r)\;,\qquad\qquad
\pi^{ij}(u, {\bm w})=\delta^{ij}p(r)\;.\en
Moreover, we assume that the internal distribution of matter does not depend on the local coordinate time $u$
that excludes radial oscillations of the body from consideration. Radial oscillations can be easily included
in our version of the PPN formalism but we postpone treatment of this problem for future work
\footnote{Notice that radial oscillations are irrelevant for consideration of this problem in general relativity
due to the Birkhoff's theorem \cite{ll,mtw}}.

Spherically-symmetric distribution of matter must generate a spherically-symmetric gravitational field.
Therefore, the
multipolar expansion of the Newtonian gravitational potential of the body must have in the local coordinates only
a monopole term
\be\label{as8}
\hat U_{\sss B}(u,{\bm w})=G\int_{V_{\sss B}}\frac{\rho^*(u, {\bm w}')d^3w'}{|{\bm w}-{\bm w}'|}=
\frac{G{\cal M}_{*{\sss B}}}{r}\;,
\en
where the baryon (Newtonian) mass ${\cal M}_{*{\sss B}}$ is defined in Eq. (\ref{a}).
Strictly speaking, this monopole expansion will be violated at some order of approximation because the external
tidal force of the background gravitational field acts on the body and deforms its spherically-symmetric
distribution of matter. This tidal distortion is proportional to some numerical coefficient
\footnote{Love's number $k_2$ \cite{zhtr}.} which characterizes elastic properties of the body under
consideration. Assuming that the body is made of matter with sufficiently low elasticity one can reduce
the tidal deformation of the body to a negligibly small value, at least in the first post-Newtonian
approximation \footnote{Analytic estimate of the magnitude of the tidal deformation comparatively with
the magnitude of the post-Newtonian forces has been done in \cite{dam1983,kop1985}.}.

We shall consider the case of rigidly rotating bodies for
which the internal velocity of matter (as defined in the local coordinates) is a vector product of the
angular velocity $\Omega_{\sss
B}^i$, referred to the local frame, and the radius-vector $w^i$, that is
\br\label{qma}
\nu^i&=&\varepsilon^i_{\;jk}{\Omega}^j_{\sss
B}w^k\\\label{pa3}
\nu^2&=&\frac{2}{3}\Omega^2_{\sss B} r^2-\Omega^j_{\sss B}\Omega^k_{\sss B} w^{<jk>}\;,
\er
in the local frame of the body B. Again, one should remind that rotation causes rotational deformation
of the body and distorts spherical symmetry. However, the rotational deformation is proportional to the
same Love's number, $k_2$, \cite{zhtr} and by assuming that the body is rigid enough and rotates sufficiently slow,
one can make the rotational deformation to be negligibly small. In what follows we shall use this assumption and
neglect the rotational deformation.

Spherical symmetry of each body assumes that one can use the following (pure geometric) properties which are valid
for any function $f(r)$, depending on radial coordinate $r$ only \cite{thorne}:
\br
\label{9.6.1}
    \int_{V_{\sss B}}f(r)w^{i_1i_2...i_{2l}}d^3w&=&\frac{1}{2l+1}\,
\delta_{(a_1a_2}...\delta_{a_{2l-1}a_{2l)}}\int_{V_{\sss B}}f(r)r^{2l}d^3w\;,\\\nonumber\\\label{apk}
    \int_{V_{\sss B}}f(r)w^{i_1i_2...i_{2l+1}}d^3w&=&0\;,
\er
where $\delta_{(a_1a_2}...\delta_{a_{2l-1}a_{2l)}}$ is the fully symmetric linear combination of the Kronecker
delta symbols \cite{thorne}.
In particular, for any $l\ge1$ one has
\br
\label{pqf}
    \int_{V_{\sss B}}f(r)w^{<i_1i_2...i_{l}>}d^3w&=&0\;.
\er

One will also need several other equations for performing integration over sphere in the local coordinates of
body B. They are as follows:
\br\label{zpo1}\hspace{-1cm}
A_{<iL>}B_{<N>}\int_{V_{\sss B}}\rho^*w^{<L>}w^{<N>}d^3w&=&\left\{\begin{array}{c}\displaystyle\frac{l!}{(2l+1)!!}
\,A_{<iL>}B_{<L>}{\cal I}^{(2l)}_{\sss B}\;,\quad(n=l)\\0\;,\qquad\qquad\qquad\qquad\qquad\qquad(n\not=l)\end{array}\right.
\\\nonumber\\\label{zpo2}
A_{<iL>}\int_{V_{\sss B}}\rho^*\nu^2w^{<L>}d^3w&=&\left\{\begin{array}{c}-\displaystyle\frac{2}{15}\,A_{<ijk>}
\Omega_{\sss B}^j\Omega_{\sss B}^k {\cal I}^{(4)}_{\sss B}\;,\qquad(l=2)\\0\;,\qquad\qquad\qquad\qquad\qquad\quad(l>2)
\end{array}\right.
\er
where we used Eq. (\ref{pa3}), $A_{<L>}$ and $B_{<L>}$ are arbitrary STF tensors, and
\be {\cal I}^{(2l)}_{\sss B}=\int_{V_{\sss B}}\rho^*r^{2l}d^3w,                \label{9.6.3}
\en
is $2l$-th order rotational moment of inertia of the body B \footnote{Notice that all odd rotational moments
${\cal I}^{(2l+1)}=0$.}.

Eqs. (\ref{9.6.1}) -- (\ref{9.6.3}) will be used for calculation of multipolar expansions of various gravitational
potentials entering translational equations of motion of the bodies.

\subsection{Coordinate Transformation of Multipole Moments}

Multipolar expansion of the Newtonian potential in the global coordinates, $x^\alpha=(ct,{\bm x})$, introduces
multipole moments of a body defined as integrals over the body's volume taken on hypersurface of constant time
$t$, that is
\be\label{aq9}\mathbb{I}_{\sss B}^L=\int_{V_{\sss B}}\rho^*(t,{\bm x})R_{\sss B}^{i_1}R_{\sss B}^{i_2}...
R_{\sss B}^{i_l}d^3x\;,\en
where $R_{\sss B}^i=x^i-x^i_{\sss B}$, and $x^i_{\sss B}$ is the origin of the local coordinates coinciding with
the center of mass of the body. We have postulated that the density and other structure-dependent functions
inside the body have a spherical-symmetric distribution in the local coordinates $w^\alpha=(cu,{\bm w})$,
so that according to Eq. (\ref{pqf}) the following relationship must held for any $l\ge1$
\be\label{zah}
\int_{V_{\sss B}}\rho^*(r')w'^{<L>}d^3w'=0\;,\qquad\qquad(l\ge1)
\en
where the integration is over a hypersurface of constant local coordinate time $u$.
However, Eq. (\ref{zah}) does not assume that the multipole moments of the body B defined in the global
coordinates, $\mathbb{I}_{\sss B}^L$, equal zero, in fact $\mathbb{I}_{\sss B}^L\not=0$ for $l\ge1$.
One can calculate $\mathbb{I}_{\sss B}^L$ directly from Eq. (\ref{zah}) after making use of transformation
formula, Eq. (\ref{sot}), from the local to global coordinates \footnote{This kind of transformation of
the multipole moments turns out to be important for calculation of the 3-d post-Newtonian equations of
motion \cite{itoh}.}
\br\label{zap2}
w'^i&=&R'^i_{\sss B}+\epsilon^2\left[\left(\frac{1}{2}v^i_{\sss B}v^j_{\sss B}+{F}_{ij}+{D}_{ij}\right)
R'^j_{\sss B}+{D}_{ijk}R'^j_{\sss B}R'^k_{\sss B}\right]\\\nonumber\\\nonumber&&\qquad+\epsilon^2
\left(v'^i-v^i_{\sss B}\right)\left(R'^j_{\sss B}-R^j_{\sss B}\right)v^i_{\sss B}+O\left(\epsilon^4\right)\;,
\er
where $R'^i_{\sss B}=x'^i-x^i_{\sss B}$, $R^i_{\sss B}=x^i-x^i_{\sss B}$, $v'^i=dx'^i/dt$, $v^i=dx^i/dt$, and
\br\label{to3}
{F}_{ij}&=&-\varepsilon_{ijk}{\cal F}^k\;,
\\\label{to1}
{D}_{ij}&=&\gamma\delta_{ij}\bar{U}({\bm x}_{\sss B})\;,\\\label{to2}
{D}_{ijk}&=&\frac{1}{2}\biggl(a^j_{\sss B}\delta^{ik}+a^k_{\sss B}\delta^{ij}-a^i_{\sss B}\delta^{jk}\biggr)\;,\er
and function ${\cal F}^k$ is defined in Eq. (\ref{3.43}).

Eq. (\ref{zap2}) must be used for transforming integrals shown in Eq. (\ref{zah}) from the local to global
coordinates. This is because the integration in Eq. (\ref{zah}) is performed over the hypersurface of constant
(local) time coordinate $u$ while similar integrals in the multipolar decomposition of the Newtonian
gravitational potential in the global coordinates are defined on the hypersurface of constant (global)
time coordinate $t$. Transformation of space coordinates from the space-like hypersurface of constant
time $u$ to that of time $t$ depends on space coordinates, ${\bm x}$, of the point at which matching of
the local and global coordinates is done so that the two space coordinates: ${\bm x}'$ -- the point of
integration, and, ${\bm x}$ - the matching point, appear in Eq. (\ref{zap2}) simultaneously and they
both belong to the hypersurface of constant time $t$.

Substitution of Eq. (\ref{zap2}) into Eq. (\ref{zah}) yields \cite{kop1987}
\br\label{po4}
\int_{V_{\sss B}}\rho^*(u,{\bm w}')w'^{<L>}d^3w'&=&\mathbb{I}_{\sss B}^{<L>}+\epsilon^2\left(\frac{l}{2}
v_{\sss B}^jv_{\sss B}^{<i_l}\mathbb{I}_{\sss B}^{L-1>j}-lF^{j<i_l}\mathbb{I}_{\sss B}^{L-1>j}\right.\\
\nonumber\\\nonumber
&+&\left.
lD^{j<i_l}\mathbb{I}_{\sss B}^{L-1>j}+l\mathbb{I}_{\sss B}^{jk<L-1}D^{i_l>jk}+v_{\sss B}^j
\dot{\mathbb{I}}_{\sss B}^{j<L>}\right.\\\nonumber\\\nonumber
&-&\left.\left(v_{\sss B}^jR_{\sss B}^j\right)\dot{\mathbb{I}}_{\sss B}^{<L>}-v_{\sss B}^j
\int_{V_{\sss B}}\rho^*(u,{\bm w}')\nu'^jw'^{<L>}d^3w'\right)+O\left(\epsilon^4\right)\;.
\er
Taking into account Eq. (\ref{zah}) one concludes that only the dipole, $\mathbb{I}^i_{\sss B}$,
and the quadrupole, $\mathbb{I}^{ij}_{\sss B}$, moments differ from zero. More specifically, one has
\br\label{pa5}
\mathbb{I}^i_{\sss B}&=&\frac{\epsilon^2}{3}{\cal I}^{(2)}_{\sss B}\left(\varepsilon_{ijk}v^j_{\sss B}
\Omega^k_{\sss B}+\frac{1}{2}a^i_{\sss B}\right)+O\left(\epsilon^4\right)\;,\\\nonumber\\\label{pa6}
\mathbb{I}^{<ij>}_{\sss B}&=&-\frac{\epsilon^2}{3}{\cal I}^{(2)}_{\sss B}v^{<i}_{\sss B}v^{j>}_{\sss B}+
O\left(\epsilon^4\right)\;,\\\nonumber\\\label{pa7}
\mathbb{I}^{<L>}_{\sss B}&=&O\left(\epsilon^4\right)\;,\qquad(l\ge3)
\er

The same expressions for the multipole moments can be obtained in a different way by making use of
multipolar expansions of the Newtonian potential of body B in the local and global coordinates and their
subsequent comparison with the help of transformation formula shown in Eq. (\ref{zal}). We have checked
that both derivations are self-consistent and yield identical expressions for multipole moments given
in Eqs. (\ref{pa5}) -- (\ref{pa7}). Transformation of multipole moments of Earth's gravitational field
from the global to local coordinates were used in our papers \cite{bk-kin,bk-nc} in order to derive the
post-Newtonian equation of motion of Earth's artificial satellites (valid also for Moon) in the geocentric
frame with taking into account relativistic corrections due to the presence of Earth's quadrupole field.

\subsection{Multipolar Decomposition of Gravitational Potentials in Global Coordinates}\label{mdgco}

In order to derive equations of motion of the bodies in global coordinates one will need to know the
multipolar decomposition of gravitational potentials ${U}_{\sss B}(t,{\bm x})$, ${U}^i_{\sss (B)}(t,{\bm x})$,
${\Phi}_{\sss (B)}(t,{\bm x})$, and ${\chi}_{\sss (B)}(t,{\bm x})$ in these coordinates. The potentials under
discussion are defined in Eqs. (\ref{12.10}) -- (\ref{12.16}). For the Newtonian potential one has
\br\label{re4}
{U}_{\sss B}(t,{\bm x})&=&\frac{G{\cal M}_{*{\sss B}}}{R_{\sss B}}+\nn{l=1}\frac{(-1)^l}{l!}G\mathbb{I}^{<L>}
\frac{\partial^L}{\partial x^L}\left(\frac{1}{R_{\sss B}}\right)\\\nonumber\\\nonumber&=&\frac{G{\cal M}_*}
{R_{\sss B}}-G\mathbb{I}^{i}\frac{\partial}{\partial x^i}\left(\frac{1}{R_{\sss B}}\right)+\frac{1}{2}G
\mathbb{I}^{<ij>}\frac{\partial^2}{\partial x^i\partial x^j}\left(\frac{1}{R_{\sss B}}\right)+O\left(\epsilon^4\right)\;,
\er
where the dipole, $\mathbb{I}^i$, and quadrupole, $\mathbb{I}^{<ij>}$, moments are given by Eqs. (\ref{pa5})
and (\ref{pa6}).

Vector-potential ${U}^i_{\sss (B)}(t,{\bm x})$ is decomposed as follows
\be\label{ar8}
{U}^i_{\sss (B)}(t,{\bm x})=\frac{G{\cal M}_{*{\sss B}}v^i_{\sss B}}{R_{\sss B}}-\frac{1}{3}G{\cal I}^{(2)}_{\sss B}
\varepsilon^i_{\;jk}\Omega^j_{\sss B}\frac{\partial}{\partial x^k}\left(\frac{1}{R_{\sss B}}\right)+O\left(\epsilon^2
\right)\;,
\en
where we have used the fact that inside the body, $v^i=v^i_{\sss B}+\nu^i+O\left(\epsilon^2\right)$,
and the internal velocity, $\nu^i$, is defined in Eq. (\ref{qma}).

Superpotential ${\chi}_{\sss (B)}(t,{\bm x})$ has the following multipolar decomposition
\be\label{ar2}
{\chi}_{\sss (B)}(t,{\bm x})=-G{\cal M}_{*{\sss B}}R_{\sss B}-\frac{1}{3}\frac{G{\cal I}^{(2)}_{{\sss B}}}
{R_{\sss B}}+O\left(\epsilon^2\right)\;.
\en

Potential ${\Phi}_{\sss (B)}(t,{\bm x})$ consists of a linear combination of four functions as shown in
Eq. (\ref{12.9ex}). For each of these functions one has
\br\label{bv1}
{\Phi}^{\sss (B)}_1(t,{\bm x})&=&G\int_{V_{\sss B}}\frac{\rho^*(t,{\bm x}')v'^2d^3x'}{|{\bm x}-{\bm x}'|}=
\frac{G{\cal M}_{*{\sss B}}v^2_{\sss B}}{R_{\sss B}}+\frac{G}{R_{\sss B}}\int_{V_{\sss B}}\rho^*(r)\nu^2d^3w\\
\nonumber&&+\frac{2G}{3}\frac{\varepsilon_{ijk}R^i_{\sss B}v^j_{\sss B}\Omega^k_{\sss B}{\cal I}^{(2)}_{\sss B}}
{R^3_{\sss B}}-\frac{G}{5}\frac{\Omega^{<i}\Omega^{j>}R^i_{\sss B}R^j_{\sss B}{\cal I}^{(4)}_{\sss B}}
{R^5_{\sss B}}+O\left(\epsilon^2\right)\;,
\\\nonumber\\\nonumber\\    \label{bv2}
{\Phi}^{\sss (B)}_2(t,{\bm x})&=&G\int_{V_{\sss B}}\frac{\rho^*(t,{\bm x}'){U}(t,{\bm x}')d^3x'}{|{\bm x}-{\bm x}'|}=
\frac{G}{R_{\sss B}}\int_{V_{\sss B}}\rho^*(r)\hat{U}_{\sss B}(r)d^3w\\\nonumber
&&+G^2\sum_{A\not=B}\sum_{l=0}^\infty\frac{(-1)^l}{(2l+1)l!}\frac{{\cal M}_{*{\sss A}}{\cal I}^{(2l)}_{\sss B}
R^{<L>}_{BA}}{R^{2l+1}_{BA}}\frac{\partial^L}{\partial x^L}\left(\frac{1}{R_{\sss B}}\right)+O\left(\epsilon^2\right)\;,
\\\nonumber\\\nonumber\\\label{bv3}
{\Phi}^{\sss (B)}_3(t,{\bm x})&=&G\int_{V_{\sss B}}\frac{\rho^*(t,{\bm x}')\Pi(t,{\bm x}')d^3x'}{|{\bm x}-{\bm x}'|}=
\frac{G}{R_{\sss B}}\int_{V_{\sss B}}\rho^*(r)\Pi(r)d^3w+O\left(\epsilon^2\right)\;,\\\nonumber\\\nonumber\\\label{bv4}
{\Phi}^{\sss (B)}_4(t,{\bm x})&=&G\int_{V_{\sss B}}\frac{\pi^{kk}(t,{\bm x}')d^3x'}{|{\bm x}-{\bm x}'|}=\frac{3G}
{R_{\sss B}}\int_{V_{\sss B}}p(r)d^3w+O\left(\epsilon^2\right)\;.
\er
This concludes the set of equations describing the multipolar decomposition of the gravitational potentials
in the global coordinates.

\subsection{Translational Equations of Motion}

Both conditions of spherical symmetry and rigid rotation, Eqs. (\ref{as7}) and (\ref{qma}), allow us to
simplify equation (\ref{9.5.1}) of translational motion of body B drastically.
For example, Eq. (\ref{pqf}) assumes that all multipole moments of the body ${\cal I}^L=O\left(\epsilon^2\right)$
for all $l\ge1$, and ${\cal S}^L=O\left(\epsilon^2\right)$ for all $l\ge2$. Therefore, calculation of the tidal
Newtonian force for body B, taken from Eq. (\ref{9.5.1}), yields
\br\label{qk8}
\mathbb{F}^i_N&=&\epsilon^2\left\{\frac{2\gamma+1}{30}Q_{ijk}\Omega^j_{\sss B}
\Omega^k_{\sss B}{\cal I}^{(4)}_{\sss B}-\nn{l=1}\frac{2(1-\beta)P_L-Q_L}{l!(2l+1)!!}\,Q_{iL}
{\cal I}^{(2l)}_{\sss B}\right\}\;,
\er
which has the post-Newtonian, ($\sim \epsilon^2$), order of magnitude.

The post-Newtonian gravitomagnetic tidal force from Eq. (\ref{9.5.1}) in case of a spherically-symmetric
body is reduced to
\be\label{pn6}
{\mathbb{F}}^i_{pN}=-\frac34 C_{ij}{\cal S}^j_{\sss B}\;,
\en
where the external (gravitomagnetic-type) quadrupole
\br\label{pn7}
C_{ij}=&-&\frac{20(1+\gamma)G}{3}\sum_{C\neq B}{\cal I}^{(2)}_{\sss C}\frac{\Omega^p_{\sss C}R_{\sss{BC}}^{<ijp>}}
{R_{\sss{BC}}^7}
\\\nonumber
&+&2(1+\gamma)G\sum_{C\neq B}\frac{{\cal M}_{\sss C}(v^p_{\sss C}-v^p_{\sss B})}{R_{\sss{BC}}^5}
\biggl(\varepsilon_{ipq}R^{<jq>}_{\sss{BC}}+\varepsilon_{jpq}R^{<iq>}_{\sss{BC}}  \biggr)\;,
\er
${\cal M}_{\sss B}$ is the mass, and
\be\label{sp4}
{\cal S}^i_{\sss B}=\frac23{\cal I}^{(2)}_{\sss B}\,\Omega^i_{\sss B}\;,
\en
is the spin of the body B.

All other terms in Eq. (\ref{9.5.1}) depending on ${\cal I}^L$ are equal to zero. Hence, Eq. (\ref{9.5.1}) is
drastically simplified for spherically-symmetric bodies and reads
\br \label{mnj}
{\tilde{\cal M}}_{\sss B}\nnaa{i} &=& {\cal M}_{\sss B}\bar
V_{,i}\nnxe-\mathbb{F}^i_N\\\nonumber & +\epsilon^2&{\cal M}_{\sss B}\Biggl\{\biggl[\gamma
\delta_{ik}\nnv{2}-\nnv{i}\nnv{k} -2(\gamma+\beta)\delta_{ik}\bar
U\nnxe\biggr]\bar U_{,k}\nnxe \\\nonumber &&
+2(1+\gamma)\dot{\bar
U}^i\nnxe-2(1+\gamma)\nnv{k}\bar U^{k,i}\nnxe
-(1+2\gamma)\nnv{i}\dot{\bar U}\nnxe\Biggr\}\\\nonumber &&+\epsilon^2\biggl[\frac12(1-\gamma)\varepsilon_{ipk}
{\cal S}^p\dot{\bar U}_{,k}\nnxe+\frac34 C_{ij}{\cal S}^j\biggr]
 + O(\epsilon^4)\;,
\er
where the {\it conformal}, $\tilde{\cal M}_{\sss B}$, and {\it active}, ${\cal M}_{\sss B}$,masses od the body B
are related to each other via Eq. (\ref{acmass}), that is
\br\label{amz1}
\tilde{\cal M}_{\sss B}&=&{\cal M}_{\sss B}
+\slch^2\left[\frac{\eta}{2}\int_{V_{\sss B}}\rho^*\hat{U}_{\sss B}d^3w
+2(\beta-1)\sum_{C\neq B}
\frac{G{\cal M}_{\sss C}{\cal M}_{\sss B}}{R_{\sss{CB}}}\right]\;,
\er
and the gravitational potential
\be\label{oy4}
\bar V({\bm x})=\bar
U({\bm x})+\epsilon^2\biggl[\bar\Phi({\bm x})-\frac{1}{2}\bar{\chi}_{,tt}({\bm x})\biggr]\;.
\en The tidal force
$\mathbb{F}^i_N$ is given by Eq. (11.4.1) and $C_{ij}$ is shown in Eq. (\ref{pn7}).

We are to calculate all terms in the right side of Eq. (\ref{mnj}) explicitly in terms of body's mass, rotational
moment of inertia, and spin. Among them the most complicated is the first one, that is $\bar V_{,i}\nnxe$.
By making use of Eqs. (\ref{re4}) -- (\ref{bv4}) we obtain
\br\label{xc7}\hspace{-1cm}
\bar V(t, \bm x)&=& \sum_{C\neq
B}\frac{G{\cal M}_{\sss C}}{R_{\sss C}}\left\{1+\slch^2\left[(\gamma+1)v^2_{\sss C}-\frac{1}{2}a_{\sss C}^k
R_{\sss C}^k-\frac{\left(
v_{\sss C}^k R_{\sss C}^k\right)^2}{2R_{\sss C}^2}-\gamma\sum_{D\neq C}\frac{G{\rm M}_{\sss D}}{R_{\sss{CD}}}
\right]\right\}\\\nonumber &
+\slch^2&G\sum_{C\neq B}
\Biggl\{\frac13{\cal I}^{(2)}_{\sss C}\frac{R^k_{\sss C}}{R_{\sss C}^3}\biggl[2(1+\gamma)\varepsilon_{kpq}
v^p_{\sss C}\Omega^q_{\sss C}+a^k_{\sss C}\biggr]
-\frac{1+2\gamma}{10}\frac{R^j_{\sss C}R^k_{\sss C}}{R_{\sss C}^5}
\Omega^{<j}_{\sss C}\Omega^{k>}_{\sss C}{\cal I}^{(4)}_{\sss C}
\\\nonumber &&
+(1-2\beta)\nn{l=1}\frac{(2l-1)!!}{(2l+1)l!}{\cal I}^{(2l)}_{\sss C}
\frac{R^{<L>}_{\sss C}}{R_{\sss C}^{2l+1}}\sum_{D\neq
C}G{\rm M}_{\sss D}\frac{R^{<L>}_{\sss{CD}}}{R^{2l+1}_{\sss{CD}}}
\Biggr\}+O(\slch^4),
\er
where
\br\label{mzq}
{\cal M}_{\sss C}&=&{\rm M}_{\sss C}
-\slch^2\left[\frac{\eta}{2}\int_{V_{\sss C}}\rho^*\hat{U}_{\sss C}d^3w
+(2\beta-\gamma-1)\sum_{D\neq C}
\frac{G{\rm M}_{\sss C}{\rm M}_{\sss D}}{R_{\sss{CD}}}\right]\;,
\er
is the active mass of the body C, ${\rm M}_{\sss C}$ is the general relativistic mass of the body C defined
by Eq. (\ref{ij3}) where for the sake of simplicity we assumed $Y=0$, and $\eta=4\beta-\gamma-3$ is the
Nordtvedt parameter.

After calculating derivatives from potentials $\bar V(t, \bm x)$, $\bar U(t, \bm x)$ and substituting them into
Eq. (\ref{mnj}) one obtains the following expression for acceleration of the center of mass of the body B:
\br\label{9.6.11}
{\rm M}_{\sss B}\nnaa{i}&=&
F^i_{\sss N} + \epsilon^2\left\{ F^i_{\sss{EIH}} +
+F^i_{\sss{\cal S}} + F^i_{\sss{\cal I}GR}+\delta F^i_{\sss{\cal I}GR}\right\} + O(\epsilon^4)\;,
\er
where $F^i_{\sss N}$ is the Newtonian force and $F^i_{\sss{EIH}}$,
$F^i_{\sss\Omega}$, $F^i_{\sss{\cal I}}$ are the post-Newtonian relativistic corrections.
Gravitational forces in the right side of this equation are given by the following expressions
\br\label{gko1}
F^i_{\sss N} &=&
\sum_{\sss{C\neq B}}\frac{G\mathfrak{M}_{\sss B}\mathfrak{M}_{\sss C}R^i_{\sss{BC}}}
{R^3_{\sss{BC}}}\;, \\\nonumber\\\nonumber\\\label{gko2}
F^i_{\sss{EIH}} &=&
\sum_{\sss{C\neq B}}\frac{G{\rm{M}}_{\sss B}{\rm{M}}_{\sss C}R^i_{\sss{BC}}}
{R^3_{\sss{BC}}}\Biggl\{\gamma v^2_{\sss B}-2(1+\gamma)(\bm v_{\sss B}
\cdot\bm v_{\sss C})+(1+\gamma)v^2_{\sss C}
\\\nonumber&&
-\frac32\left(\frac{\bm R_{\sss{BC}}\cdot\bm v_{\sss C}}{R_{\sss{BC}}}\right)^2
-(1+2\gamma+2\beta)\frac{G{\rm M}_{\sss B}}{R_{\sss{BC}}}
-2(\gamma+\beta)\frac{G{\rm M}_{\sss C}}{R_{\sss{BC}}}
\\\nonumber&&
+\sum_{\sss{D\neq B,C}}\left[(1-2\beta)\frac{G{\rm M}_{\sss D}}{R_{\sss{CD}}}
-2(\gamma+\beta)\frac{G{\rm M}_{\sss D}}{R_{\sss{BD}}}
+\frac{G{\rm M}_{\sss D}(\bm R_{\sss{BC}}\cdot\bm R_{\sss{CD}})}{2R^3_{\sss{CD}}}\right]
\Biggr\}\\\nonumber&&
+\sum_{\sss{C\neq B}}\Biggl\{
\frac{G{\rm M}_{\sss B}{\rm M}_{\sss C}(v^i_{\sss C}-v^i_{\sss B})}{R^3_{\sss{BC}}}
\biggl[2(1+\gamma)(\bm v_{\sss B}\cdot\bm R_{\sss{BC}}) -(1+2\gamma)
(\bm v_{\sss C}\cdot\bm R_{\sss{BC}})\biggr]
\\\nonumber&&
+\frac{3+4\gamma}{2}\frac{G{\rm M}_{\sss B}{\rm M}_{\sss C}}{R_{\sss{BC}}}
\sum_{\sss{D\neq B,C}}\frac{G{\rm M}_{\sss D}R^i_{\sss{CD}}}{R^3_{\sss{CD}}}\Biggr\}\;,
\\\nonumber\\\nonumber\\\label{gko3}
F^i_{\sss{\cal S}}&=&G\sum_{\sss{C\neq B}}\Biggl\{
\frac{{\rm M}_{\sss C}{\cal S}^p_{\sss B}
(v^k_{\sss C}-v^k_{\sss B})}{2R^5_{\sss{BC}}}
\biggl[3(1+\gamma)\left(\varepsilon_{ikq}R^{<pq>}_{\sss{BC}}-
\varepsilon_{kpq}R^{<iq>}_{\sss{BC}}\right)\\\nonumber&&+(1-\gamma)\varepsilon_{ipq}R^{<kq>}_{\sss{BC}}\biggr]
+3(1+\gamma)\frac{{\rm M}_{\sss B}
{\cal S}^p_{\sss C}(v^k_{\sss C}-v^k_{\sss B})}{R^5_{\sss{BC}}}\biggl[ \varepsilon_{ipq}R^{<kq>}_{\sss{BC}}-
\varepsilon_{kpq}R^{<iq>}_{\sss{BC}}\biggr]\\\nonumber&&
-\frac{15(1+\gamma)}{2}\frac{{\cal S}^j_{\sss B}{\cal S}^k_{\sss C} R^{<ijk>}_{\sss{BC}}}{R^{7}_{\sss{BC}}}
-\left(\gamma+\frac{1}{2}\right)\frac{R^{<ijk>}_{\sss{BC}}}{R^{7}_{\sss{BC}}}\biggl[ {\rm M}_{\sss B}
{\cal I}^{(4)}_{\sss C}\Omega^j_{\sss C}\Omega^k_{\sss C} +{\rm M}_{\sss C}{\cal I}^{(4)}_{\sss B}
\Omega^j_{\sss B}\Omega^k_{\sss B}  \biggr]
\Biggr\}\;,
\\\nonumber\\\nonumber\\
\label{gko4}
F^i_{\sss{\cal I}GR}&=&-
G^2\sum_{\sss{C\neq B}}\! \nn{l=2}\nna\!\left[(-1)^{l}
 {\rm M}_{\sss B}
{\cal I}^{(2l)}_{\sss C}\frac{R^{<iL>}_{\sss{BC}}}{R^{2l+3}_{\sss{BC}}}\!
\sum_{\sss{D\neq C}}
\frac{{\rm M}_{\sss D}R^{<L>}_{\sss{CD}}}{R^{2l+1}_{\sss{CD}}}
\right.\\\nonumber&&\left.
+{\rm M}_{\sss C}
{\cal I}^{(2l)}_{\sss B}\frac{R^{<L>}_{\sss{BC}}}{R^{2l+1}_{\sss{BC}}}\!
\sum_{\sss{D\neq B}}\frac{{\rm M}_{\sss D}R^{<iL>}_{\sss{BD}}}{R^{2l+3}_{\sss{BD}}}\right]\!,
\\\nonumber\\\nonumber\\
\label{gko5}
\delta F^i_{\sss{\cal I}GR}&=&2(1-\beta)G^2
\sum_{\sss{C\neq B}}\left\{
{\rm M}_{\sss C}{\cal I}^{(2)}_{\sss B}\frac{R^{k}_{\sss{BC}}}{R^3_{\sss BC}}
\sum_{\sss{D\neq B}}\frac{{\rm M}_{\sss D}R^{<ik>}_{\sss{BD}}}{R^5_{\sss{BD}}}\right.\\\nonumber&&\left.
+{\rm M}_{\sss B} {\cal I}^{(2)}_{\sss C}\frac{R^{<ik>}_{\sss{BC}}}{R^5_{\sss{BC}}}
\sum_{\sss{D\neq C}}\frac{{\rm M}_{\sss D}R^k_{\sss{CD}}}{R^3_{\sss{CD}}}
+\nn{l=2}\nna\biggl[ (-1)^l{\rm M}_{\sss B}
{\cal I}^{(2l)}_{\sss C}\frac{R^{<iL>}_{\sss{BC}}}{R^{2l+3}_{\sss{BC}}}
\sum_{\sss{D\neq C}}
\frac{{\rm M}_{\sss D}R^{<L>}_{\sss{CD}}}{R^{2l+1}_{\sss{CD}}}
\right.\\\nonumber&&\left.
+ {\rm M}_{\sss C}
{\cal I}^{(2l)}_{\sss B}\frac{R^{<L>}_{\sss{BC}}}{R^{2l+1}_{\sss{BC}}}
\sum_{\sss{D\neq B}}\frac{{\rm M}_{\sss D}R^{<iL>}_{\sss{BD}}}{R^{2l+3}_{\sss{BD}}}\biggr]\right\}\;,
\er
where we have defined $R^i_{\sss{BC}}=x^i_{\sss C}-x^i_{\sss B},$
 $R_{\sss{BC}}=|\bm x_{\sss C}-\bm x_{\sss B}|$, and spin ${\cal S}^i_{\sss B}$ of body B relates to
 the angular speed of the body's rotation $\Omega^i_{\sss B}$ via Eq. (\ref{sp4}).

Eq. (\ref{9.6.11})  elucidates that {\it inertial} mass ${\rm M}_{\sss B}$ of the body B is simply its
general relativistic mass given by Eqs. (\ref{ij3}). This mass is conserved (constant) for spherically-symmetric
bodies as follows from Eq. (\ref{ij4}).
The  {\it Nordtvedt gravitational} mass $\mathfrak{M}_{\sss B}$ of the body B
depends on the gravitational defect of mass multiplied with the Nordtvedt parameter $\eta=4\gamma-\beta-3$
\be\label{dm}
\mathfrak{M}_{\sss B}={\rm M}_{\sss
B}-\frac{1}{2}\epsilon^2\eta\int_{V_{\sss B}}\rho^*\hat{U}_{\sss B}d^3w\;.
\en
The Newtonian gravitational force $F^i_{\sss N}$, Eq. (\ref{gko1}), depends in the scalar-tensor theory only
on the Nordtvedt gravitational masses of the bodies. Will \cite{will} distinguishes the "active" and "passive"
gravitational masses which depend on the entire set of the PPN parameters. In our approach used in the present
paper only two PPN parameters, $\beta$ and $\gamma$, exist. In this case the "active" and "passive"
gravitational masses coincide and reduce to one and the same expression given by Eq. (\ref{dm}).
Inertial and gravitational masses of the body are not equal in the scalar-tensor theory of gravity
\cite{dicke1,n-1,n-2}. This inequality violates the strong principle of equivalence for massive bodies.
This violation can be also explained from the point of view of interaction of the gravitational field of the body
under consideration with the scalar field generated by external bodies. This interaction leads to a local force
which brings about a non-zero value of the time derivative of the body's linear momentum. Indeed, assuming that
the body under consideration has finite size, is non-rotating, and spherically-symmetric, one obtains
from Eq. (\ref{6.2d})
\br\label{sdo2}
\dot{\cal P}^i&=&{\cal M}Q_i\left(1+
\frac{1}{2{\cal M}}\epsilon^2\eta\int_{V_{\sss B}}\rho^*\hat{U}^{(\sss B)}d^3w\right)-
\\\nonumber&&\frac{1}{2}\epsilon^2\eta P_i\int_{V_{\sss B}}\rho^*\hat{U}^{(\sss B)}d^3w -{\rm\bf F}^i_N+O(\epsilon^4) \;,
\er
where $P_i$ is a gradient of the scalar field of the external bodies.
If one keeps the body's center of mass at the origin of the local coordinate system ($\dot{\cal P}^i=0$),
then, the body's center of mass will experience an anomalous acceleration $Q_i\not=0$. This anomalous
acceleration is due to the interaction of the gravitational energy ("gravitational charge") of the body
under consideration with the gradient of the external scalar field. The coupling constant of this interaction
is the dimensionless Nordtvedt parameter $\eta$.

The post-Newtonian forces, Eqs. (\ref{gko2})--(\ref{gko5}), depend only on the general relativistic masses of
the bodies which coincide with the Newtonian definition of mass, Eq. (\ref{a}) in the approximation under
consideration.  The post-Newtonian force (\ref{gko2}) is known as the (Lorentz-Droste) Einstein-Infeld-Hoffmann
(EIH) force \cite{soffel-book} presently used as a basis of JPL ephemerides \cite{sman}. It was derived in
general relativity by Lorentz and Droste \cite{ldr} and later by Einstein, Infeld and Hoffmann \cite{eih},
Petrova \cite{petr}, and Fock \cite{fock}. In the Brans-Dicke theory this force was derived by Estabrook
\cite{esbk} in case of $\beta=1, \gamma\not=1$ and by Dallas \cite{dall} in the case $\beta\not=1,
\gamma\not=1$ (see also \cite{vin}). These derivations assumed that the bodies have negligible ratio
of their radii to the characteristic distance between the bodies (a point-like body approximation) as
well as that they are non-rotating and move along geodesic world lines.

Corrections to the EIH force are given by Eqs. (\ref{gko3}) --(\ref{gko5}).
The force $F^i_{\sss{\cal S}}$ given by Eq. (\ref{gko3}) describes the relativistic post-Newtonian correction
to the EIH force due to the coupling of the body's spin with orbital angular momentum and rotational spins
of other bodies. It depends on the PPN parameter $\gamma$ only. If one takes $\gamma=1$ in Eq. (\ref{gko3}),
the force $F^i_{\sss{\cal S}}$ is reduced exactly to its general relativistic expression obtained earlier by
other researchers \cite{dsx1992,brr,barcon,xws}.  Our Eq. (\ref{gko3}) for the PPN force $F^i_{\sss{\cal S}}$
coincides with that derived by Klioner and Soffel \cite{kls}.

The force (\ref{gko4}) describes general relativistic correction to the EIH force due to the finite size of
the bodies. This correction is proportional to the forth-order rotational moments of inertia of the bodies,
${\cal I}^{(4)}$, while all terms, which are proportional to the second-order body's rotational moments of
inertia, ${\cal I}^{(2)}$, cancelled mutually out. Nordtvedt \cite{n-3} considered the problem of translational
motion of extended bodies in the general class of scalar-tensor theories of gravity. He came to the conclusion
that covariant formulation of the variational principle requires the second-order moment of inertia of
extended body to be coupled with the Ricci tensor of the background gravitational field generated by the
external bodies.  Hence, such coupling must disappear in general theory of relativity by virtue of the
vanishing of the background Ricci tensor in vacuum.  However, body's moments of inertia of higher order
couple with the full Riemann tensor and its derivatives, and for this reason they can present in general
relativistic equations of motion of spherically-symmetric bodies as demonstrated in Eq. (\ref{gko4}).

Nordtvedt's calculation \cite{n-3} of the equations of motion of extended spherically-symmetric bodies
agrees with our derivation of these equations based on the implementation of the matched asymptotic
expansion technique and separate solution of the internal and external problems for gravitational field of
the N-body system.  However, vanishing of all terms depending on the second order rotational moment of inertia
in general relativity is in disagreement with calculations of Brumberg \cite{brr}, Spyrou \cite{spyr1},
Dallas \cite{dall} and Vincent \cite{vin} who came to the conclusion that the general relativistic Lagrangian
for the system of N spherically-symmetric bodies must depend on the second-order moments of inertia of these
bodies, ${\cal I}^{(2)}$. Brumberg's expression for the force due to the finite size of the bodies is \cite{brr}
\br\label{brum}\hspace{-1cm}
F^i_{\sss{\rm Brumberg}} &=& \epsilon^2G\sum_{\sss{C\neq B}}\left\{
\frac{5\left({\rm {M}}_{\sss B}{\cal I}^{(2)}_{\sss C}+{\rm {M}}_{\sss C}{\cal I}^{(2)}_{\sss B}\right)
v^j_{\sss C}v^k_{\sss C}R^{<ijk>}_{\sss{BC}}}{2R^7_{\sss{BC}}}\right.\\\nonumber&&\left.
+\frac{G\left({\rm {M}}_{\sss B}{\rm {M}}_{\sss C}{\cal I}^{(2)}_{\sss B}-
{\rm {M}}^2_{\sss B}{\cal I}^{(2)}_{\sss C}-2{\rm {M}}^2_{\sss C}{\cal I}^{(2)}_{\sss B}\right)R^i_{\sss{BC}}}
{3R^6_{\sss{BC}}}
\right.\\\nonumber&&\left.
-\frac{G^2}{2}\sum_{\sss{D\neq B,C}}{\rm {M}}_{\sss C}\Biggl[
\frac{{\rm {M}}_{\sss B}{\cal I}^{(2)}_{\sss D}R^k_{\sss{CD}}R^{<ik>}_{\sss{BD}}}
{R^3_{\sss{CD}}R^5_{\sss{BD}}}\right.\\\nonumber&&\left.
+\frac{{\rm {M}}_{\sss D}{\cal I}^{(2)}_{\sss B}}{R^3_{\sss{BC}}}\biggl(
\frac{R^k_{\sss{CD}}R^{<ik>}_{\sss{BC}}}{R^3_{\sss{CD}}R^2_{\sss{BC}}}
+\frac{R^k_{\sss{BD}}R^{<ik>}_{\sss{BC}}}{R^3_{\sss{BD}}R^2_{\sss{BC}}}
+\frac{R^k_{\sss{BC}}R^{<ik>}_{\sss{BD}}}{R^5_{\sss{BD}}}\biggr)\Biggr]\right\}\;,
\er
We have analyzed and pinned down the origin of the disagreement between Nordtvedt's \cite{n-3} and Brumberg's
calculations. Nordtvedt \cite{n-3} used covariant approach while Brumberg \cite{brr} followed Fock's method
\cite{fock} and operated with the coordinate-dependent definitions of the multipole moments. In fact, Brumberg
defined multipole moments of the bodies in the global (barycentric) frame of the N-body system. Spherical
symmetry of the bodies was also defined by Brumberg in the global frame. Such definition of the spherical
symmetry does not comply with the relativistic law of transformation between local and global frames. Hence,
the bodies can sustain their spherically-symmetric shape in the global frame if and only if there are internal
stresses in the body's matter which compensate for the Lorentz and gravitational contractions of the body's
shapes \cite{kop1987}.  Existence of such coordinate-dependent internal stresses inside the body under
consideration is unnatural. Furthermore, assumption about spherical symmetry of the bodies in the global
barycentric frame brings about the force $F^i_{\sss{\rm Brumberg}}$ which is a pure coordinate effect.

In order to prove that the force  $F^i_{\sss{\rm Brumberg}}$ has no physical origin we have considered translational
equation of motion for body B defined in the global frame as follows
\br\label{for}
\int_{V_{\sss B}}\rho^*\frac{dv^i}{dt}d^3x&=&\int_{V_{\sss B}}\left(\frac{\partial\pi_{ij}}{\partial x^j} -
\rho^*\frac{\partial U}{\partial x^i}\right)d^3 x+O\left(\epsilon^2\right)\;,
\er
where the post-Newtonian corrections (not shown explicitly) include the relativistic point-like effects and
Brumberg's force (\ref{brum}). Barycentric velocity $v^i$ of the body's matter is obtained by differentiation
with respect to time of both sides of Eq. (\ref{zap2}) and can be decomposed to the sum of the barycentric
velocity of the body's center of mass, $v^i_{\sss B}$, and the internal velocity's field in local coordinates
\br\label{pkg}
v^i&=&v^i_{\sss B}(t)+\nu^i(u,{\bm w})+\epsilon^2\Delta\nu^i(u,{\bm w})\;,\er
where $\Delta\nu^i$ is the relativistic correction to the local velocity $\nu^i$. This correction is a quadratic
function of the local coordinates $w^i$ of the body. Hence, subsequent calculation of the time derivative of $v^i$
and calculation of the integral in the left side of Eq. (\ref{for}) bring about terms which actually depend on
the moment of inertia of body B. This moment of inertia is reduced to the rotational moment of inertia
${\cal I}^{(2)}_{\sss B}$ in case of a spherically-symmetric body. Calculation of the integral from the
Newtonian potential in the right side of Eq. (\ref{for}) has to be done by splitting the potential in
two parts - internal and external (see Eq. (\ref{3.1})), and applying Eqs. (\ref{re4}) and (\ref{pa5}),
(\ref{pa6}) for calculation of the integrals from the external potential. This again gives a number of
terms depending on the rotational moment of inertia  ${\cal I}^{(2)}_{\sss B}$ of the body B. Summing up
all these terms one obtains exactly the same expression as in Eq. (\ref{brum}) but with opposite sign.
It means that these terms cancel out with the force $F^i_{\sss{\rm Brumberg}}$. This completely agrees with our
calculations of the force $F^i_{\sss{\cal I}GR}$ which does not depend on the body's rotational moments
of inertia of the second order. We conclude that the origin of the coordinate-dependent force $F^i_{\sss{\rm Brumberg}}$
is directly associated with an inappropriate choice of the body's center of mass and the property of its spherical
symmetry which must be defined with respect to the local coordinate frame co-moving with the body under consideration.
Brumberg attempted to make more physical calculation of the force $F^i_{\sss{\cal I}GR}$ in his another
book \cite{brberg} but he did not arrive to any definite conclusion regarding whether the
force $F^i_{\sss{\cal I}GR}$ depends on the body's rotational moments of inertia of the second order or not.
Our calculations resolve the problem and demonstrate that force $F^i_{\sss{\rm Brumberg}}$  does not exist
\footnote{In the sense that $F^i_{\sss{\rm Brumberg}}$ has no physical impact on the motion of the bodies having
finite size.} and that general relativistic correction due to the finite size of the moving bodies
is proportional to the forth and higher order rotational moments of inertia of the bodies. These corrections
are extremely small for the bodies comprising the solar system and can be neglected in treatment of the solar
system gravitational experiments. However, finite size effects can become important during the final stage of
coalescence of binary neutron stars so that they should be included in the precise calculation of templates of
the gravitational waveforms.

The force $\delta F^i_{\sss{\cal I}GR}$ describes relativistic correction due to the finite size of the bodies
in the scalar-tensor theory of gravity. This force is proportional to the parameter $\beta-1$ only and, in contrast
to general relativity, depends on the second order rotational moments of inertia, ${\cal I}^{(2)}_{\sss B}$.
 This dependence was noticed by Nordtvedt \cite{n-3,n-4} who has found that in the case of weakly self-gravitating
  bodies the finite-size effects are proportional to $\eta=4\beta-\gamma-3$. This is in disagreement with our
  calculations of the force $\delta F^i_{\sss{\cal I}GR}$ but one can easily reconcile the two formulations.
  The matter is that Nordtvedt \cite{n-4} worked in harmonic coordinate system defined by the condition
  $\partial_\alpha(\sqrt{-g}g^{\alpha\beta})=0$ while we worked in the quasi-harmonic coordinates defined
  by the Nutku condition  $\partial_\alpha(\phi\sqrt{-g}g^{\alpha\beta})=0$ (see Eq. (\ref{11.3}) in the
  present paper).  This leads to two different forms of the transformation between spatial global and local
  coordinates. In harmonic coordinates this transformation reads \cite{kls,n-4}
\br\label{mjn}
w^i_{\rm harmonic}&=&\nnr{i}+\epsilon^2\left[
   \biggl(\frac{1}{2}\nnv{i}\nnv{k}+\gamma
   \delta^{ik}\bar U\nnxe+{F}^{ik}\biggr)\nnr{k}\right.\\\nonumber&&\left.
   +\gamma
   \nnaa{k}\nnr{i}\nnr{k}-\frac{\gamma}{2}\nnaa{i}\nnr{2}
   \right]+O(\epsilon^4)\;,
\er
while in the quasi-harmonic coordinates, used in the present paper, we have
\br\label{mjn2}
w^i&=&\nnr{i}+\epsilon^2\left[
   \biggl(\frac{1}{2}\nnv{i}\nnv{k}+\gamma
   \delta^{ik}\bar U\nnxe+{F}^{ik}\biggr)\nnr{k}\right.\\\nonumber&&\left.+
   \nnaa{k}\nnr{i}\nnr{k}-\frac{1}{2}\nnaa{i}\nnr{2}
   \right]+O(\epsilon^4)\;.
\er
The two transformations have different dependence on $\gamma$ in terms proportional to the acceleration so that
\br\label{gak}
w^i_{\rm harmonic}=w^i+(\gamma-1)\epsilon^2\left(\nnaa{k}\nnr{i}\nnr{k}-\frac{1}{2}\nnaa{i}\nnr{2} \right)\;.\er
 It is due to this difference the parameter $\gamma$ had appeared in Nordtvedt's calculations of the finite size
 effects and made the parameter $\eta$ characterizing the magnitude of the finite-size effects in the harmonic
 coordinates. However, dependence of the magnitude of the finite-size effects on the parameter $\gamma$
 in Nordtvedt's calculations is a pure coordinate effect which has no physical meaning. Parameter $\gamma$
 can be eliminated from the force $\delta F^i_{\sss{\cal I}GR}$ if one works in the quasi-harmonic coordinates
 defined by the Nutku condition (\ref{11.3}) of this paper.

\subsection{Rotational Equations of Motion}

Derivation of rotational equations of motion for spherically-symmetric bodies requires calculation of the
multipole moments ${\cal R}_L$ of the body under consideration. One has
\be\label{op-1}
{\cal R}_L=O\left(\epsilon^2\right)\;.
\en
All other multipole moments have been calculated in the previous section. Performing calculations of the torques
and body's spin given by Eqs. (\ref{spin-5})--(\ref{spin-7}) one obtains
\br\label{op-3}
 {\cal T}^i &=&
\slch^2\left[\frac{2\gamma+1}{15}\varepsilon_{ijk}Q_{jn}\Omega^k_{\sss{(B)}}\Omega^n_{\sss{(B)}}
{\cal I}^{\sss{(4)}}_{\sss{(B)}}+\varepsilon_{ijk}{\cal
S}^jC_k\right]\,,
\\\nonumber\\\label{op-4}
\Delta{\cal T}^i &=& 0\,,
\\\nonumber\\\label{op-5}
\Delta{\cal S}^i &=& \frac{2}{3}{\cal
I}^{\sss{(2)}}_{\sss{(B)}}C_i+ \left[Q-Y+(\gamma-1)\bar
U\nnxe\right]{\cal S}^i\,. \er
Consequently, the rotational equation of motion for the body's spin is
\be\label{op-6} \frac{d{\cal S}^i_{+}}{du} =
\epsilon^2\left(\frac{2\gamma+1}{15}\varepsilon_{ijk}Q_{jn}\Omega^k_{\sss{(B)}}\Omega^n_{\sss{(B)}}
{\cal I}^{\sss{(4)}}_{\sss{(B)}}+\varepsilon_{ijk}{\cal S}^jC_k\right)
+ O(\slch^4)\,, \en
where $C_k$ is angular velocity of rotation of the local coordinate frame with respect to that which axes
are subject to the Fermi-Walker transport.

Our Eq. (\ref{op-6}) has one extra term comparatively with the corresponding equation (9.75) from the
paper \cite{kls} by Klioner and Soffel. This term depends on the forth-order rotational moment of
inertia, ${\cal I}^{(4)}_{\sss B}$, of the body B and has pure general relativistic origin. This
term was not taken into account by Klioner and Soffel because they neglected finite size of the rotating body.
Contribution of the  the forth-order rotational moment of inertia, ${\cal I}^{(4)}_{\sss B}$ to the rotational
torque is negligibly small for the theory of Earth's rotation. However, it may become significant during last
several orbits of a coalescing binary neutron star. It would be interesting to study the impact of this
term on the form of gravitational waves emitted such binaries.

\appendix

\section {Solution of the Laplace Equation for Scalar, Vector and Tensor Fields}\label{ap1}

In this appendix we find solutions of the Laplace equations for
scalar $F(t,{\bm x})$, vector $F_i(t,{\bm x})$, and tensor
$F_{ij}(t,{\bm x})$ fields. These equations are \be \label{7.a}
\triangle F(t,{\bm x})=0\qquad\quad\;,\quad\qquad
\triangle F_i(t,{\bm x})=0\qquad\quad\;,\qquad\quad \triangle
F_{ij}(t,{\bm x})=0\;. \en The procedure of finding solutions of
equations (\ref{7.a}) is based on the approach developed in
\cite{gelf} (see also \cite{thorne}, \cite{bldam2} and references
therein).

Basic spherical functions are
\be
   Y^{lm}(\theta,\,\phi)=\frac{1}{\sqrt{2\pi}}                 \label{7.1}
   e^{im\phi}P^m_l(cos\theta)\,,\quad (-l\le m\le l),
\en where $P^m_l(\cos \theta)$ are the associated Legendre
polynomials. According to the general theorem any arbitrary
function $f(\theta,\phi)$ such that its square is integrable over
the sphere, can be expanded in the convergent series \be
   F(\theta,\,\phi)=\nn{l=0}\sum\limits_{m=-l}^{l}             \label{7.2}
   F_{lm}Y^{lm}(\theta,\,\phi)\,.
\en
Making use of transformation from spherical to Cartesian
coordinates  one can obtain \cite{thorne} a one-to-one mapping
between the spherical harmonics and the symmetric trace-free (STF)
tensors with rank $l$
\be
   Y^{lm}(\theta,\,\phi)={\cal Y}^{lm}_{<K_l>}N^{K_l}\,,       \label{7.3}
\en where $N^{K_l}=n^{k_1}...n^{k_l}$ are products of components
of unit vector ${\bm n}={\bm r}/r.$ Tensors ${\cal
Y}^{lm}_{<K_l>}$ with $-l \le m \le l$ form a basis in
$(2l+1)$-dimensional space of symmetric and trace-free tensors
with $l$ indices, that is any STF tensor of rank $l$ can be
represented as
 \be
   {\cal F}_{<K_l>}=\sum\limits_{l=-m}^{m}                    \label{7.4}
   F^{lm}{\cal Y}^{lm}_{<K_l>}\,.
\en Hence, equation (\ref{7.2}) can be recast to the following
form
 \be
   f(\theta,\,\phi)=\nn{l=0}{\cal F}_{<K_l>}N^{K_l}\,.        \label{7.5}
\en Spherical functions are the eigenfunctions of the orbital
angular momentum operator \be
   {\rm\bf L}^2Y^{lm}\equiv \Bigl[
   \partial_r(r^2\partial_r)-r^2\nabla^2\Bigr]                \label{7.6}
   Y^{lm}=l(l+1)Y^{lm}\,,
\en that is a consequence of definition of canonical basis in the
sub-space in which the irreducible representation with weight $l$
is realized.

Thus, equations (\ref{7.5}) and (\ref{7.6}) reveal that any scalar
function $F(t,{\bm x})$ that is solution of the Laplace equation
(\ref{7.a}) is given by
 \be
   F(t,\nnx)=\nn{l=0}\left[A_{<L>}
   \Bigl(\frac1r\Bigr)_{,L}+B_{<L>}x^L\right]\,,            \label{7.7}
\en
where $A_{<L>}$ and $B_{<L>}$ are STF multipole moments depending on time $t$ only.

Vector and tensor spherical harmonics are obtained from the direct
product of two irreducible representations of the rotation group
with weights $l'$ and $l''$ which can be expanded into irreducible
representations with weights $|l'-l''|\le l \le |l'+l''|$.
Canonical orthonormal basis  in the sub-space of the vector
spherical functions in which the irreducible representation has
weight $l$, is the set of $3(2l+1)$ functions
\be
\label{7.8}
   {\bf Y}^{l',\,lm}(\theta,\,\phi)=
   \sum\limits_{m'=-l'}^{l'}\sum\limits_{m''=-1}^1
   (1l'\,m''m'\,|\,l m)
   {\bm \xi}_{m''}Y^{l'm'}(\theta,\,\phi)\,,
\en where $l'$ can take either of three values $l-1$, $l$, $l+1$,
notation $(l''l'\,m''m'\,|\,lm)$ stands for the Clebsch-Gordan
coefficients \cite{gelf}, and three vectors \be
   {\bm\xi}_{-1}=\frac{{\bf e}_x-i{\bf e}_y}{\sqrt{2}}\quad\,,
   \quad
   {\bm\xi}_0={\bf e}_z\quad\,,                                      \label{7.9}
   \quad
   {\bm\xi}_1=\frac{-{\bf e}_x-i{\bf e}_y}{\sqrt{2}}
\en represent canonical basis of the main matrix representation.

Canonical basis in the three-dimensional space of constant tensors
of second rank is made of 9 tensors such that five of them,
 \be \label{7.10}
   {\rm \bf t}^m=
   \sum\limits_{m'=-1'}^{1}\sum\limits_{m''=-1}^1
   (11\,m'm''\,|\,2m)
   {\bm\xi}_{m'}\otimes {\bm\xi}_{m''}\;,
\en are symmetric trace-free tensors, three tensors,
 \be \label{7.11}
   {\rm \bf P}^m=
   \sum\limits_{m'=-1'}^{1}\sum\limits_{m''=-1}^1
   (11\,m'm''\,|\,1m)
   {\bm\xi}_{m'} \otimes {\bm\xi}_{m''}\;,
\en are fully antisymmetric, and one,
 \be \label{7.12}
   { \bm \delta}=
   \sum\limits_{m'=-1'}^{1}\sum\limits_{m''=-1}^1
   (11\,m'm''\,|\,0m)
   {\bm \xi}_{m'} \otimes {\bm \xi}_{m''}\;,
\en
is the unit tensor. We shall abandon the antisymmetric part of
the basis as we are interested only in the symmetric tensors.
Hence, the canonical basis in the sub-space of such tensors of the
second rank with irreducible representation with weight $l$ is
formed from $6(2l+1)$ tensor harmonics
 \be
\label{7.13}
   {\rm \bf T}^{2l',\,lm}=
   \sum\limits_{m'=-l'}^{l'}\sum\limits_{m''=-2}^2
   (l'2\,m'm''\,|\,l m)
   Y^{l'm'}{\rm\bf t}^{m''}\,,
\en where $l-2\le l'\le l+2$, and $2l+1$ scalar spherical
harmonics
 \be
  {\rm\bf T}^{0l,\,lm}=Y^{lm}{\bm \delta}\,.                 \label{7.14}
\en Finally, solutions of the Laplace equations for vector, $F_i$, and tensor, $F_{ij}$, functions
(\ref{7.a}) are given as follows
 \br
 F_i(t,{\bm x})&= & \label{7.15}
\nn{l=1}\left[ C_{<iL-1>}\Bigl(\frac{1}{r}\Bigr)_{,L-1}+
D_{<iL-1>}x^{L-1}\right]\\\nonumber && + \nn{l=0}\left[
G_{<L>}\Bigl(\frac{1}{r}\Bigr)_{,iL}+
H_{<L>}x^{iL}\right]\\\nonumber && +  \nn{l=1}\varepsilon_{ipq}
\left[ E_{<qL-1>}\Bigl(\frac{1}{r}\Bigr)_{,pL-1}+
F_{<qL-1>}x^{pL-1}\right] \;,
\\\nonumber\\\nonumber\\\nonumber\\
 F_{ij}(t,{\bm x})&= & \delta_{ij}                               \label{7.16}
\nn{l=0}\left[ I_{<L>}\Bigl(\frac{1}{r}\Bigr)_{,L}+
J_{<L>}x^{L}\right]\\\nonumber &&+ \nn{l=0}\left[ K_{<L>}\Bigl(\frac{1}{r}\Bigr)
_{,ijL}+ M_{<L>}x^{<ijL>}\right] \\\nonumber &&+ \nn{l=2}\left[
V_{<ijL-2>}\Bigl(\frac{1}{r}\Bigr)_{,L-2}+
W_{<ijL-2>}x^{L-2}\right] \\\nonumber &&+ \nn{l=1}\left[
N_{<iL-1>}\Bigl(\frac{1}{r}\Bigr)_{,jL-1}+
P_{<iL-1>}x^{<jL-1>}\right]^{{\rm Sym}(ij)}\\\nonumber &&+
\nn{l=1}\left[\varepsilon_{ipq}\Bigl( Q_{<qL-1>}\Bigl(\frac{1}{r}\Bigr)
_{,jpL-1}+ R_{<qL-1>}x^{<jpL-1>}\Bigr)\right]^{{\rm Sym}(ij)}+
\\\nonumber && + \nn{l=2}\left[\varepsilon_{ipq}\Bigl(
S_{<qjL-2>}\Bigl(\frac{1}{r}\Bigr)_{,pL-2}+
T_{<qjL-1>}x^{<pL-2>}\Bigr)\right]^{{\rm Sym}(ij)}\,, \er
where the symbol ${\rm Sym}(ij)$ denotes symmetrization, and $C_L$, $D_L$,..., $T_L$ represent STF multipole moments
depending on time $t$ only.

\section{The Christoffel Symbols and the Riemann Tensor}\label{ap2}
\subsection{The Christoffel Symbols}
In this appendix we give formulas for the Christoffel symbols and
the Riemann tensor which elucidates the physical meaning of
external multipole moments in the expression for the local metric
tensor (\ref{1.24b}) -- (\ref{1.26b}). Christoffel symbols are
defined by standard expression \be\label{b1}
   \Gamma^{\alpha}_{\beta\gamma}=\ffrac{1}{2}
    g^{\alpha\delta}\left(g_{\beta\delta,\gamma}+
     g_{\gamma\delta,\beta}-g_{\beta\gamma,\delta}\right)=\frac12\left(\eta^{\alpha\delta}-h^{\alpha\delta}
     \right)\left(h_{\beta\delta,\gamma}+
     h_{\gamma\delta,\beta}-h_{\beta\gamma,\delta}\right)\;,
\en where the metric tensor components are taken from equations (\ref{11.4})
-- (\ref{11.6}). Calculation results in \br\label{b2}
   \Gamma^{0}_{00}&=&-\ffrac{\epsilon^3}{2}\left(\nne{2}{00}{,0}+\nne{1}{0i}{}\nne{1}{0i}{,0}\right)+O(\epsilon^4)\;,
   \\\nonumber   \\                                                                               \label{b3}
   \Gamma^{0}_{0i}&=&-\ffrac{\epsilon^2}{2}\left[\nne{2}{00}{,i}-\nne{1}{0j}{}\left(\nne{1}{0i}{,j}-\nne{1}{0j}
   {,i}\right)\right]+O(\epsilon^3)\;,                                                                 \\\nonumber \\
   \label{b4}
   \Gamma^{i}_{00}&=&\epsilon^2\left(\nne{1}{0i}{,0}-\ffrac{1}{2}\nne{2}{00}{,i}\right)
   \\\nonumber&&+
                    \epsilon^4\left(\nne{3}{0i}{,0}-
                    \ffrac{1}{2}\nne{4}{00}{,i}-\nne{2}{i j}{}\nne{1}{0j}{,0}+\frac12\nne{1}{0i}{}\nne{2}{00}{,0}-
                    \ffrac{1}{2}\nne{2}{i}{k}
                    \nne{2}{00}{,k}\right)+O(\epsilon^5)\;,
                                                 \\\nonumber           \\ \label{b5}
   \Gamma^{i}_{0k}&=&\epsilon\left(\nne{1}{0i}{,k}-
                   \nne{1}{0k}{,i}\right)\\\nonumber&&+
                   \ffrac{\epsilon^3}{2}\left[\nne{3}{0i}{,k}-
                   \nne{3}{0k}{,i}+\nne{2}{i k}{,0}+\nne{2}{i j}{}\left(\nne{1}{0k}{,j}-\nne{1}{0j}{,k}\right)+
                   \nne{1}{0i}{}\nne{2}{00}{,k}\right]+O(\epsilon^4)\;,
                                           \\\nonumber                 \\ \label{b6}
   \Gamma^{0}_{ik}&=&-\epsilon\left(\nne{1}{0i}{,k}+
                   \nne{1}{0k}{,i}\right)
                   \\\nonumber&&-\ffrac{\epsilon^3}{2}\left[\nne{3}{0i}{,k}+
                   \nne{3}{0k}{,i}-\nne{2}{i k}{,0} +\nne{2}{00}{}\left(\nne{1}{0i}{,k}+
                   \nne{1}{0k}{,i}\right)\right.\\\nonumber&&\left.-\nne{1}{0j}{}\left(\nne{2}{i j}{,k}+\nne{2}{k j}{,i}
                   -\nne{2}{i k}{,j}\right)\right]+O(\epsilon^4)\;,
                                                          \\\nonumber  \\ \label{b7}
   \Gamma^{i}_{jk}&=&\ffrac{\epsilon^2}{2}\left[
                   \nne{2}{i j}{,k}+\nne{2}{i k}{,j}-
                   \nne{2}{j k}{,i}+\nne{1}{0i}{}\left(\nne{1}{0j}{,k}+\nne{1}{0k}{,j}\right)\right]+O(\epsilon^3)\;.
\er We neglect in our calculations all terms which are
quadratic with respect to the angular speed of rotation $\Omega^i$
and linear velocity ${\cal V}^i$ in $\nne{1}{0i}{}$. In addition, we note many terms in Eqs. (\ref{b1})--(\ref{b7})
equal zero due to the specific structure
of $\nne{1}{0i}{}$. Thus, after simplification the expressions for the
Christoffel symbols are reduced to  \br\label{bb2}
   \Gamma^{0}_{00}&=&-\ffrac{\epsilon^3}{2}\,\nne{2}{00}{,0}+O(\epsilon^4)\;,       \\\nonumber\\
   \label{bb3}
   \Gamma^{0}_{0i}&=&-\ffrac{\epsilon^2}{2}\,\nne{2}{00}{,i}+O(\epsilon^3)\;,
   \\\nonumber\\                   \label{bb4}
   \Gamma^{i}_{00}&=&\epsilon^2\left(\nne{1}{0i}{,0}-\ffrac{1}{2}\nne{2}{00}{,i}\right)+
                    \epsilon^4\left(\nne{3}{0i}{,0}-
                    \ffrac{1}{2}\nne{4}{00}{,i}-
                    \ffrac{1}{2}\nne{2}{i}{k}
                    \nne{2}{00}{,k}\right)+O(\epsilon^5)\;,
                                                            \\\nonumber\\ \label{bb5}
   \Gamma^{i}_{0k}&=&\epsilon\left(\nne{1}{0i}{,k}-
                   \nne{1}{0k}{,i}\right)+
                   \ffrac{\epsilon^3}{2}\left(\nne{3}{0i}{,k}-
                   \nne{3}{0k}{,i}+\nne{2}{i k}{,0}\right)+O(\epsilon^4)\;,
                                                            \\\nonumber\\ \label{bb6}
   \Gamma^{0}_{ik}&=&-\ffrac{\epsilon^3}{2}\left(\nne{3}{0i}{,k}+
                   \nne{3}{0k}{,i}-\nne{2}{i k}{,0}\right)+O(\epsilon^4)\;,
                                                            \\\nonumber\\ \label{bb7}
   \Gamma^{i}_{jk}&=&\ffrac{\epsilon^2}{2}\left(
                   \nne{2}{i j}{,k}+\nne{2}{i k}{,j}-
                   \nne{2}{j k}{,i}\right)+O(\epsilon^3)\;.
\er
These expressions have been used in the present paper.
\subsection{The Riemann Tensor of External Gravitational Field in Local Coordinates}\label{rten}

Components of the Riemann tensor computed by making use of the external metric tensor only, are
\br                                                                 \label{b8}
   R^{\sss(E)}_{0i0j}&= &
                  -\epsilon^2\nn{l=0}\ffrac{1}{l!}\,Q_{ijL}w^{L}\;,\\\nonumber
                                                        \\                  \label{b9}
   R^{\sss(E)}_{0ijk}&= &\epsilon^3
                  \nn{l=1}\ffrac{l}{(l+2)\,l!}
                   \left[\delta_{ij}\dot{Q}_{kL} -
       \delta_{ik}\dot{Q}_{jL}\right]w^{L}
       \\\nonumber&&
       +2\nn{l=0}\ffrac{1}{(l+3)\,l!}
                   \left[\dot{Q}_{ijL}w^{kL}-
                   \dot{Q}_{ikL}w^{jL}\right]\\ \nonumber
                &&- (1-\gamma)\biggl\{\nn{l=0}\ffrac{l+1}{(l+2)\,l!}
                   \left[\delta_{ij}\dot{P}_{kL}-
                    \delta_{ik}\dot{P}_{jL}\right]w^{L}
                   \\\nonumber&&
                    +\nn{l=0}\ffrac{1}{(l+3)\,l!}
                   \left[\dot{P}_{ijL}w^{kL}-
                    \dot{P}_{ikL}w^{jL}\right]\biggr\}      \\ \nonumber
                && +\nn{l=0}\ffrac{l+3}{(l+2)!}\varepsilon_{jpk}
                   \left[(l+1)C_{ipL}+
                    \ffrac{1}{4}\dot{G}_{ipL}\right]w^{L}
                     \;,                             \\ \nonumber
                                                        \\                  \label{b10}
   R^{\sss(E)}_{ijkn}&= &\epsilon^2
                   \nn{l=0}\ffrac{1}{l!}\biggl[
                   \delta_{in}Q_{jkL}+\delta_{jk}Q_{inL}-
                   \delta_{ik}Q_{jnL}-\delta_{jn}Q_{ikL}
                   \biggr]w^{L}                          \\ \nonumber
                && -(1-\gamma)\nn{l=0}\ffrac{1}{l!}\biggl[
                   \delta_{in}P_{jkL}+\delta_{jk}P_{inL}-
                   \delta_{ik}P_{jnL}-\delta_{jn}P_{ikL}
                   \biggr]w^{L}\,.
\er
Here, all multipole moments $P_L$ are caused by the presence of scalar field.

\section{Comparison with the Klioner-Soffel Approach to Reference Frames in the PPN Formalism}\label{ksfor}
Klioner and Soffel \cite{kls} have worked out an independent approach aimed to amend the standard PPN formalism
\cite{nw,wn,will} with a mathematical procedure allowing us to construct the local PPN coordinates in the
vicinity of moving massive bodies (Earth, Moon, etc.) comprising an N-body system. Klioner-Soffel approach is
a straightforward extension of the standard PPN formalism and, thus,
is not based on the field equations of a particular parametrized theory of gravity in order to preserve that
generality which
the original PPN formalism was trying to reach in the global PPN coordinates. Our point of
view is that such general PPN formulation is not possible without having recourse to the field equations of a
particular theory of gravity (or a set of theories). This is because the goal of gravitational experiments is to
explain physics of gravitational field and exclude the field equations of untenable theories of gravity.
Formal procedure for extension of the PPN formalism, proposed by Klioner and Soffel, is not able to solve this task.
Even in the case of two parameters, $\gamma$ and $\beta$, one can not say what physical theory is hiding behind
mathematical manipulations present in the Klioner-Soffel approach \cite{kls}.

The two alternative points of view on one and the same problem led naturally to
a number of differences in many equations which have been derived. In this appendix we summarize the differences
we have found between the Klioner-Soffel formulation \cite{kls} of the local reference frames in the PPN formalism
and that based directly on the scalar-tensor
theory of gravity used in the present paper. These differences are as follows:
\begin{itemize}
\item Klioner and Soffel restricted their attention to the case of two PPN parameters, $\beta$ and $\gamma$
like we did. However, Klioner and Soffel did not rely upon any specific theory of gravitation, thus, leaving
open the question about what kind of field equations is compatible and can be used along with the PPN coordinate
transformations and the equations of motion they have obtained. Calculations in the present paper have been done
in the framework of a general class of scalar-tensor theories of gravity with one scalar field which is
also parametrized
by the PPN parameters $\gamma$ and $\beta$. In our formulation we know exactly the field equations which are
compatible with our coordinate
transformations and equations of motion. Hence, measuring these parameters restricts domain of applicability
of a particular physical theory of gravity.
\item Klioner and Soffel worked in harmonic coordinates defined by the condition
$\partial_\alpha(\sqrt{-g}g^{\alpha\beta})=0$ while we worked in the quasi-harmonic coordinates defined by
the Nutku condition  $\partial_\alpha(\phi\sqrt{-g}g^{\alpha\beta})=0$ \footnote{See Eq. (\ref{11.3}) in the
present paper.}. Spatial transformation from local to global coordinates given by Eqs. (\ref{5.13}) in our
approach does not contain the PPN parameter $\gamma$ in the quadratic terms with respect to $R^i_{\sss B}$ while
that of Klioner and Soffel does.
\item Klioner and Soffel postulated (guessed) a specific form of the metric tensor in the local frame with one free
function -- potential $\Psi$ entering their Eqs. (3.33) and (4.25). Then, they postulated specific rules for matching
of the local and global frames and used them in order to determine the post-Newtonian coordinate transformation along
with the a priory unknown potential $\Psi$ in the local frame. We have used the field equations in order to find the
metric tensor in the local frame, so that it gets fully determined, and matched it with the metric tensor in the global
frame by making use of the same procedure as that used in general theory of relativity \cite{iau2000,skpw}.
Hence, our matching procedure is not voluntary but is in a complete agreement with the field equations.
\item Klioner and Soffel noticed two distinguished properties of the metric tensor in the local frame valid
in general relativity:
\begin{itemize}
\item [{\bf A}.] Gravitational field of external bodies is represented only in the form of a relativistic tidal
potential being at least of the second order in the local spatial coordinates. It coincides with the usual
Newtonian tidal potential in the Newtonian limit of the PPN formalism;
\item [{\bf B}.] The internal gravitational field of a body (or sub-system of bodies) coincides with the
gravitational field of a corresponding isolated source provided that the tidal influence of the external matter
is neglected.
\end{itemize}
Klioner-Soffel approach can not keep the two properties of the metric tensor in the local frame simultaneously
while in our approach to the problem under consideration these two properties are retained similar to general
relativity.
\item Klioner and Soffel do not distinguish {\it active}, {\it conformal}, and {\it scalar} multipole moments
characterizing different properties of the gravitational field of an isolated body. In fact, they operate only
with {\it active} multipole moments. We use two types of the multipole moments in our calculations and show that
the {\it conformal} dipole moment has more advanced mathematical properties than the {\it active} one for definition
of body's center of mass. It allows us to simplify translational equations of motion of the bodies drastically.
\item Klioner and Soffel derived equations of motion of a body in its local frame and discovered that in their
approach these equations depend explicitly not only on the set of multipoles being inherent to general theory
of relativity but contain one more family of the multipole moments which they denoted as ${\cal N}_L$
\footnote{See our Eq. (\ref{aer}) for its precise definition.}. In our approach the equations of motion of
the body in the local frame do not explicitly depend on such multipole moments except for the second time
derivative of the monopole term, ${\cal N}\equiv{\cal I}^{(2)}$, entering Eq. (\ref{fopm}). The reader is
invited to compare equations (9.37)--(9.40) from the paper \cite{kls} with equations (\ref{6.2d}) -- (\ref{fopm})
of the present paper to decide which type of equations is more economic.
\item Klioner and Soffel defined the center of mass of a body by making use of {\it active} dipole moment
of the body (see their Eq. (9.24)). They have found that if the Nordtvedt parameter $\eta\not=0$
and/or $\gamma\not=0$, the second time derivative of this dipole does not vanish for the case of one
isolated body unless specific physical conditions are met (secular stationarity). Thus, Klioner and Soffel
admit the existence of a self-accelerated motion of the body in their formalism which violates the third
(action-counteraction) Newton's law. On the other hand, we defined the center of mass of each body by making
use of {\it conformal} dipole moment of the body (see our Eq. (\ref{6.2a}). The second time derivative of the
{\it conformal} dipole moment vanishes perfectly  in the case of a single body and no self-accelerated terms
in motion of the body comes about in our formalism.
\item Equations of rotational motions in Klioner-Soffel formalism contain torques which depend on the Nordtvedt
parameter $\eta=4-3\beta-1$. We confirm this observation. However, we noticed that Klioner and Soffel did not
take into account finite-size effects of the bodies and they worked in a different gauge. This led to appearance
of several terms in our equations which are not present in Klioner-Soffel's analysis \cite{kls}. These terms
can be important in some particular situations, for example, in rotational equations of motion of coalescing
binaries.
\end{itemize}

\end{document}